\documentclass[preprint]{aastex}
\pdfoutput=1 
\usepackage[usenames,dvipsnames]{color}
\usepackage{natbib,graphicx,latexsym,lscape,pdflscape,url,pdfpages}
\usepackage[caption=false]{subfig}

\newcommand{\Hipparcos}{{\sl Hipparcos}}
\newcommand{\Gaia}{{\sl Gaia}}
\newcommand{\HST}{{\sl HST}}

\newcommand{\Msun}{\mbox{$M_{\sun}$}}

\newcommand{\Lsun}{\mbox{$L_{\sun}$}}
\newcommand{\Rsun}{\mbox{$R_{\sun}$}}
\newcommand{\Mjup}{\mbox{$M_{\rm Jup}$}}
\newcommand{\Rjup}{\mbox{$R_{\rm Jup}$}}
\newcommand{\degree}{\mbox{$^{\circ}$}}

\newcommand{\kms}{\mbox{km\,s$^{-1}$}}
\newcommand{\masyr}{\hbox{mas\,yr$^{-1}$}}

\newcommand{\Hc}{\mbox{$H_{\rm cont}$}}
\newcommand{\Kc}{\mbox{$K_{\rm cont}$}}
\newcommand{\Kp}{\mbox{$K^{\prime}$}}
\newcommand{\Ks}{\mbox{$K_S$}}

\newcommand{\CHs}{\mbox{$CH_4s$}}
\newcommand{\Kn}{\mbox{$K_{\rm H2}$}}
\newcommand{\Mtot}{\mbox{$M_{\rm tot}$}}

\newcommand{\Lbol}{\mbox{$L_{\rm bol}$}}

\newcommand{\Teff}{\mbox{$T_{\rm eff}$}}
\newcommand{\logg}{\mbox{$\log(g)$}}

\newcommand{\Lp}{\mbox{${L^\prime}$}}

\newcommand{\aphot}{\mbox{${a_{\rm phot}}$}}
\newcommand{\mur}{\mbox{${\mu_{\alpha^*}}$}}
\newcommand{\mud}{\mbox{${\mu_{\delta}}$}}
\newcommand{\vrad}{\mbox{${v_{\rm rad}}$}}

\newcommand{\intg}{\textsc{int-g}}
\newcommand{\fldg}{\textsc{fld-g}}

\shortauthors{Dupuy \& Liu}

\begin{document}

\title{Individual Dynamical Masses of Ultracool Dwarfs\altaffilmark{*\dag\ddag}}

\author{Trent J.\ Dupuy\altaffilmark{1} and
        Michael C.\ Liu\altaffilmark{2}}

      \altaffiltext{*}{Data presented herein were obtained at the
        W.M.\ Keck Observatory, which is operated as a scientific
        partnership among the California Institute of Technology, the
        University of California, and the National Aeronautics and
        Space Administration. The Observatory was made possible by the
        generous financial support of the W.M.\ Keck Foundation.}

      \altaffiltext{\dag}{Based on data obtained with WIRCam, a joint
        project of CFHT, Taiwan, Korea, Canada, France, at the
        Canada-France-Hawaii Telescope, which is operated by the
        National Research Council of Canada, the Institute National
        des Sciences de l'Univers of the Centre National de la
        Recherche Scientifique of France, and the University of
        Hawaii.}

      \altaffiltext{\ddag}{Based on observations made with the
        NASA/ESA \textsl{Hubble Space Telescope}, obtained at the
        Space Telescope Science Institute, which is operated by the
        Association of Universities for Research in Astronomy, Inc.,
        under NASA contract NAS~5-26555. These observations are
        associated with programs GO-11593, GO-12317, and GO-12661.}

      \altaffiltext{1}{The University of Texas at Austin, Department
        of Astronomy, 2515 Speedway C1400, Austin, TX 78712, USA}

      \altaffiltext{2}{Institute for Astronomy, University of Hawai`i,  
        2680 Woodlawn Drive, Honolulu, HI 96822, USA}

\begin{abstract}

  We present the full results of our decade-long astrometric
  monitoring programs targeting 31 ultracool binaries with component
  spectral types M7--T5. Joint analysis of resolved imaging from Keck
  Observatory and \textsl{Hubble Space Telescope} and unresolved
  astrometry from CFHT/WIRCam yields parallactic distances for all
  systems, robust orbit determinations for 23~systems, and photocenter
  orbits for 19~systems. As a result, we measure 38 precise individual
  masses spanning 30--115\,\Mjup. We determine a model-independent
  substellar boundary that is $\approx$70\,\Mjup\ in mass ($\approx$L4
  in spectral type), and we validate \citet{2015A&A...577A..42B}
  evolutionary model predictions for the lithium-depletion boundary
  (60\,\Mjup\ at field ages).  Assuming each binary is coeval, we test
  models of the substellar mass--luminosity relation and that find in
  the L/T transition, only the \citet{2008ApJ...689.1327S} ``hybrid''
  models accounting for cloud clearing match our data. We derive a
  precise, mass-calibrated spectral type--effective temperature
  relation covering 1100--2800\,K. Our masses enable a novel direct
  determination of the age distribution of field brown dwarfs spanning
  L4--T5 and 30--70\,\Mjup. We determine a median age of 1.3\,Gyr, and
  our population synthesis modeling indicates our sample is consistent
  with a constant star formation history modulated by dynamical
  heating in the Galactic disk. We discover two triple-brown-dwarf
  systems, the first with directly measured masses and eccentricities.
  We examine the eccentricity distribution, carefully considering
  biases and completeness, and find that low-eccentricity orbits are
  significantly more common among ultracool binaries than solar-type
  binaries, possibly indicating the early influence of long-lived
  dissipative gas disks. Overall, this work represents a major advance
  in the empirical view of very low-mass stars and brown dwarfs.

\end{abstract}



\section{Introduction}

Mass is the most important property governing the destiny of
self-gravitating gaseous objects in hydrostatic equilibrium.  Mass
determines whether an object becomes a star, generating sufficient
energy through nuclear fusion of hydrogen to stabilize itself against
gravitational collapse, or a brown dwarf, which primarily supports
itself by degeneracy pressure.  The dichotomy in the way these two
classes of objects satisfy hydrostatic equilibrium results in
drastically different evolutionary outcomes.  Sufficiently massive
objects maintain a relatively stable luminosity and temperature,
becoming stars on the main sequence for millions to trillions of
years.  Meanwhile, objects of substellar mass steadily and drastically
decrease in luminosity and temperature.  One key observational
consequence is that the coldest gaseous objects can be confidently
identified as brown dwarfs \citep[e.g.,][]{1995Sci...270.1478O}.
However, the masses of most ultracool dwarfs are indeterminate because
observables like luminosity and temperature are degenerate between
mass and age, such that the mass could be lower or higher if the
object is younger or older, respectively.

The evolution of brown dwarfs and the substellar boundary itself are
largely understood based on theoretical work over the last two decades
\citep[e.g.,][]{1994ApJ...424..333S, 1995ApJ...446L..35B,
  1997ApJ...491..856B, 1997A&A...327.1039C, 1999ApJ...519..793L,
  2000ApJ...542..464C, 2001RvMP...73..719B, 2003A&A...402..701B,
  2008ApJ...689.1327S, 2009ApJ...700..387M, 2015A&A...577A..42B}.
Broadly speaking, the interior physics of evolutionary models over
this time has remained the same, while the treatment of the surface
boundary conditions has advanced greatly due to improved molecular
line lists, chemistry, and cloud modeling \citep[see a recent review
by][]{2015ARA&A..53..279M}.  
Clouds in particular have long thought to be the key to explaining
major variations in the emergent flux of brown dwarfs, especially in the
change from L to T spectral types (the L/T transition). Understanding
clouds in theoretical modeling is imperative given that they can
exert great influence on the atmosphere's opacity, but the physical
processes governing cloud formation, particle size distribution, and
sedimentation are all effectively free parameters that can only be
constrained loosely by observations.
Over the past decade, model atmospheres have been largely developed
with guidance from observations of colors, absolute magnitudes, and
spectra, so they can reproduce the observed surface properties of
ultracool dwarfs reasonably well.  However, as atmospheres are a key
input to evolutionary models, there arises an entirely different
question of whether predictions of fundamental properties (luminosity,
radius, etc.) are improved by the adoption of the latest cloud
prescriptions and molecular opacities.

Direct mass measurements are central to tests of evolutionary models.
Previous work on ultracool dwarfs (spectral types $\geq$M7) has
focused on \emph{total} dynamical masses that can be readily
determined from \emph{relative} astrometric orbits because the narrow
field-of-view afforded by adaptive optics (AO) imaging rarely contains
any reference stars that could be used for absolute astrometry.  To
date there are total dynamical masses for more than a dozen ultracool
binaries \citep[e.g.,][]{2001ApJ...560..390L, 2001A&A...367..183L,
  2004A&A...423..341B, 2008ApJ...689..436L, 2009ApJ...706..328D,
  2009ApJ...692..729D, 2009ApJ...699..168D, 2010ApJ...721.1725D,
  2014ApJ...790..133D, 2009AIPC.1094..509C, 2010ApJ...711.1087K}.  In
a few special cases, individual masses have been determined: for two
single companions to main-sequence stars \citep{2008ApJ...678..463I,
  2012ApJ...751...97C}; for the companion AB~Dor~C by assuming a mass
for AB~Dor~A \citep{2005Natur.433..286C, 2006A&A...446..733G}; and for
three ultracool binary systems where mass ratios are measured using
radial velocities or absolute astrometry \citep{2004ApJ...615..958Z,
  2006ApJ...644.1183S, 2008A&A...484..429S, 2010ApJ...711.1087K,
  2012A&A...541A..29K, 2016ApJ...827...23D}.  However, most of the
objects in this small sample in the literature are likely to be stars
and not brown dwarfs.  For additional context, our individual mass
measurements for the L6.5+T1.5 binary SDSS~J1052+4422AB marked the
first individual masses for any field L or T dwarfs
\citep{2015ApJ...805...56D}.  This stands in stark contrast to the
situation for individual dynamical masses of earlier-type M~dwarfs,
where dozens of measurements over the years have now reached very high
precision \citep[e.g.,][]{1993AJ....106..773H, 2000A&A...364..217D,
  2016AJ....152..141B}.  Thus, even though significant progress has
been made on ultracool dwarfs using mostly total dynamical masses, the
individual masses needed for the strongest tests of models have been
lacking.

We present here orbit and mass determinations for a large sample of
ultracool binaries with component spectral types of M7--T5. Results
here are based on a uniform (re-)analysis of Keck AO imaging and
masking data from the past 10 years, as well as \HST\ imaging from the
past 20~years, and nearly a decade of absolute astrometric monitoring
at the Canada-France-Hawaii Telescope (CFHT). The combined data set
not only provides total dynamical masses from the relative orbits and
CFHT parallaxes but also yields precise mass ratios from measurements
of photocenter motion in our CFHT data. We report observations for our
entire sample of 31 binaries, from which we determine robust orbits
for 23~systems and individual masses for 19~systems (two of which turn
out to be previously unknown triple systems). We critically examine
some of the basic predictions of evolutionary models, such as the mass
limits for hydrogen and lithium fusion, and we establish empirical
relations between parameters such as mass, luminosity, effective
temperature, absolute magnitude, and spectral type. We use the brown
dwarfs in our sample as clocks (age-dated from models using mass and
luminosity) to directly determine the age distribution of the field
population. Finally, we re-visit the eccentricity distribution of
ultracool dwarfs \citep[initially discussed in][]{2011ApJ...733..122D}
and the potential implications for the earliest stages of evolution.


\section{Sample Selection for Keck \& CFHT Orbit Monitoring \label{sec:sample}}

The orbit sample presented here was drawn from the very low-mass
visual binaries ($\Mtot \lesssim 0.2$\,\Msun; using published mass
estimates) known to us in early 2008.  In addition to the total mass
criterion, we also excluded binaries with integrated light spectral
types of M6 or earlier and published distance estimates
$\gtrsim$40\,pc.  For the work presented here we also only consider
binaries with types earlier than T6.  As the starting point for
defining this sample we used the summary table of
\citet{2007prpl.conf..427B}\footnote{We note the following significant
  updates to Table~1 of \citet{2007prpl.conf..427B}:
  2MASS~J0652307+471034AB is listed as a 2.0~AU binary (Reid et al.,
  in prep.), but \citet{2006ApJ...639.1114R},
  \citet{2010ApJ...711.1087K}, and Keck LGS AO observations of our own
  confirm that it is single; SDSS~J233558.51$-$001304.1AB is much more
  distant ($160\pm30$\,pc) and thus wider ($9.0\pm1.7$\,AU) than
  previously estimated given that its spectral type is M7
  \citep{2008AJ....135..785W} not early- to mid-L.} to which we added
other binaries from the literature and from our own proprietary Keck
LGS AO data in hand at the time. From this whole sample of binaries,
we selected only those with estimated periods that suggested
$\gtrsim$30\% of their orbit could be complete by 2010. We computed
these initial period estimates using the estimated total mass from the
literature and a range of semimajor axes corresponding to the
projected separation at discovery \citep[e.g.,][]{1999PASP..111..169T,
  2011ApJ...733..122D}, where we adopted the 1$\sigma$ minimum period
for our estimates.

In Table~\ref{tbl:sample} we list all 33 binaries in our Keck+CFHT
astrometric monitoring sample, along with three other binaries that
have published orbit and parallax measurements.
We began obtaining resolved Keck AO astrometry in 2007--2008, and we
combined our new astrometry with available data in the literature or
public archives (e.g., \HST\ and Gemini) to refine our orbital period
estimates and thereby our prioritizaton for Keck observations. As
described in \citet{2009arXiv0912.0738D}, we performed a Monte Carlo
analysis of this multi-epoch data that provided period estimates based
on actual orbital motion, not simply projected separations. We
subsequently deprioritized Keck observations of the systems with
longer expected periods, focusing our observational efforts on the
shorter-period systems. We did occasionally obtain Keck astrometry for
the systems with the longest expected periods in our sample.

For most of the binaries in our sample, we began integrated-light
astrometric monitoring with CFHT in the second half of 2007 or in 2008
and continued collecting data until parallaxes were determined.  We
did not include in our CFHT program the three binaries with
\Hipparcos\ parallaxes (Gl~417BC, HD~130948BC, Gl~569Bab), but we did
include other systems with published parallaxes at the time (11
binaries) in order to improve the distance accuracy.  Indeed, for
seven of these binaries we reduce the uncertainty by a factor of
1.9--6.7.  Most of these initial parallax determinations were reported
in \citet{2012ApJS..201...19D}.  We have continued monitoring this
sample with CFHT up to the present in order to place constraints on
photocenter motion and thereby the binary mass ratios.

Our orbit sample selection is essentially based on spectral type and
projected separation, but for some purposes, like studies of the
eccentricity distribution \citep[e.g.,][]{2011ApJ...733..122D}, it is
important to consider whether our target selection and prioritization
has introduced biases in our sample beyond those intrinsic to the
population of ultracool binaries known in late 2007 and early
2008. Although our sample was initially defined in terms of the
probability that a given binary would yield an orbit determination by
2010, we can retroactively determine over what range of observable
properties our sample is complete. To the best of our knowledge, our
initial sample included all observable binaries known at the time with
integrated-light spectral types of M6.5--T5.5,\footnote{Our spectral
  type cuts excluded the late type binaries 2MASSW~J1225543$-$273947AB
  (T6) and 2MASS~J15530228+1532369AB (T7). At earlier types, we
  excluded LSPM~J1314+1320AB based on an estimated spectral type of M6
  from \citet{2006MNRAS.368.1917L}, but note that it has since been
  updated to M7 \citep{2009AJ....137.4109L, 2016ApJ...827...23D}.
  Other notable systems near the boundary of our spectral type cutoff
  and thus not included here are L~726-8AB a.k.a.\ Gl~69AB (M5.5+M6;
  \citealp{1988AJ.....95.1841G}, \citealp{1991ApJS...77..417K}),
  L~789-6ABC a.k.a.\ Gl~866ABC (M5+M5.5:+M6.5:;
  \citealp{1994AJ....108.1437H}, \citealp{2000A&A...364..217D}), and
  GJ~1245ABC (M5.5+M5.5+M6.5:; \citealp{2012ApJ...753..156K,
    2016AJ....152..141B}).}  projected separations $\leq$6\,AU at
discovery, and distances $\lesssim$40\,pc (based on the
spectrophotometric distance estimate in the absence of a
parallax).\footnote{By focusing on nearby systems, the only binaries
  with projected separations $\leq$6\,AU that we excluded were
  2MASS~J16000548+1708328AB \citep[61\,pc;][]{2003AJ....126.1526B} and
  binaries in Taurus and Upper Scorpius \citep{2005ApJ...633..452K,
    2006ApJ...649..306K}.  The most distant binary in our sample is
  LP~415-20AB at $39.7\pm1.1$\,pc, and its original estimated distance
  was actually $30\pm5$\,pc \citep{2003ApJ...598.1265S}.}
2MASSI~J1426316+155701AB is the only binary meeting these criteria for
which we never ultimately obtained any CFHT or Keck LGS AO data
because we de-prioritized it based on archival and published
astrometry \citep{2008A&A...481..757B} that showed its projected
separation had increased from 4.1\,AU at discovery to 6.9\,AU
projected separation by 2006~June.  Although we have attempted to
resolve 2MASSI~J0856479+223518AB on multiple occasions with Keck LGS
AO, we have never been successful, so this binary does not appear in
any of the following discussion.

Practical observational considerations imposed some limitations on our
sample that should not correspond to any physically meaningful
selection biases.
The largest subset of targets meeting the above criteria but excluded
from our sample are those not accessible to Keck AO because they are
too far south: GJ~1001BC \citep[L5;][]{2004AJ....128.1733G},
DENIS~J035726.9$-$441730AB \citep[L0;][]{2003AJ....126.1526B,
  2003AJ....125.3302G}, SCR~J1845$-$6357AB
\citep[M8.5;][]{2006ApJ...641L.141B}, $\epsilon$~Ind~Bab
\citep[T2.5;][]{2004A&A...413.1029M}, and 2MASS~J22551861$-$5713056AB
\citep[L6;][]{2008AJ....135..580R}.
Another practical limitation is that not all ultracool binaries have a
nearby star bright enough for tip-tilt correction as needed for LGS AO
imaging at Keck ($R\lesssim18.5$\,mag within $\approx$65\arcsec). This
excluded DENIS~J100428.3$-$114648AB (M8), 2MASSW~J1239272+551537AB
(L5), and 2MASS~J14304358+2915405AB (L2). This tip-tilt star
limitation also excluded three of the most promising T~dwarf binaries,
SDSS~J0423$-$0414AB (T0), SDSS~J0926+5847AB (T4.5), and
2MASS~J0518$-$2828AB (T1p), so we conducted \HST\ monitoring of these
binaries instead.

Some of our sample binaries already have published orbital monitoring
results, either from our own work \citep{2008ApJ...689..436L,
  2009ApJ...706..328D, 2009ApJ...692..729D, 2009ApJ...699..168D,
  2010ApJ...721.1725D, 2011ApJ...733..122D, 2014ApJ...790..133D,
  2015ApJ...805...56D} or from others \citep{2001ApJ...560..390L,
  2004ApJ...615..958Z, 2006ApJ...644.1183S, 2010ApJ...711.1087K}. In
the following, we perform a complete, uniform analysis of all data on
these systems, even in cases where we have previously published
results.  2MASS~J0746+2000AB had a published orbit
\citep{2004A&A...423..341B} before we began our observing program, and
\citet{2010ApJ...711.1087K} provided significant additional astrometry
and a refined orbit.  We only obtained Keck astrometry for
2MASS~J0746+2000AB at one epoch, but we have been monitoring its
absolute astrometry from CFHT since 2008.  Therefore, we have
re-analyzed published \HST\ and Keck data of 2MASS~J0746+2000AB from
public archives in order to include it in our sample of joint
Keck+CFHT orbital analysis.
LHS~1070BC is the only other binary that would have been in our sample
if it did not have a previously published orbit
\citep{2001A&A...367..183L, 2008A&A...484..429S, 2012A&A...541A..29K}.
Since the published orbit of LHS~1070BC is not based on Keck
astrometry, and we did not observe it as part of our CFHT astrometry
program, we simply use the orbit parameters quoted in the literature
and perform no orbital analysis of our own.
We also include the recently published orbit for LSPM~J1314+1320AB
\citep{2016ApJ...827...23D}, which is based on the same orbital
analysis presented here.
The $\epsilon$~Ind~Bab system is too far south to observe from
Maunakea, but it also has a published orbit determination from
\citet{2009AIPC.1094..509C}, so we include the published parameters in
our discussion as well.


\section{Observations \label{sec:obs}}

\subsection{Relative Astrometry from High-Angular Resolution Observations}

\subsubsection{Keck/NIRC2 AO Imaging \& Masking  \label{sec:keck}}

We present here new Keck/NIRC2 AO imaging and non-redundant aperture
masking observations, both in natural guide star and laser guide star
modes, in addition to a re-analysis of our own previously published
data and publicly available archival data for our sample binaries.
Our approach for reducing NIRC2 imaging and obtaining the binary
parameters separation, position angle (PA), and flux ratio is well
established in our previous work \citep{2006ApJ...647.1393L,
  2008ApJ...689..436L, 2009ApJ...706..328D, 2009ApJ...692..729D,
  2009ApJ...699..168D, 2010ApJ...721.1725D, 2015ApJ...803..102D}.
Briefly, we apply standard calibrations (dark subtraction, flat
fielding) and then fit an analytic, three-component Gaussian model to
each point source in the images.  In cases where the binary components
are spatially well separated, we perform PSF-fitting using StarFinder
\citep{2000A&AS..147..335D}.  In other cases where a third,
unsaturated star is in the field (e.g, Gl~569A and Gl~569Bab), we use
the third single star as an empirical PSF to fit the binary.
Analysis of our NIRC2 masking data was performed using a pipeline
similar to previous papers \citep[e.g.,][]{2008ApJ...678..463I,
  2008ApJ...678L..59I} and is described in detail in Section~2.2 of
\citet{2009ApJ...699..168D}.
For NIRC2 narrow camera images obtained before 2015~April~13~UT, we
use the astrometric solution of \citet{2010ApJ...725..331Y} to correct
our measured $(x,y)$ image coordinates for nonlinear optical
distortion, using their derived pixel scale of
$9.952\pm0.002$\,mas\,pixel$^{-1}$ and $+0\fdg252\pm0\fdg0.009$
correction for the orientation given in the NIRC2 image headers.  For
narrow camera imaging obtained after 2015~April~13~UT, we use an
updated distortion solution from \citet{2016PASP..128i5004S} that
accounts for the first ever realignment of the AO system.  The
post-realignment pixel scale and orientation are
$9.971\pm0.004$\,mas\,pixel$^{-1}$ and $+0\fdg262\pm0\fdg0.020$,
respectively.  For NIRC2 wide camera images, which represent a very
small subset of our observations, we used the distortion solution of
Fu et al.\ (2012, priv.\
comm.)\footnote{\url{http://astro.physics.uiowa.edu/~fu/idl/nirc2wide/}}
that assumes a pixel scale of 39.686\,mas\,pixel$^{-1}$ and the Yelda
et al.\ PA offset of +0\fdg252.

We have made some small improvements to our astrometric analysis
compared to our previous work, as described in
\citet{2016ApJ...817...80D}.  For data obtained in vertical angle
mode, where the sky rotation of the images relative to the detector
axes is constantly changing, we corrected the rotator angles reported
in the header to correspond to the midpoint instead of the start of
the exposure.  We also apply corrections for differential aberration
and atmospheric refraction.  The refraction correction requires
knowledge of the air temperature, pressure, and humidity on Maunakea
during our observations, for which we used the weather data archived
by CFHT.\footnote{\url{http://mkwc.ifa.hawaii.edu/archive/wx/cfht/}}
We generally adopt errors that are the rms of measurements from
individual dithers at a given epoch.  In some cases where we have
previously published astrometry we instead use estimates of errors
from Monte Carlo simulations of fitting artificial binaries.

Table~\ref{tbl:keck} gives our measured astrometry and flux ratios for
all Keck AO data used in our orbital analysis.  In total there are 339
distinct measurements (unique bandpass and epoch for a given target),
where 302 of these are direct imaging and 37 are non-redundant
aperture masking.  The median Keck separation error is 0.6\,mas with
90\% of measurements having errors between 0.06--1.9\,mas.  The median
PA error is 0\fdg3 with 90\% of measurement errors between 0\fdg02 and
1\fdg3.  (We caution that there are possible systematic effects, e.g.,
uncertainty in the distortion solution and tip-tilt jitter, that are
difficult to quantify and may impact the few measurements with
astrometric errors well below 0.1\,mas.  The fact that none of these
data points show up as outliers in our orbit fitting analysis
described later could be due to the fact that not all phases of all
orbits are overconstrained by numerous degrees of freedom.)  Eight of
the imaging measurements are from six unpublished archival data sets.
For HD~130948BC we used data sets from 2005~Feb~25~UT (PI Prato) and
2011~Mar~25~UT (PI Bowler).  For Gl~569Bab we used data sets from
2003~Apr~15~UT (PI Simon), 2004~Aug~10~UT (PI Kulkarni), and
2011~Mar~25~UT (PI Bowler).  For LSPM~J1735+2634AB we used a $JHK$
data set from 2007~Aug~1~UT (PI N.~Law).  Fourteen other measurements
are from our re-analysis of ten previously published data sets in the
archive: LP~349-25, 2MASS~J0920+3517AB, 2MASS~J0746+2000AB,
HD~130948BC, 2MASS~J1847+5522AB, and 2MASS~J2206$-$2047AB \citep[PI
Ghez;][]{2010ApJ...711.1087K} and SDSS~J2052$-$1609AB \citep[PI
C.~Gelino;][]{2015AJ....150..163B}.

\subsubsection{\HST/ACS-WFC Monitoring \label{sec:acs}}

In addition to our Keck AO monitoring, we also obtained data for three
T~dwarf binaries lacking suitable LGS tip-tilt stars over a 3-year
\HST\ program\footnote{Programs GO-11593, GO-12317, and GO-12661 (PI
  Dupuy).} using the Advanced Camera for Surveys (ACS) Wide Field
Camera (WFC).  ACS-WFC has a pixel scale of $\approx$50\,mas, and we
used the $F814W$ bandpass for all of our observations to optimize
between S/N and angular resolution.  SDSS~J0423$-$0414AB (T0) and
SDSS~J0926+5847AB (T4.5) were both resolved in all of our images.
2MASS~J0518$-$2828AB (T1p) was not resolved in any of our six
observations spanning 2009~Dec~11~UT to 2012~Oct~27~UT.  Based on
simulations we performed for our observation proposal, we expected to
readily resolve 2MASS~J0518$-$2828AB if it was at a similar separation
as its discovery \citep[51\,mas;][]{2006ApJS..166..585B}.  Given that
the flux ratio of 2MASS~J0518$-$2828AB in the $F814W$ band is unknown
(and difficult to extrapolate from the available infrared data), and a
larger flux ratio of $\approx$2\,mag could make it quite difficult to
clearly resolve, we conservatively estimate an upper limit of
$<$50\,mas on the putative companion's separation during our
observations.

For the resolved binaries, we performed PSF fitting of our ACS-WFC
images in a similar manner to our previous work on data from other
\HST\ cameras \citep[e.g.,][]{2008ApJ...689..436L,
  2009ApJ...706..328D, 2009ApJ...692..729D}.  We used TinyTim
\citep{2011SPIE.8127E..0JK} to generate a PSF model for each pipeline
reduced (\texttt{flt}) FITS image that we downloaded from the \HST\
archive.  The only inputs to this model are the target's position on
the detector and a template spectrum of comparable spectral type to
the target.  For the latter, we used optical-to-infrared spectra from
\citet{2002ApJ...564..466G}, adopting the integrated-light spectrum of
SDSS~J0423$-$0414AB\footnote{\url{http://staff.gemini.edu/~sleggett/spectra/T0_SDSS0423-04.txt}}
for each of its components and the spectrum of
2MASS~J0559$-$1404\footnote{\url{http://staff.gemini.edu/~sleggett/spectra/T4.5_2MASS0559-14.txt}}
for the components of SDSS~J0926+5847AB.  To test the impact of
assuming the same spectrum for each component, we also tried L5 and T5
templates for both components of SDSS~J0423$-$0414AB and found the
astrometry changed by $<$0.6\,mas ($<$1$\sigma$).  We use the same PSF
model for both components and shift, scale, and add them to each other
to create a model binary image.  We used the \texttt{amoeba} function
in IDL, based the routine from \citet{1992nrca.book.....P}, to find
the best-fit parameters for the binary model: position and flux of the
primary ($x_0$, $y_0$, $f_0$) and the binary separation, PA, and flux
ratio.  We used the information contained in the FITS files to correct
our fitted $(x,y)$ positions on the detector to sky coordinates.  This
accounts for detector defects from the \texttt{D2IMARR} FITS
extension, a polynomial distortion correction contained in FITS header
keywords, and non-polynomial distortion corrections from the
\texttt{WCSDVARR} FITS extension \citep{2015acs..rept....6K}.  We then
applied the \texttt{CD} matrix in the FITS header and a tangent
projection to convert the corrected $x$ and $y$ values to RA and Dec.

Table~\ref{tbl:other} reports the mean and rms of our best-fit binary
parameters derived from the individual dithered ACS-WFC images for
SDSS~J0926+5847AB.  While the first five observations show the binary
moving outward and then inward slightly (67\,mas to 73\,mas to
63\,mas), at the final epoch it had moved in significantly and was
only marginally resolved.  Over the first five epochs we measured flux
ratios consistent with being constant at $\Delta{F814W} =
0.560\pm0.024$\,mag.  We therefore fixed the flux ratio to be
0.56\,mag in our fitting of the final imaging observations, giving a
separation of $33\pm5$\,mas and a very uncertain PA of
$283\pm30$\degree.  We note that there is nominally an ambiguity in
this measured PA, typical for marginally resolved binaries, such that
it could be $103\pm30$\degree\ instead.  However, $283\pm30$\degree\
is in better agreement with the orbit fit, so we adopt this value.  

At separations of 104--151\,mas, SDSS~J0423$-$0414AB was much wider in
our ACS-WFC imaging than SDSS~J0926+5847AB.  We found that using the
rms of our SDSS~J0423$-$0414AB measurements as the uncertainties
produced a somewhat high $\chi^2$ for the final orbit fit, implying
that the rms does not fully capture all systematic errors for this
well resolved binary.  We therefore examined the rms about the orbit
fit of our ACS-WFC data and used this as the common uncertainty in
separation (0.6\,mas) and PA (0.7\degree) for all epochs of
SDSS~J0423$-$0414AB imaging.  The $F814W$ flux ratio we measured
across multiple epochs shows some variation within the rms errors, but
it would be consistent with being constant if there are 0.05\,mag
systematic errors in our measured SDSS~J0423$-$0414AB flux ratios.

\subsubsection{Other Published \& Archival \HST\ Imaging \label{sec:hst}}

Many of our sample binaries have \HST\ imaging data in the public
archive. These are often the observations that discovered the binaries
initially, providing the first epoch of astrometry that we use in our
orbit fits.  As in our previous work, we have re-analyzed the
available archival data using our own TinyTim PSF-fitting routine as
described in Section~\ref{sec:acs}.  These data come from the WFPC2
Planetary Camera (WFPC2-PC1), ACS High Resolution Channel (ACS-HRC),
and NICMOS Camera 1 (NICMOS-NIC1).  For all three of these cameras,
numerous images of single ultracool dwarfs are also available in the
archive from various imaging surveys.  We collected these images to
make a library of observed single PSFs for each relevant camera and
filter combination and used them to perform Monte Carlo simulations of
our PSF fitting.  For a given binary observation, we replicated the
binary properties as closely as possible to the actual fractional
pixel separation in $x$ and $y$ by co-adding two different library
PSFs centered at different subpixel locations, and scaling the two
PSFs to match the flux ratio and S/N of the actual binary data.  We
ran 100 simulations for each observation, applied the mean systematic
offset as a correction to our measurement, and used the rms as the
error.

Table~\ref{tbl:other} gives the resulting binary parameters from our
analysis of archival WFPC2-PC1, ACS-HRC, and NICMOS-NIC1 imaging.
WFPC2-PC1 data come from programs GO-6345 (PI Kirkpatrick), GO-8146
(PI Reid), GO-8581 (PI Reid), GO-8563 (PI Kirkpatrick), GO-8720 (PI
Brandner), GO-9157 (PI Mart\'{i}n), GO-9345 (PI Mart\'{i}n), and
GO-9968 (PI Mart\'{i}n).  ACS-HRC data come from programs GO-9451 (PI
Brandner) and GO-10559 (PI Bouy).  NICMOS-NIC1 data come from programs
GO-7952 (PI Mart\'{i}n), GO-9833 (PI Burgasser), GO-9843 (PI Gizis),
GO-10143 (PI Reid), and GO-11136 (PI Liu).

\subsection{Absolute Astrometry from CFHT/WIRCam  \label{sec:cfht}}

We obtained high-precision unresolved astrometry of our sample to
determine parallaxes, proper motions, and to constrain the photocenter
motion due to the binary orbits.
We present here an updated analysis of our data from the Hawaii
Infrared Parallax Program that uses the CFHT facility infrared camera
WIRCam \citep{2004SPIE.5492..978P}.  Our observing strategy and custom
astrometry pipeline are described in detail in
\citet{2012ApJS..201...19D}.  Briefly, for a given target we obtain
data sets on multiple nights each season, with each data set typically
comprising 20--30 dithered frames in $J$ band, or in a narrow $K$-band
filter (\Kn) if the target would saturate in $J$.  Queue observing
constraints at CFHT are used to require good seeing and that all
$J$-band observations occur near transit, i.e., at similar airmass, to
guard against systematic astrometric errors due to differential
chromatic refraction (DCR).  We measure Gaussian-windowed centroids by
running \texttt{sextractor~v2.19.5} \citep{1996A&AS..117..393B} on the
detrended images provided by CFHT.  We cross-match detections at a
given epoch, using our custom distortion solution and simple linear
transformations, and adopt the standard error on the mean as the
astrometric measurement error for any given star at each epoch.  We
then solve for the linear transformations between epochs, iteratively
solving for and masking high proper motion and/or parallax sources.
We finally cross-match to an astrometric reference catalog (2MASS or
SDSS-DR9) to provide absolute calibration of the linear terms and then
solve for the proper motion and parallax of every star in the field.

Unlike our previous work, we do not use a stand-alone analysis of our
CFHT data to compute the parallax and proper motion of our target
binaries.  This approach was used by \citet{2012ApJS..201...19D},
where we then applied evolutionary model-based corrections to account
for the expected orbital motion in the CFHT photocenter.  As we have
continued monitoring our sample binaries from 2007 or 2008 up to the
present, we now have data spanning a much longer fraction of many of
our targets' orbital periods.  It therefore becomes even more
important to freely fit for photocenter motion (as described in
Section~\ref{sec:orbit}), without imposing assumptions about
mass--magnitude relations from evolutionary models or potentially
unresolved multiplicity.  Indeed, the fact that we are now in a
position to detect significant photocenter motion allows us to place
strong empirical constraints on the relationship between mass and
magnitude, as we have recently demonstrated with SDSS~J1052+4422AB
\citep{2015ApJ...805...56D}, and possibly detect higher-order
multiplicity.  We have also made some small changes to improve the
performance of our pipeline compared to \citet{2012ApJS..201...19D},
such as using the mean instead of the median measured position of a
given source at each epoch.  In addition, when computing the
correction from relative to absolute parallax and proper motion using
the Besan\c{c}on model of the Galaxy \citep{2003A&A...409..523R}, we
now exclude model stars with proper motions larger than the cutoff
value we use in our iterative solution of the epoch-to-epoch linear
terms.

Table~\ref{tbl:cfht} gives all of the absolute astrometry from CFHT
for the sample of binaries for which we have orbit-monitoring data
from Keck and \HST.  Many of these CFHT observations were originally
published in \citet{2012ApJS..201...19D}, but the data presented here
are nonetheless distinct from what was published previously since we
no longer attempt to remove any photocenter motion from the reported
absolute astrometry.


\section{Joint Orbit \& Parallax Analysis \label{sec:orbit}}

We performed a Markov Chain Monte Carlo (MCMC) analysis to
simultaneously fit a relative orbital solution and an absolute
astrometric solution to our resolved (Keck, \HST, etc.) and unresolved
(CFHT) astrometry, respectively.  This joint fitting approach was
originally motivated by binaries that show significant photocenter
motion in our CFHT data.  However, this approach also has the
advantage of appropriately marginalizing over the uncertainty in CFHT
photocenter motion when determining parallaxes and proper motions.
Fitting for photocenter motion, even if it ends up being consistent
with zero, also allows us to constrain mass ratios given that we have
independently measured flux ratios.  For example, most binaries in our
sample have flux ratios near unity, so if their mass ratios are not
near unity due to unresolved triple components, then they would
display photocenter motion commensurate with the amount of unresolved
mass present.

Our joint orbit and parallax MCMC analysis method is similar to our
previous work on SDSS~J1052+4422AB \citep{2015ApJ...805...56D}.  Six
orbital parameters are shared in common between the resolved and
absolute astrometric solutions: period ($P$), eccentricity ($e$),
inclination ($i$), PA of the ascending node ($\Omega$), argument of
periastron ($\omega$), and mean longitude at the reference epoch
($\lambda_{\rm ref}$).\footnote{Here, in the absence of radial
  velocity information, there is a 180\degree\ ambiguity in $\Omega$,
  and consequently $\omega$ and $\lambda_{\rm ref}$, such that one of
  $\Omega$ or $\Omega+180\degree$ is actually the ascending node while
  the other is the descending node.  Therefore, we follow the standard
  convention that $0\degree \leq \Omega \leq 180\degree$.  If future
  radial velocities show that $\Omega+180\degree$ is actually the
  ascending node then 180\degree\ should be added to our reported
  values of $\Omega$, $\omega$, and $\lambda_{\rm ref}$.} The
reference epoch ($t_{\rm ref}$) is defined to be
2010~January~1~00:00~UT (2455197.5~JD) and is related to the time of
periastron passage $T_0 = t_{\rm ref} - P\frac{\lambda_{\rm
    ref}-\omega}{360\degree}$.  For the resolved orbit we fit for the
total semimajor axis ($a = a_1 + a_2$, where $a_1$ and $a_2$ are the
semimajor axes of the primary and secondary components about the
system barycenter).  For the unresolved orbit we fit for a photocenter
semimajor axis \aphot.  For the absolute astrometry we also fit for
the usual five parameters of parallax ($\pi$), position in RA
($\alpha$) and Dec ($\delta$) at a reference epoch, which we choose to
be the same as $t_{\rm ref}$ above, and proper motion in RA ($\mur
\equiv \mu_{\alpha}\cos\delta$) and Dec ($\mud$).  For a circular
orbit the parameter $\omega$ has no physical meaning.  Therefore, in
order to allow our MCMC to robustly explore parameter space for nearly
circular orbits we chose to step in $\omega+\Omega$ and
$\omega-\Omega$.  Our priors are uniform in $e$, $\omega$, $\Omega$,
$\lambda_{\rm ref}$, and the ratio of $\aphot/a$.  We adopt log-flat
priors in $P$ and $a$ by multiplying likelihoods by $1/P$ and $1/a$.
We allow for inclinations to be randomly distributed in space by
multiplying likelihoods by $\sin(i)$.  Our prior on parallax assumes a
uniform density in space volume, and thus we multiply our likelihood
by $1/\pi^2$.

To determine the starting point of our chains, we made an initial
estimate of the best-fit parameters.  For the relative orbit
parameters, we performed a grid search over the parameters $P$, $e$,
and $T_0$.  Once these three parameters are specified, then the
eccentric anomaly can be computed and the best-fit orbit can be
directly analytically determined, as described in detail by
\citet{2014A&A...563A.126L}.  We searched across 10$^4$ randomly drawn
values between $0 \leq e < 1$, $0 \leq T_0/P < 1$, and $\log(P)$
initially from 10$^3$ to 10$^6$~days.  At subsequent iterations we
refined the range of $P$ to center on the best-fit orbital period from
the previous trials.  After three iterations, we passed the single set
of orbit parameters with the lowest $\chi^2$ to our custom
least-squares minimization routine for orbit fitting
\citep{2010ApJ...721.1725D}, which is based on the MPFIT package in
IDL \citep{2009ASPC..411..251M}, in order to optimize our starting
position for the MCMC.  For the parallax parameters, including the
photocenter orbit size \aphot, we used the same method as described in
Section~2.4.1 of \citet{2012ApJS..201...19D} to find the best-fit
values.  Briefly, we used our best-fit orbit parameters, the binary
flux ratio for the CFHT bandpass (either $J$ or $K$ band), and an
estimated mass ratio from Cond evolutionary models
\citep{2003A&A...402..701B}.  Combining the flux ratio and mass ratio
gives the scale factor by which to multiply $a$ in order to derive
\aphot.  We then subtract the estimated photocenter orbit offsets from
the CFHT astrometry and find the best-fit parallax, proper motion, and
RA and Dec.

We used the Python implementation of the \citet{affine-mcmc}
affine-invariant ensemble sampler \texttt{emcee~v2.1.0}
\citep{2013PASP..125..306F} to perform our MCMC analysis.  We
initialized  10$^3$ walkers with our best-fit parameters after
adding a small amount of Gaussian noise having standard deviations of
$P \times 10^{-4}$ in $P$; $a \times 10^{-4}$ in $a$; $10^{-4}$ in
$e$; $10^{-4}$\,radians in $i$, $\omega-\Omega$, $\omega+\Omega$, and
$\lambda_{\rm ref}$; $10^{-6}$\,degrees in RA and Dec;
$10^{-8}$\,degrees\,yr$^{-1}$ in RA and Dec proper motion; $\pi \times
10^{-3}$ in $\pi$; and $10^{-3}$ in $\aphot/a$.  We allowed our $10^3$
walkers to run for $10^5$ steps, saving results every 500 steps. After
removing the first 50\% of the chains as burn-in, our final chains
possess a total of $1.0\times10^5$ values for each parameter.

Our CFHT astrometry gives a measurement of the parallax and proper
motion relative to the grid of reference stars.  In order to determine
distances and absolute proper motions, we estimate the parallax and
proper motion of our reference grids using the Besan\c{c}on model of
the Galaxy
\citep{2003A&A...409..523R}\footnote{\url{http://model.obs-besancon.fr/}}
in a similar fashion to our previous work \citep{2012ApJS..201...19D,
  2015ApJ...805...56D}.  The only change we have adopted here is to
exclude Besan\c{c}on model stars with large proper motions and
parallaxes.  This accounts for the fact that our astrometric pipeline
always excludes such stars from our reference grid, using cuts of
10\,mas in parallax and 30\,\masyr\ in proper motion.  Therefore, our
relative-to-absolute corrections reported here are slightly smaller in
amplitude than our past work.  For example, in
\citet{2015ApJ...805...56D} we derived corrections of
$\Delta\pi = 1.7\pm0.3$\,mas, $\Delta\mur = -6\pm3$\,\masyr, and
$\Delta\mud = -7\pm3$\,\masyr\ for SDSS~J1052+4422AB, but here we
derive $1.33^{+0.13}_{-0.16}$\,mas, $-3.0\pm1.4$\,\masyr, and
$-4.2\pm1.2$\,\masyr, respectively.  This results in a $<$0.5$\sigma$
change in our absolute parallax for SDSS~J1052+4422AB
($38.4\pm0.7$\,mas previously, $38.1\pm0.6$\,mas here).

Tables~\ref{tbl:mcmc-LP349-25}--\ref{tbl:mcmc-2M2252-17} present for
each binary the median, best-fit, and 1$\sigma$ and 2$\sigma$ credible
intervals for the 13 parameters in our joint orbit and parallax
analysis.  These tables also report various properties of interest
that can be derived from our fitted parameters, such as $T_0$,
distance ($d$), and semimajor axis in AU.  We also report the total
system mass for each binary,
$\Mtot/\Mjup = 1047.93 \times (a/{\rm AU})^3 (P/{\rm yr})^{-2}$, both
with and without including the uncertainty in the parallax so that
future improvements in distances for our sample binaries can be easily
adapted into improved masses.
Table~\ref{tbl:chisq} gives the $\chi^2$ values of the best-fit
solutions as well as the final mean acceptance fraction for each MCMC
chain.
Figure~\ref{fig:orbit} shows all of the relative and absolute
astrometry used in our joint orbit/parallax analysis for each binary
along with the resulting astrometric solution, and
Figure~\ref{fig:orb-hist} shows the resulting posterior distributions
from our MCMC analysis.

\subsection{Assessing Reliability of Orbit Determinations \label{sec:bad}}

Although we have applied our MCMC analysis to all of our astrometric
data, not all binaries will have orbit determinations sufficiently
reliable for astrophysical use.  The primary quality metric for
assessing our orbit fits should be the precision with which they
constrain the system mass, as very large mass uncertainties will be of
little use for constraining models in the following analysis.  In
addition, if orbital parameters are poorly determined, then the
resulting posteriors will be strongly influenced by our adopted
priors.  While we have attempted to choose priors that are as
uninformed as possible, we should not rely on them to constrain
physical parameters ($P$, $a$, $e$) that are not sufficiently
constrained by our data.  Finally, some of our orbit determinations
indicate that our observations span only a small fraction of the full
orbital period.  Therefore, we also consider this as a possible
indicator of whether our data has provided reliable constraints on
orbital parameters.

Table~\ref{tbl:quality} provides a summary of the orbit quality
metrics that we use to assess the reliability of our orbit fits.  The
metrics $\delta\log\Mtot$ and $\delta{e}$ are computed as the
difference between the maximum and minimum credible interval
(1$\sigma$) of the total mass at fixed distance and eccentricity,
respectively, and the metric $\Delta{t_{\rm obs}}/P$ is the fraction
of the orbital period (median of the posterior) covered by our
resolved astrometry.  We now consider our orbit fits, beginning with
the apparently least reliable, and moving toward binaries with better
reliability metrics until we finally determine which orbits to carry
forward in our analysis.

SDSS~J0926+5847AB has the worst mass precision
($\delta\log\Mtot = 0.38$\,dex) and eccentricity constraint
($\delta{e} = 0.38$).  Even though it has a good time baseline
($\Delta{t_{\rm obs}}/P = 0.68$), the actual coverage of the orbit on
the sky is poor, partly due to the fact that is seen nearly edge on
($i = 91.7^{+0.6}_{-0.8}$\degree).  The next worse mass precisions are
$\Delta\log\Mtot = 0.23$\,dex for SDSS~J2052$-$1609AB and 0.20\,dex
for 2MASS~J1750+4424AB.  The latter has the smallest time baseline of
any binary here ($\Delta{t_{\rm obs}}/P = 0.04$) owing to its very
long, quite uncertain orbital period ($210^{+40}_{-60}$\,yr), although
it only has a modestly large $\delta{e} = 0.15$.  SDSS~J2052$-$1609AB
has a better time baseline ($\Delta{t_{\rm obs}}/P = 0.19$) but a
poorly constrained eccentricity ($\delta{e} = 0.21$).
2MASS~J0850+1057AB has relatively poor mass precision
($\Delta\log\Mtot = 0.12$\,dex) as well as a poorly constrained
eccentricity ($\delta{e} = 0.11$) despite a marginally acceptable time
baseline ($\Delta{t_{\rm obs}}/P = 0.31$).  All of the above orbits we
do not consider sufficiently reliable to be used in the following
analysis.  SDSS~J1534+1615AB and SDSS~J1021$-$0304AB nominally have
much better mass precision than any of the above
($\delta\log\Mtot = 0.09$\,dex and 0.06\,dex, respectively).  However,
both have poor time baselines with $\Delta{t_{\rm obs}}/P = 0.17$ and
0.10, respectively, and poorly constrained eccentricities with
$\delta{e} = 0.43$ and 0.14, respectively.  (For reference, our input
prior on eccentricity alone would correspond to $\delta{e} = 0.68$.)
We therefore also chose not to use these orbits.

DENIS~J1228$-$1557AB and 2MASS~J1847+5522AB are both marginal cases.
DENIS~J1228$-$1557AB has a modestly large mass error
($\delta\log\Mtot = 0.12$\,dex) but a better constrained eccentricity
($\delta{e} = 0.06$) and longer observational baseline
($\Delta{t_{\rm obs}}/P = 0.29$).  On the other hand,
2MASS~J1847+5522AB has good mass precision
($\delta\log\Mtot = 0.035$\,dex) but a poorly constrained eccentricity
($\delta{e} = 0.14$) and an observational baseline of only
$\Delta{t_{\rm obs}}/P = 0.22$.  We conservatively choose to place our
cutoffs in orbit quality metrics to exclude these two marginal cases.
The next worse mass precision in our sample is LP~415-20AB
($\delta\log\Mtot = 0.06$\,dex), but it appears to have a reliably
determined orbit, with $\delta{e} = 0.023$ and most of the orbital
period covered by observations ($\Delta{t_{\rm obs}}/P = 0.67$).  The
worst eccentricity constraint for any of the remaining binaries is for
2MASS~J1728+3948AB ($\delta{e} = 0.028$), but it has much better mass
precision ($\delta\log\Mtot = 0.0089$\,dex) and better orbital
coverage ($\Delta{t_{\rm obs}}/P = 0.34$) compared to our excluded
orbits.  LSPM~J1735+2634AB has the worst orbital coverage of our
remaining sample ($\Delta{t_{\rm obs}}/P = 0.29$), but its orbit is
very well determined ($\delta\log\Mtot = 0.0017$\,dex,
$\delta{e} = 0.0075$) thanks to our observations serendipitously
bracketing its periastron passage.

In summary, we do not use the orbit determinations for
2MASS~J0850+1057AB, SDSS~J0926+5847AB, DENIS~J1228$-$1557AB,
2MASS~J1750+4424AB, 2MASS~J1847+5522AB, and SDSS~J2052$-$1609AB in our
following analysis.  The basis for their exclusion is poor mass
precision, poor observational coverage of the orbit, and/or poorly
constrained eccentricity that could make our results overly dependent
on our uniform eccentricity prior.  The remaining sample comprises 23
binaries with orbit quality metrics ranging from
$\delta\log\Mtot = 0.0017$--0.06\,dex, $\delta{e} = 0.0007$--0.028,
and $\Delta{t_{\rm obs}}/P = 0.29$--5.01.

\subsection{Comparison to Published Orbits \label{sec:compare-orbit}}

Some of the binaries in our sample have published orbital parameter
determinations, either from work of our own or others.  
By far, the largest overlap of our 31-binary sample is with 14
binaries from \citet{2010ApJ...711.1087K}, followed closely by our own
work (13 binaries).  In most of these overlapping cases, our new orbit
determinations agree with and improve upon previously published work
as expected given our longer time baseline and more numerous epochs.
For example, compared to \citet{2010ApJ...711.1087K} our orbital
period uncertainties are much smaller, with a median difference in
errors of a factor of 7.
We now discuss the cases where our new results for astrophysical
parameters (i.e., mass, period, and eccentricity but not viewing
angles) differ significantly ($\gtrsim$2$\sigma$) from published work.

Gl~569Bab has had a number of published orbit determinations, and
despite being a very well studied system the reported semimajor axes
of the orbit have varied significantly compared to the reported
errors.  As discussed in Section~3.1 of \citet{2010ApJ...721.1725D},
this is likely mostly due to astrometric calibration issues in data
sets that used an early generation of cameras behind the Keck AO
system \citep[KCAM and SCAM;][]{2001ApJ...560..390L} that we did not
use here or in our previous work.  Our analysis here uses some
archival NIRC2 imaging not included in our previous work, which
improves the coverage of this short period orbit
($P = 2.3707\pm0.0005$\,yr), and results in a semimajor axis of
$93.64\pm0.14$\,mas.  This sits in between values derived in our prior
work \citep[$95.6^{+1.1}_{-1.0}$\,mas][]{2010ApJ...721.1725D} and
values published by others that largely relied on KCAM and SCAM data:
$91.8\pm1.0$\,mas \citep{2004ApJ...615..958Z}, $90.4\pm0.7$\,mas
\citep{2006ApJ...644.1183S}, and $90.8\pm0.8$\,mas
\citep{2010ApJ...711.1087K}.  In spite of these different semimajor
axes, the dynamical total mass ($138\pm7$\,\Mjup\ here) varies by
$<$1$\sigma$ because the error is dominated by the \Hipparcos\
parallax uncertainty ($\sigma_{\pi}/\pi = 0.016$) not our semimajor
axis measurement ($\sigma_a/a = 0.0016$).  The improved precision of
of our new orbit fit is due to more measurements (19 from
astrometrically well calibrated Keck/NIRC2 or \HST\ imaging) that now
span 11.87\,yr (i.e., 5.006 orbital periods).

HD~130948BC has a very well determined orbit, so small differences in
orbital parameters are more statistically significant. In our most
recent previous analysis we found $e = 0.176\pm0.006$
\citep{2011ApJ...733..122D} compared to $0.1627\pm0.0017$ here, a
2.1$\sigma$ difference.  This difference is not astrophysically
significant, and perhaps it is simply a statistical fluctuation due to
the significant improvement in other orbital parameters (e.g., both
$a$ and $P$ are improved by nearly a factor of 20 compared to our
original orbit in \citealp{2009ApJ...692..729D}).  Both our current
and past values are consistent at $<$2$\sigma$ with the eccentricity
of $0.16\pm0.01$ from \citet{2010ApJ...711.1087K}.

Our orbit for LP~415-20AB agrees with our previous work but disagrees
somewhat with the analysis of \citet{2010ApJ...711.1087K}.  This is
discussed in detail in Section~2 of \citet{2011ApJ...733..122D} as
likely being jointly due to the small number of degrees of freedom in
the \citet{2010ApJ...711.1087K} analysis (1~dof) and one of their
measurements being a significant outlier with respect to the rest of
the available data.  Compared to their orbital parameters, our period
of $14.82\pm0.24$\,yr is longer (compared to $11.5\pm1.2$\,yr), our
semimajor axis of $96.5^{+1.1}_{-1.4}$\,mas is smaller (compared to
$108\pm24$\,mas), and our eccentricity of $0.706^{+0.011}_{-0.012}$ is
smaller (compared to $0.9\pm0.1$).  Our inferred total dynamical mass
is quite discrepant from that of \citet{2010ApJ...711.1087K}, given
that their fitted distance of $21\pm5$\,pc (derived by combining
astrometry and radial velocities) is 2.7$\sigma$ different from our
parallactic distance of $38.6\pm1.1$\,pc.  This is likely due both to
our different orbit determinations and the fact that their radial
velocity measurement ($\Delta$RV = $-0.7\pm1.4$\,\kms) disagrees with
the prediction from our relative orbit ($\Delta$RV =
$-2.96\pm0.16$\,\kms).  With 12 measurements spanning 9.97\,yr, our
new orbit fit has many more degrees of freedom (17~dof) for the
relative orbit fit than previous work, making the resulting orbit
parameters more robust and more precise by a factor of 5--17.

Our orbit for 2MASS~J1534$-$2952AB agrees well with
\citet{2010ApJ...711.1087K} but disagrees with
\citet{2008ApJ...689..436L}.  The cause of the discrepancy with
\citet{2008ApJ...689..436L} is the choice of eccentricity prior.  In
that work we adopted a prior of $p(e) = e$, which deweights smaller
eccentricities and rules out circular orbits entirely, whereas we
adopt here a more conservative uniform prior $p(e) = 1$.  Our
eccentricity posterior here piles up at zero with a 2$\sigma$ interval
of $e = 0.000$--0.014, and $a$ and $P$ are both correlated with $e$ in
the sense that larger $e$ corresponds to smaller $a$ and $P$.
According to Figure~6 of \citet{2008ApJ...689..436L}, that posterior
distribution of $P$ as $e$ approaches zero would agree with our
orbital period of $P = 20.29\pm0.07$\,yr.  According to Figure~8 of
\citet{2008ApJ...689..436L}, our values of $P$ and
$a = 213.7\pm0.5$\,mas agree well with that $P$--$a$ posterior
distribution in spite of the difference in eccentricity priors.
However, both \citet{2008ApJ...689..436L} and
\citet{2010ApJ...711.1087K} used the parallax of $73.6\pm1.2$\,mas
from \citet{2003AJ....126..975T} to convert their angular semimajor
axes into physical units and thereby compute dynamical masses.  As we
discuss in Appendix~\ref{sec:2M1534-29}, our parallax of
$63.0\pm1.1$\,mas leads to very different dynamical masses than in
previous work.

2MASS~J2206$-$2047AB also has an eccentricity posterior that piles up
at zero, which differs from the results of our previous work on this
system.  In \citet{2009ApJ...706..328D} we used a prior of $p(e) = e$
that suppressed low eccentricity orbit solutions and resulted in
$e = 0.25\pm0.08$.  Our new posterior has a 2$\sigma$ interval of
$e = 0.000$--0.027. As a result of this difference,
\citet{2009ApJ...706..328D} found very different values for semimajor
axis and period ($a = 213^{+24}_{-18}$\,mas, $P = 35^{+6}_{-5}$\,yr)
compared to our new analysis ($a = 167.7\pm0.5$\,mas,
$P = 23.96^{+0.23}_{-0.21}$\,yr).  However, thanks to the strong
correlation between $a$ and $P$, our new values follow the posterior
distribution of \citet{2009ApJ...706..328D} as shown in their
Figure~6.  Therefore, the resulting total dynamical mass was
$184\pm4$\,\Mjup\ from \citet{2009ApJ...706..328D} but is
$188.3^{+2.9}_{-3.1}$\,\Mjup\ here (assuming a fixed distance of
28.0\,pc for the purpose of this comparison).  Therefore, the choice
of eccentricity prior does not change the best-fit total mass,
although it would have affected the uncertainty in the mass, where the
uniform $e$ prior used here is more conservative.

SDSS~J2052$-$1609AB has a preliminary orbit determination from
\citet{2015AJ....150..163B} based on combining their data with
published astrometry from \citet{2011A&A...525A.123S} and 3 epochs of
our data retrieved from the NIRC2 archive.  Their posterior
eccentricity distribution ($e = 0.014^{+0.023}_{-0.010}$) is
significantly different from what we find in our analysis
($e = 0.20^{+0.09}_{-0.11}$, with a 2$\sigma$ interval of 0.08--0.50).
Their quoted uncertainties in other parameters are also generally much
smaller than ours, e.g., they find
$\Mtot = 86.2^{+3.9}_{-1.8}$\,\Mjup\ while we find
$\Mtot = 69^{+14}_{-20}$\,\Mjup, despite our analysis using more data
over a longer time baseline.  We also performed a cursory analysis of
their astrometry using our MCMC fitter and were unable to reproduce
their posterior distributions.  Therefore, we speculate that the
difference in our results is most likely due to differing MCMC
analysis methods.  They report that acceptance fractions for their
single Metropolis-Hastings sampled chain were typically 0.5\%--1\%,
whereas our analysis using the affine-invariant sampler of
\texttt{emcee} had much higher acceptance fractions of 8.7\%
(Table~\ref{tbl:chisq}).  Their MCMC also included distance as an
eighth parameter in addition to the seven orbit parameters that are
constrained by the relative astrometry.
Since distance is not constrained by the relative astrometry, it
appears that its use as a parameter was intended to marginalize over
the uncertainty in the \citet{2012ApJS..201...19D} parallax.  It is
therefore curious that the posterior distribution on distance from
\citet{2015AJ....150..163B} had smaller errors
($30.7^{+0.2}_{-0.4}$\,pc) than the input parallactic distance of
$29.5\pm0.7$\,pc.  Regardless, it is not clear how this would explain
the small uncertainties on the orbital parameters or why their MCMC
analysis apparently avoided the part of parameter space preferred by
our \texttt{emcee} analysis.  Our new orbit fit is based on 8 epochs,
rather than 6 epochs, spanning 8.58\,yr instead of 4.55\,yr, in
addition to the fact that our joint Keck+CFHT analysis properly
marginalizes over the uncertainty in orbital photocenter motion when
determining the parallax.  Therefore, we conservatively choose to use
our own orbit fitting results in the following analysis.

Kelu-1AB has an unrefereed orbit determination by
\citet{2008arXiv0811.0556S} based on 9 epochs of astrometry spanning
2.84\,yr.  Our orbit based on 13 epochs spanning 16.66\,yr agrees at
$\lesssim$1$\sigma$ with their eccentricity ($e = 0.82\pm0.10$) but
not their semimajor axis ($a = 339^{+129}_{-66}$\,mas) or orbital
period ($P = 38^{+8}_{-6}$\,yr).  We find
$a = 227.9^{+0.9}_{-1.1}$\,mas and $P = 24.98\pm0.19$\,yr.  Assuming a
fixed distance of 20.8\,pc for both orbits, we find a total mass of
$\Mtot = 180.1\pm1.1$\,\Mjup\ compared to their
$\Mtot = 244^{+156}_{-76}$\,\Mjup\ that agrees within 1$\sigma$
because of their large uncertainties.

\subsection{Comparison to Published Parallaxes \label{sec:compare-plx}}

In Table~\ref{tbl:compare-plx} we provide a comparison of the
parallaxes we measure here with other values in the literature,
including our own past work \citep{2012ApJS..201...19D,
  2015ApJ...805...56D}.  Our previously published parallaxes are not
statistically independent of the values given here, but they were
derived from somewhat different analysis methods.  Namely, in
\citet{2012ApJS..201...19D} we did not marginalize over the
uncertainty in the orbital parameters or mass ratio in the same way as
we have done here, and we used a slightly different method for
computing the correction from relative to absolute parallax.  It is
therefore not surprising that one of the largest discrepancies is for
2MASS~J0920+3517AB (2.0$\sigma$ different from our published
parallax), because we previously assumed a model-based mass ratio
estimated from the measured flux ratio, but as we discuss in
Section~\ref{sec:disc} and Appendix~\ref{sec:2M0920+35} this system is
likely a hierarchical triple where the fainter component is actually
much more massive than the brighter component.  The only other
$\gtrsim$2$\sigma$ difference is for 2MASS~J1728+3948AB, where our new
data spans 9.1\,yr (instead of 3.3\,yr) and with somewhat better
parallax phase coverage resulting in a parallax that is 6\%
(2.5$\sigma$) smaller, and we suggest this is likely a natural
statistical variation given the new better constraints on proper
motion in both RA and Dec (the parallax amplitude in Dec is almost as
large as RA for this object).

Among other literature parallaxes, the largest discrepancy is for
2MASS~J1534$-$2952AB (3.9$\sigma$).  As we have previously discussed
in \citet{2012ApJS..201...19D}, this is likely due to the uncertainty
in the \citet{2003AJ....126..975T} value of $73.6\pm1.2$\,mas being
somewhat underestimated.  Our value of $63.0\pm1.1$\,mas significantly
changes the inferred dynamical mass since $\Mtot \propto d^3$,
resulting in a factor of 1.6 increase in mass.  The next largest
discrepancies are 1.3--1.9$\sigma$ for LP~349-25AB,
SDSS~J0423$-$0414AB, 2MASS~J0700+3157AB, and 2MASS~J0746+2000AB and
the remaining 19 comparisons agree within $\leq$0.6$\sigma$.
Two binaries have parallaxes in the literature more precise than our
CFHT values.  Kelu-1AB has a parallax of $53.6\pm2.0$\,mas from
\citet{2002AJ....124.1170D} and $51.75\pm1.16$\,mas from
\citet{2016AJ....152...24W} that are both more precise but likely less
conservative than our value of $49.8\pm2.2$\,mas because we
marginalize over photocenter orbital motion.  All of these values
agree with each other at 0.8--1.3$\sigma$, but using the different
parallaxes results in quite different dynamical masses, and we
therefore exclude Kelu-1AB from our further mass analysis (see
Appendix~\ref{sec:KELU-1} for a discussion).
2MASS~J0746+2000AB has a remarkably precise parallax of
$81.9\pm0.3$\,mas from the USNO optical program
\citep{2002AJ....124.1170D}.  An updated USNO analysis that accounts
for the photocenter orbital motion of 2MASS~J0746+2000AB
\citep{2015csss...18..413H} results in a parallax of
$81.24\pm0.25$\,mas (H.\ Harris 2015, private communication).  Given
the longer time baseline and higher precision of the USNO parallax, we
use their value in our following analysis.


\section{Empirically Determined Properties}

\subsection{Spectral Types \& Magnitudes \label{sec:spt}}

Most binaries in our sample have spectral types derived through
spectral decomposition from \citet{2012ApJS..201...19D}, using
integrated-light NIR spectra and resolved NIR photometry.  For most of
these we simply adopt the same spectral types as in that work, but a
few binaries were not in that sample or have updated resolved
photometry that warrants new analysis.  We use the same method
described in Section~5.2 of \citet{2012ApJS..201...19D}.  Briefly, we
pair spectra from a template library and find the optimal scaling
ratio for each pair that best matches the observed integrated-light
spectrum.  We compute synthetic photometry for these pairs and then
cull all pairings that do not agree with the observed resolved
photometry, $p(\chi^2) < 0.05$.  We then examine the remaining pairs,
ranked by how well they match the observed spectrum, and assign types
and errors on types that best represent the results.  A summary of all
available magnitudes for our sample binaries, both integrated-light
and resolved, is given in Table~\ref{tbl:mags}.  We report magnitudes
on both the 2MASS and MKO photometric systems if data exist in either
system by using the empirical relations between photometric system
conversion and absolute magnitude that we derived in
Appendix~\ref{app:phot-corr}.

LP~415-20AB and 2MASS~J1047+4026AB were not included in the
\citet{2012ApJS..201...19D} sample.  LP~415-20AB has an
integrated-light optical spectral type of M7.5
\citep{2000AJ....120.1085G}, but we find that primary templates as
early as M5 and as late as M7 provide good matches to the infrared
SpeX SXD spectrum and resolved photometry, while the best secondary
templates range from M7.5--M8.5.  Therefore, we adopt types of
M$6.0\pm1.0$ for LP~415-20A and M$8.0\pm0.5$ for LP~415-20B.
2MASS~J1047+4026AB has an integrated-light optical type of M8
\citep{2000MNRAS.311..385G}, and we find types of M$8.0\pm0.5$ and
L$0.0\pm1.0$ in our spectral decompostion.  In this work we have added
new flux ratios in $CH4_S$ and $K$ bands for 2MASS~J1404$-$3159AB that
allow us to refine our analysis, and we find the same spectral types
as we did in \citet{2012ApJS..201...19D}, L$9\pm1$ and T$5.0\pm0.5$.
We have added a $J$-, $H$-, and $K$-band flux ratios for
2MASS~J2140+1625AB resulting in the same primary type (M$8.0\pm0.5$)
but an updated secondary type of L$0.5\pm1.0$ (was M$9.5\pm0.5$).
Finally, for DENIS~J2252$-$1730AB we have new flux ratios in $Y$, $J$,
$H$, $CH4_S$, and $K$ bands.  This allows us to improve the spectral
types to L$4.0\pm1.0$ (was L$4.5\pm1.5$) and T$3.5\pm0.5$ (was
T$3.5\pm1.0$).

\subsection{Individual Masses \label{sec:indiv}}

Our joint analysis of resolved relative astrometry and unresolved
absolute astrometry provides direct measurements of the relative
orbit, parallax, and photocenter orbit.  The first two of these give
the total system mass directly.  When the flux ratio of the binary is
known, then the photocenter orbit can also be used to directly measure
the individual component masses.  Our absolute astrometry from
CFHT/WIRCam is all in either $J$ or \Kn\ bands, and we have resolved
$J$- and $K$-band photometry for all of our binaries.  Therefore, we
can derive individual masses for any binary with a sufficiently well
constrained photocenter orbit semimajor axis (\aphot).  As noted
above, we assume a uniform prior in the ratio of the photocenter
semimajor axis to the true semimajor axis ($\aphot/a$).  If we define
the ratio of the secondary-to-total mass as $\mu \equiv M_2/(M_1+M_2)$
and the ratio of secondary to total flux as
$\beta \equiv F_2/(F_1+F_2)$, then $\aphot/a = \mu - \beta$.  An
unbounded and uniform prior on $\aphot/a$ and $\beta$, as we have
assumed, can in principle allow for unphysical values of $\mu < 0$.
This is simply a consequence of not using any information about the
flux ratio in our astrometric analysis, motivated by the fact that
improved flux ratios could easily be obtained in the future.
Therefore, not all astrometric solutions result in meaningful
constraints on the individual mass ratios, but in our sample of
well-determined orbits only Kelu-1AB has an \aphot\ uncertainty
sufficiently large to encompass a wide range of unphysical values.

In order to derive flux ratios, we used the values of $\Delta{J}$ and
$\Delta{K}$ reported in Table~\ref{tbl:mags}.  We use the MKO $J$-band
data directly, while for the narrow-band \Kn\ filter on WIRCam we must
derive a correction to be applied to our broadband $K$ measurements.
As described in Appendix~\ref{app:narr-corr}, we derived third order
polynomial relations between $K$-band absolute magnitude and both
$K_{\rm MKO}-\Kn$ color and $K_{\rm 2MASS}-\Kn$ color.  These
relations have an rms scatter of 0.021\,mag and 0.025\,mag,
respectively.  When applying the offset from these relations we added
these values of the rms scatter in quadrature to the flux ratio
uncertainty in any case where our observed $\Delta{K}$ was more than
2$\times$ larger than the scatter.  In other words, for binaries with
flux ratios within $\lesssim$0.05\,mag of unity we did not add this
uncertainty in quadrature.

Table~\ref{tbl:orbsumm} summarizes all of the photocenter orbit sizes
and flux ratios measured in our analysis, including systems for which
the dynamical masses are not reliable.  Table~\ref{tbl:propsumm} gives
the resulting individual masses we derive from our dynamical total
masses and photocenter mass ratios (19 systems) or from other
information (3 systems).  Two of the three binaries in our relative
orbit sample that do not have absolute astrometry from CFHT have
stellar companions with \Hipparcos\ distances (Gl~417BC and
HD~130948BC). We provide the model-derived individual masses in
Table~\ref{tbl:propsumm} for reference, but we do not consider them as
part of the following analysis of individual masses. The third binary
with an orbit but no CFHT data is Gl~569Bab, and for this system we
use the mass ratio of $M_1/M_2 = 1.4\pm0.3$ derived from a joint
analysis of all available radial velocity data by \citet[][see their
Section~5.4.6]{2010ApJ...711.1087K}.  Gl~569Bab along with LHS~1070BC
\citep{2012A&A...541A..29K} are the only two ultracool binary systems
that previously had individual mass determinations in the literature
not from our own work.  Our sample of 38 objects with individual
masses therefore increases the sample size by an order of magnitude.

Figure~\ref{fig:spt-m} shows the all our our individual masses, as
well as those from the literature, as a function of spectral type.
In the analysis that follows, we will use this information to discuss
the substellar boundary.
Here we simply note that in most cases our individual masses do not
grossly deviate from rough astrophysical expectations, e.g., nearly
equal flux systems have mass ratios near unity and vice versa.  There
are a few remarkable exceptions, such as the secondary
2MASS~J0920+3517B (L$9.0\pm1.5$) that has a mass of
$116^{+7}_{-8}$\,\Mjup, well in excess of the expected hydrogen fusion
limit ($\approx$70\,\Mjup; Section~\ref{sec:hlimit}).

Figure~\ref{fig:cmd-mass} shows our dynamical mass sample on the
color--magnitude diagram (CMD) with data points colored by the
measured masses.  Overall, mass broadly decreases through the CMD
sequence, although there is not a one-to-one correspondence between
mass and CMD location, as expected for our sample that is drawn from
the field population of ultracool dwarfs spanning a wide range of
ages. The most clear illustration of this is that the lowest mass
objects happen to be located roughly in the middle of the L/T
transition, while the latest-type, bluest objects are somewhat more
massive.  This more likely reflects the distribution of ages in our
sample rather than indicating that later type objects tend to be more
massive.  Figure~\ref{fig:spt-abs} shows our measured absolute
magnitudes as a function of spectral type, colored according to mass,
in order to assess the reliability of both plotted quantities.  Our
sample is broadly consistent with the field sequence, indicating the
accuracy of our spectral types, with the largest outlier being
2MASS~J0920+3517B, which is consistent with our dynamical masses
showing that this is an unresolved pair of brown dwarfs in a triple
system.

To provide a simple summary of our results, Table~\ref{tbl:avgmass}
reports the mean mass in seven spectral type bins selected to contain
3--7 objects per bin, excluding unusual objects (unresolved binaries
and the pre--main-sequence system LP~349-25AB).  As expected, the mean
mass broadly decreases with spectral type.  We caution that this table
is simply meant to provide a guide to the typical masses of objects in
the field population and is not intended to be used for any
quantitative astrophysical purpose.  At earlier spectral types,
Table~\ref{tbl:avgmass} may be useful for providing model-independent
mass estimates for very low-mass stars.
Next, we briefly compare these results to previous work and defer
discussion of the astrophysical interpretation of our individual
masses to later sections.

\subsubsection{Comparison to Literature Mass Ratios \label{sec:compare-q}}

Four of our 19 systems with individual masses have mass ratios derived
from radial velocities by \citet{2010ApJ...711.1087K}.  They quote
mass ratios as $M_1/M_2$, rather than the standard $q \equiv M_2/M_1$,
so for the purposes of comparison here we quote our measurements as
$1/q$ as well.  For 2MASS~J0746+2000AB we find $1/q = 1.05\pm0.03$,
which agrees within the large uncertainties of $4.0^{+0.1}_{-3.8}$ as
found by \citet{2010ApJ...711.1087K}.  They report a similar mass
ratio for 2MASS~J2140+1625AB with much smaller uncertainties
($4.0^{+0.0}_{-0.1}$), which disagrees at $\approx$8$\sigma$ with our
result ($1.65^{+0.19}_{-0.23}$).  This seems to confirm the suggestion
in Section 5.4.5 of \citet{2010ApJ...711.1087K} that their mass ratio
for this binary was likely more uncertain than the quoted errors
implied.  One of our most precise mass ratios is for LHS~2397aAB
($1/q = 1.42^{+0.05}_{-0.06}$), and it agrees well with the value of
$1.5^{+7.1}_{-1.4}$ reported by \citet{2010ApJ...711.1087K}.  Finally,
we find $1/q = 1.06\pm0.03$ for LP~349-25AB, which disagrees with the
value of $0.5\pm0.3$ found by \citet{2010ApJ...711.1087K}.  As
discussed in Section~4.2 of \citet{2010ApJ...721.1725D}, a mass ratio
that agrees with our value can be accommodated by the radial
velocities of \citet{2010ApJ...711.1087K}, suggesting that their
quoted errors may be underestimated.

The only other directly measured mass ratio in the literature for a
system in our sample is from \citet{2015csss...18..413H} for
2MASS~J0746+2000AB.  Their USNO astrometry gives a value of
$q = 0.925$, which agrees well with our value of
$0.952^{+0.026}_{-0.027}$.

\subsection{Bolometric Luminosities \label{sec:lbol}}

We have derived luminosities for all objects in our sample using our
resolved photometry along with the empirical relations between
luminosity and absolute magnitude that are described in
Appendix~\ref{app:lbol-mag}.  This is somewhat different from our past
work where we relied on bolometric correction--spectral type relations
to derive individual luminosities.  Our new approach obviates the need
to reference spectral types, as it links absolute magnitude directly
to luminosity.  Given that $J$- and $H$-band absolute magnitudes get
brighter and plateau across the L/T transition, respectively, only
$K$-band absolute magnitudes are suitable for the entire range of
luminosity of our sample.  The luminosity scatter about the $K$-band
polynomial relation is somewhat higher across the L/T transition
(0.07\,dex; $M_K>13.0$\,mag) than at brighter magnitudes (0.04\,dex).
Interestingly, if we only consider the late-M and earlier L dwarfs
with $M_H<13.3$\,mag, the scatter in luminosity about the $H$-band
relation is significantly lower (0.023\,dex). It is not obvious why
this should be the case from an astrophysical perspective, but we
nonetheless use this to our advantage.  For binaries where both
components have $M_H<13.3$\,mag, we use the $H$-band relation with
luminosity, and for the remainder we use the $K$-band relation.  In
all cases we used whichever photometry was more precise between 2MASS
and MKO. We report the resulting component luminosities in
Table~\ref{tbl:propsumm}, and Figure~\ref{fig:m-l} shows our derived
luminosities as a function of our individual masses.

As a check on our method, we compare our results to published work
using integrated-light spectra and photometry to derive luminosities.
For objects in common with \citet{2004AJ....127.3516G},
\citet{2014AJ....147...94D}, and \citet{2015ApJ...810..158F}, all of
whom use somewhat different methods and generations of models for
deriving luminosities, we find that our total luminosities agree well
within the errors after accounting for the different distances
assumed in each work.  For example, summing our luminosities for the
components of 2MASS~J0746+2000AB we find
$\log(\Lbol/\Lsun) = -3.375\pm0.020$\,dex, which agrees well with
published values of $-3.41\pm0.02$\,dex \citep{2004AJ....127.3516G},
$-3.413\pm0.009$\,dex \citep{2014AJ....147...94D}, and
$-3.391\pm0.003$\,dex \citep{2015ApJ...810..158F}.  Other objects in
common include Kelu-1, LHS~2397a, SDSS~J0423$-$0414, and
2MASS~J1728+3948.


\section{Evolutionary Model-Derived Properties \label{sec:model}}

Directly measured masses and luminosities are just two of the many
physical properties that are of interest in characterizing our sample
of very low-mass stellar and brown dwarf binaries.  In principle, all
physical properties of both stars and brown dwarfs can be predicted
from just a few fundamental parameters: mass, age, and composition.
(Other properties such as the entropy of formation, initial angular
momentum, and magnetic fields may also be important as initial
conditions.)  Our dynamical mass sample consists of binaries that may
be conservatively presumed to be coeval to within a few Myr and to be
composed of the same material.  It is therefore an ideal sample for
pairing with evolutionary models, since the three most fundamental
parameters are all constrained at some level.

We consider two families of evolutionary models here, the Lyon group's
models and the \citet{2008ApJ...689.1327S} models
(Figure~\ref{fig:models-t-l} shows all model grids used here).  The
most recent grid from the Lyon group comes from
\citet{2015A&A...577A..42B}, hereinafter BHAC, and extends over masses
of 0.01--1.4\,\Msun.  The published BHAC grid samples masses in the
range 0.02--0.1\,\Msun\ in increments of 0.01\,\Msun\ (with two extra
models at 0.072\,\Msun\ and 0.075\,\Msun) and only tracks evolution
down to $\Teff=1500$\,K.  We have obtained BHAC model tracks that
extend down to 1300\,K (only for fundamental properties, not absolute
magnitudes), as well as more finely gridded tracks over the mass range
0.05--0.07\,\Msun\ in 0.001\,\Msun\ increments.  This enhanced BHAC
evolutionary model grid was kindly provided by I.\ Baraffe (2016,
private communication) so that we could more accurately sample the
hydrogen-fusion and lithium-fusion mass limits.
Many of our sample binary components have luminosities too low to be
covered by the BHAC grid, and when this is the case we use the Lyon
Cond models instead \citep{2003A&A...402..701B}.  We do not consider
Lyon Dusty models \citep{2000ARA&A..38..337C} as these have been
supplanted by BHAC models over the luminosity range for which they
would be appropriate (namely late-M and L~dwarfs).  The second family
of models we consider are from \citet{2008ApJ...689.1327S},
hereinafter SM08, which conversely to BHAC only cover cooler
temperatures ($\Teff\lesssim2400$\,K;
$\log(\Lbol/\Lsun) \lesssim -3.3$\,dex) and thus many of our more
luminous objects are not covered by the SM08 grid.
These hard limits on the model grids means that we cannot always
report model-derived properties for a given component because we do
not extrapolate beyond the range of any model grid.

In order to perform tests of models, we first use the observed
properties of mass and luminosity to infer other properties from
evolutionary models.
For our entire sample we have individual luminosities and precise
total masses.  In most cases (19 of 22 systems) we also have
individual masses, but the precision of our mass ratios causes most of
these individual masses to be less precise than the total mass.
Therefore we take a two-pronged approach to inferring physical
properties from models, either using the more precise total mass alone
or the full individual mass information.

The total-mass analysis we use here is distinct from methods that we
have used in our previous work \citep[e.g.,][]{2008ApJ...689..436L,
  2009ApJ...692..729D} as well as other approaches in the brown dwarf
literature (see Section~4.5 of \citealp{2010ApJ...721.1725D} for a
discussion of different methods).  Our past Monte Carlo methods
relied on a two-step interpolation, first using the individual
luminosities to compute a model-predicted total mass as a function of
age, and then applying our observational constraint on the total mass
to determine the age distribution.  Such a method is susceptible to
numerical problems where more than one mass corresponds to a given
luminosity either at a given age (e.g., due to deuterium burning) or
over a range of ages (e.g., even very low-mass stars become slightly
more luminous at old ages than they were on the zero age main
sequence).  More importantly, measurement errors can cause Monte Carlo
samples to fall in regions of parameter space not covered by models,
e.g., a luminosity that scatters low so that models predict a star of
that mass would never be that faint, which causes a simple
interpolation approach to fail.  Such interpolation problems can also
occur in the case of individual-mass analysis methods applied to
stars, even though they have worked well for brown dwarfs and young
stars in our previous studies \citep{2015ApJ...805...56D,
  2016ApJ...827...23D}.

Therefore, we have developed a new approach for deriving physical
properties from models that is based on the statistical technique of
rejection sampling.  This approach works well when the number of
parameters of interest is small, and in the individual-mass analysis
we only need to find the probability distribution for one unknown
parameter---age---because age combined with mass allows all other
properties to be directly interpolated from models.  Our approach
begins with a uniform distribution of Monte Carlo drawn ages spanning
the minimum to maximum age of a given model grid, and through
rejection sampling we end up with a distribution of ages that match
the observed luminosities at the measured masses.  For the case of
individual masses this is conceptually straightforward, but when using
only the total mass we must simultaneously try both random ages and
component masses (i.e., essentially allowing for one more unknown
parameter, mass ratio).  In both cases, we use the final individual
mass and age samples remaining at the end of the rejection sampling to
derive other properties from models (\Lbol, radius, \Teff, etc.).  
We now describe each case in more detail.

In our individual-mass analysis, we begin with $10^6$ Monte Carlo
trials of age that are randomly paired with masses from our MCMC
results.  The input distribution of ages defines the prior on age, and
we choose uniformly distributed values, i.e., a flat prior.  Each mass
and age uniquely corresponds to a luminosity from models, which we
determine from bilinear interpolation of a uniform 2-d grid of mass
and age constructed using Delaunay triangulation as implemented in the
\texttt{trigrid} function in IDL.  We compute a $\chi^2$ for each
sample given our measured median \Lbol\ and its uncertainty.  We
account for the fact that our luminosity and mass measurements are
correlated (due to the common parallax) by fitting a line to
$\log(\Lbol)$ as a function of $\log(M)$.  At each Monte Carlo-drawn
mass, the effective luminosity ($L_{\rm bol}^{\prime}$) is given by
the coefficients of this linear fit.  The observational error in
luminosity at fixed mass ($\sigma_{L_{\rm bol}^{\prime}}$) is given by
the rms scatter about the fit, which is equal to or smaller than the
actual \Lbol\ error depending on the degree of covariance.  Thus, we
compute
$\chi^2 = (L_{\rm bol, model}-L_{\rm bol}^{\prime})^2/\sigma_{L_{\rm
    bol}^{\prime}}^2$
and a corresponding probability of $p = e^{-(\chi^2-\min(\chi^2))/2}$,
normalized by the Monte Carlo sample with the lowest $\chi^2$.  We
then draw random, uniformly distributed variates $u$, and we rejected
samples where $p < u$.  In order to efficiently sample the most
relevant ages, we iterated this process gradually narrowing the trial
age distribution based on the results of the previous iteration.  We
found that after three iterations the number of successful Monte Carlo
trials stabilized.  After determining our final age distribution, we
then determined other properties (\Teff, lithium depletion, etc.) from
2-d grids constructed with \texttt{trigrid} in the same way as for
luminosity.

In our total-mass analysis we have two unknown parameters, age and
mass ratio.  Therefore, we not only draw random, uniformly distributed
ages as above, but also uniformly distributed values of $\log(M_1)$
for one component.  The mass of the other component can then be
computed from the total mass in that Monte Carlo sample,
$M_2 = \Mtot - M_1$.  We initially sample all ages covered by the
model tracks and masses for $M_1$ ranging from the lowest mass model
to the maximum total mass in the MCMC chain.  We interpolate the
luminosity from models for each component mass and age pair and
compute a total $\chi^2$ by comparing the interpolated luminosities
for both components to our measurements.  As in our individual-mass
analysis, we account for correlation between our measured total mass
and component luminosities by fitting lines to $\log(\Lbol_{,1})$ and
$\log(\Lbol_{,2})$ as a function of $\log(\Mtot)$.  We reject Monte
Carlo samples based on the probabilities computed from the sum of the
components' $\chi^2$ and then iteratively narrow the range of age and
component mass searched, again finding three iterations sufficient in
all cases.  We use the final remaining samples of age and component
masses to interpolate other properties (\Teff, etc.)  from the 2-d
model grids.

Tables~\ref{tbl:props-LP349-25}--\ref{tbl:props-2M2252-17} show the
results of our analyses using total mass and individual luminosities
as well as, when possible, individual masses and luminosities.  Our
rejection sampling analysis naturally produces output distributions of
individual luminosities and masses in all cases.  Conceptually, this
is because any inputs to the rejection sampling (even measured
quantities) can be thought of as priors on those properties, so the
output distributions may be somewhat different (narrower or slightly
shifted).  The final output properties will always be fully consistent
with model predictions.  For example, in the case of a star on the
main sequence, combining our measured luminosity and age prior
typically results in a much narrower range of masses according to
models because only certain masses are predicted to correspond to the
input luminosity.  In most cases, the median mass and luminosity
output by the rejection sampling analysis agrees within
$\lesssim$1.5$\sigma$ of the input values, as expected, but
exceptional cases are discussed in the following sections.
Figure~\ref{fig:model-derive} shows our input measurements compared to
model tracks and the resulting age distributions from both
individual-mass and total-mass analyses.

Our new approach to deriving properties from models via rejection
sampling gives essentially identical results as our past approach for
cases where a given mass and luminosity measurement lie well within
the bounds of model isochrones, i.e., for brown dwarfs and young
stars.  The main advantage to rejection sampling is that it properly
handles stars on the main-sequence and that it explicitly specifies a
prior on age.  For example, in our past work on LHS~1901AB we quoted
median and $\pm$1$\sigma$ values from our age posterior from Lyon
models of 0.28$^{+9.72}_{-0.08}$\,Gyr \citep{2010ApJ...721.1725D}.
This low median age was a consequence of directly interpolating our
measured luminosity that is slightly higher than, but still well
within the 1$\sigma$ errors of, the model-predicted main-sequence
luminosity for objects of that mass.  Using our new method we report a
more sensible result of 5.2$^{+1.9}_{-4.7}$\,Gyr, thanks to both
allowing measured luminosities to scatter above and below the main
sequence value and imposing an explicit age prior.  In addition, in
our new approach we always interpolate directly from model grids in
mass and age, which circumvents issues related to the fact that
luminosity can be double-valued at a given mass as a function of age.

Finally, we note that the models used in our analysis all assume solar
metallicity.  Different assumptions about metallicity would impact
opacities and likely also influence cloud formation and evolution,
both of which would change how luminosity evolves with time.  In our
work on Gl~417BC \citep{2014ApJ...790..133D}, we discussed this impact
of changing opacities on cooling, inferring from the cloudless SM08
models that a change of $\pm$0.3\,dex in metallicity changes
luminosity by no more than $\pm$0.05\,dex (super-solar models are more
luminous at a given mass and age).  Thus, if our sample is on average
significantly offset from solar metallicity for some reason, this
would result in only a slight systematic shift in our results.  For
reference, \citet{2008A&A...480..889S} report a mean and rms
metallicity of $-0.10\pm0.24$\,dex for stars in the solar
neighborhood. Even for such relatively large metallicity differences
as $\pm$0.3\,dex, the corresponding difference in luminosity is
comparable to our typical luminosity uncertainties, so we do not
expect metallicity to strongly affect our conclusions.


\section{Discussion \label{sec:disc}}

Our large sample of dynamical masses enables a broad array of
empirical tests of substellar evolution, many of which have not been
possible before without precise individual masses.  We now discuss a
number of key topics, both in testing model predictions and
establishing empirical relations.  In the following analysis we will
mostly deal with our results as whole, but a detailed discussion of
each individual system is provided in Appendix~\ref{app:objects}.
When discussing our results below, we generally do not explicitly
exclude anomalous objects (e.g., the unresolved components of triple
systems discussed in Section~\ref{sec:triple}) every time they might
be relevant.

\subsection{Empirical Test of the Substellar Boundary \label{sec:hlimit}}

Our work represents the first sample of individual masses for
spectrally classified L and T~dwarfs.  This allows us to examine the
maximum mass of the latest-type objects, which is the mass of the
boundary between stars and brown dwarfs.  Figure~\ref{fig:m-l} shows
our 38 individual mass and luminosity measurements along with all
other published model-independent masses, including the spectrally
unclassified objects Gl~802B
\citep[$80\pm14$\,\Mjup;][]{2008ApJ...678..463I} and HR~7672B
\citep[$68.7^{+2.4}_{-3.1}$\,\Mjup;][]{2012ApJ...751...97C}, as well
as the M8.5+M9 binary Gl~569Bab \citep[$80^{+9}_{-8}$\,\Mjup\ and
$58^{+7}_{-9}$\,\Mjup;][]{2010ApJ...711.1087K}, the M9.5+L0 binary
LHS~1070BC \citep[$81\pm5$\,\Mjup\ and
$74\pm4$\,\Mjup;][]{2012A&A...541A..29K}, and the M7+M7 binary
LSPM~J1314+1320AB \citep[$92.8\pm0.6$\,\Mjup\ and
$91.7\pm1.0$\,\Mjup;][]{2016ApJ...827...23D}.  None of the components
violate the most basic predictions of models within the observational
uncertainties, except for systems suspected to be higher-order
multiples based on our analysis independent of models.  No objects
have a lower luminosity than expected given their mass and the finite
age of the Universe.

Figure~\ref{fig:spt-m} shows our individual masses as a function of
spectral type.  The lack of objects at high mass and late spectral
type is empirical evidence for a mass limit to hydrogen fusion.
Moreover, examining the maximum mass of the latest-type objects (or
equivalently the lowest luminosity objects in Figure~\ref{fig:m-l})
allows us to constrain the upper limit on the mass of the substellar
boundary, and correspondingly a lower limit of the mass of main
sequence stars.  In fact, our dynamical mass sample is (somewhat
unfortunately) ideal for this test because we are biased toward high
masses.  The main observational limitation in measuring masses to date
has been achieving the time baseline needed for robust orbit
determinations of binaries with typical periods of 20--30\,yr.  The
highest mass late-L or T~dwarfs in our sample are 2MASS~J1728+3948B
(L$7.0\pm1.0$; $67\pm5$\,\Mjup;
$\log(\Lbol/\Lsun) = -4.49\pm0.04$\,dex) and 2MASS~J1404$-$3159A
(L$9.0\pm1.0$; $65\pm6$\,\Mjup;
$\log(\Lbol/\Lsun) = -4.52\pm0.05$\,dex).  The next most massive
late-type object is 2MASS~J2132+1341B (L$8.5\pm1.5$; $60\pm4$\,\Mjup;
$\log(\Lbol/\Lsun) = -4.50^{+0.05}_{-0.04}$\,dex).  Going to earlier
types, there are three objects that have masses and spectral types
that are consistent among each other within the errors:
2MASS~J0920+3517A (L$5.5\pm1.0$; $71\pm5$\,\Mjup;
$\log(\Lbol/\Lsun) = -4.28\pm0.03$\,dex), 2MASS~J1728+3948A
(L$5.0\pm1.0$; $73\pm7$\,\Mjup;
$\log(\Lbol/\Lsun) = -4.29^{+0.04}_{-0.05}$\,dex), and
2MASS~J2132+1341A (L$4.5\pm1.5$; $68\pm4$\,\Mjup;
$\log(\Lbol/\Lsun) = -4.22\pm0.05$\,dex).
Despite the fact that we are more sensitive to massive systems, even
our most massive late-type objects are only consistent with having
masses as high as 71\,\Mjup\ within 1$\sigma$, and even at spectral
types as early as L4 the most massive objects are consistent with
70\,\Mjup.  Therefore, we estimate an empirical substellar boundary of
$\approx$70\,\Mjup.  

The theoretical mass limit of hydrogen fusion quoted in the literature
varies widely.  One review by \citet{2001RvMP...73..719B} places the
minimum mass for H fusion between 0.070--0.075\,\Msun\ (73--79\,\Mjup)
for solar metallicity and at 0.092\,\Msun\ (96\,\Mjup) for zero
metallicity.  More recent models from \citet{2011ApJ...736...47B} also
show a range of 0.070--0.075\,\Msun\ for the H-fusion mass assuming
different helium fractions (their Figure~6).
A review by \citet{2000ARA&A..38..337C} gives the minimum mass of
H~fusion as 0.070\,\Msun\ (73\,\Mjup) when cloud opacity is included
and 0.072\,\Msun\ (75\,\Mjup) for cloudless atmospheres.  If the
H-fusion mass limit is indeed as high as $\approx$75\,\Mjup, then it
is surprising that we have not uncovered objects with masses closer to
this limit in our sample.  Figure~\ref{fig:models-t-l} shows more
recent tracks from BHAC that includes a 0.072\,\Msun\ (75\,\Mjup)
track that stabilizes in luminosity after $\approx$2\,Gyr and thus we
would consider a star.  Some of the next lower mass tracks appear more
star-like in that sense, but by 0.067\,\Msun\ (70\,\Mjup) the
luminosity appears to be on course to monotonically decrease for the
age of the Universe.
Finally, SM08 quote a minimum mass of 0.070\,\Msun\ (73\,\Mjup) from
their cloudy models, because that track ultimately reaches 1550\,K
where clouds are very significant in the atmosphere.
Thus, the lack of any $\gtrsim$70\,\Mjup\ brown dwarfs in our sample
is consistent with the lower range of predictions in the literature
for the minimum mass of H~fusion (70--73\,\Mjup).

If we adopt our empirical substellar boundary of $\approx$70\,\Mjup,
we can then determine the latest spectral type of objects above this
boundary (i.e., objects that may be stars at the very bottom of the
main sequence).  The most precise stellar mass in our sample is for
2MASS~J0746+2000B (L$1.5\pm0.5$; $78.4\pm1.4$\,\Mjup;
$\log(\Lbol/\Lsun) = -3.777^{+0.028}_{-0.027}$\,dex), but this is
likely well above the H-fusion limit.  Going to later spectral types
we find 2MASS~J0700+3157A (L$3.0\pm1.0$; $68.0\pm2.6$\,\Mjup;
$\log(\Lbol/\Lsun) = -3.95\pm0.04$\,dex) and 2MASS~J1017+1308B
(L$3.0\pm1.0$; $75\pm7$\,\Mjup;
$\log(\Lbol/\Lsun) = -3.84\pm0.04$\,dex).  After this is the cluster
of $\approx$L4--L6 objects with masses consistent with 70\,\Mjup\
mentioned above.  We therefore conclude that the end of the main
sequence occurs within the range of spectral types of L3--L5 and
luminosities of $10^{-4.3}$ to $10^{-3.9}$\,dex, independent of model
assumptions.  Objects with spectral types later than this, or
luminosities lower than this, are brown dwarfs, while objects with
earlier types or higher luminosities could either be stars or brown
dwarfs.  Our result is consistent with early work on the first samples
of L~dwarfs that showed the substellar boundary was likely within this
spectral type range \citep[e.g.,][]{1999ApJ...519..802K}.  A larger
sample and more precise component spectral types would enable us to
refine the location of this empirically defined substellar boundary.

Finally, we discuss our results in the context of the work of
\citet{2014AJ....147...94D} on the substellar boundary. Their approach
uses the minimum radius (determined via luminosity and a temperature
derived from model atmospheres) on the Hertzsprung-Russell
diagram. The physical motivation is that the transition to
degeneracy-dominated interiors marks the beginning of the substellar
regime, as degeneracy pressure is what prevents objects below a
critical mass from reaching sufficiently high core temperatures to
generate a significant amount of energy by fusing hydrogen. To perform
their test, they start at the late-M dwarfs and examine progressively
cooler objects that have progressively smaller radii, until the trend
reverses and cooler objects no longer have significantly smaller
radii. \citet{2014AJ....147...94D} find that this occurs at a radius
of $0.086\pm0.003$\,\Rsun\ for the object 2MASS~J0523$-$1403, which
has a spectral type of L2.5 and $\log(L/\Lsun) = -3.9$\,dex, and they
conclude that this marks the end of the main sequence. Compared to the
boundary derived here using mass, their result is somewhat earlier in
type and more luminous than our result but broadly consistent within
the uncertainties. \citet{2014AJ....147...94D} discuss the fact that
models predict a locus for their radius test at somewhat cooler
temperatures and suggest that lowering the metal abundances adopted in
current models could explain the discrepancy; however, they also note
that such a change in the abundances would alter the mass limit of the
boundary to higher masses. The fact that we find a rather low mass
limit for the substellar boundary would therefore seem to indicate
that abundances alone will not rectify the tensions between models and
the results of \citet{2014AJ....147...94D}. We speculate that one
possibility could be systematic errors in temperatures derived from
model atmospheres akin to those identified in our previous work on
late-M dwarfs \citep{2010ApJ...721.1725D}, where model atmospheres
gave higher temperatures than implied by evolutionary model radii and
measured luminosities.

\subsection{Lithium Fusion Mass Limit \label{sec:lithium}}

Theoretical predictions of lithium depletion in ultracool dwarfs as a
function of mass and age are thought to be one of the most reliable
methods for age-dating young clusters
\citep[e.g.,][]{1996ApJ...459L..91C, 1997ApJ...482..442B,
  2014MNRAS.438L..11B, 2014AJ....147..146K}.  They have also long been
used as a method of confirming ultracool dwarfs as substellar
\citep[e.g.,][]{1992ApJ...389L..83R, 1996ApJ...458..600B}.  The brown
dwarfs in our sample span masses both above and below the predicted
lithium depletion boundary, and many of them are in systems with
published optical, integrated-light spectroscopy that constrains the
presence or absence of lithium.  One potential complication is that
the \ion{Li}{1} doublet at 6708\,\AA, which provides the most
sensitive observational probe of lithium, is expected to disappear in
cool objects where monatomic lithium becomes incorporated into LiCl
and LiOH molecules or at cooler temperatures condenses into LiF and
Li$_2$S \citep[e.g.,][]{1999ApJ...519..793L}.  This chemical depletion
at low temperatures has not previously been tested with objects of
known mass.  The latest-type object known to display \ion{Li}{1}
absorption is WISE~J1049$-$5319B \citep[T0;][]{2014ApJ...790...90F,
  2015A&A...581A..73L}, implying that any of the L~dwarfs in our
sample might plausibly display \ion{Li}{1} absorption if they are not
depleted by Li~fusion.

Figure~\ref{fig:m-age} shows our individual masses as a function of
age for the 13 systems with well constrained model-derived ages, i.e.,
systems containing at least one brown dwarf, as well as the
pre--main-sequence binary LP~349-25AB.  (The low-mass main-sequence
binaries in our sample have unconstrained ages, and, as expected, none
are known to display \ion{Li}{1} absorption.)  Only two of these 13
binaries have published lithium detections, SDSS~J0423$-$0414AB
\citep[EW = 11\,\AA;][]{2008ApJ...689.1295K} and Gl~417BC \citep[EW =
11.5\,\AA;][]{2001AJ....121.3235K}.  Seven more systems have published
spectra that exclude lithium absorption in integrated light:
LP~349-25AB \citep[$<$0.5\,\AA;][]{2009ApJ...705.1416R},
2MASS~J0700+3157AB \citep[$<$0.3\,\AA;][]{2003PASP..115.1207T},
LHS~2397aAB \citep[$<$0.5\,\AA;][]{2009ApJ...705.1416R},
2MASS~J1728+3948AB \citep[$<$4\,\AA;][]{2000AJ....120..447K},
2MASS~J2132+1341AB \citep{2007AJ....133..439C}, and
DENIS~J2252$-$1730AB \citep{2008AJ....136.1290R}.  (See
Appendix~\ref{app:objects} for a detailed discussion of all of these
binaries and their lithium constraints.)

Also shown in Figure~\ref{fig:m-age} are the lithium-depletion mass
limits predicted by the BHAC models and by the Tucson models
\citep{1997ApJ...491..856B}. While we do not use the Tucson models in
our other model analysis, mainly because in addition to their non-gray
atmosphere models they used gray atmospheres at the higher
temperatures (including some temperatures relevant for our sample) in
order to cover the full range of evolution, we include them here
because they are the only other available evolutionary tracks that
report lithium depletion.
As expected, lithium nondetections in the most massive objects,
LHS~2397aA (M8, $93\pm4$\,\Mjup) and the young
($271^{+22}_{-29}$\,Myr, BHAC) components of LP~349-25AB (M7+M8,
$85\pm4$\,\Mjup\ and $80\pm3$\,\Mjup), are all consistent with
model-predicted lithium depletion.
Likewise, the two systems with strong lithium detections comprise some
of the lowest mass components in the sample ($\lesssim$50\,\Mjup), so
they are consistent with both sets of models that predict they should
be abundant in lithium.

The lowest mass objects with lithium not detected are the
$59\pm5$\,\Mjup\ and $41\pm4$\,\Mjup\ components of
DENIS~J2252$-$1730AB (L4+T3.5; $1.10^{+0.15}_{-0.18}$\,Gyr). Within
the 1$\sigma$ uncertainties the primary is consistent with being
strongly depleted according to BHAC models,
log(Li/Li$_{\rm init}) = -2.1^{+0.7}_{-1.5}$\,dex (Cond), and the
late-type secondary is likely too cool to possess monatomic lithium.
In contrast, the Tucson models predict that DENIS~J2252$-$1730A should
have retained most of its lithium,
log(Li/Li$_{\rm init}) = -0.04^{+0.04}_{-0.06}$\,dex, and are only
consistent with full depletion at 2.0$\sigma$.  Five other objects in
four systems with masses below 70\,\Mjup\ (2MASS~J2132+1341AB,
2MASS~J0700+3157A, LHS~2397aB, 2MASS~J1728+3948B) are all consistent
at $\leq$1$\sigma$ with full lithium depletion according to BHAC
models. In contrast, Tucson models predict that all of these objects
could possess a substantial fraction (up to 100\%) of their initial
lithium within the 1$\sigma$ errors.  This can be seen visually in
Figure~\ref{fig:m-age} as the cluster of lithium nondetections around
60--70\,\Mjup\ and 2\,Gyr that mostly lie above the BHAC 99.9\%
depletion curve but that lie entirely below the Tucson 99.9\%
depletion curve.

We have noted this tension between predictions of Lyon and Tucson
models in our previous work on HD~130948BC
\citep{2009ApJ...692..729D}.  Our large sample of individual masses
below the substellar boundary suggests that the lithium is destroyed
via fusion at lower masses than the Tucson models predict but at
masses consistent with BHAC models.  The fact that the Tucson models
predict a higher mass limit for lithium fusion indicates that their
central temperatures are lower, which can be understood as a
consequence of their atmospheres being less opaque as discussed in
detail in Section~3.2 of SM08.  According to BHAC models, at field
ages of $\gtrsim$1\,Gyr the lithium depletion boundary is at a mass of
60\,\Mjup. Some of the most interesting systems for testing lithium
depletion are the young (400--800\,Myr) binaries Gl~569Bab,
HD~130948BC, and Gl~417BC that span masses of $\approx$50--70\,\Mjup\
but that do not yet have directly measured individual masses and/or
optical spectra.  Individual masses are attainable in the future with
absolute astrometry, either relative to their stellar hosts
(Gl~569Bab, HD~130948BC) or from wide-field \HST\ imaging (Gl~417BC).
In addition, resolved optical spectra from \HST/STIS would benefit the
entire sample, enabling stricter tests of models than is currently
possible with integrated-light spectra.

\subsection{Coevality Tests \& the Mass--Luminosity Relation in the L/T Transition \label{sec:coevality}}

In principle, every binary in our sample can be thought of as a
``mini-cluster'' of coeval, co-compositional objects that can be used
to test model isochrones \citep[e.g.,][]{2010ApJ...722..311L}.  The
most straightforward approach is to frame this as a test of coevality.
Given our directly measured individual masses and luminosities, we
derive from models an age posterior distribution for each component
($t_1$ and $t_2$).  We then examine the posterior distribution of the
difference in ages, and the extent to which it is different from zero
corresponds to a discrepancy in the mass--luminosity relation
predicted by model isochrones.  Because we use a Monte Carlo approach,
covariances due to parameters in common between components (distance
and total mass) are naturally accounted for in our analysis.

Figure~\ref{fig:mb-dt} displays the median and 1$\sigma$ intervals of
the posterior distributions for $\Delta\log(t) \equiv \log{t_2} -
\log{t_1}$.  To examine how this coevality criterion depends on the
underlying physical properties, we plot these values as a function of
the secondary component mass (i.e., component that is less luminous).
At the highest masses, where the binaries are composed of two
main-sequence stars that each have unconstrained age posteriors, the
resulting coevality values are all $\Delta\log{t} \approx
0.0\pm0.5$\,dex at 1$\sigma$.  In other words, both stars are
consistent with the full range of main sequence ages (a few hundred
Myr, depending on mass, to 10\,Gyr, the maximum age of models).

The more interesting coevality tests are for binaries composed of two
brown dwarfs.  All but two of these systems give consistent ages for
the primary and secondary at $\lesssim$1$\sigma$ across all models.
The exceptions are two binaries spanning the L/T transition that have
the lowest-mass secondaries of any objects in our sample.  According
to Lyon Cond models, SDSS~J0423$-$0414AB and SDSS~J1052+4422AB are
3.0$\sigma$ and 1.5$\sigma$ discrepant with coevality, respectively.
In contrast, the SM08 models predict a shallower mass--luminosity
relation in the L/T transition and thus yield coeval ages for both
systems (consistent at 0.5$\sigma$ and 0.2$\sigma$, respectively).
The key difference between Lyon models (Cond and Dusty) and SM08
hybrid models is that SM08 prescribes an ad hoc change from a cloudy
to a cloudless photosphere as temperature drops from 1400\,K to
1200\,K.  This results in luminosity dropping less quickly during and
immediately following the transition from cloudy to cloudless
atmospheres.

We originally discussed this discrepancy between SM08 and Lyon models
for SDSS~J1052+4422AB in \citet{2015ApJ...805...56D}.  The case of
SDSS~J0423$-$0414AB (L$6.5\pm1.5$ and T$2.0\pm0.5$) is very similar to
SDSS~J1052+4422AB (L$6.5\pm1.5$ and T$1.5\pm1.0$).  The primary masses
are the same within the errors ($51.6^{+2.3}_{-2.5}$\,\Mjup\ and
$51\pm3$\,\Mjup, respectively), but SDSS~J0423$-$0414AB has a lower
mass ratio ($0.62\pm0.04$ compared to $0.78\pm0.07$) and luminosity
ratio ($\Delta\log(\Lbol) = 0.31^{+0.09}_{-0.08}$\,dex compared to
$0.13\pm0.08$\,dex).  Thus, SDSS~J0423$-$0414AB has even more leverage
in our coevality test, and indeed it confirms our previous findings at
even higher significance.  The reason for the coevality test failure
in both cases is that the secondary is more luminous (or equivalently
the primary is less luminous) than predicted for their masses at a
single age in Lyon models.  According to the SM08 hybrid models,
luminosity does not fade as quickly during and immediately following
the transition from cloudy to cloud-free atmospheres.  So according to
SM08 models it is the secondary that is more luminous than expected
(rather than a suppressed luminosity of the primary) because the
secondary has already cooled through the L/T transition and lost its
clouds.

A new physical explanation for the L/T transition has been proposed by
\citet{2016ApJ...817L..19T}, where the primary driver is a
thermo-chemical instability in the carbon chemistry (CO/CH$_4$).  Our
measurement of the mass--luminosity relation through the L/T
transition will provide a key test of this new idea after it is
implemented in evolutionary models in the future.

\subsection{Discovery of Triple Systems \label{sec:triple}}

Our individual mass and luminosity measurements have revealed that
some objects in our sample are likely triple systems where the higher
order multiplicity is unresolved by current observations.  The most
extreme case is 2MASS~J0920+3517AB (L5.5+L9), where the total mass of
$187\pm11$\,\Mjup\ is high enough that it suggests the presence of an
unresolved massive component.  Our individual masses reveal that
indeed the less luminous (by $0.06\pm0.04$\,dex) component
2MASS~J0920+3517B is much more massive ($116^{+7}_{-8}$\,\Mjup) than
the more luminous 2MASS~J0920+3517A ($71\pm5$\,\Mjup).  In fact, the
anomalous mass of 2MASS~J0920+3517B would be obvious on its own, given
that its spectral type is L$9.0\pm1.5$, and that models cannot
reproduce its luminosity at such a high mass
(Figure~\ref{fig:triple-ml}).  2MASS~J0920+3517B can plausibly be
explained by a simplistic model in which it is composed of two
equal-luminosity components with equal masses of
$58^{+3}_{-4}$\,\Mjup.  The unresolved pair must be rather tight, as
we have never observed an elongated PSF for the secondary relative to
the primary, and the astrometric residuals about the resolved orbit
fit have an rms of 0.5\,mas for the better half (13 out of 25) of our
measurements.  This implies an inner orbit that is much smaller than
the observed outer orbit ($a = 68.15\pm0.23$\,mas, $2.11\pm0.04$\,AU).
We discuss this system in more detail in Appendix~\ref{sec:2M0920+35}.

Unlike 2MASS~J0920+3517AB, the total mass of the L3+L6.5 binary
2MASS~J0700+3157AB ($141^{+4}_{-5}$\,\Mjup) is not unusually high.
Indeed, our total-mass model analysis readily finds a self-consistent
way to apportion the mass of the two components
($76.8^{+1.5}_{-1.3}$\,\Mjup\ and $66^{+4}_{-3}$\,\Mjup, Cond)
according to their quite different luminosities
($\Delta{\log(\Lbol/\Lsun)} = 0.50\pm0.06$\,dex) at an age of
$2.1^{+0.4}_{-0.6}$\,Gyr.  However, our directly measured masses are
$68.0\pm2.6$\,\Mjup\ and $73.3^{+2.9}_{-3.0}$\,\Mjup.  The less
luminous component is again the more massive one, implying that
2MASS~J0700+3157B is in fact an unresolved binary.  The model-derived
ages for 2MASS~J0700+3157A are $0.76^{+0.09}_{-0.14}$\,Gyr (SM08) and
$1.01^{+0.13}_{-0.19}$\,Gyr (Cond).  If we assume that
2MASS~J0700+3157B is composed of equal-luminosity components with
equal masses of $36.7^{+1.4}_{-1.5}$\,\Mjup, their model-derived ages
are somewhat inconsistent with the primary's
($1.15^{+0.11}_{-0.12}$\,Gyr and $0.83^{+0.08}_{-0.09}$\,Gyr,
respectively), suggesting that this simplistic scenario is not likely.
As with 2MASS~J0920+3517AB, we have never observed evidence for PSF
elongation of 2MASS~J0700+3157B in our Keck LGS AO imaging, and the
rms about the relative orbit fit is 0.22\,mas for the better half (11
out of 22) of our measurements.  Thus, the inner orbit in this system
is likely very small relative to the outer orbit
($a = 377^{+5}_{-6}$\,mas, $4.25\pm0.08$\,AU).  We discuss this system
in more detail in Appendix~\ref{sec:2M0700+31}.

We have identified a third candidate triple system.  The total mass of
LP~415-20AB ($248^{+26}_{-29}$\,\Mjup) is very high for such late
spectral type components (M6+M8).  Our individual masses show that
this is because the primary is very massive
($156^{+17}_{-18}$\,\Mjup), while the secondary's mass
($92^{+16}_{-18}$\,\Mjup) is consistent with its luminosity according
to BHAC models.  Given the relatively large individual mass
uncertainties, it is unclear whether the putative unresolved component
of LP~415-20A is another low-mass star or if it is a brown dwarf.  It
is less likely, but still possible, that the apparent discrepancy here
is due to the rather large observational uncertainties.  An improved
parallax for this system, which is also notable as a possible member
of the Hyades, will help clarify the situation as we discuss in more
detail in Appendix~\ref{sec:LP415-20}.

2MASS~J0700+3157 and 2MASS~J0920+3517 bring the tally of high-order
multiples composed entirely of likely brown dwarfs to five, and these
are the first such triple systems with directly measured masses,
semimajor axes, and eccentricities.  The first triple brown dwarf
identified was DENIS~J020529.0$-$115925 \citep{2005AJ....129..511B},
where the primary (L5) is orbited by a L/T transition binary (L8+T0).
\citet{2013ApJ...778...36R} discovered the triple T~dwarf system
2MASS~J08381155+1511155, where the brightest component (T3) is orbited
by two slightly fainter components (T3+T4.5).
\citet{2016ApJ...818L..12S} discovered that the primary in the young
system VHS~J125601.92$-$125723.9 \citep{2015ApJ...804...96G} is
actually a binary; making this the only one of these hierarchical
triples in which the inner pair contains the most massive components.
(Note that the distance of VHS~J1256$-$1257 is not secure, and the
more massive inner pair could be composed of low-mass stars and not
brown dwarfs.) The hierarchical mass ordering of the
DENIS~J0205$-$1159 and 2MASS~J0838+1511 systems appear to be similar
to 2MASS~J0700+3157 and 2MASS~J0920+3517, but the relative sizes of
inner and outer orbits may be different.  The inner and outer pairs of
DENIS~J0205$-$1159 have projected separations of 1.3\,AU and 7\,AU,
respectively, and 2MASS~J0838+1511 has projected separations of
2.5\,AU and 27\,AU.  If the inner pairs of 2MASS~J0700+3157 and
2MASS~J0920+3517 were only a factor of 7--10$\times$ smaller than
orbit, then we would have either resolved them directly or possibly
seen perturbations in our relative astrometry.  Thus, it is likely
that the semimajor axis ratios of 2MASS~J0700+3157 and
2MASS~J0920+3517 are much smaller than for DENIS~J0205$-$1159 and
2MASS~J0838+1511.  Finally, we note that the eccentricity of the outer
orbit is very low for 2MASS~J0700+3157 ($0.017^{+0.005}_{-0.007}$) and
is also rather low for 2MASS~J0920+3517 ($0.180^{+0.006}_{-0.007}$).
This would seem to rule out a violent dynamical origin for these
triple systems.

\subsection{Effective Temperature Relations \label{sec:teff}}

Without directly measured radii for field brown dwarfs of known
spectral type, all previous work on spectral type--\Teff\ relations
has relied on assumptions about age combined with model radii
\citep[e.g.,][]{2004AJ....127.3516G, 2004AJ....127.2948V,
  2009ApJ...702..154S, 2015ApJ...810..158F}.  Our sample of visual
binaries with directly measured luminosities and precise model-derived
ages and radii allows us to examine spectral type--\Teff\ relations
without uncertainties in radius due to the unknown ages or masses.  We
consider the effective temperatures derived from BHAC and Cond models
as a single ``Lyon'' relation, using BHAC results when available.  We
used results from our individual-mass analysis when possible, relying
on our total-mass analysis for the three systems without astrometric
mass ratios (Gl~417BC, Gl~569Bab, and HD~130948BC). We excluded the
three likely unresolved binaries LP~415-20A, 2MASS~J0700+3157B, and
2MASS~J0920+3517B.  LHS~2397aB is also excluded because it lacks a
spectral type determination.  This results in a sample of 40 objects
with Lyon temperatures and 22 objects with SM08 temperatures.  (SM08
results are only for the coolest objects; our highest SM08
model-derived \Teff\ is $2090\pm50$\,K for 2MASS~J1017+1308A.)

Figure~\ref{fig:spt-teff} shows our model-derived effective
temperatures as a function of spectral type.  We calculated
second-order polynomial fits of \Teff\ as a function of spectral type,
weighting by the quadrature average of the +$1\sigma$ and $-1\sigma$
uncertainties in \Teff.  The coefficients of these fits are given in
Table~\ref{tbl:poly}.  The Lyon fit covers spectral types of M7--T5
and $\Teff\approx1100$--2800\,K, and the residuals have an rms scatter
of 90\,K.  The SM08 fit covers spectral types of L1.5--T5 and
$\Teff\approx1100$--2100\,K, and the residuals have an rms scatter of
80\,K.  The two fits agree reasonably well in the overlapping spectral
type range, with the main difference being that the SM08 fit gives
$\approx$100\,K cooler temperatures for L4--L7 dwarfs.  This seems to
be a reflection of the fact that the SM08 model-derived temperatures
for these objects are indeed systematically lower, due to SM08 radii
being $\approx$10\% larger than Cond radii at these temperatures, and
not simply a quirk of the polynomial fit.  This is expected from the
fact that SM08 models have clouds while Cond models do not
\citep[e.g., see discussion of cloudy versus clear models
by][]{2011ApJ...736...47B}. There are no significant outliers in the
SM08 fit, and in the Lyon fit only Gl~569Bb appears be be
$\gtrsim$1.5$\sigma$ discrepant.  It has a model-derived temperature
that is 310\,K cooler than expected given its M9 spectral type and
more consistent with L0.5--L1 types.  Gl~569Ba is also slightly
discrepant for its M8.5 spectral type (140\,K cooler than the fit).
Therefore, Gl~569Bb may appear discrepant because the system as whole
is a slight outlier in \Teff\ due to the shared distance uncertainty
in addition to the spectral type of Gl~569Bb perhaps being slightly
underestimated.

Figure~\ref{fig:spt-teff} also shows literature polynomial relations
from \citet{2004AJ....127.3516G} and \citet{2015ApJ...810..158F}.  Our
mass-calibrated relation broadly agrees with this past work, except
for the temperature of the L/T transition from
\citet{2004AJ....127.3516G} which is systematically high, likely due
to the presence of unrecognized binaries in that early work
\citep[e.g.,][]{2005ApJ...634L.177B, 2006ApJ...647.1393L}. The
relation of \citet{2015ApJ...810..158F} is systematically lower than
our Lyon relation by $\approx$90\,K, which appears to be due to the
fact that they used SM08 model radii.  Our SM08 relation agrees very
well with \citet{2015ApJ...810..158F}, except at the extreme late-type
end where we find a 150\,K warmer \Teff\ for spectral type T5.  This
seems to be due to their polynomial not capturing the near-plateau in
temperatures through the L/T transition that in our data extends to at
least a type of T5.  While our sample does not span exactly the same
range in spectral type, the scatter about our polynomial relation
(90\,K for M7--T5) is somewhat smaller than previous relations
spanning M6--T8 (124\,K for \citealp{2004AJ....127.3516G}; 113\,K for
\citet{2015ApJ...810..158F}).

Interestingly, all six T dwarfs in our sample, ranging from T1.5--T5,
have effective temperatures that are consistent with each other within
the uncertainties, i.e., $p(\chi^2)\approx0.5$.  The weighted average
and rms of the model-derived temperatures is $1200\pm60$\,K for Lyon
and $1190\pm60$\,K for SM08.  The batch of objects at slightly earlier
types are five L6.5--L8.5 dwarfs that also have internally consistent
temperatures of $1475\pm30$\,K (Lyon) and $1400\pm25$\,K (SM08).  In
comparison, the only object in our sample without a spectral type
determination is LHS~2397aB, which has model-derived temperatures of
$1520\pm40$\,K (Lyon) and $1440\pm40$\,K (SM08), consistent with the
warmer L6.5--L8.5 dwarfs.  This is the first significant sample of
objects with individually measured masses spanning the L/T transition.
Their model-derived radii combined with our luminosities imply that
the spectral type range of L6.5--T5 corresponds to a temperature range
of only 200--300\,K, with Lyon models favoring the slightly larger
temperature range.  The narrower \Teff\ range derived from the SM08
models is likely due to the fact that \Lbol\ does not drop as steeply
through the L/T transition, as discussed in the previous subsection.

Figure~\ref{fig:cmd-teff} shows our sample on the CMD with the data
points colored according to their model-derived temperatures.  As
expected, temperature correlates strongly with the location along the
CMD sequence, consistent with temperature being a primary driver of
the spectral energy distributions of ultracool dwarfs.  Of course, our
derived \Teff\ is directly correlated with the plotted absolute
magnitudes that are also used to derive \Lbol.  However, in the L/T
transition, objects have similar absolute magnitudes and yet they span
a relatively wide range of temperatures ($\approx$1500\,K to
$\approx$1100\,K), with objects bluer in $J-K$ having systematically
cooler model-derived temperatures.

\subsection{Age Distribution \label{sec:age-dist}}

The age distribution of the field population provides information
about the star formation history of the galaxy and is a key component
in studies of substellar mass function in the solar neighborhood
\citep[e.g.,][]{2010MNRAS.406.1885B, 2012ApJ...753..156K}.
Without directly measured fundamental properties, the age distribution
of ultracool dwarfs has previously only been constrained in a
statistical sense from kinematic or population synthesis studies
\citep[e.g.,][]{2004ApJS..155..191B, 2005ApJ...625..385A,
  2007ApJ...666.1205Z, 2009AJ....137....1F}.  However, precise age
information is potentially available for brown dwarfs because, unlike
stars, they are continuously changing their most easily observable
properties, namely luminosity and temperature.  The most precise
age-dating method within current capabilities is to combine mass and
luminosity, two properties that can be measured very accurately as we
have shown here and in our past work, and infer an age from
evolutionary models \citep[for a review of various brown dwarf
age-dating techniques see][]{2009IAUS..258..317B}.  Note that, as in
all stellar astrophysics, age determinations for brown dwarfs are
model dependent, so the accuracy of the derived ages will depend on
how well evolutionary models predict substellar luminosity evolution.
As we have found in our past work, this is not yet a solved problem,
with potential issues at the factor of $\approx$2 level
\citep[e.g.,][]{2009ApJ...692..729D, 2014ApJ...790..133D,
  2015ApJ...805...56D}.  However, given that our derived ages span
nearly two orders of magnitude ($\sim$250\,Myr to $\sim$10\,Gyr), such
inaccuracies should not have significant influence on the broad trends
in our sample.

Figure~\ref{fig:age-hist} shows the distribution of system ages for
the 10~binaries in our sample with at least one substellar component
and thereby a well-determined age. When possible we use SM08
model-derived ages from our total-mass analysis; we only use our
individual-mass analysis for the special cases of 2MASS~J0700+3157AB,
where the secondary is an unresolved binary, and LHS~2397aAB, where
the primary is main-sequence star.  We also tested ages derived from
Cond models and found that they tended to give slightly
($\approx$10\%) older ages.  We exclude two systems from our analysis
here because their presence in our sample is not independent of their
age.  HD~130948BC and Gl~569Bab were both discovered in targeted
surveys of young stars, and so it would not be correct to include them
in this sample of field objects that have different observational
selection effects with respect to age.  Unlike these two companion
systems, Gl~417BC was originally identified on its own in 2MASS and
was only later associated with the young star Gl~417 by
\citet{2001AJ....121.3235K}.  Likewise, the pre--main-sequence binary
LP~349-25AB is not included here because it only has a precise age
determination by virtue of its youth.

The median age of our sample of field brown dwarfs is 1.3\,Gyr
(2.3\,Gyr mean), and the age interval containing 90\% of the joint
posterior distribution of all 10~systems is 0.4--4.2\,Gyr.  This age
distribution is broadly consistent with or somewhat younger than
previous statistical age distributions.  From a kinematic analysis of
21 L and T dwarfs, \citet{2007ApJ...666.1205Z} found a statistical age
of $1.2^{+1.1}_{-0.7}$\,Gyr, in good agreement with our age
distribution.  (Their sample includes just one of our age-dated
systems, 2MASS~J1728+3948AB.)  In contrast,
\citet{2009AJ....137....1F} found an older statistical age range from
a kinematic analysis of proper motions for a much larger sample of 184
L0--T8 dwarfs, ranging from 3--8\,Gyr for the whole sample or
2--4\,Gyr excluding high tangential velocity ($>$100\,\kms) objects.
(None of our systems have such high tangential velocities.)  The
\citet{2009AJ....137....1F} results are thus systematically older than
but marginally consistent with our age distribution.

\citet{2005ApJ...625..385A} found that brown dwarfs with spectral
types ranging from L5 to early-T are expected to have a mean age of
3\,Gyr, given a uniform prior on age from 0--10\,Gyr and a nominal
mass function of $\Psi(m) \propto m^{-0.8}$.
The fact that we find a younger mean age could imply that the solar
neighborhood comprises objects with a slightly younger age
distribution.  To investigate this possible discrepancy we performed a
population synthesis simulation using a power-law mass function that
is consistent with recent work
\citep[$\Psi(m) \propto m^{0.5}$;][]{2012ApJ...753..156K,
  2013MNRAS.433..457B} for a range in mass of 30--70\,\Mjup, i.e., the
lowest to highest masses covered by our age-dated sample.  Rather than
assume a constant input age distribution, we adopted the Besan\c{c}on
model for the solar neighborhood that assumes a constant star
formation rate and accounts for Galactic dynamics in a self-consistent
way \citep{2003A&A...409..523R}.\footnote{To be clear about the
  terminology we are using, ``star formation rate'' refers to the
  number of stars and brown dwarfs that were formed in the Galactic
  disk as a function of time.  ``Age distribution'' refers to the
  local, present day number of stars and brown dwarfs as a function of
  age.}  Because the Sun lies near the Galactic midplane, dynamical
heating skews the age distribution toward younger ages as older stars
are preferentially scattered away from the midplane over time.  In our
simulation we assume that ages are distributed uniformly among each of
the Besan\c{c}on model age bins, where each bin contains a fraction of
the total population as follows: 20.0\% for 0.15--1\,Gyr, 16.1\% for
1--2\,Gyr, 11.9\% for 2--3\,Gyr, 16.6\% for 3--5\,Gyr, 12.9\% for
5--7\,Gyr, and 14.4\% for 7--10\,Gyr.  We drew random ages according
to this distribution, restricting our population synthesis simulation
to 0.3--10\,Gyr in order to accurately represent our observed sample.

After assigning masses and ages to each simulated object, we
interpolated \Teff\ from the SM08 models, keeping only simulated
objects that fall within the temperature range of our sample
(1100--2100\,K).  As a check, the fractional breakdown in spectral
types between $\leq$L6, L6.5--L8.5, and L9--T5 dwarfs was
0.22/0.36/0.42, very similar to the actual breakdown of objects in our
age-dated sample (0.28/0.33/0.39).  Also note that by performing our
simulation with the same evolutionary models used to derive ages for
our sample, the two age distributions are self-consistent by
construction.  In other words, the simulation output is directly
comparable to our age distribution, even if the models were to have
large systematic uncertainties in the absolute ages.

The age distribution resulting from our simple population synthesis
simulation is shown in Figure~\ref{fig:age-hist}. It has a median of
1.7\,Gyr, a mean of 2.3\,Gyr, and the interval 0.3--5.0\,Gyr contains
90\% of the distribution.  The general shape of the distribution is
quite similar to our observations, piling up at younger ages and
tailing off at older ages.  We checked if the input mass function
might change this results, but using $\alpha = 0$ or $-1$ instead of
$-0.5$ only changed the final median age by $\approx$5\%.
The simulation predicts that we would have found $\approx$1 old system
($>$5\,Gyr) in our sample of 10 binaries, and the fact that none of
our sample is definitively that old is the main cause of the slight
discrepancy between the population synthesis and our observations.
Overall however, our age distribution is remarkably consistent with
our input assumption of a constant star formation rate.

The simulated population accounts for the main selection effects in
our sample, namely a limited spectral type range and the fact that we
can only age-date brown dwarfs and not stars (i.e., objects with mass
$\lesssim$70\,\Mjup).  The spectral type selection should bias our
sample against old ages, and indeed the simulation indicates that we
are the most complete at young ages.
There is one remaining selection effect that is difficult to quantify,
the fact that the youngest binaries at a given spectral type will have
the lowest masses, and lower system masses correspond to longer
orbital periods for a given semimajor axis range.  Our dynamical mass
sample relies on robust orbit determinations and thus suffers
incompleteness at long periods (and thereby low masses and young ages)
because there has been insufficient time to constrain the longest
period orbits. (See Section~\ref{sec:lowest} for a detailed discussion
of this selection effect.)  The fact that our sample is biased in this
way against young ages, and yet we still find an age distribution
skewed slightly younger than our simulation suggests that this
selection effect has only a marginal influence.

Overall, our population synthesis simulation demonstrates that our
observed age distribution is consistent with a constant star formation
rate in the Galactic disk.  We find that the age distribution of
M7--T5 dwarfs in the solar neighborhood is somewhat younger than in
previous work, but we determine that this can be explained by
accounting for dynamical heating in the disk that boosts the scale
height of older stars, removing them from the local volume.  As a
test, we performed a second simulation with uniform age input (i.e.,
constant star formation rate without dynamical heating) and this
skewed the output median age older by almost a factor of two.
In the future it will be possible to extend to older ages this
empirical determination of the age distribution in the solar
neighborhood with dynamical mass measurements for even cooler brown
dwarfs \citep[e.g.,][]{2015ApJ...803..102D}.  

\subsection{Eccentricity Distribution \label{sec:ecc}}

We have previously examined the eccentricities of ultracool binaries
and the implications for formation models in
\citet{2011ApJ...733..122D}.  The sample we used previously included
eleven visual binaries, four spectroscopic binaries, and one eclipsing
binary, and we had to contend with potential selection effects in this
sample due to discovery bias (eccentric visual binaries are easier to
discover with poor angular resolution), our own bias in selecting
which binaries to monitor, and whether eccentricity might make some
binaries more difficult to yield orbit determinations.  Our new sample
is not only much larger (ten new orbits), but we are also in a better
position to address issues regarding observer and orbit-fitting
biases.  This is because we present here our entire monitoring sample
with orbit fits for all binaries.  The only selection is therefore
based on the quality of the resulting fits, and this allows us to
robustly quantify the completeness of our sample.

As described in detail in Section~\ref{sec:sample}, our input sample
of 31~binaries is, to the best of our knowledge, complete for binaries
with projected separations at discovery of $\leq$6\,AU, given our
other non-orbit related selection criteria (spectral types M6.5--T5.5,
$d\leq40$\,pc, observable from Maunakea with Keck LGS AO).
Figure~\ref{fig:sep-p} shows all orbit fitting results for our sample,
including binaries with marginally or poorly determined orbits, as
well as visual binary orbits from the literature.  As expected, the
binaries with the widest separations at discovery have turned out to
have longer orbital periods.  Because we are hampered by the available
observational time baseline for longer period binaries, many of these
do not have well determined orbits.  Only one binary with $P>30$\,yr
has a well determined orbit (2MASS~J1728+3948AB); all the rest are
marginal or poor.  The only binary at $P<30$\,yr that does not have a
well determined orbit is SDSS~J0926+5847AB ($e = 0.00$--0.58 at
2$\sigma$).  This is a pathological case where the orbit is very close
to being viewed edge on, and the phase coverage from our 3-year \HST\
program does not allow us to uniquely determine the orbit.  It is
likely that if SDSS~J0926+5847AB were observable with Keck LGS AO,
then we would have better phase coverage over a longer time baseline
that would enable an orbit fit.
Therefore, we conclude that for periods $<$30\,yr, our sample of orbit
determinations is effectively complete, including 22 binaries from our
sample and 3 binaries from the literature.

A complete sample can still be biased, with the key effect here being
``discovery bias,'' which is explored in detail in Section~3.1.1 of
\citet{2011ApJ...733..122D}. Briefly, every imaging survey has an
inner working angle (IWA) inside of which binary companions are not
detectable.  Given a binary with semimajor axis $a$, the projected
separation will sometimes be inside of this limit, depending on the
inclination, eccentricity, and epoch of observation.  When
${\rm IWA} \ll a$, the probability of detecting any binary approaches
unity.  When ${\rm IWA} > a$, circular binaries are completely
undetectable while more eccentric binaries are more favorable to
detect because the apoastron distance is $a\times(1+e)$ (see Figure~1
of \citealp{2011ApJ...733..122D}).  By excluding binaries discovered
very near the IWA of a survey we can minimize the impact of this
discovery bias.  Figure~\ref{fig:p-iwa} shows the ratio of IWA/$a$ for
binaries in our sample.  The highest IWA/$a$ ratios are 1.58 and 1.24
for 2MASS~J1047+4026AB and LP~415-20AB, respectively, and indeed both
of these binaries are quite eccentric ($0.7485\pm0.0013$ and
$0.706^{+0.011}_{-0.012}$).  The next highest are IWA/$a = 0.88$ for
2MASS~J0920+3517AB ($e = 0.180^{+0.006}_{-0.007}$) and IWA/$a = 0.69$
for LP~349-25AB ($e = 0.0468^{+0.0019}_{-0.0018}$).  Based on the
simulations of \citet{2011ApJ...733..122D}, we adopt a cutoff of
IWA/$a < 0.75$ to mitigate discovery bias in our sample, and this
includes 2MASS~J1047+4026AB, LP~415-20AB, and the more marginal case
of 2MASS~J0920+3517AB.  Eccentricities for our entire sample, along
with the three other published visual binary orbits, are shown in
Figure~\ref{fig:a-e} as a function of semimajor axis.

After making the cuts described above, the final complete, de-biased
sample comprises 22 ultracool visual binaries.  In addition, we
examine published orbits for various unresolved binaries: three
astrometric orbits \citep{2015A&A...577A..15S, 2015A&A...579A..61S,
  2016AJ....151...57K}, three spectroscopic orbits
\citep{1999AJ....118.2460B, 2010A&A...521A..24J, 2016ApJ...827...25B},
and one double-lined eclipsing orbit for a very young brown dwarf
binary in Orion \citep{2006Natur.440..311S}.  Figure~\ref{fig:p-e}
shows all of these results, spanning more than three orders of
magnitude in period.  The unresolved binaries all have modest
eccentricities (0.2--0.6), whereas our sample has a number of very
eccentric ($>$0.7) and nearly circular (0.0--0.1) systems.  The lack
of low eccentricities ($<$0.2) in the seven unresolved binaries is the
most striking, as nearly half (10 of 22) of the visual binaries have
$e<0.2$.  More short-period binaries are needed to determine the
significance of this apparent difference, given the current small
sample with short periods.

Figure~\ref{fig:ecc-hist} shows the eccentricity distribution of our
de-biased visual binary sample. The abundance of low eccentricity
orbits is the most clear result, reinforcing the conclusion of
\citet{2011ApJ...733..122D} that very-low mass binaries are very
inconsistent with the high eccentricity orbits found in simulations
from \citet{2009MNRAS.392..413S} but consistent with the more modest
eccentricities predicted from the simulations of
\citet{2009MNRAS.392..590B}.  In fact, our larger sample here is less
consistent with the \citet{2009MNRAS.392..590B} results that do not
predict as many nearly circular orbits.  For both sets of simulations,
it is possible that the influence of gas over timescales longer than
the simulated durations would damp eccentricities further and bring
the simulations into better agreement with our observations.

Compared to the extensive orbital information available for solar-type
binaries spanning $\approx$6 orders of magnitude in period
\citep[e.g.,][]{1991A&A...248..485D, 2010ApJS..190....1R}, our visual
binary sample covers a relatively narrow range of periods ($\lesssim$1
order of magnitude).  However, one striking feature of the recent
large \citet{2010ApJS..190....1R} compilation of solar-type multiple
systems (127~eccentricities) from is that for a wide range of periods
($\sim$10$^2$--10$^6$\,d), there are no binary orbits with $e<0.1$ and
only three orbits in triple or quadruple systems with $e<0.1$.  In
contrast, our de-biased visual binary sample has 5 out of 22 orbits
with $e<0.1$.  Two of these are orbits in triple systems, where one is
an ``inner'' orbit (i.e., the orbit of the tighter pair in the
hierarchical triple; HD~130948BC) and one is an ``outer'' outer
(2MASS~J0700+3157AB).  The other three are all seem to be binaries
with no other companions (LP~349-25AB, 2MASS~J1534$-$2952AB, and
2MASS~J2206$-$2047AB).  This discrepancy between the very low-mass and
and solar-type orbits samples was first noted in
\citet{2011ApJ...733..122D}, and it is even stronger now after roughly
doubling the sample size of visual binaries.  For example, the Fisher
exact test comparing the number of binaries above and below $e=0.1$ in
our sample to \citet{2010ApJS..190....1R} gives a $p$-value of 0.0019,
implying the two samples are inconsistent at 3.1$\sigma$.
This difference in low-$e$ orbits could be due to differences in
initial conditions (i.e., binary formation is different at very low
masses) and/or early evolution (i.e., different physical processes at
work after formation).  In the latter case, if dissipative gas disks
are longer lived for very low-mass objects than solar type stars, this
could potentially result in a higher fraction of nearly circular
orbits even if the initial conditions are the same for both samples.

\subsection{The Lowest Mass Brown Dwarfs in the Sample? \label{sec:lowest}}

As described in Section~\ref{sec:bad}, we have excluded some binaries
from our analysis because of their uncertain orbits.  This is largely
due to poor orbital coverage given their apparently long orbital
periods relative to our observational time baseline.  Since our sample
was selected to have the smallest semimajor axes possible, some of
these binaries are likely to be the lowest mass systems in our sample.
Indeed, if we take our orbit determinations at face value, then
2MASS~J0850+1057AB, SDSS~J1021$-$0304AB, SDSS~J1534+1615AB, and
SDSS~J2052$-$1609AB all have lower total masses than the lowest mass
object used in our analysis (SDSS~J0423$-$0414AB,
$\Mtot = 83\pm3$\,\Mjup).

2MASS~J0850+1057AB (L6.5+L8.5) has a total mass of only
$54\pm8$\,\Mjup, and thus perhaps component masses of 20--30\,\Mjup.
Given the component luminosities, this would correspond to an age of
150--400\,Myr according to both Lyon and SM08 models.
The 2MASS~J0850+1057 system has previously been suggested to be young
based on having an M~dwarf companion (NLTT~20346; 4$\farcm$13 or
8000\,AU away) with a young X-ray activity age
\citep{2011AJ....141...71F} and based on 2MASS~J0850+1057A being
bright at $J$-band for its spectral type \citep{2011AJ....141...70B}.
However, the physical association of NLTT~20346 to the
2MASS~J0850+1057 system is questionable \citep[see Section~6.4
of][]{2012ApJS..201...19D}, and the discovery and characterization of
young late-L dwarfs indicates that they tend to be fainter at
$J$-band, not brighter \citep{2012ApJ...752...56F,
  2013AN....334...85L, 2016ApJ...833...96L}.  Our improved
proper motion measurement for the 2MASS~J0850+1057 system is
consistent with our previous findings, discrepant by 45\,\masyr\
(6.3$\sigma$) with the \citet{2011AJ....141...71F} proper motion for
NLTT~20346, implying that the wide pair is not likely to be physically
associated.
\citet{2011AJ....141...70B} also suggested that 2MASS~J0850+1057A
could be an unresolved binary based on its unusually bright $J$- and
$K$-band absolute magnitudes given its spectral type of L7, making
this a triple system of ultracool dwarfs.  Our distance
($31.8\pm0.6$\,pc instead of $38\pm6$\,pc) and spectral type of
L$6.5\pm1.0$ give absolute magnitudes comparable to other L5--L7
dwarfs (see Table~15 of \citealp{2012ApJS..201...19D}), reducing the
reason to suspect 2MASS~J0850+1057A is an unresolved binary.
Moreover, if our low total mass is accurate, then the component masses
(if a triple system) would be 10--15\,\Mjup, making the system even
younger.  This seems unlikely given that the integrated-light spectrum
lacks spectral similarities to other comparably young late-L dwarfs
\citep[e.g.,][]{2013ApJ...777L..20L, 2015ApJ...799..203G}.

SDSS~J1021$-$0304AB (T0+T5, $\Mtot = 52^{+6}_{-7}$\,\Mjup) and
SDSS~J1534+1615AB (T0+T5.5, $\Mtot = 46^{+6}_{-7}$\,\Mjup) have the
most precise masses of any of the systems with poorly determined
orbits.  For plausible mass ratios the implied component masses of
these binaries would be 20--30\,\Mjup.  With orbital period posteriors
of $86^{+13}_{-17}$\,yr and $58^{+39}_{-24}$\,yr, respectively, it is
unlikely that we will achieve significant improvement in their orbit
determinations in the near future. SDSS~J2052$-$1609AB (L8.5+T1.5,
$\Mtot = 69^{+14}_{-20}$\,\Mjup) has a very uncertain total mass but
the best chance of yielding a robust orbit in the near term, as its
derived period is $44^{+10}_{-14}$\,yr and our observations already
cover 20\% of this.


\section{Conclusions}

We present astrometric monitoring of a well-defined sample of 31
ultracool dwarf visual binaries (component spectral types M7--T5) from
our ground- and space-based observing programs spanning nearly a
decade in time baseline.  We determine robust orbits using resolved
astrometry by combining our Keck laser guide star adaptive optics
imaging and aperture masking interferometry observations, \HST\
imaging, and other published archival data.  Our unresolved infrared
imaging from CFHT/WIRCam allows us to measure precise parallactic
distances and constrain the photocenter orbital motion of our
binaries.  We perform a simultaneous MCMC analysis of both resolved
and unresolved astrometry in order to accurately determine posterior
distributions of all 13 astrometric parameters.  By careful
consideration of our posteriors, we propose orbit quality metrics that
separate robustly determined orbits (23 systems) from marginal and
poorly determined orbits that typically have long orbital periods
($\gtrsim$50\,yr) and thus lack sufficient observational time
baseline.  Our combined orbit and parallax results directly yield
dynamical total masses (median precision 5\%, as good as 1.1\%), and
combining our measured infrared flux ratios with the photocenter orbit
constraints enables determination of individual masses for 19 systems
(median precision 7\%, as good as 1.7\%).

We perform empirical tests of evolutionary models based solely on the
observed properties of our sample.  Also, we critically examine the
fundamental properties of objects from the bottom of the main sequence
through the L/T transition by deriving precise ages, radii, etc.\ from
the models.  For this analysis, we develop new relations for deriving
bolometric luminosity from $H$- and $K$-band absolute magnitudes, as
well as a method for inferring parameters from models given our mass
and luminosity constraints.  Unlike our past work that used simple
interpolation of models, our new approach is based on the Monte Carlo
technique of rejection sampling and allows for explicit definitions of
age priors as well as proper treatment of cases where an object of a
given mass might have the same luminosity at multiple times during its
evolution (e.g., a star on the main sequence or a deuterium-burning
brown dwarf).  We summarize our findings as follows.
\begin{enumerate}

\item We perform a novel empirical test of the mass limit for hydrogen
  fusion in stars by examining the maximum mass of the coolest objects
  in our sample.  Among all late-L and T~dwarfs, we find no objects
  more massive than 70\,\Mjup, implying a substellar boundary that is
  lower in mass than found in some previous work (75\,\Mjup\ has often
  been quoted) but reasonably consistent with the latest BHAC
  evolutionary tracks.  Among objects more massive than 70\,\Mjup\ the
  latest spectral types are L3--L5, implying that any object with a
  later spectral type than this is likely to be a brown dwarf
  independent of any model assumptions.

\item We examine the mass limit for lithium fusion in brown dwarfs.
  Our sample has 13 systems with at least one substellar component,
  and two of these have integrated-light \ion{Li}{1} absorption
  reported in the literature while seven have published nondetections
  of lithium.  After considering the potential for chemical depletion
  of monatomic lithium into molecular form (e.g., LiCl, LiOH), the
  observations are fully consistent with predictions of lithium
  depletion from Lyon evolutionary models \citep{2015A&A...577A..42B}.
  In contrast, Tucson models \citep{1997ApJ...491..856B} predict that
  lithium should be present at detectable levels in some of our more
  massive binary components (60--65\,\Mjup), but this is inconsistent
  with numerous \ion{Li}{1} nondetections.  The most massive objects
  in systems with lithium detected (SDSS~J0423$-$0414A and Gl~417B)
  have masses of $\approx$52\,\Mjup, consistent with (in fact, much
  lower than) the BHAC model-predicted lithium depletion boundary of
  60\,\Mjup\ at ages $\gtrsim$1\,Gyr.

\item We perform coevality tests, deriving ages for each binary
  component independently and checking for consistency to test
  evolutionary model isochrones.  All binaries are consistent with
  predictions from the various models we tested, except for two
  binaries spanning the L/T transition that have the lowest mass
  secondary components in our sample (SDSS~J0423$-$0414AB and
  SDSS~J1052+4422AB).  For these two systems, our observations show a
  smaller difference in luminosity between the binary components given
  their difference in mass than predicted by Lyon models (3.0$\sigma$
  and 1.5$\sigma$, respectively).  However, both binaries are
  consistent with SM08 hybrid models that assume a gradual loss of
  cloud opacity as objects cool from 1400\,K to 1200\,K, which has the
  side effect of making the mass--luminosity relation shallower
  through the L/T transition. This reinforces the results of
  \citet{2015ApJ...805...56D}.

\item We discover two new brown dwarf triple systems, where the inner
  pairs are identified via direct mass measurements but not spatially
  resolved.  2MASS~J0700+3157B and 2MASS~J0920+3517B are each the
  fainter component of a visual binary but are significantly more
  massive than the brighter component, implying that they are each
  unresolved, roughly equal-mass binaries. A third system LP~415-20AB
  has an unusually massive primary ($156^{+17}_{-18}$\,\Mjup) given
  its M6 spectral type, making this a third candidate triple in our
  sample. 2MASS~J0700+3157 and 2MASS~J0920+3517 are only the fourth
  and fifth known triple brown dwarf systems known, and they are the
  only such systems with directly measured masses, semimajor axes, and
  eccentricities, information that could be used to understand their
  origins.

\item We construct a spectral type--effective temperature relation for
  ultracool dwarfs that is based on model radii free of assumptions
  about mass or age.  We fit a second order polynomial to temperatures
  derived from Lyon models (a sample that spans M7--T5 and
  1100--2800\,K) as well as SM08 models (valid over a narrower range
  of L1.5--T5 and 1100--2100\,K).  The scatter about the fit is
  90\,K for Lyon models and 80\,K for SM08 models.  Examining
  individual subsets of objects, we find that all six T1.5--T5 dwarfs
  have consistent temperatures within their uncertainties (weighted
  average and rms of $1190\pm60$\,K using SM08 models), and the five
  L6.5--L8.5 dwarfs likewise have internally consistent temperatures
  ($1400\pm25$\,K using SM08 models).

\item The 10 systems in our sample with precisely determined ages (and
  provenances free of explicit age bias) allow us to directly measure
  the age distribution of brown dwarfs in the solar neighborhood for
  the first time.  The median age of our sample is 1.3\,Gyr, and the
  mean is 2.3\,Gyr.  Compared to previous statistical age constraints
  on the field population, our age distribution is either consistent
  \citep{2007ApJ...666.1205Z} or systematically younger
  \citep{2009AJ....137....1F}, and it is also younger than previous
  modeling of the solar neighborhood
  \citep[e.g.,][]{2005ApJ...625..385A}.  We performed a population
  synthesis simulation designed to mimic our sample in mass and
  spectral type range, using an input constant star formation history.
  We find good agreement with our observations only after accounting
  for dynamical heating in the Galactic disk, which preferentially
  removes older stars from the solar neighborhood.  Thus, our sample
  is consistent with a constant star formation rate in the Galactic
  disk, at least over the relatively young ages that we probe
  ($\lesssim$4\,Gyr).
  
\item We have revisited the eccentricity distribution of very low-mass
  binaries.  In general, our results reinforce the conclusions of
  \citet{2011ApJ...733..122D}, now with a larger sample and more
  robustly quantified completeness.  Perhaps the most striking feature
  of our visual binaries is the preponderance of low-eccentricity
  orbits. Out of 22 orbits in our de-biased sample, ten have $e<0.2$
  and five have $e<0.1$.  This is stark contrast to solar-type
  binaries and possibly also the relatively small sample of seven
  short-period, unresolved ultracool binaries with spectroscopic or
  astrometric orbits.  We speculate that low-eccentricity orbits could
  reflect more extensive tidal damping during the early evolution of
  ultracool binaries if their gas disks are longer lived than more
  massive stars.

\end{enumerate}

Our individual masses provide a direct window into substellar
astrophysics, empirically quantifying the mass limits for hydrogen
(and lithium) fusion.  The unique ability of binaries to serve as
``mini-clusters'' has provided the strongest tests of the
mass--luminosity relation in the poorly understood L/T transition.
Indeed, recent theoretical work by \citet{2016ApJ...817L..19T} has
called into question the primacy of clouds in driving changes in the
most readily observable properties (colors, magnitudes, and spectral
types) across the L/T transition.  Our results highlight how the
modeling of surface properties can have a significant influence on the
bulk evolution of brown dwarfs, as the only available models that
match our mass--luminosity relation data are those that treat the L/T
transition as a change in cloud properties.  Evolutionary models
adopting the alternative hypothesis that the L/T transition is driven
by a CO/CH$_4$ thermo-chemical instability are needed to test this
idea against our observations.

Our sample lays the groundwork for future tests of models using
objects of known mass.  For example, objects from our sample will have
a significant advantage for testing model atmospheres as they provide
a joint constraint on \Teff\ and \logg\ (via radius) given the
directly measured masses and luminosities. To enable such tests,
resolved spectroscopy of our binaries will need to be obtained, and
this is within reach of existing and future high angular resolution
facilities.  Moreover, resolved optical spectroscopy will be key in
strengthening our tests of lithium depletion, as our current tests
rely on integrated-light spectroscopy alone.  In the farther future,
dynamical masses for late-T and Y~dwarf binaries
\citep[e.g.,][]{2011ApJ...740..108L, 2015ApJ...803..102D} will open
the door to new tests of models.  Such measurements will reveal both
the older analogs of the $\gtrsim$30\,\Mjup\ brown dwarfs in our
current sample, as well as pushing to much lower masses, perhaps even
to the mass range shared by directly imaged planets (5--15\,\Mjup).

\clearpage

\acknowledgments

We are deeply grateful to Michael Ireland, especially for his early
contributions to this project that included training the author on
analysis of masking data from NIRC2, as well as very helpful comments
on this manuscript.
We also are very thankful to Andrew Mann for assistance in validating
our methods for determining bolometric luminosities.
This work would not have been possible without the excellent, long
running support of CFHT staff past and present, including Lo\"{\i}c
Albert, Todd Burdullis, Pascal Fouqu\'e, Nadine Manset, Karun
Thanjavur, Kanoa Withington, and all service observers who have
executed our program over the last nine years.
It is a pleasure to thank the Keck Observatory staff for assistance
with the Keck LGS AO observing, including Joel Aycock, Randy Campbell,
Al Conrad, Scott Dahm, Greg Doppmann, Heather Hershley, Marc Kassis,
Jim Lyke, Jason McIlroy, Carolyn Jordan, Gary Punawai, Julie Riviera,
Luca Rizzi, Terry Stickel, Hien Tran, and Cynthia Wilburn.
We thank
Isabelle Baraffe, Didier Saumon, and Mark Marley for providing
expanded model grids and helpful discussions about substellar
evolution;
and the anonymous referee for a thoughtful and timely review.
T.J.D.\ is indebted to many people for providing useful discussions
over the course of this project, including Katelyn Allers, Josh
Carter, Rio Jennings, Adam Kraus, and Aaron Rizzuto.
This work was supported by multiple NASA Keck PI Data Awards,
administered by the NASA Exoplanet Science Institute.
Support for our long-term \HST\ program (GO-11593, GO-12317, and
GO-12661) was provided by NASA through a grant from the Space
Telescope Science Institute, which is operated by the Association of
Universities for Research in Astronomy, Inc., under NASA contract
NAS~5-26555.
T.J.D. also acknowledges support from NASA Exoplanet Research Program
grant NNX15AD23G.
M.C.L.\ acknowledges support from NSF grants AST09-09222 and
AST15-18339.
Our research has employed 
NASA's Astrophysical Data System;
Posse East;
the SIMBAD database operated at CDS, Strasbourg, France;
the 2MASS data products;
and the SpeX Prism Spectral Libraries, maintained by Adam Burgasser at
\url{http://www.browndwarfs.org/spexprism}.
Finally, the authors wish to recognize and acknowledge the very
significant cultural role and reverence that the summit of Maunakea
has always had within the indigenous Hawaiian community.  We are most
fortunate to have the opportunity to conduct observations from this
mountain.

{\it Facilities:} \facility{HST (WFC3)} \facility{Keck:II (LGS AO, NIRC2)} \facility{CFHT (WIRCam)} \facility{IRTF (SpeX)}

\clearpage

\appendix

\section{Empirical Relations with Absolute Magnitude \label{app:absmag-rel}}

While there are many useful empirical relations in the literature
based on spectral type, many tight binary components do not have
directly measured spectra.  However, we do have accurate near-infrared
absolute magnitudes for all of our binaries from parallactic distances
and resolved photometry.  We will soon have many more late-M and L
dwarfs with distances from \Gaia\ that will not immediately have
spectra available.  Therefore, we have derived empirical relations of
various useful quantities with respect to absolute magnitude using the
existing sample of ultracool dwarfs with parallaxes.  Because $J$- and
$H$-band absolute magnitudes do not change monotonically through the
L/T transition \citep[e.g.,][]{2012ApJS..201...19D}, we primarily
derive these relations using $K$-band absolute magnitudes on the 2MASS
and MKO photometric systems.  We also examined some relations as a
function of $H$-band absolute magnitude for $M_H < 13.3$\,mag objects
for which the relations have lower scatter in luminosity and spectral
type and color changes monotonically.

\subsection{2MASS-to-MKO Photometric Conversion \label{app:phot-corr}}

In order to report complete photometry across both MKO and 2MASS
photometric systems for our sample binaries, we compute photometric
system conversions based solely on $K$-band absolute magnitude.  
We used our compilation of ultracool dwarf parallaxes available
online,\footnote{\url{http://www.as.utexas.edu/~tdupuy/plx}} selecting
only single objects with a published spectrum, a parallax error of
$<$15\%, and a ``null'' object flag, meaning that there is nothing
unusual about the object.  We use synthesized 2MASS/MKO photometric
conversions from \citet{2012ApJS..201...19D} for each of the 46
objects in each of $J$, $H$, and $K$ bands.  For $J$ and $H$ bands, we
found that a third-order polynomial provided a good fit to the data.
The $K$-band conversion undergoes a steep jump between
$M_K \approx 13$--14\,mag that cannot be well approximated by a single
polynomial.  Therefore, we computed a three-part piecewise linear fit
to the $K$-band conversion as a function of $M_K$.  In addition to
these relations as a function of $M_K$, we also computed an $H$-band
conversion as a function of $M_H$.  The polynomial coefficients are
given in Table~\ref{tbl:poly} and the fits are displayed in
Figure~\ref{fig:poly-corr}.  The typical rms about the fits ranges
from 0.003\,mag for $H$-band to 0.015\,mag for $J$ band.

\subsection{Broadband $K$ to Narrowband CFHT/WIRCam \Kn\
  Conversion \label{app:narr-corr}}

Determining mass ratios from photocenter motion requires knowing the
flux ratios in the imaging bandpass.  Our absolute astrometry for the
brightest targets observed with CFHT/WIRCam was obtained in a narrow
$K$-band filter, \Kn, centered on 2.122\,\micron\ with a bandwidth of
0.032\,\micron.  We do not have resolved photometry for these binaries
in this specialized filter, so we must synthesize it from our
broadband $K$ photometry in order to determine the flux ratios of our
binaries in our CFHT imaging.

We used the same sample of objects selected from our online parallax
compilation as described in the previous section.  We used the filter
curve provided by CFHT to synthesize \Kn\
photometry,\footnote{\url{http://www.cfht.hawaii.edu/Instruments/Filters/curves/cfh8304.dat}}
and we assume that the difference between this scan done in the lab
and the transmission at colder operational temperatures is negligible
for our purposes.  We computed $\Kn-K$ and $\Kn-\Ks$ colors for these
objects and found a smooth relation for objects with
$M_K < 13.1$\,mag, corresponding roughly to spectral types $<$L9.
(The smooth relations break down for later types as the \Kn\ filter
samples part of the $K$-band flux peak of T~dwarfs while their
broadband $K$ flux plummets.)  We fit a second-order polynomial as a
function of $M_K$ for 25 objects.  The coefficients are given in
Table~\ref{tbl:poly} and the fits are displayed in
Figure~\ref{fig:poly-corr}.  The rms about these fits were 0.025\,mag
for $K$ and 0.021\,mag for \Ks.

\subsection{Bolometric Luminosity \label{app:lbol-mag}}

Determining the component luminosities using our resolved photometry
is a critical part of our model tests.  We have used the large sample
of luminosities from \citet{2015ApJ...810..158F} to derive polynomial
relations between luminosity and absolute magnitude.  We used only
objects with parallactic distances, nominal ages $\gtrsim$0.5\,Gyr,
and normal properties (i.e., no subdwarfs, young objects, or objects
dubbed peculiar in their spectral type).  We supplemented the observed
magnitudes reported by \citet{2015ApJ...810..158F} by applying our
2MASS/MKO conversions from above.  For any object with either a 2MASS
or MKO measurement, we derived the complementary magnitude using the
$K$-band absolute magnitude.  For objects with both 2MASS and MKO
measurements, we used the more precise magnitude to derive the
complementary magnitude, supplanting the directly observed value. This
results in a sample of 126 normal field objects with luminosities and
$K$-band absolute magnitudes.  (Since our 2MASS/MKO relations are only
valid between $M_K = 9.1$--17.0\,mag, there are a few objects with only
2MASS or MKO photometry that changes the input sample slightly between
2MASS and MKO.)  A subset of 76 of these objects had magnitudes of
$M_H<13.3$\,mag, which we used in our $H$-band relations.

We fit luminosity as a function of $K$-band absolute magnitude with a
fifth-order polynomial, since there was clear structure in the
residuals using only a third order polynomial.  For the more limited
$H$-band magnitude range, a third-order polynomial was sufficient.  In
$K$-band, it was visually apparent that the scatter at brighter
magnitudes was less than at fainter magnitudes.  The fact that the
nominal errors on the magnitudes and luminosites are not significantly
larger for the fainter objects implies that this increased scatter is
due to either unquantified systematic errors in the luminosities of
the lower luminosity objects or real astrophysical scatter in the
relationship between $K$-band absolute magnitude and luminosity.
Regardless, we quantify this effect by separately computing the rms
scatter at the bright end ($M_K \leq 13.0$\,mag; rms of 0.04\,dex for
90 objects) and faint end (rms of 0.07\,dex for 36 objects).  The
coefficients for all relations are given in Table~\ref{tbl:poly}, and
the fits and residuals are displayed in Figure~\ref{fig:poly-lbol}.

\section{Discussion of Individual Objects \label{app:objects}}

\subsection{LP~349-25AB (M7+M8) \label{sec:LP349-25}}

In our previous work, we used a parallax of $75.8\pm1.6$\,mas from
\citet{2009AJ....137..402G} to compute luminosities and a total mass
that gave a Lyon Dusty model-derived age of $127^{+21}_{-17}$\,Myr
\citep{2010ApJ...721.1725D}.  Our new more accurate parallax of
$69.2\pm0.9$\,mas from CFHT results in somewhat higher masses and
luminosities but still the youngest model-derived age of any object in
our sample, $271^{+22}_{-29}$\,Myr (BHAC).  We now have directly
measured individual masses, and the ages derived from the component
masses and luminosities are coeval within the uncertainties
($\Delta{t}=50\pm50$\,Myr).  Our measured mass ratio
($q = 0.941^{+0.029}_{-0.030}$) is consistent within the uncertainties
with the model-derived mass ratio from our total-mass analysis
($0.88^{+0.03}_{-0.04}$).  Such a young age implies that LP~349-25AB
is a pair of pre--main-sequence stars with masses of $85\pm4$\,\Mjup\
and $80\pm3$\,\Mjup.  At a distance of only
$14.45^{+0.18}_{-0.19}$\,pc, this is the nearest pre--main-sequence
system containing very low-mass stars ($<$0.1\,\Msun), with the next
closest being LSPM~J1314+1320AB \citep[M7+M7;][]{2016ApJ...827...23D}
at 17.25\,pc.

According to the BHAC tracks, a 0.075\,\Msun\ object comparable to
LP~349-25B requires $\approx$130\,Myr to deplete 99.9\% of its initial
supply lithium, while 0.075\,\Msun\ object comparable to LP~349-25B
would need $\approx$110\,Myr.  Therefore, BHAC models predict both
components are fully depleted in lithium, in agreement with the
nondetection of \ion{Li}{1} absorption ($<$0.5\,\AA) reported by
\citet{2009ApJ...705.1416R}.  LP~349-25AB therefore provides a
critical test of the lithium depletion boundary in low-mass stars and
brown dwarfs at young ages that is used to age-date young clusters at
$\sim$20--200\,Myr \citep[e.g.,][]{1997ApJ...482..442B,
  2014MNRAS.438L..11B, 2014AJ....147..146K}.  It joins
LSPM~J1314+1320AB as the only other pre--main-sequence system where
objects of known mass empirically constrain the model-predicted
lithium depletion boundary \citep{2016ApJ...827...23D}.  Like
LSPM~J1314+1320AB, LP~349-25AB is a luminous radio source
\citep{2007ApJ...658..553P, 2009ApJ...700.1750O}, which makes it a
rare benchmark for characterizing radio emission in ultracool dwarfs.
High-precision VLBI astrometry may be possible for LP~349-25AB, like
\citet{2016ApJ...827...22F} recently showed for LSPM~J1314+1320AB,
which would refine the individual mass precision and enable even
stronger constraints on models.

LP~349-25AB has not previously been reported as having spectral
signatures of low surface gravity.  However, examination of our own
integrated light SpeX SXD spectrum ($R \approx 1200$) indicates that
it has a classification of M8~\intg\ on the
\citet{2013ApJ...772...79A} system.  Given the gap in known young
associations between the ages of the AB~Doradus ($\sim$150\,Myr) and
Ursa~Majoris ($\sim$400\,Myr) groups, LP~349-25AB (250--300\,Myr)
provides unique evidence that spectral signatures of low gravity in
the infrared can persist to at least 250\,Myr for late-M dwarfs.
There is only one other system of ultracool dwarfs with spectral
signatures of youth that have directly measured masses,
LSPM~J1314+1320AB \citep[$92.8\pm0.6$\,\Mjup\ and
$91.7\pm1.0$\,\Mjup;][]{2016ApJ...827...23D}.  Despite
LSPM~J1314+1320AB having more marginal signatures of low gravity, its
components actually have lower model-derived surface gravities
($\logg \simeq 4.8$\,dex) than the older and somewhat less massive
LP~349-25AB (5.1\,dex).  This implies that while the current gravity
classification system does indeed pick out young objects
($\lesssim$300\,Myr), there is a limit to the granularity by which
relative strengths of spectral features can be used to distinguish
surface gravity and thus age.

\subsection{LP~415-20AB (M6+M8) \label{sec:LP415-20}}

LP~415-20AB has the largest total mass in our sample
($248^{+26}_{-29}$\,\Mjup) and also the largest fractional uncertainty
in its mass ($^{+10}_{-12}$\%).  Given the component spectral types of
M$6.0\pm1.0$ and M$8.0\pm0.5$, and the integrated-light optical type
of M7.5 found by \citet{2000AJ....120.1085G}, a mass as large as
250\,\Mjup\ would be quite puzzling.  The implied component masses
would be $\approx$120--130\,\Mjup, and objects that massive are not
known at such late spectral types \citep[e.g., L~726-8AB and
L~789-6ABC are both M5.5 in integrated light;][]{2000A&A...364..217D}.
The size of our mass uncertainty suggests that LP~415-20AB could
simply be a $\gtrsim$2$\sigma$ outlier, but we also consider the
possibility that it is an unresolved triple system.  Our directly
measured mass ratio of $0.59^{+0.10}_{-0.12}$ is rather far from unity
despite a modest luminosity ratio of
$\Delta\log(\Lbol/\Lsun) = 0.25^{+0.04}_{-0.03}$\,dex.  This is also
consistent with higher order multiplicity, but the mass ratio
uncertainty is too large to make a definitive statement.

In our total-mass analysis, rejection sampling eschewed masses at the
high end of the input distribution, resulting in
$\Mtot = 201^{+4}_{-3}$\,\Mjup, 1.7$\sigma$ lower than our measurement
(Table~\ref{tbl:props-LP415-20}).  This is because the models do not
allow the higher masses to correspond to the luminosity we measure for
LP~415-20A (Figure~\ref{fig:triple-ml}).  The individual-mass analysis
confirms that this mass discrepancy is found to be due entirely to the
primary.  The mass of LP~415-20A after rejection sampling is
$109.8^{+3.1}_{-2.9}$\,\Mjup, 2.6$\sigma$ lower than the input
$156^{+17}_{-18}$\,\Mjup, whereas the input and output secondary
masses are essentially unchanged.  Therefore, both sets of analyses
suggest that if our unusually high mass is due to an unresolved
component of LP~415-20A, its mass is expected to be
$\approx$50\,\Mjup\ (assuming that the unresolved component
contributes negligibly to the luminosity).

As a test, we also performed our analysis for a hypothetical scenario
where LP~415-20A is an equal-mass, equal-flux binary with each
component having $M = 78^{+8}_{-9}$\,\Mjup\ and
$\log(\Lbol/\Lsun) = -3.32\pm0.03$\,dex.  This scenario is the
simplest to implement because we do not need to invoke a
mass--luminosity relation to apportion flux between the hypothetical
components.  We found reasonable agreement with models, though the
resulting LP~415-20A component masses after rejection sampling
preferred a somewhat higher mass of $92.3^{+2.2}_{-1.6}$\,\Mjup\
(Table~\ref{tbl:props-LP415-20PRI2}).  This is consistent with the
idea that if LP~415-20A is an unresolved binary, its components are
somewhat unequal in mass.
However, we reiterate that this discrepancy could also simply be a
statistical outlier, e.g., due to the somewhat uncertain parallax
($\sigma_{\pi}/\pi = 3$\%) that dominates the mass uncertainty.  As
one of the brighter targets in our sample, LP~415-20AB should be well
detected by \Gaia, which would resolve the current mass discrepancy.

In all cases above, the masses and luminosities of the components of
LP~415-20 are consistent with being coeval stars, most likely on the
main sequence but also consistent with ages of a few hundred Myr (the
BHAC model-derived age for LP~415-20B is $5.0^{+1.9}_{-4.7}$\,Gyr).
Thus, from our model-derived age distribution alone we do not rule out
the possibility that LP~415-20 could be a member of the Hyades
\citep[$750\pm150$\,Myr;][]{2015ApJ...807...58B}, as was originally
noted by \citet{1992ApJ...388L..23B}.
We can however use the kinematic information, both from our own
astrometry and the literature, to reassess its potential membership.
We have projected the space velocity of the Hyades,
$(U,V,W) = (42.3, -19.1, -1.5)$\,\kms\ \citep{2001A&A...367..111D},
into proper motion and radial velocity using our RA, Dec, and parallax
of LP~415-20.  We simulate the 0.3\,\kms\ velocity dispersion of the
Hyades and our parallax uncertainty in a Monte Carlo fashion, finding
$\mur = 132.2\pm4.4$\,\masyr, $\mud = -42.8\pm2.1$\,\masyr, and
$\vrad = 38.8\pm0.3$\,\kms.  In comparison, our measured absolute
proper motion is $126.1\pm0.7$\,\masyr\ and $-38.2\pm0.8$\,\masyr,
which is only $8.5\pm3.4$\,\masyr\ ($1.6\pm0.5$\,\kms) away from the
predicted proper motion assuming Hyades kinematics.  Thus, under these
assumptions, the motion of LP~415-20AB appears fairly inconsistent
(2.5$\sigma$) with membership.  The mean radial velocity of the two
components from \citet{2010ApJ...711.1087K} is $40.8\pm1.4$\,\kms,
only $2.0\pm1.4$\,\kms\ different from the predicted Hyades value.
The full 3-d distance in $UVW$ between the proper motion and RV
measurements and the predicted Hyades values is $2.7\pm1.1$\,\kms,
indicating that it is only a marginal candidate for membership in the
Hyades.  However, if the velocity dispersion of the Hyades were larger
than currently thought, e.g., 1\,\kms\ as is seen for other young
clusters, then LP~415-20AB would be a stronger candidate member.

\subsection{SDSS~J0423$-$0414AB (L6.5+T2) \label{sec:SD0423-04}}

SDSS~J0423$-$0414AB has the lowest total mass in our sample
($83\pm3$\,\Mjup), and it also has a well determined mass ratio from
our CFHT data ($0.62\pm0.04$), yielding individual masses of
$51.6^{+2.3}_{-2.5}$\,\Mjup\ and $31.8^{+1.5}_{-1.6}$\,\Mjup.  Our
total-mass analysis gives an SM08 model-derived mass ratio of
$0.65^{+0.05}_{-0.06}$, in good agreement with our measurement, and a
system age of $0.81^{+0.07}_{-0.09}$\,Gyr.  In contrast, Lyon Cond
models give a mass ratio of $0.80\pm0.05$ (2.8$\sigma$ discrepant with
our measurement) but a very similar age ($0.80^{+0.07}_{-0.08}$\,Gyr).
Our measured mass ratio of $0.62\pm0.04$ is much lower than previous
estimates, e.g., 0.8--1.0 from \citet{2005ApJ...634L.177B} and
$0.80\pm0.03$ from \citet{2010ApJ...722..311L}.

The case of SDSS~J0423$-$0414AB is similar to SDSS~J1052+4422AB
(L$6.5\pm1.5$+T$1.5\pm1.0$, SM08 age of $1.04^{+0.14}_{-0.15}$\,Gyr),
with a nearly identical primary mass but with mass and luminosity
ratios further from unity.  Like we find for SDSS~J0423$-$0414AB,
\citet{2015ApJ...805...56D} found a similar discrepancy between the
measured and the Lyon model-derived mass ratios for SDSS~J1052+4422AB
but agreement with SM08 models.  Using the individual mass and
luminosity of each component independently, in both cases the Lyon
models predict a younger age for the secondary as compared to the
primary, $\Delta{t} = -0.44\pm0.15$\,Gyr for SDSS~J0423$-$0414AB and
$-0.35^{+0.25}_{-0.22}$\,Gyr for SDSS~J1052+4422AB.  This is caused by
the secondary being more luminous than expected according to Lyon
models, resulting in a younger model-derived age.  In contrast, SM08
models give a higher luminosity for the secondary because they predict
that luminosity does not fade as quickly during and immediately
following the transition from cloudy to cloud-free atmospheres, which
they prescribe ad hoc to occur as temperature drops from 1400\,K to
1200\,K.  The components of SDSS~J0423$-$0414AB have SM08
model-derived temperatures straddling this transition,
$1430^{+30}_{-40}$\,K and $1200\pm40$\,K.

Both components of SDSS~J0423$-$0414AB have masses so low that they
should never significantly deplete their initial supply of lithium
\citep[$\lesssim$55\,\Mjup;][]{2000ARA&A..38..337C}.  However, the
monatomic lithium that is readily probed by the \ion{Li}{1} doublet at
6708\,\AA\ can be removed from the atmosphere chemically.  Models from
\citet{1999ApJ...519..793L} indicate that monatomic lithium ceases to
be the dominant lithium-bearing species below $\Teff \approx 1500$\,K
at pressures near the photosphere ($\sim$1\,bar).  Detailed spectral
synthesis modeling shows that \ion{Li}{1} absorption can in fact
persist to much lower temperatures
\citep[e.g.,][]{2000A&A...355..245P, 2001ApJ...556..357A}, and indeed
it has been detected in objects as cool as WISE~J1049$-$5319B
\citep[T0;][]{2014ApJ...790...90F, 2015A&A...581A..73L}.
\citet{2008ApJ...689.1295K} reported strong \ion{Li}{1} absorption (EW
= 11\,\AA) in the integrated-light spectrum of SDSS~J0423$-$0414AB.
This is consistent with our Lyon Cond model analysis that predicts
SDSS~J0423$-$0414A should still possess all or most of its initial
lithium, log(Li/Li$_{\rm init}) = -0.04^{+0.04}_{-0.20}$\,dex, along
with the previous detection of lithium in even cooler objects than
SDSS~J0423$-$0414A.

\subsection{2MASS~J0700+3157AB (L3+L6.5) \label{sec:2M0700+31}}

Our measured mass ratio of $q = 1.08\pm0.05$ indicates that
2MASS~J0700+3157B is in fact slightly more massive than
2MASS~J0700+3157A, despite the fact that it is $0.50\pm0.06$\,dex less
luminous.  Combined with our total mass of $141^{+4}_{-5}$\,\Mjup, we
find individual masses for 2MASS~J0700+3157A and 2MASS~J0700+3157B of
$68.0\pm2.6$\,\Mjup\ and $73.0^{+2.9}_{-3.0}$\,\Mjup, respectively.
These masses and luminosities result in very different SM08
model-derived ages, $0.76^{+0.09}_{-0.14}$\,Gyr for 2MASS~J0700+3157A
and $7.8^{+2.8}_{-5.2}$\,Gyr for 2MASS~J0700+3157B, or in other words,
a difference of $\Delta\log{t} = 1.00^{+0.29}_{-0.20}$\,dex.

The most natural explanation is that 2MASS~J0700+3157B is an
unresolved binary itself composed of two brown dwarfs.  We performed a
simple test, dividing the mass and luminosity of 2MASS~J0700+3157B in
two ($36.7\pm1.5$\,\Mjup, $-4.75\pm0.04$\,dex) and repeating our model
analysis.  In this hypothetical scenario, we found model-derived ages
that agreed better between 2MASS~J0700+3157A and 2MASS~J0700+3157B
($\Delta\log{t} = 0.18^{+0.09}_{-0.07}$\,dex for SM08,
$-0.09^{+0.09}_{-0.08}$\,dex for Cond).

Interestingly, the total mass of 2MASS~J0700+3157AB is not high enough
to betray its unresolved higher-order multiplicity without additional
mass-ratio information.  When we performed our total-mass analysis,
evolutionary models are able to self-consistently apportion the total
mass between the two components at their appropriate luminosities at a
single coeval age ($1.9^{+0.6}_{-0.7}$\,Gyr for SM08,
$2.1^{+0.4}_{-0.6}$\,Gyr for Cond).  But the model-derived mass ratio,
e.g., $0.86^{0.04}_{-0.03}$ from Cond, is 3.4$\sigma$ lower than we measure.
This case highlights the important role directly measured mass ratios
play in the substellar regime, where a given mass does not correspond
to a particular luminosity or spectral type unlike for main-sequence
stars.

\citet{2003PASP..115.1207T} reported a lack of \ion{Li}{1} absorption
in the integrated-light spectrum of 2MASS~J0700+3157AB (EW
$<0.3$\,\AA).  This is consistent with the prediction from Lyon models
that the $68.0\pm2.6$\,\Mjup\ primary should deplete 99.9\% of its
lithium within $\approx$200\,Myr.  Given the unknown nature of the
unresolved secondary's individual components, it is not clear whether
or not they should cause lithium absorption to the integrated-light
spectrum.  When present, lithium is routinely detected at EW $\sim
10$\,\AA\ in late-L dwarfs, so it seems plausible that if one of the
unresolved components of 2MASS~J0700+3157B were lithium-bearing it
would have been detectable even when diluted by up to a factor of
$\sim$20 in flux given typical detection limits of EW $\sim0.5$\,\AA.
The lack of a \ion{Li}{1} detection therefore hints at two possible
scenarios for the unresolved components. (1)~A lithium-depleted higher
mass object is paired with a lower mass object that either has lithium
but is too diluted in flux or has lithium in molecular form only
(LiCl, LiOH) because it is very cool.  (2)~2MASS~J0700+3157B is
composed of two cool, chemically depleted objects.  In the equal-flux,
equal-mass scenario, the absolute magnitudes of the components would
be $M_J = 15.06\pm0.04$\,mag and $M_K = 13.44\pm0.04$\,mag (MKO),
roughly consistent with the faint end of field objects with spectral
type L7--T0 \citep[see Table~15 of][]{2012ApJS..201...19D}.  This is
the same range of spectral type over which lithium becomes
undetectable in field objects, with later-type objects being more
likely to be chemically depleted in lithium.  Later-type objects would
also display some methane absorption, and this could be detectable in
the spectrum of 2MASS~J0700+3157B even if it were somewhat diluted by
an earlier type component.  No evidence of methane is seen in the
integrated light spectrum of 2MASS~J0700+3157 \citep[see Figure~13
of][]{2012ApJS..201...19D}, so resolved spectroscopy with AO or
spectroastrometry will be needed to further examine the possible
presence of an unresolved methane-bearing component.

2MASS~J0700+3157 was originally discovered by
\citet{2003PASP..115.1207T} as an unresolved single L3.5 dwarf
displaying parallax and proper motion during the course of an
astrometry program targeting cataclysmic variables.  Therefore, unlike
most other known ultracool dwarfs, the discovery of 2MASS~J0700+3157
did not involve a traditional magnitude-selected sample, which would
have been subject to Malmquist bias favoring the discovery of such
high-order multiples.  However, our own dynamical-mass sample is more
complete for the most massive systems as they orbit faster at a given
semimajor axis ($P \propto \Mtot^{-1/2}$), and our orbit
determinations are mostly limited by having sufficient time baseline.
Therefore, our dynamical-mass sample is slightly biased toward more
massive systems.  As example of the potential amplitude of such bias,
the period we measure for 2MASS~J0700+3157AB is $23.9\pm0.5$\,yr, but
if this system contained only $\frac{2}{3}$ of the mass the period
would have been 29.3\,yr.

\subsection{LHS~1901AB (M7+M7) \label{sec:LHS1901}}

In our previous work, we used a parallax of $77.8\pm3.0$\,mas from
\citet{2009AJ....137.4109L} to compute a total mass of
$203^{+26}_{-22}$\,\Mjup\ \citep{2010ApJ...721.1725D}.  Our new more
accurate parallax from CFHT ($76.4\pm1.1$\,mas) and updated orbit
determination results in a consistent but more precise mass of
$213^{+9}_{-10}$\,\Mjup.  We also now have directly measured
individual masses of $113\pm8$\,\Mjup\ and $99\pm7$\,\Mjup, based on
our mass ratio of $0.87^{+0.09}_{-0.11}$.  In our total-mass analysis,
the BHAC model-derived mass ratio is somewhat higher but consistent
within the errors ($0.97^{+0.03}_{-0.04}$), as expected given that the
luminosity ratio is near unity
($\Delta\log(\Lbol) = 0.04\pm0.03$\,dex).  Given our measured masses
and luminosities, LHS~1901AB is consistent with being a main-sequence
stellar binary, with a correspondingly unconstrained age (BHAC models
only give a 3$\sigma$ lower limit of $>$0.3\,Gyr from our total mass
analysis).  As noted by \citet{2010ApJ...721.1725D}, this system is
likely to be old given that it is lacking H$\alpha$ emission, which is
unusual for late-M dwarfs \citep[e.g.,][]{2008AJ....135..785W}.
Finally, we note that in our latest CFHT imaging from 2016 that we do
not use here, LHS~1901AB has become blended with a background source
that had previously been well separated, making these images unusable
for absolute astrometry.

\subsection{2MASS~J0746+2000AB (L0+L1.5) \label{sec:2M0746+20}}

The masses of 2MASS~J0746+2000A ($82.4^{+1.4}_{-1.5}$\,\Mjup) and
2MASS~J0746+2000B ($78.4\pm1.4$\,\Mjup) have the highest fractional
precision (2\%) of any individual masses in our sample.  Given our
determination of the substellar boundary of $\approx$70\,\Mjup, the
masses indicate that this binary is pair of stars and that neither is
a brown dwarf, in contrast to the analysis of
\citet{2004A&A...423..341B} but in agreement with
\citet{2006AJ....131..638G}.  Our measured mass ratio of
$0.952^{+0.026}_{-0.027}$ agrees well with the BHAC model-derived mass
ratio from our total mass analysis ($0.957^{+0.012}_{-0.010}$),
indicating that our masses and luminosities follow BHAC isochrones
well.  However, because the components of 2MASS~J0746+2000AB are both
main-sequence stars, the age is essentially unconstrained by models,
which only give 3$\sigma$ lower limits of $>$0.9\,Gyr for
2MASS~J0746+2000A and $>$1.0\,Gyr for 2MASS~J0746+2000B.

One of the components of 2MASS~J0746+2000AB is a radio emitter with a
rotation period of $2.0720\pm0.0018$\,hr \citep{2009ApJ...695..310B}.
In addition, \citet{2013ApJ...779..101H} report optical variability in
integrated light with a period of $3.32\pm0.15$\,hr.  Unfortunately,
it is not known which component corresponds to which period.
\citet{2012ApJ...750...79K} report $v\sin{i_{\star}} = 19\pm2$\,\kms\
and $33\pm3$\,\kms\ for the primary and secondary components,
respectively, suggesting that the secondary may be radio emitter.  We
consider this possibility (Case~I) and the alternative (Case~II) and
compare the implied stellar radii from combining the rotation periods
with projected rotational velocities to the radii predicted from
evolutionary models.  In Case~I, we compute minimum radii of the
primary and secondary of $R\sin{i_{\star}} = 0.50\pm0.06$\,\Rjup\ and
$0.55\pm0.05$\,\Rjup, respectively.  In Case~II we compute minimum
radii of the primary and secondary of
$R\sin{i_{\star}} = 0.32\pm0.03$\,\Rjup\ and $0.89\pm0.09$\,\Rjup,
respectively.

BHAC models predict radii of $0.959^{+0.007}_{-0.008}$\,\Rjup\ and
$0.914^{+0.007}_{-0.008}$\,\Rjup\ for the primary and secondary,
respectively.  Both possibilities that we consider are consistent with
these model radii, i.e., no minimum radii are larger than these
values, but they would correspond to very different projected
alignments of the stellar spin axes.  In Case~I, inclinations of
$i_{\star} = 32\pm4$\degree\ and $37\pm4$\degree\ for the primary and
secondary, respectively, would be needed to match the model radii.
This is consistent with co-alignment of stellar spin axes, as well as
rough alignment with the orbital plane
($i = 138.56^{+0.20}_{-0.21}$\degree\ means that the spin axes would
need to be $i_{\star} = 138.56$\degree\ or 41.44\degree\ to be
aligned).  In contrast, for Case~II projected stellar inclinations of
$i_{\star} = 19\pm2$\degree\ and $67\pm4$\degree\ for the primary and
secondary, respectively, would be needed to match the model radii.
This would rule out either of the stellar spin axes from being aligned
with each other or with the orbital plane.  We conclude that Case~I is
more likely because of the remarkable coincidence of the stellar spin
axes and orbital inclination all being within $\pm$5\degree\ of each
other.  However, we note that this is still only a probabilistic
argument and that the stellar spin axis angles are only seen in
projection, and we cannot rule out determine if they are aligned or
misaligned out of the plane of the sky.

\subsection{2MASS~J0920+3517AB (L5.5+L9) \label{sec:2M0920+35}}

The integrated-light spectral type of 2MASS~J0920+3517AB is L6.5 in
the optical \citep{2000AJ....120..447K} and T0p in the infrared
\citep{2006ApJ...637.1067B}.  We determined component types of
L$5.5\pm1.0$ and L$9.0\pm1.5$ from spectral deconvolution
\citep{2012ApJS..201...19D}.  It is therefore surprising that we find
a mass of $116^{+7}_{-8}$\,\Mjup\ for the fainter 2MASS~J0920+3517B,
much higher than the substellar boundary ($\approx$70\,\Mjup).
Moreover, the total system mass $187\pm11$\,\Mjup\ suggests that there
is more mass in this system than two brown dwarfs.  The most plausible
explanation is that 2MASS~J0920+3517B is itself an unresolved binary.

Using only the primary mass ($71\pm5$\,\Mjup) and luminosity, SM08
models give an essentially unconstrained age of $7^{+3}_{-5}$\,Gyr and
Cond models give a somewhat more tightly constrained age of
$3.1^{+1.5}_{-1.7}$\,Gyr.  If we arbitrarily divide 2MASS~J0920+3517B
into two equal-mass ($58^{+3}_{-4}$\,\Mjup), equal-luminosity
components, then we find ages from SM08 ($2.3^{+0.3}_{-0.4}$\,Gyr) and
Cond ($1.99^{+0.25}_{-0.37}$\,Gyr) that are in reasonable agreement
with the primary.  Therefore, the model-derived ages are consistent
with the hypothesis that 2MASS~J0920+3517B is an unresolved, nearly
equal-mass binary.

The nondetection of \ion{Li}{1} absorption in the integrated light
spectrum \citep[$<$0.5\,\AA;][]{2000AJ....120..447K} is consistent
with the fact that the primary mass ($71\pm5$\,\Mjup) is well above
the 55--60\,\Mjup\ lithium depletion boundary at field ages
(Section~\ref{sec:lithium}).  In principle, a lithium-bearing
component in 2MASS~J0920+3517B could still be detectable even when
diluted by the somewhat brighter 2MASS~J0920+3517A
($\Delta{F814W} = 0.30\pm0.10$\,mag).  In our hypothetical equal-mass
binary scenario for 2MASS~J0920+3517B, our interpolation of Cond
models predicts only modest lithium depletion of
log(Li/Li$_{\rm init}) = -0.0215^{+0.0019}_{-0.0018}$\,dex for the
components.  The case of 2MASS~J0920+3517B is therefore quite similar
to that of 2MASS~J0700+3157B, which we discussed in detail above.  We
note that while the unresolved components of 2MASS~J0920+3517B are
likely both substellar, the primary 2MASS~J0920+3517A could be a star
or brown dwarf within the measurement uncertainties.

\subsection{2MASS~J1017+1308AB (L1.5+L3) \label{sec:2M1017+13}}

2MASS~J1017+1308AB (L1.5+L3) has one of the larger fractional
uncertainties in total mass ($^{+9}_{-12}$\%) in our sample, owing to
a somewhat large parallax uncertainty (3.4\%).  Our directly measured
mass ratio ($0.92\pm0.08$) results in component masses of
$81^{+10}_{-11}$\,\Mjup\ and $75\pm7$\,\Mjup.  Our measured mass ratio
agrees well with the model-derived mass ratios from our total mass
analysis ($0.987^{+0.017}_{-0.015}$ from BHAC, and SM08 gives
essentially the same result).  This means that our individual masses
and luminosities agree with both model mass--luminosity relations,
although this system does not offer a very strong test of models given
the near-unity mass and luminosity ratios.  Because the components of
2MASS~J1017+1308AB are both likely main-sequence stars, the system age
is essentially unconstrained by models, which only give 3$\sigma$
lower limits of $>$0.6\,Gyr (BHAC) and $>$0.5\,Gyr (SM08) from our
total-mass analysis.

\subsection{2MASS~J1047+4026AB a.k.a.\ LP~213-68AB (M8+L0) \label{sec:2M1047+40}}

We measure a total mass of $178^{+11}_{-12}$\,\Mjup\ for
2MASS~J1047+4026AB and individual masses of $97^{+6}_{-7}$\,\Mjup\ and
$80\pm6$\,\Mjup.  Our measured mass ratio of $0.82\pm0.06$ agrees
somewhat lower than but consistent within the uncertainties of the
BHAC model-derived mass ratio from our total-mass analysis
($0.947\pm0.029$), indicating that our masses and luminosities are
consistent with BHAC isochrones.  However, because the components of
2MASS~J1047+4026AB are both main-sequence stars, the age is
essentially unconstrained by models, which only give a 3$\sigma$ lower
limits of $>$0.5\,Gyr from our total mass analysis.

The M8 dwarf 2MASSW~J1047138+402649 (a.k.a.\ LP~213-68) is a proper
motion companion to the M6.5 dwarf LP~213-67 (2MASSW~J1047126+402643,
14$\farcs$4 away) as described by \citet{2000MNRAS.311..385G} in their
paper presenting the discovery of both objects as ultracool dwarfs.
\citet{2003ApJ...587..407C} reported the discovery of several binaries
from their Gemini AO survey, but they report conflicting information
regarding this source.  \citet{2003ApJ...587..407C} use the 2MASS name
of the M6.5 dwarf LP~213-67 but refer to the object they observed as
an M8 dwarf.  We believe that they indeed observed the M8 dwarf
LP~213-68, given that we have observed more than a full orbital period
of this binary with Keck and the separation and PA reported by
\citet{2003ApJ...587..407C} agree very well with our orbit.  If
\citet{2003ApJ...587..407C} observed LP~213-67, then it would have to
be a binary that happened to have a separation and PA matching our
ephemeris for LP~213-68 at the epoch of their observations, which
would be a remarkable coincidence.  \citet{2008A&A...481..757B}
present Lick AO observations with very clear details indicating that
they observed LP~213-68 (despite referring to it by the 2MASS name of
LP~213-67 in their tables), and their reported separation and PA
($106\pm14$\,mas and $319.3\pm1.0$\degree) agree reasonably well with
our ephemeris for LP~213-68 at that epoch ($109\pm2$\,mas and
$307.3\pm1.9$\degree).  In contrast, \citet{2010ApJ...711.1087K}
report observations that do not appear to be consistent with our
ephemeris of LP~213-68.  They detected a binary at $32\pm2$\,mas on
2006~Jun~21~UT but nothing outside of 47\,mas on 2006~Nov~27~UT and
2007~Dec~2~UT.  Our ephemeris predicts separations and PAs on those
respective dates of: $87\pm5$\,mas and $292\pm3$\degree;
$101\pm3$\,mas and $302\pm2$\degree, and $117.5\pm1.0$\,mas and
$316.6\pm1.6$\degree.  It is possible that \citet{2010ApJ...711.1087K}
observed the M6.5 dwarf LP~213-67 instead of the M8 dwarf LP~213-68
because their tables use the 2MASS name and coordinates of LP~213-67.
We conclude that this is the most likely explanation for their
measurements.  Interestingly, this implies that the M6.5 companion is
also a binary, even tighter than LP~213-68, making this a hierarchical
quadruple system.  Unlike other known quadruples containing ultracool
dwarfs (e.g., Gl~337, 2MASS~J04414565+2301580) the LP~213-67/LP~213-68
system seems to be entirely composed of $>$M6 dwarfs.

\subsection{SDSS~J1052+4422AB (L6.5+T1.5) \label{sec:SD1052+44}}

The properties of SDSS~J1052+4422AB are discussed in detail in
\citet{2015ApJ...805...56D}.  Our new homogeneous analysis here
results in slightly revised properties for both mass and luminosity,
but the key results are unchanged.  Our measured mass ratio of
$0.78\pm0.07$ is further from unity than expected given the near-unity
luminosity ratio $\Delta\log(\Lbol) = 0.13\pm0.08$\,dex.  The mass
ratio derived from SM08 models ($0.82^{+0.09}_{-0.11}$, 0.3$\sigma$
different) provides a better match to our measurement than Cond models
($0.91\pm0.05$, 1.5$\sigma$ different).  Or put in terms of age, our
directly measured individual masses and luminosities yield ages that
agree within the errors from SM08 models
($\Delta\log{t} = -0.02\pm0.12$\,dex) but somewhat discrepant ages
from Cond models ($\Delta\log{t} = -0.15\pm0.10$\,dex).

\subsection{Gl~417BC (L4.5+L6) \label{sec:GL417B}}

The properties of Gl~417BC and corresponding astrophysical
interpretation are discussed in detail in \citet{2014ApJ...790..133D}.
Our new homogeneous analysis here results in slightly revised
properties for both mass and luminosity, but the key results are
unchanged.

Gl~417 is the only system in our sample for which a \Gaia\ DR1
parallax is available from the Tycho-Gaia Astrometric Solution
\citep{2016A&A...595A...4L}.  Unfortunately, this parallax
($43.86\pm0.34$\,mas) is 3.1$\sigma$ discrepant with the \Hipparcos\
value ($45.61\pm0.44$\,mas) used in our previous work and the analysis
presented here.  If we adopt the \Gaia-DR1 parallax, the total mass
would be higher ($111.6_{-3.1}^{+2.9}$\,\Mjup, compared to
$99.2_{-3.3}^{+3.0}$\,\Mjup), the component luminosities would be
0.03\,dex higher, and the model-derived ages would be correspondingly
older.  Using the \Hipparcos\ parallax, both BHAC and SM08 models give
an age of $490^{+30}_{-40}$\,Myr, but using the \Gaia-DR1 parallax the
BHAC and SM08 models would give ages of $620\pm40$\,Myr and
$600\pm40$\,Myr, respectively.  Such older ages are more consistent
with the gyrochronology-derived age of $740^{+150}_{-120}$\,Myr for
the primary star Gl~417A and would consequently reduce the
significance of the substellar ``luminosity problem'' for this
system.  The quality of \Gaia-DR1 parallaxes have only just begun to
be assessed by the community, and we defer judgment on the discrepancy
for this system until the Tycho-Gaia Astrometric Solution errors are
better understood.

\subsection{LHS~2397aAB (M8+L~dwarf) \label{sec:LHS2397a}}

We originally presented an analysis of this system based on the total
mass alone in \citet{2009ApJ...699..168D}, but we noted that the
photocenter orbit should be readily measurable given a long enough
time baseline.  Indeed, we detect the photocenter motion of
LHS~2397aAB at the highest significance of any binary in our sample
(37$\sigma$), allowing us to measure a mass ratio of
$0.706^{+0.027}_{-0.028}$.  Only the Cond models cover the mass and
luminosity of both the primary and secondary, and they predict a mass
ratio ($0.75^{+0.07}_{-0.04}$) that is in excellent agreement with our
much more precise measurement.  As expected, the primary is a star
($93\pm4$\,\Mjup), and the secondary is consistent with being
substellar ($66\pm4$\,\Mjup).  The Cond model-derived age of the
secondary is $2.6^{+0.6}_{-1.0}$\,Gyr, while the age of the primary is
essentially unconstrained but consistent with the secondary.  The SM08
models give a similar age for the secondary of
$2.8^{+2.1}_{-1.5}$\,Gyr.

Despite the fact that the components of LHS~2397aAB are different in
luminosity by $1.14\pm0.06$\,dex, further from unity than any other
binary in our sample, the fact that the components are a star and a
brown dwarf means that this system on its own does not actually
provide a strong test of the mass--luminosity relation predicted by
models.  Such a wide range of ages are allowed for the primary given
that it is on the main sequence that models are free to match the mass
and luminosity of the secondary exactly by fine-tuning the age.
Models predict that LHS~2397aB is strongly depleted in lithium,
log(Li/Li$_{\rm init}) = -1.8^{+0.7}_{-0.6}$\,dex, which is consistent
with the nondetection (EW $<0.5$\,\AA) in the integrated-light
spectrum from \citet{2009ApJ...705.1416R}.

\subsection{Kelu-1AB (L2+L4) \label{sec:KELU-1}}

Kelu-1AB is the only one of our sample binaries with a CFHT parallax
but without a measured mass ratio, owing to the relatively short time
baseline (2.26\,yr) of the available CFHT observations.  Moreover, our
parallax ($48.0\pm2.2$\,mas) disagrees somewhat with other comparably
precise published values of $53.6\pm2.0$\,mas
\citep{2002AJ....124.1170D} and $51.75\pm1.16$\,mas
\citep{2016AJ....152...24W}.  Using our parallax, the total mass is
$180^{+22}_{-26}$\,\Mjup.  In comparison, the
\citet{2002AJ....124.1170D} parallax gives $129^{+13}_{-16}$\,\Mjup\
and the \citet{2016AJ....152...24W} parallax gives
$144^{+9}_{-10}$\,\Mjup.  Given the lack of mass ratio information and
that the published parallaxes, including our own, span such a wide
range of possible masses, we exclude Kelu-1AB from our analysis.  The
fact that \ion{Li}{1} absorption is detected in integrated light
\citep[EW = 1.7\,\AA;][]{1999ApJ...519..802K} implies that at least
one component in the system should be $\lesssim$60\,\Mjup\ according
to BHAC models, which would potentially favor the lower system mass
(larger parallax).  \citet{2005ApJ...634..616L} noted that the
components of Kelu-1AB might straddle the lithium fusion boundary,
providing a uniquely constraining test of models.  However, the
current uncertainty in the dynamical mass is too large to distinguish
whether the components of Kelu-1AB are both stars or brown dwarfs, let
alone assess whether they are lithium bearing.

In unrefereed work, \citet{2008arXiv0811.0556S} proposed that Kelu-1AB
is a triple system.  They computed a dynamical mass with large errors
($185^{+118}_{-58}$\,\Mjup)\footnote{We quote the mass implied by
  their given period and semimajor axis, because their own quoted mass
  seems to assume a conversion factor of 1000\,\Mjup/\Msun, not the
  correct 1048\,\Mjup/\Msun.} and found an $H$-band feature in their
resolved spectrum of Kelu-1A that they claimed as evidence of an
unresolved T-dwarf companion.  This feature is strong enough that it
should appear in the integrated-light spectrum of the whole system
even after being diluted by Kelu-1B, but an inspection of the
published unresolved spectrum of Kelu-1AB shows no such feature.  We
therefore conclude that there is no strong evidence that Kelu-1AB is
actually a triple system.

\subsection{2MASS~J1404$-$3159AB (L9+T5) \label{sec:2M1404-31}}

2MASS~J1404$-$3159AB is a $J$-band flux-reversal binary
\citep{2008ApJ...685.1183L}, and using our photometry here we find
component spectral types of L$9.0\pm1.0$ and T$5.0\pm0.5$ by spectral
decomposition, making 2MASS~J1404$-$3159B one of the latest-type
objects in our sample.  We have a well determined total mass
($120^{+11}_{-13}$\,\Mjup) and mass ratio ($0.84\pm0.06$), resulting
in individual masses of $65\pm6$\,\Mjup\ and $55^{+6}_{-7}$\,\Mjup.
In our total-mass analysis, both sets of models give mass ratios in
good agreement with our measurement ($0.79^{+0.08}_{-0.11}$ from SM08
and $0.85^{+0.07}_{-0.06}$ from Cond).  The model-derived ages are
also consistent with SM08 giving $3.0^{+0.8}_{-1.3}$\,Gyr and Cond
giving $2.5^{+0.6}_{-0.9}$\,Gyr.

The only other $J$-band flux-reversal binary in our final mass sample
is SDSS~J1052+4422AB, and its mass ratio of $0.78\pm0.07$ is similar
to that of 2MASS~J1404$-$3159AB. (Note that SDSS~J1534+1615AB is a
$J$-band flux-reversal binary in our input sample, but it does not
have a secure orbit determination.)  However, in that case there were
discrepancies between models and observations due to the fact that
Lyon models, either Cond here or Dusty in \citet{2015ApJ...805...56D},
predicted a much steeper drop in luminosity given the change in mass.
The lack of such discrepancies for 2MASS~J1404$-$3159AB is because
despite having a similar mass ratio, its luminosity ratio is much
further from unity, $\Delta\log(\Lbol) = 0.35\pm0.09$\,dex compared to
$0.13\pm0.08$\,dex for SDSS~J1052+4422AB.  The SM08 model-derived
temperature for 2MASS~J1404$-$3159A ($1400^{+40}_{-50}$\,K) is
comparable to that of SDSS~J1052+4422A ($1366^{+25}_{-29}$\,K), while
2MASS~J1404$-$3159B ($1190\pm50$\,K) is slightly cooler than
SDSS~J1052+4422B ($1270\pm40$\,K).

2MASS~J1404$-$3159A is consistent with being massive enough that it
would be depleted in lithium,
log(Li/Li$_{\rm init} = -1.0^{+1.0}_{-0.5}$\,dex, although within
1$\sigma$ range includes the possibility that it has not depleted any
lithium.  In contrast, Cond models predict that 2MASS~J1404$-$3159B
retains almost all of its initial lithium,
log(Li/Li$_{\rm init} = -0.018^{+0.003}_{-0.006}$\,dex.  However, the
T5 spectral type of 2MASS~J1404$-$3159B suggests that no \ion{Li}{1}
will be observable in its spectrum because monatomic lithium will be
chemically depleted into LiCl and/or LiOH.  No optical spectroscopy is
yet available to test these model predictions of lithium depletion, or
lack thereof, for this system.

\subsection{HD~130948BC (L4+L4) \label{sec:HD130948B}}

The properties of HD~130948BC and corresponding astrophysical
interpretation are discussed in detail in \citet{2009ApJ...692..729D}
and updated in \citet{2014ApJ...790..133D}.  Our new homogeneous
analysis here results in slightly revised properties for both mass and
luminosity, but the key results are unchanged.  (HD~130948A does not
appear in the \Gaia-DR1 Tycho-Gaia Astrometric Solution catalog of
parallaxes.)

\subsection{Gl~569Bab (M8.5+M9) \label{sec:GL569B}}

In our previous work, we used Keck and \HST\ data to measure the orbit
of Gl~569Bab, finding a semimajor axis of $95.6^{+1.1}_{-1.0}$\,mas
\citep{2010ApJ...721.1725D} from which we computed a system mass of
$147^{+9}_{-8}$\,\Mjup.  We discussed why this may be discrepant with
other literature measurements, the smallest of which was
$90.4\pm0.7$\,mas from \citet{2006ApJ...644.1183S}.  Our new
measurement of $93.64\pm0.14$\,mas, based on a much more extensive
data set with better orbital phase coverage, is still inconsistent
(4.1$\sigma$) with the \citet{2006ApJ...644.1183S} value, and it is
also 2.0$\sigma$ smaller than our previously published value.  This
results in a somewhat smaller system mass of $138\pm7$\,\Mjup, which
is only 0.8$\sigma$ different from our previously published value,
because the error is dominated by the \Hipparcos\ parallax
($\sigma_{\pi}/\pi = 0.016$) not our semimajor axis measurement
($\sigma_a/a = 0.0016$).  (Gl~569A does not appear in the \Gaia-DR1
Tycho-Gaia Astrometric Solution catalog of parallaxes.)

The BHAC model-derived age of Gl~569Bab is $440^{+50}_{-60}$\,Myr, in
good agreement with our past work \citep[$460^{+70}_{-110}$\,Myr from
Dusty models;][]{2010ApJ...721.1725D}.  The model-derived mass ratio
is $q = 0.85\pm0.03$, or $1/q = 1.17\pm0.05$.  This is in good
agreement with the measured mass ratio of $1/q = 1.4\pm0.3$ based on a
homogeneous analysis of radial velocities by
\citet{2010ApJ...711.1087K}.  The model-derived individual masses are
$75\pm4$\,\Mjup\ and $64\pm4$\,\Mjup.  Using our total mass and the
mass ratio from \citet{2010ApJ...711.1087K} results in individual
masses of $80^{+9}_{-8}$\,\Mjup\ and $58^{+7}_{-9}$\,\Mjup.  In both
cases, given our empirical substellar boundary of $\approx$70\,\Mjup,
Gl~569Bb is likely a brown dwarf and Gl~569Ba could be a brown dwarf
or a star.

\subsection{2MASS~J1534$-$2952AB (T4.5+T5) \label{sec:2M1534-29}}

In our previous work, \citet{2008ApJ...689..436L} used a parallax of
$73.6\pm1.2$\,mas from \citet{2003AJ....126..975T} and a different
orbit analysis method to determine a total dynamical mass of
$59\pm3$\,\Mjup.  Our CFHT parallax \citep[first published
in][]{2012ApJS..201...19D} is $63.0\pm1.1$\,mas, putting this binary
significantly farther away and thus making its semimajor axis and mass
measurements larger.  \citet{2008ApJ...689..436L} quoted a total mass
of $58.3^{+2.0}_{-1.8}$\,\Msun\ at a fixed parallax of 73.6\,mas.  We
can therefore readily compute how the total mass from that orbit
analysis would have changed if we instead use our parallax, finding
$92.9^{+3.2}_{-2.8}$\,\Mjup\ at a fixed parallax of 63.0\,mas.  The
dynamical mass we derive here with more data and a different analysis
method is $99.5^{+0.8}_{-0.6}$\,\Mjup\ at a fixed parallax of
63.0\,mas, which is somewhat higher but consistent within 2$\sigma$.
As discussed in Section~\ref{sec:compare-orbit}, this is largely due
to a difference in eccentricity prior for this nearly circular orbit.
After incorporating the parallax uncertainty, our newly determined
total mass is $99\pm5$\,\Mjup.

In our total-mass analysis, SM08 models give an age of
$3.0^{+0.4}_{-0.5}$\,Gyr, which is somewhat older than the Cond age of
$2.25^{+0.29}_{-0.32}$\,Gyr.  Models give consistent mass ratios of
$0.96\pm0.06$ (SM08) and $0.95\pm0.07$ (Cond), which agree well with
our measured mass ratio of $0.95^{+0.13}_{-0.16}$.  Thus, properties
derived from our total-mass analysis agree well with our
individual-mass analysis, and the two components are coeval according
to models.  With individual masses of $51\pm5$\,\Mjup\ and
$48\pm5$\,\Mjup, both components are predicted by Cond models to have
retained $\geq$95\% (1$\sigma$) of their initial lithium.  No
\ion{Li}{1} absorption is observed in the integrated light spectrum of
2MASS~J1534$-$2952AB \citep{2003ApJ...594..510B}, implying that
monatomic lithium is likely chemically depleted into LiCl and/or LiOH
as expected.  The component temperatures derived from Cond models are
$1170\pm50$\,K and $1110^{+40}_{-50}$\,K.  This provides an empirical
lower limit on the effective temperature at which lithium is
chemically depleted for field brown dwarfs.  Finally, we note that
despite being the latest-type binary in our sample, the components of
2MASS~J1534$-$2952AB are in fact similar in mass to some of the L and
early-T dwarfs in our sample (e.g., SDSS~J0423$-$0414A,
SDSS~J1052+4422A, Gl~417B, and Gl~417C).

\subsection{2MASS~J1728+3948AB (L5+L7) \label{sec:2M1728+39}}

Combining our total mass ($140^{+7}_{-8}$\,\Mjup) and mass ratio
($0.93^{+0.11}_{-0.13}$) gives individual masses of $73\pm7$\,\Mjup\
for 2MASS~J1728+3948A (L$5\pm1$) and $67\pm5$\,\Mjup\ for
2MASS~J1728+3948B (L$7\pm1$).  In our total-mass analysis, models give
mass ratios in good agreement with our measurement, and
correspondingly the ages derived in our individual-mass analysis are
consistent between primary and secondary.  The total-mass analysis
gives the tightest age constraints of $3.4^{+2.8}_{-2.1}$\,Gyr (SM08)
and $2.9^{+0.7}_{-1.1}$\,Gyr (Cond).  2MASS~J1728+3948A is predicted
to be strongly depleted of lithium, with
log(Li/Li$_{\rm init}) = -2.9^{+0.4}_{-0.7}$\,dex, while within
1$\sigma$ 2MASS~J1728+3948B ranges from somewhat to severely depleted
in lithium, log(Li/Li$_{\rm init}) = -1.5^{+0.7}_{-0.6}$\,dex.  Given
these predictions it would be possible for the integrated-light
spectrum to show weak lithium absorption due to 2MASS~J1728+3948B.
The lack of a strong \ion{Li}{1} detection (${\rm EW}<4$\,\AA) by
\citet{2000AJ....120..447K} does not rule out such weak absorption,
although it is also consistent with both components being fully
depleted of lithium.  In that case, this nondetection of lithium would
be an indication that the secondary mass is on the higher side of our
quoted 1$\sigma$ intervals.

\subsection{LSPM~J1735+2634AB (M7.5+L0) \label{sec:LS1735+26}}

Combining our total mass ($187\pm7$\,\Mjup) and mass ratio
($0.868^{+0.023}_{-0.025}$) gives individual masses of
$100\pm4$\,\Mjup\ for LSPM~J1735+2634A (M7.5) and $87\pm3$\,\Mjup\ for
LSPM~J1735+2634B (L0).  Having one of the more precise mass ratios,
and one that is further from unity than most, LSPM~J1735+2634AB
provides a good test of the model-predicted mass--luminosity relation.
In our total-mass analysis, BHAC models predict a mass ratio of
$0.913^{+0.019}_{-0.017}$, which is only marginally (1.6$\sigma$)
higher than we measure.  Because the components of LSPM~J1735+2634AB
are both likely main-sequence stars, the age is essentially
unconstrained by models, which only give a 3$\sigma$ lower limit of
$>$0.6\,Gyr from both total-mass and individual-mass analyses.

\subsection{2MASS~J2132+1341AB (L4.5+L8.5) \label{sec:2M2132+13}}

Combining our total mass ($128^{+7}_{-8}$\,\Mjup) and mass ratio
($0.88^{+0.04}_{-0.05}$) gives individual masses of $68\pm4$\,\Mjup\
for 2MASS~J2132+1341A (L$4.5\pm1.5$) and $60\pm4$\,\Mjup\ for
2MASS~J2132+1341B (L$8.5\pm1.5$).  In our total-mass analysis, models
give mass ratios in good agreement with our measurement, and
correspondingly the ages derived in our individual mass analysis are
consistent between primary and secondary.  The total-mass analysis
gives the tightest age constraints of $1.44^{+0.26}_{-0.37}$\,Gyr
(SM08) and $1.71^{+0.28}_{-0.38}$\,Gyr (Cond).  2MASS~J2132+1341A is
predicted to be severely depleted of lithium, while 2MASS~J2132+1341B
straddles the lithium-fusion boundary within the errors but is
predicted to mostly be above it and thus not strongly depleted in
lithium, log(Li/Li$_{\rm init}) = -1.1^{+1.1}_{-0.4}$.  If \ion{Li}{1}
were present, then \citet{2007AJ....133..439C} would likely have
detected it in their integrated-light spectrum of 2MASS~J2132+1341,
although they do not quote EW limits. Their lack of a detection is
would be more consistent with the model predictions if
2MASS~J2132+1341B is on the more massive end of our measurement
uncertainties.

\subsection{2MASS~J2140+1625AB (M8+L0.5) \label{sec:2M2140+16}}

Our measured mass ratio for 2MASS~J2140+1625AB is quite low
($0.60^{+0.07}_{-0.08}$), such that it divides up our total mass
($183^{+14}_{-17}$\,\Mjup) into a quite massive M8 primary
($114^{+10}_{-12}$\,\Mjup) and a rather low-mass L0.5 secondary
($69^{+8}_{-9}$\,\Mjup).  This is strongly disfavored by models,
mainly because the primary is much less luminous than expected for
such a high mass.  Our total-mass analysis using BHAC models predicts
a mass ratio of $0.882^{+0.028}_{-0.024}$, 3.8$\sigma$ discrepant with
our measurement.  Interestingly, both our total-mass and
individual-mass analyses end up with an essentially identical total
mass as we measure.  The individual-mass analysis demonstrates how
this can be the case despite such a discrepant model-derived mass
ratio: the primary mass after performing rejection sampling was
1.7$\sigma$ lower than we input, and the secondary mass was
2.0$\sigma$ higher than we input.

This apparent discrepancy could be jointly due to two smaller
($\approx$1.5$\sigma$) discrepancies in the measured mass ratio and
parallax.  2MASS~J2140+1625 has one of the less precise mass ratios in
our sample and a 3\% parallax uncertainty that leads to an
atypically large total mass error ($^{+8}_{-9}$\%).  Therefore,
higher-precision data would be useful in determining if
2MASS~J2140+1625AB is truly an astrophysical outlier.  

\subsection{2MASS~J2206$-$2047AB (M8+M8) \label{sec:2M2206-20}}

In our previous work, we used a parallax of $37.5\pm3.4$\,mas from
\citet{2006AJ....132.1234C} to determine a dynamical total mass and
model-derived properties \citep{2009ApJ...706..328D}.  Our CFHT
parallax of $35.7\pm1.2$\,mas was first presented in
\citet{2012ApJS..201...19D}, and the updated analysis here gives
$35.8\pm1.0$\,mas, which allows for a more precise total mass than in
\citet{2009ApJ...706..328D} as well as individual masses from our
photocenter analysis.  Our total mass is $188^{+16}_{-17}$\,\Mjup,
consistent with our previous measurement of $160^{+50}_{-30}$\,\Mjup\
\citep{2009ApJ...706..328D}.  Our measured mass ratio of
$0.84^{+0.09}_{-0.10}$, consistent with the BHAC model-derived value
of $0.986^{+0.028}_{-0.029}$, gives component masses of
$102^{+10}_{-11}$\,\Mjup\ and $86^{+8}_{-10}$\,\Mjup.  In both
total-mass and individual-mass analyses, rejection sampling results in
slightly higher secondary masses and lower primary masses in the
output distributions (96\,\Mjup\ and 95\,\Mjup) but they are
consistent with the inputs to within $\leq$1$\sigma$.  The masses and
luminosities of 2MASS~J2206$-$2047AB are consistent with this being a
pair of main-sequence stars, with BHAC models giving a 3$\sigma$ lower
limit on the age of $>$0.27\,Gyr.

\subsection{DENIS~J2252$-$1730AB (L4+T3.5) \label{sec:2M2252-17}}

Combining our total mass ($101\pm7$\,\Mjup) and mass ratio
($0.70^{+0.08}_{-0.09}$) gives individual masses of $59\pm5$\,\Mjup\
for DENIS~J2252$-$1730A (L$4.0\pm1.0$) and $41\pm4$\,\Mjup\ for
DENIS~J2252$-$1730B (T$3.5\pm0.5$).  Therefore, both components are
unambiguously substellar. In our total-mass analysis, SM08 models
prefer a somewhat lower mass ratio ($0.57^{+0.03}_{-0.05}$),
consistent with our measurement within the uncertainties, while Cond
models predict a somewhat higher mass ratio ($0.71^{+0.05}_{-0.04}$).
Both sets of models give consistent ages, with the total mass analysis
providing the tightest age constraints of $1.11^{+0.19}_{-0.22}$\,Gyr
(SM08) and $1.10^{+0.15}_{-0.18}$\,Gyr (Cond).  According to Cond
models, DENIS~J2252$-$1730B is predicted to have retained nearly all
its lithium, while DENIS~J2252$-$1730A straddles the lithium fusion
boundary within the errors,
log(Li/Li$_{\rm init}) = -2.1^{+0.7}_{-1.5}$\,dex.
DENIS~J2252$-$1730B is quite cool ($1230^{+50}_{-60}$\,K, Cond) and
thus may not be expected to display \ion{Li}{1} in absorption due to
being chemically depleted.  DENIS~J2252$-$1730A on the other hand is
an L2~dwarf that should be amenable to lithium-absorption measurement.
\citet{2008AJ....136.1290R} obtained an integrated-light optical
spectrum for this relatively bright binary and did not report lithium
absorption for this source.  This suggests that DENIS~J2252$-$1730A is
indeed above the lithium-fusion limit and thus its true mass may be in
the higher range of our quoted 1$\sigma$ mass interval.

\clearpage

\begin{figure} 
  \centerline{\includegraphics[width=6.5in,angle=0]{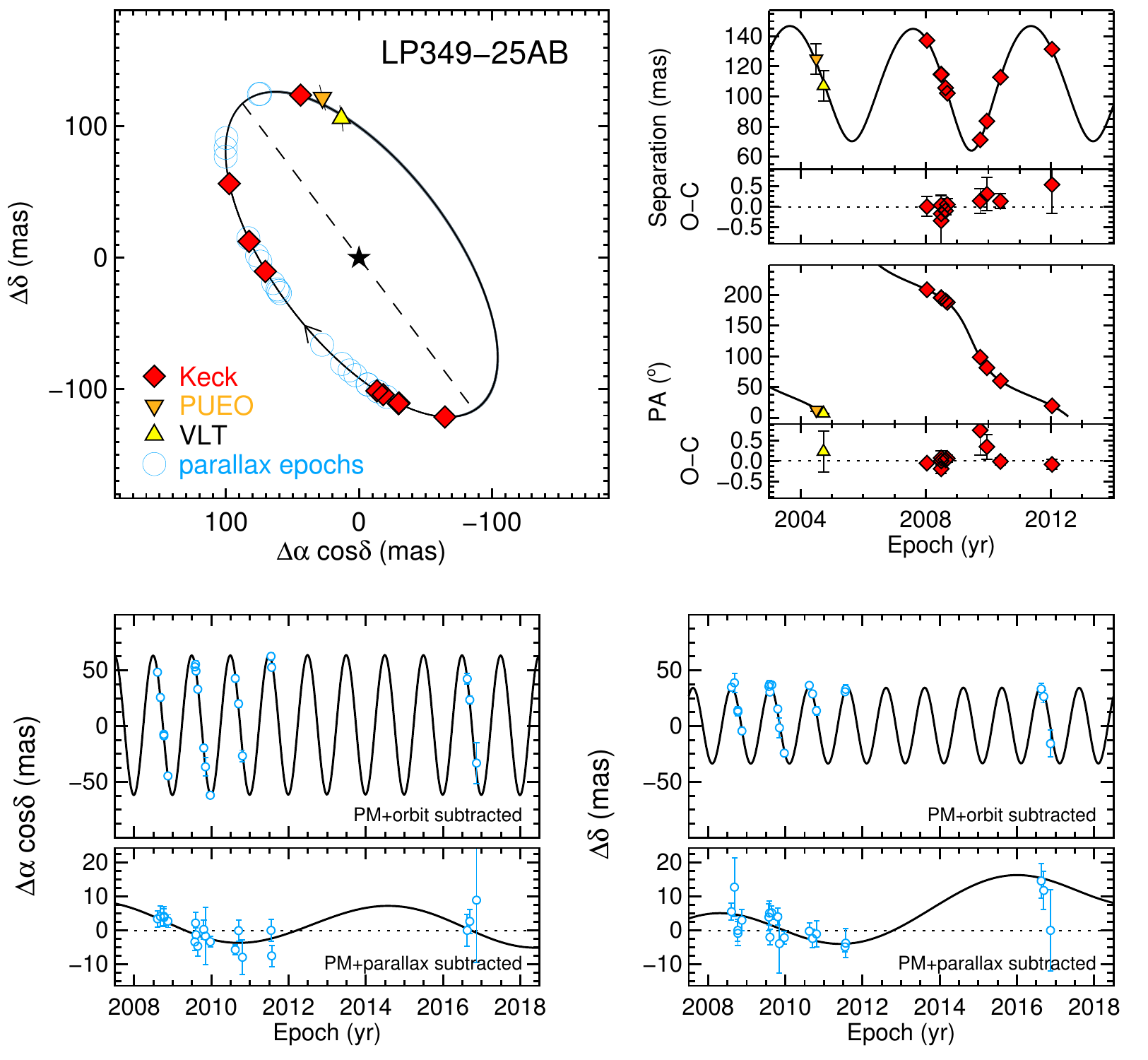}}
  \caption{\normalsize \emph{Top left:} resolved relative astrometry
    (filled symbols) shown alongside the best-fit orbit (thick black
    line) and 100 randomly drawn orbits from our MCMC chain (thin gray
    lines).  The plotting symbols typically are larger than the error
    bars.  Open blue circles indicate the epochs at which we obtained
    unresolved CFHT/WIRCam astrometry.  \emph{Top right:} our relative
    astrometry shown as a function of time (top sub-panels) and after
    subtracting the best-fit orbit solution (bottom sub-panels).
    \emph{Bottom:} CFHT/WIRCam astrometry from unresolved,
    seeing-limited imaging.  Top panels show the data with the proper
    motion and photocenter orbital motion subtracted in order to
    display our best-fit parallax solution (thick black line).  Bottom
    panels show the residuals and our best-fit astrometric solution
    (thick black line) after subtracting our best-fit parallax and
    linear motion as a function of time (including both true proper
    motion and linear orbital motion). \label{fig:orbit}}
\end{figure}
\clearpage
\begin{figure} \centerline{\includegraphics[width=6.5in,angle=0]{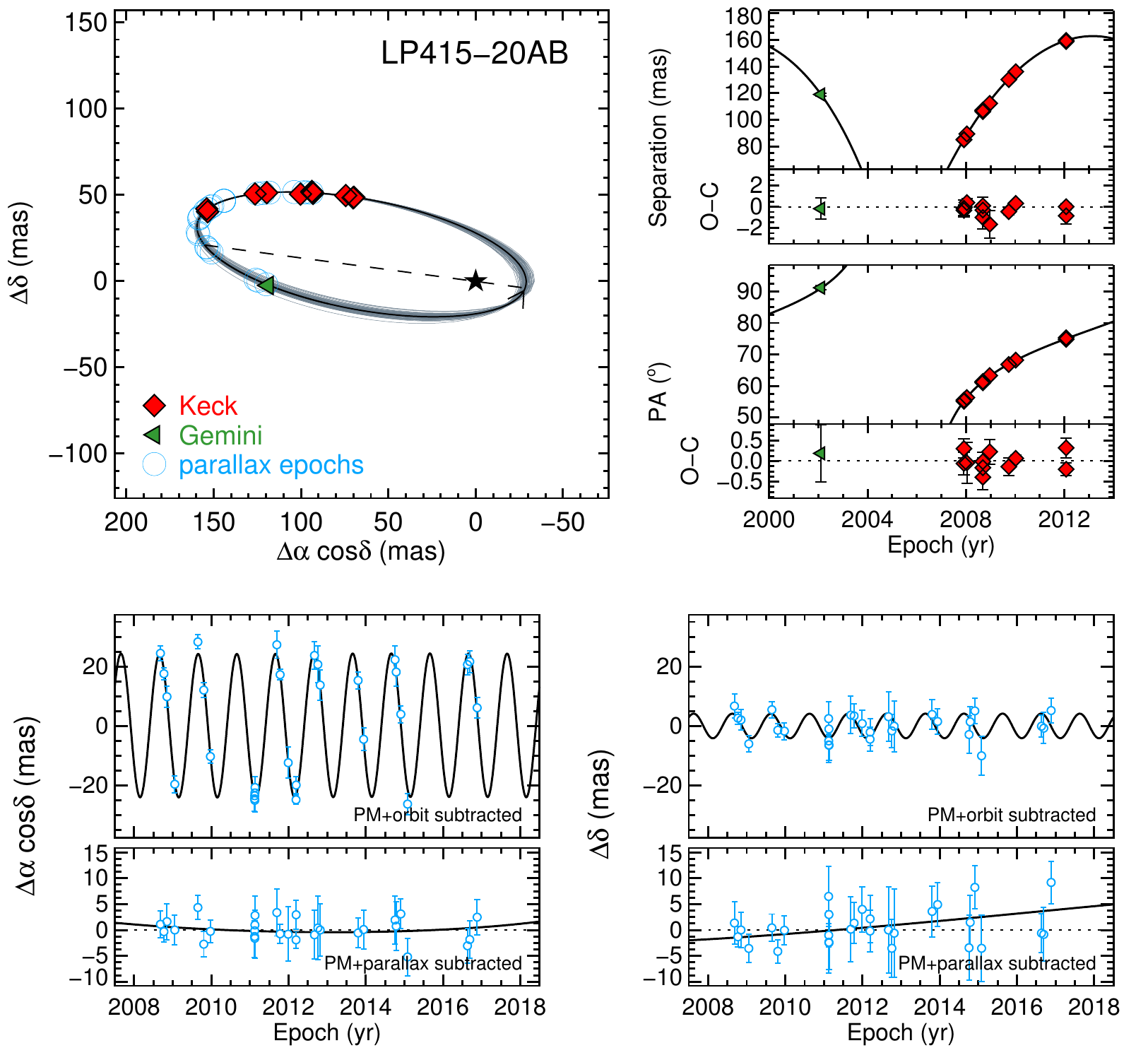}}   \ContinuedFloat  \caption{\normalsize (Continued)} \end{figure} \clearpage
\begin{figure} \centerline{\includegraphics[width=6.5in,angle=0]{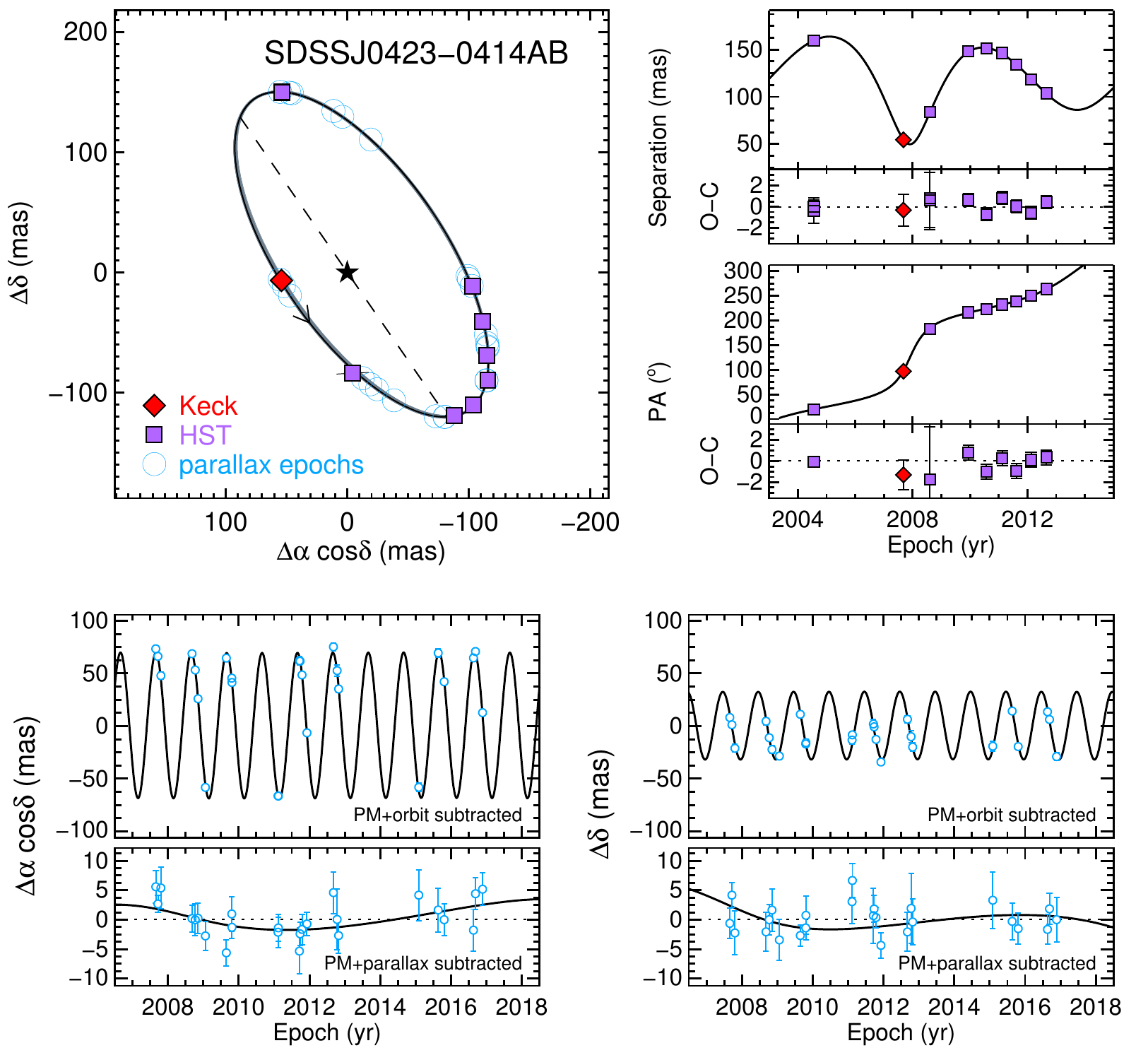}}  \ContinuedFloat  \caption{\normalsize (Continued)} \end{figure} \clearpage
\begin{figure} \centerline{\includegraphics[width=6.5in,angle=0]{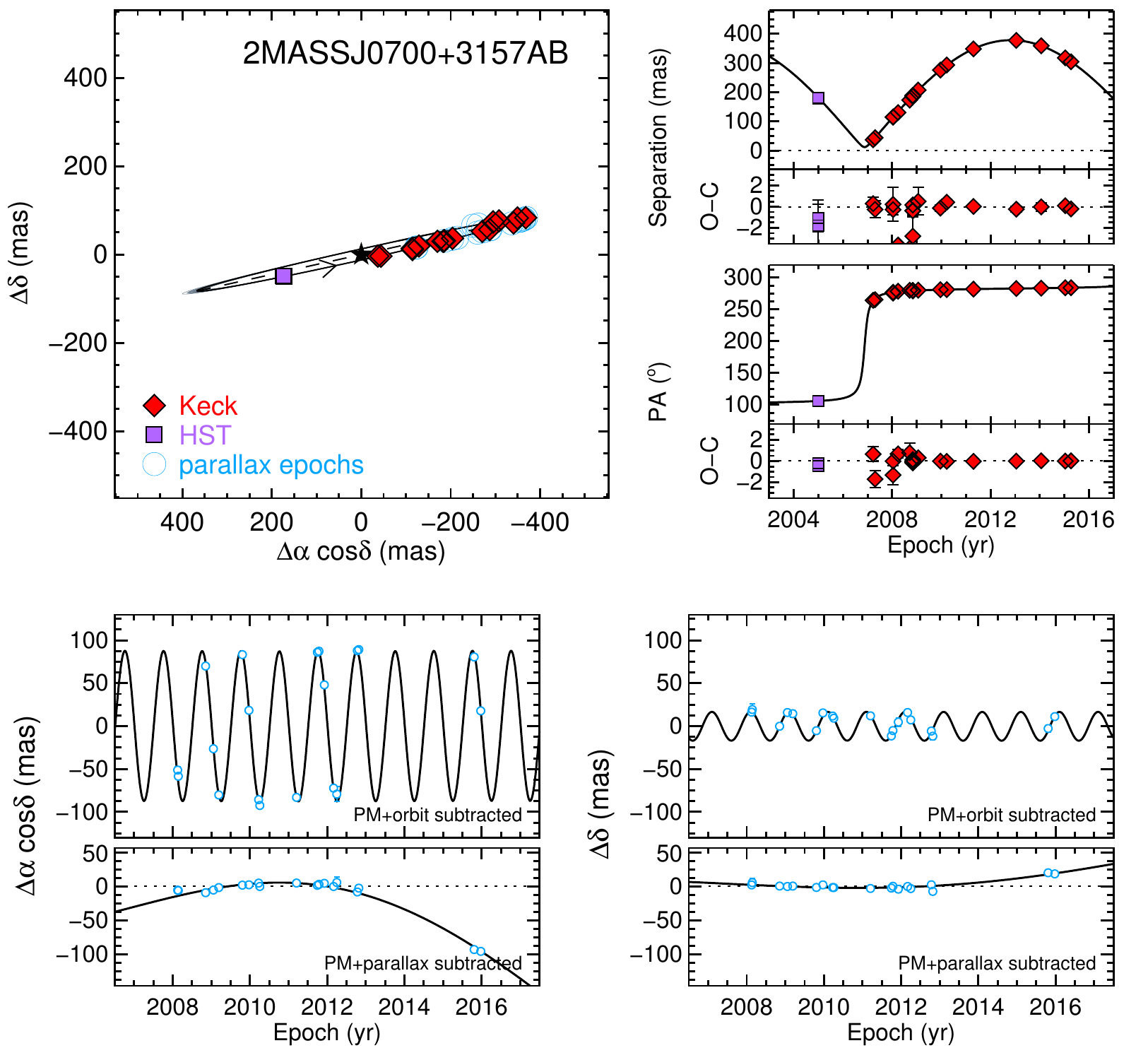}}  \ContinuedFloat  \caption{\normalsize (Continued)} \end{figure} \clearpage
\begin{figure} \centerline{\includegraphics[width=6.5in,angle=0]{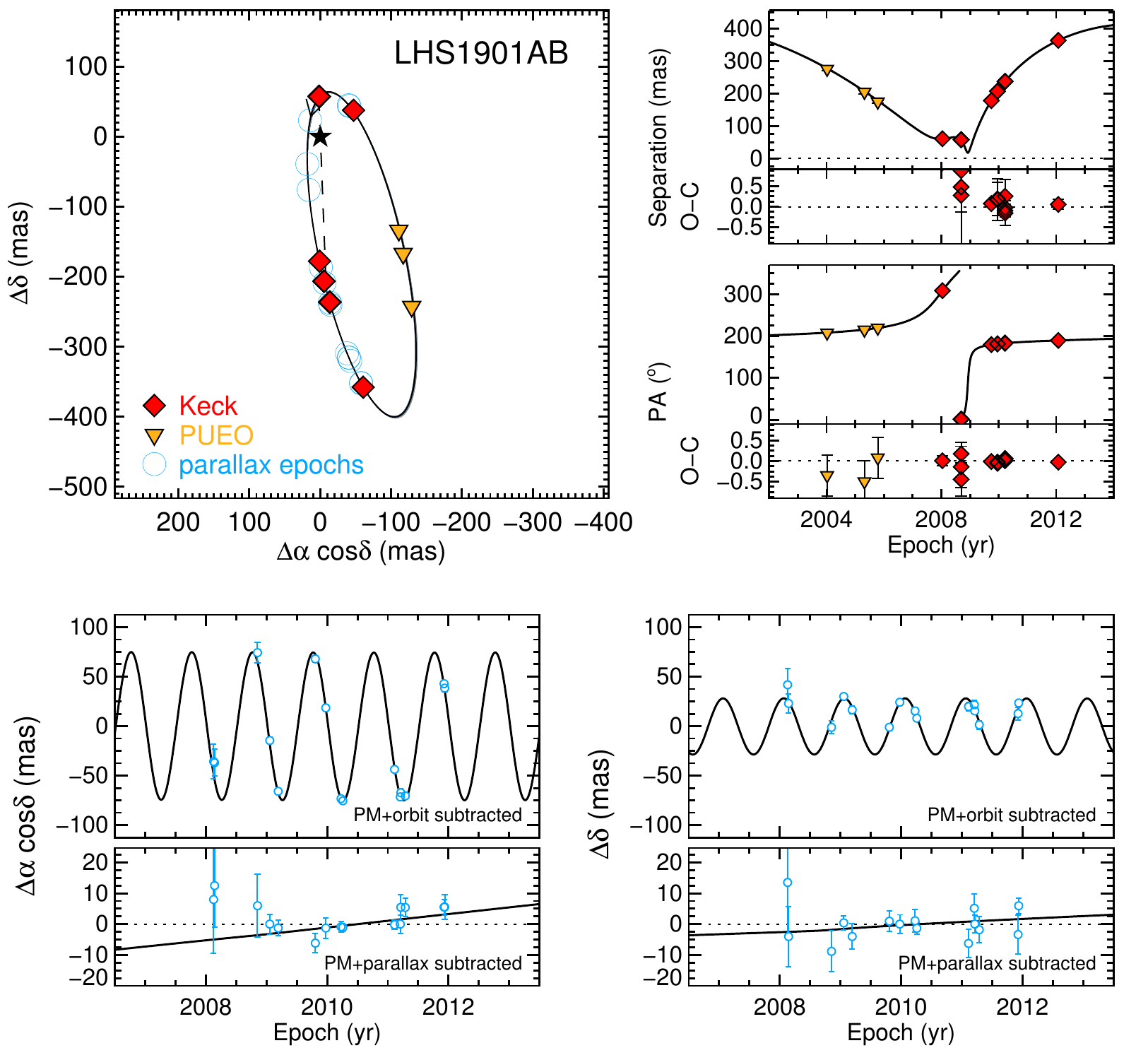}}    \ContinuedFloat  \caption{\normalsize (Continued)} \end{figure} \clearpage
\begin{figure} \centerline{\includegraphics[width=6.5in,angle=0]{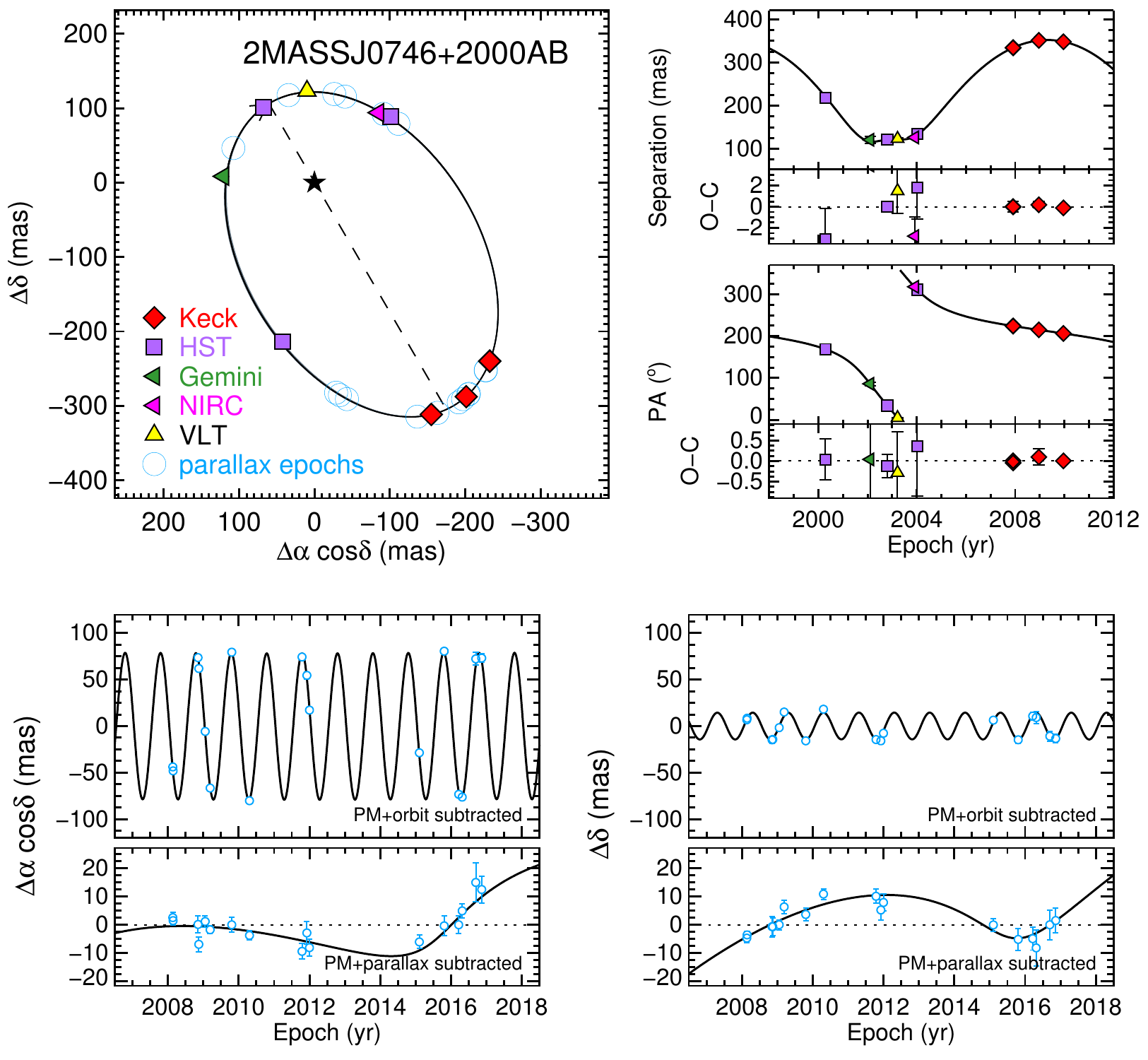}}  \ContinuedFloat  \caption{\normalsize (Continued)} \end{figure} \clearpage
\begin{figure} \centerline{\includegraphics[width=6.5in,angle=0]{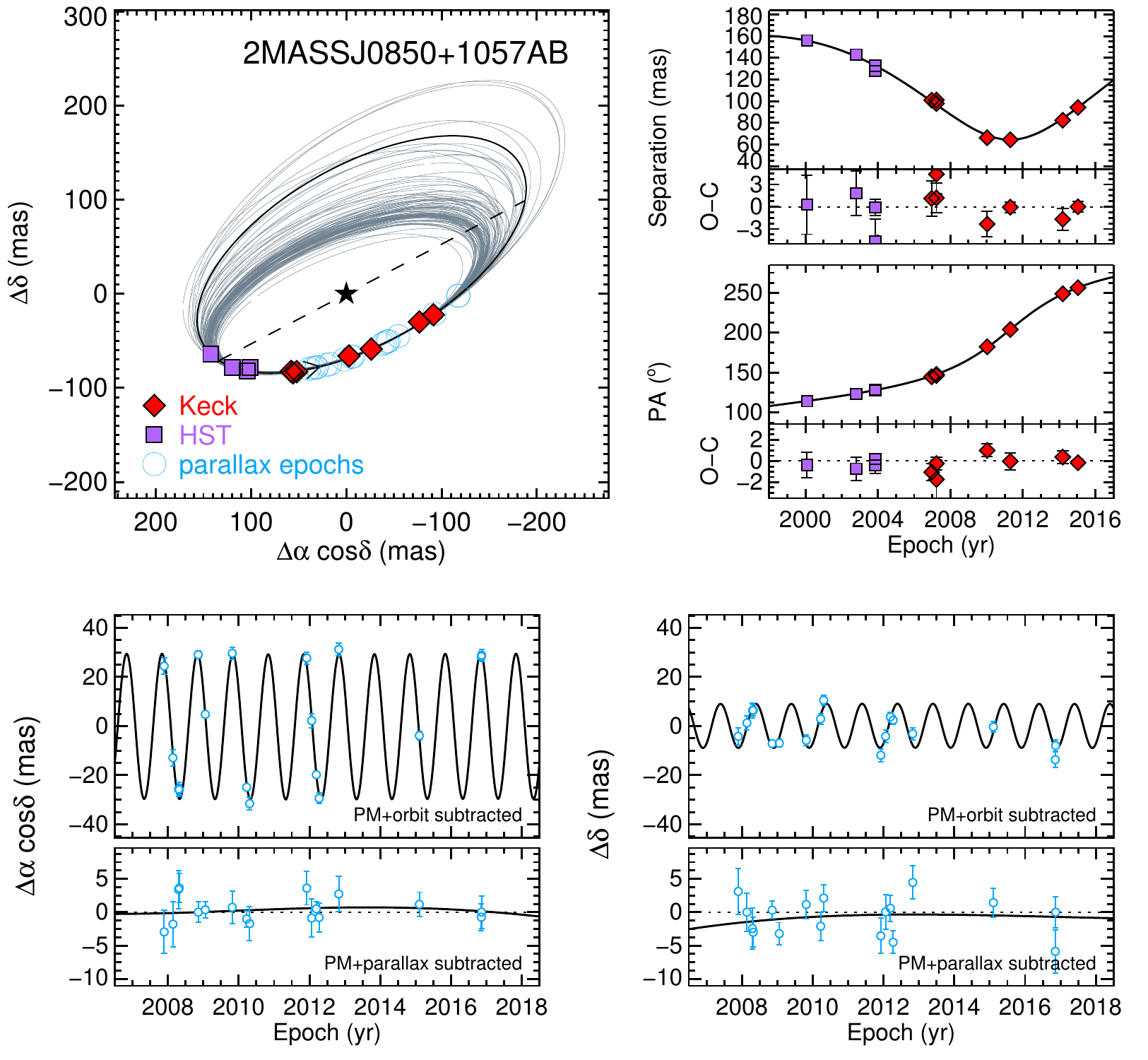}}  \ContinuedFloat  \caption{\normalsize (Continued)} \end{figure} \clearpage
\begin{figure} \centerline{\includegraphics[width=6.5in,angle=0]{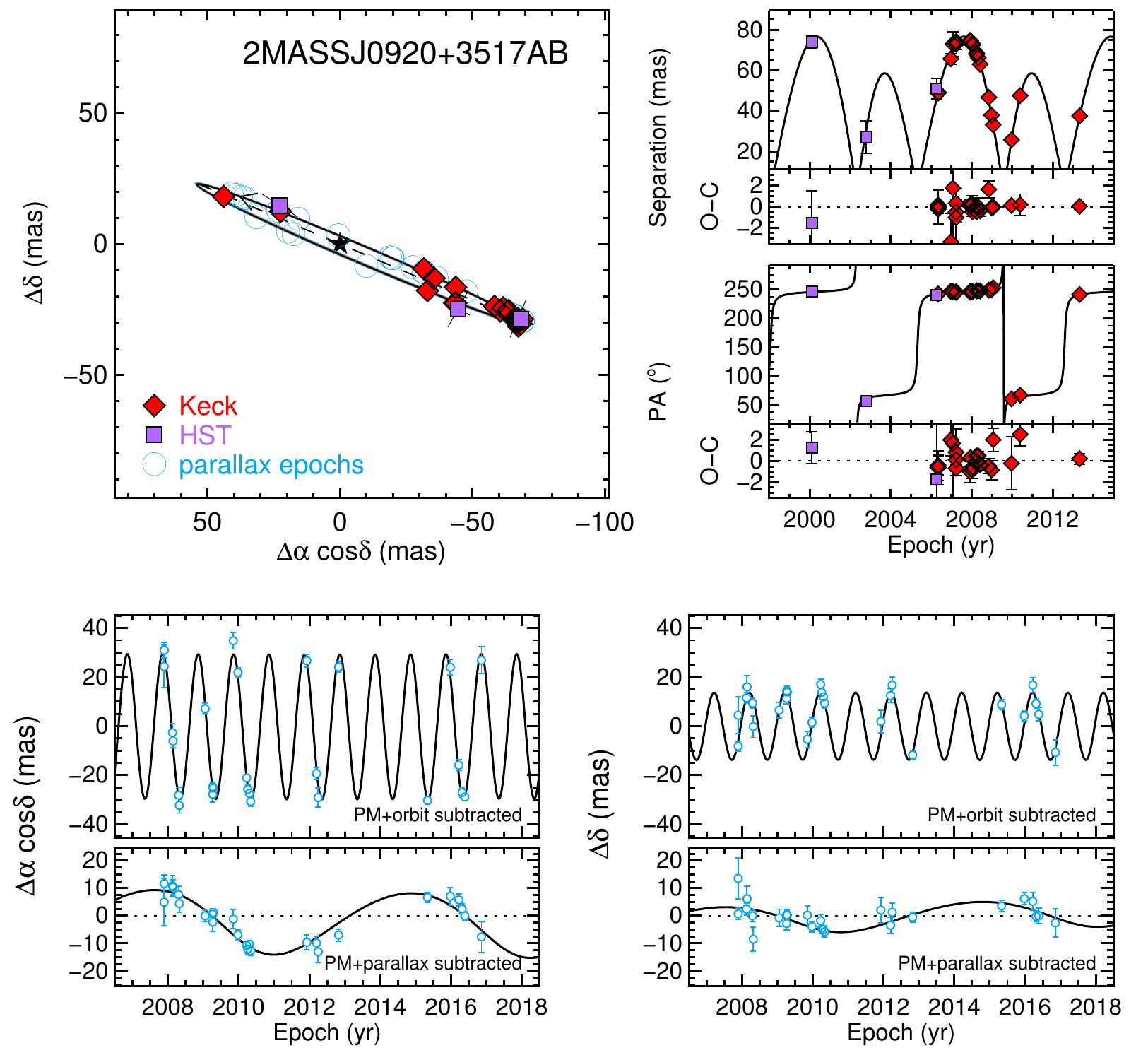}}  \ContinuedFloat  \caption{\normalsize (Continued)} \end{figure} \clearpage
\begin{figure} \centerline{\includegraphics[width=6.5in,angle=0]{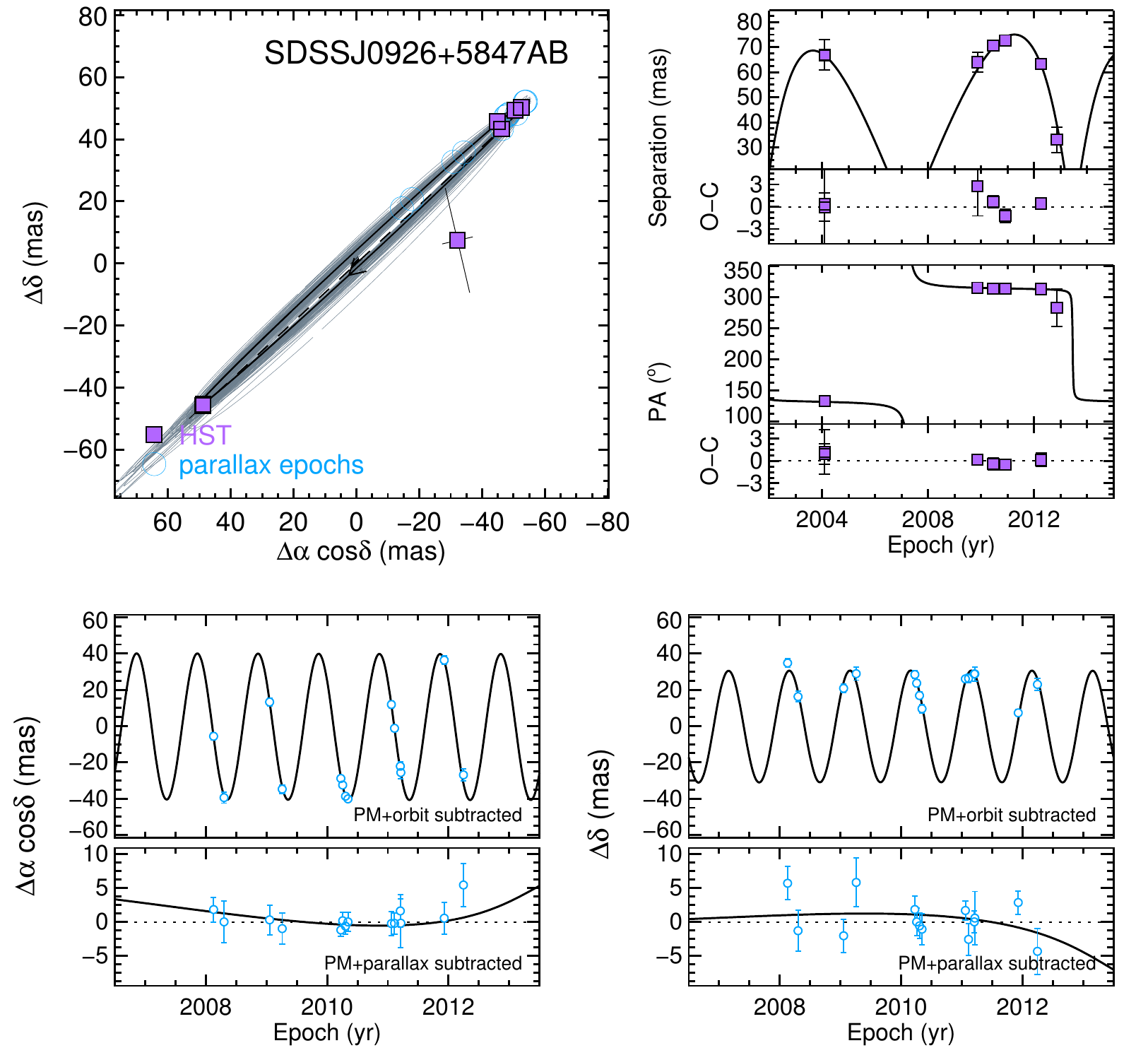}}  \ContinuedFloat  \caption{\normalsize (Continued)} \end{figure} \clearpage
\begin{figure} \centerline{\includegraphics[width=6.5in,angle=0]{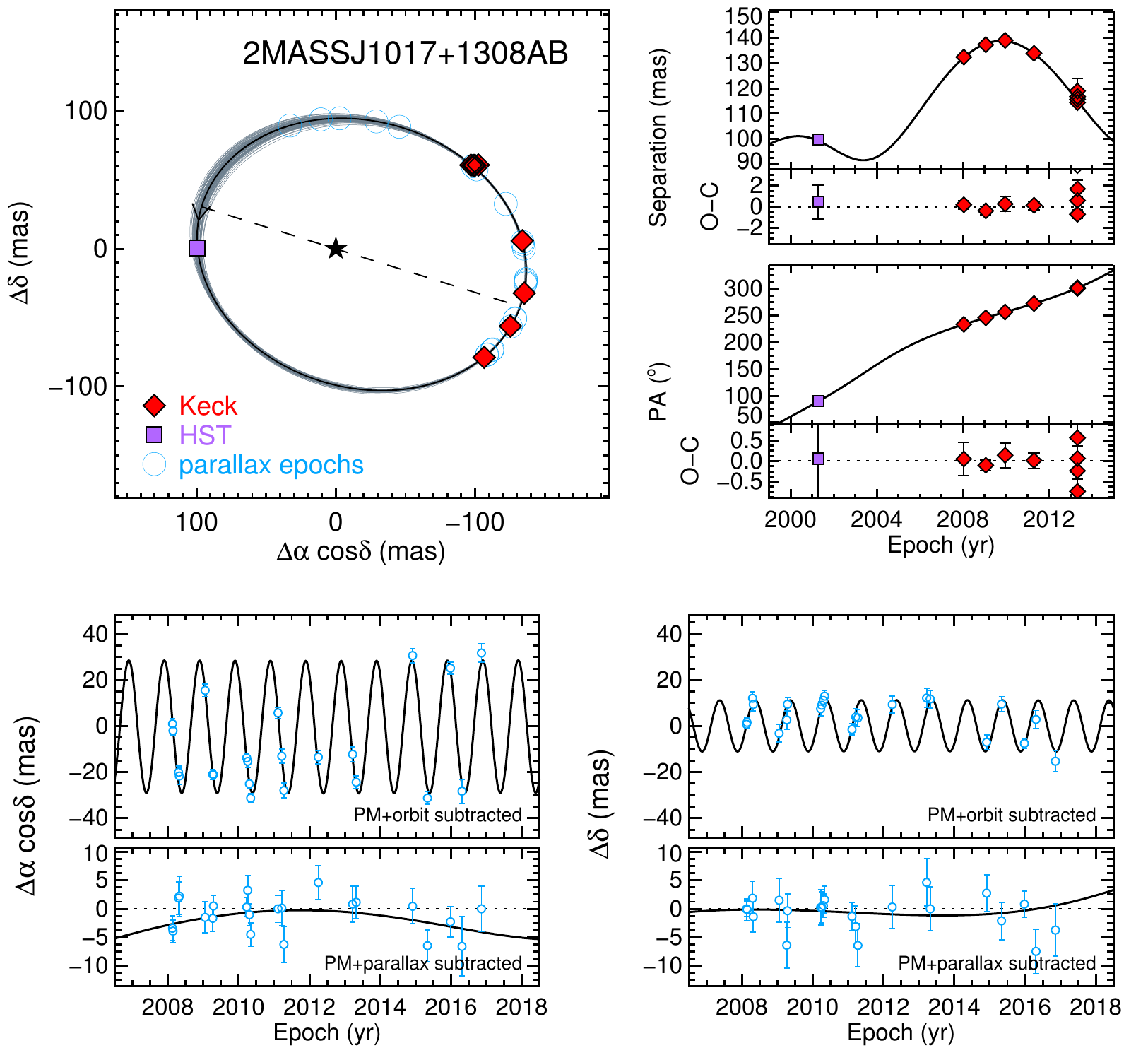}}  \ContinuedFloat  \caption{\normalsize (Continued)} \end{figure} \clearpage
\begin{figure} \centerline{\includegraphics[width=6.5in,angle=0]{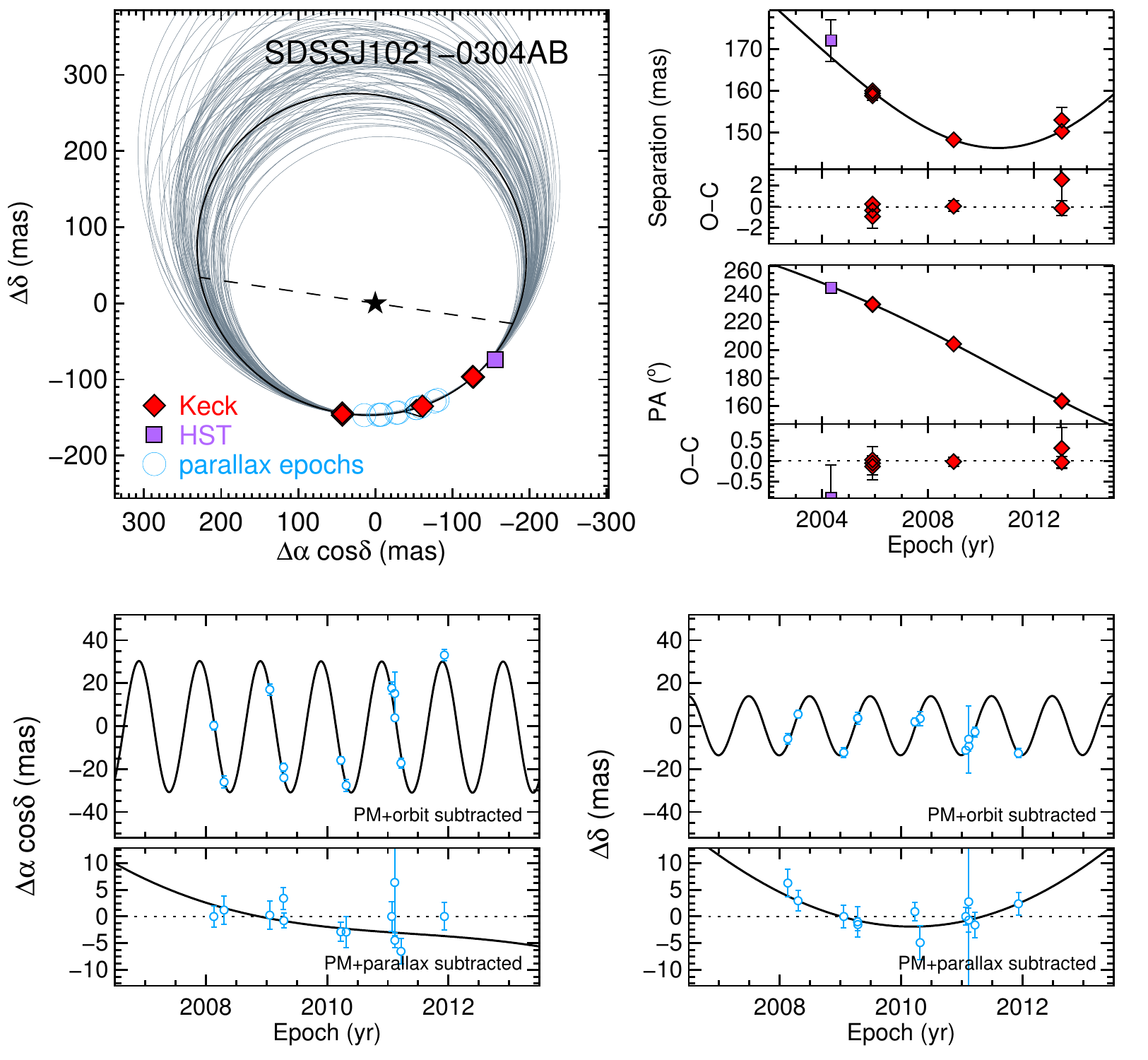}}  \ContinuedFloat  \caption{\normalsize (Continued)} \end{figure} \clearpage
\begin{figure} \centerline{\includegraphics[width=6.5in,angle=0]{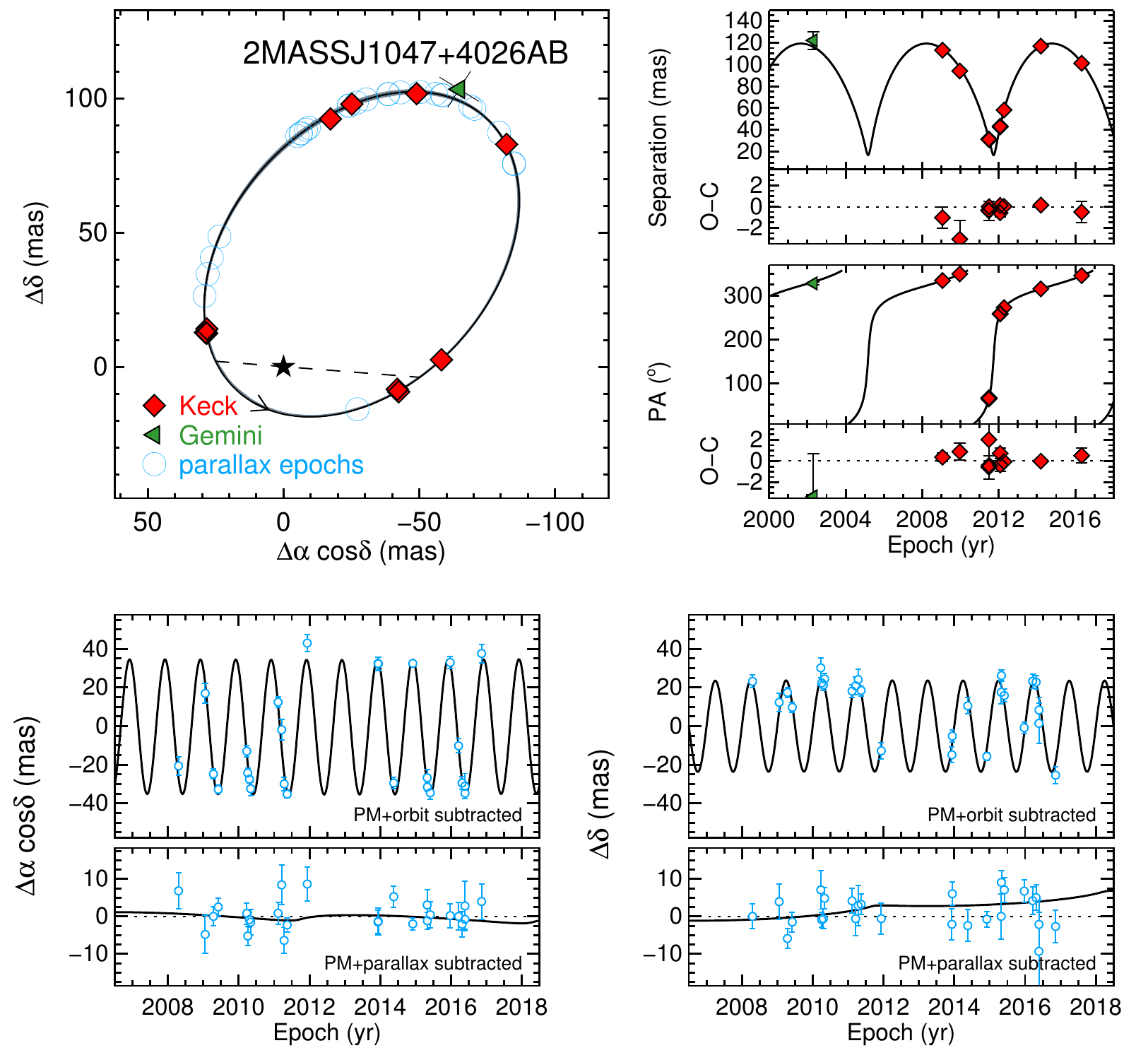}}  \ContinuedFloat  \caption{\normalsize (Continued)} \end{figure} \clearpage
\begin{figure} \centerline{\includegraphics[width=6.5in,angle=0]{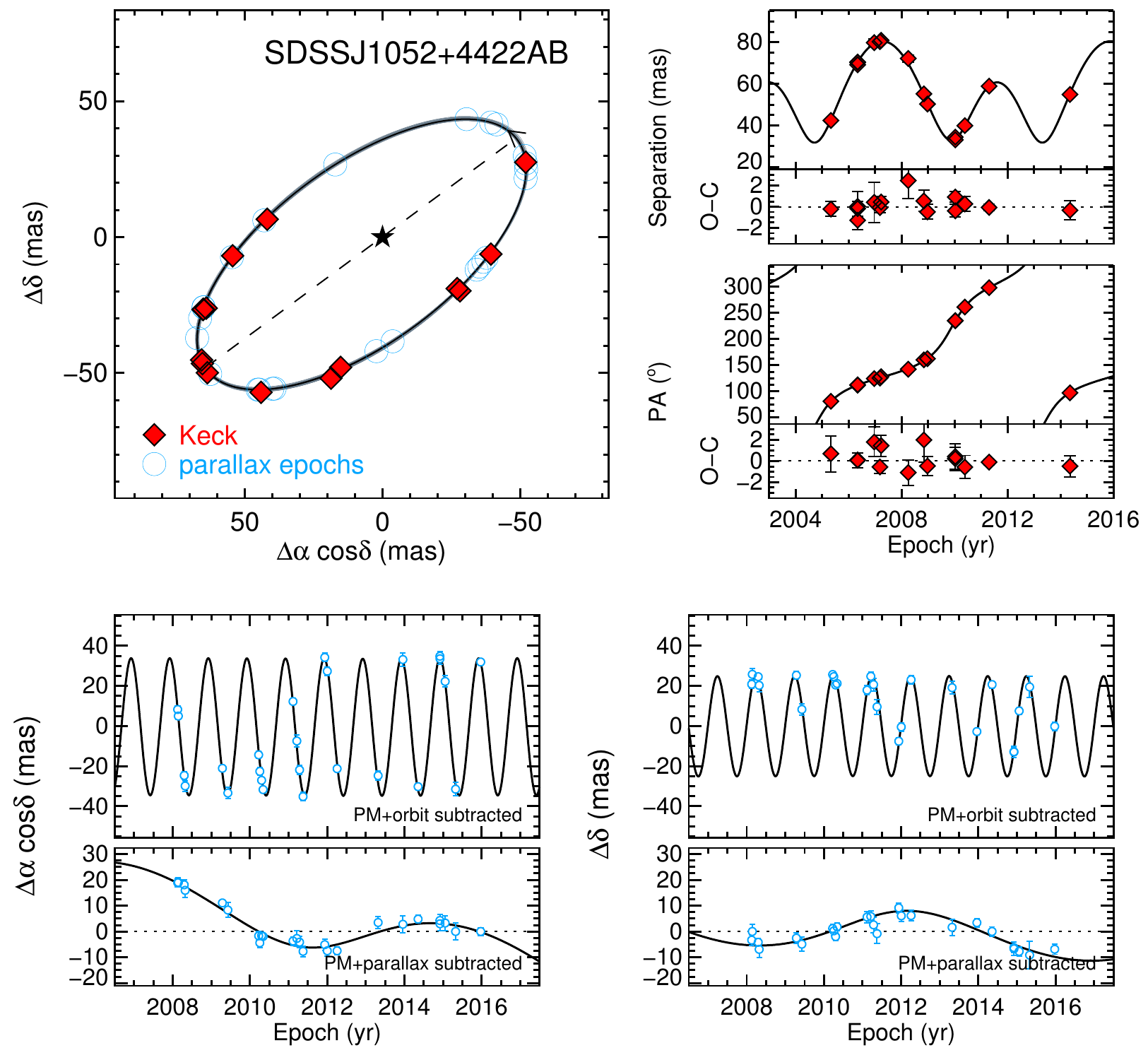}}  \ContinuedFloat  \caption{\normalsize (Continued)} \end{figure} \clearpage
\begin{figure} \centerline{\includegraphics[width=6.5in,angle=0]{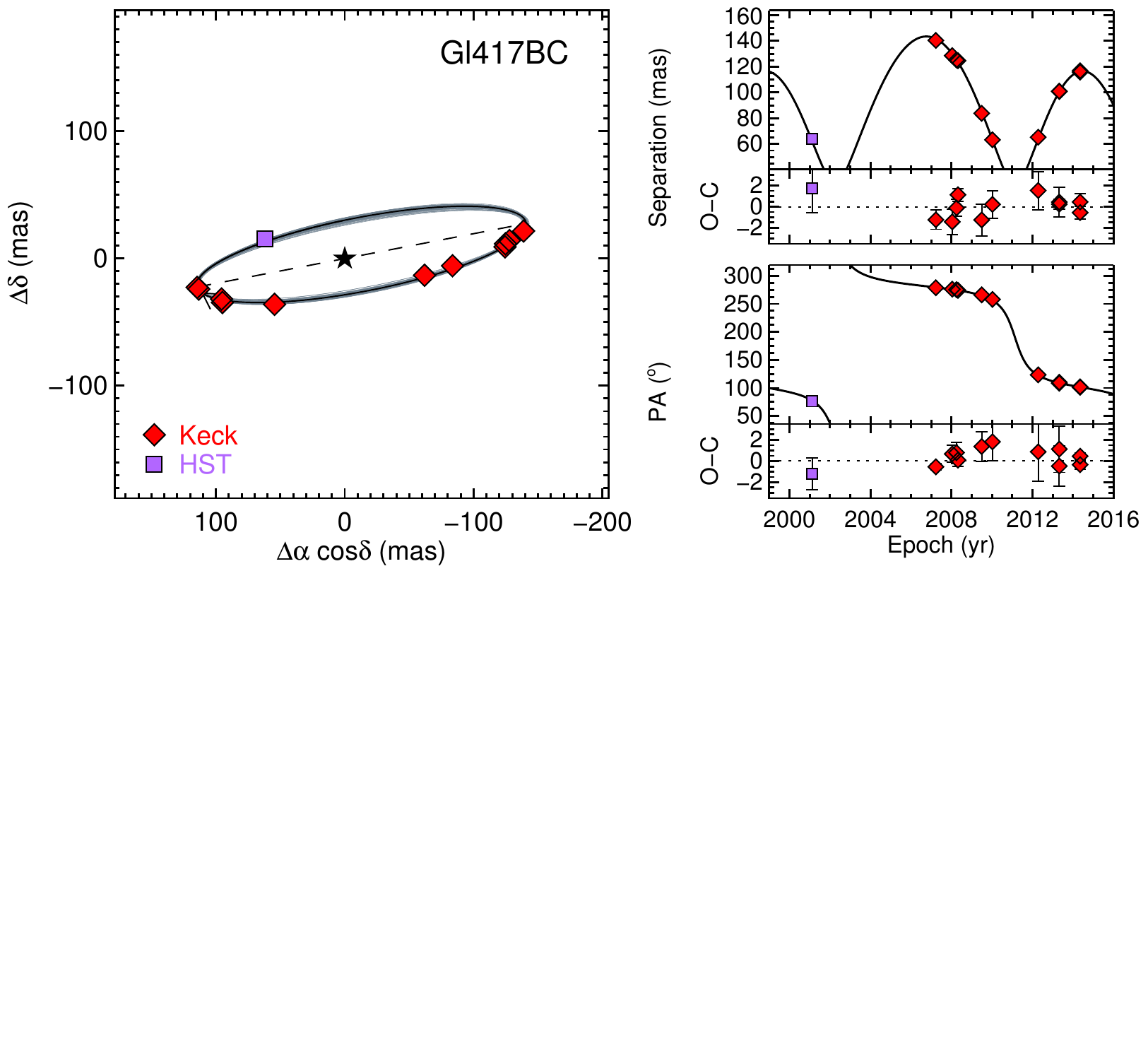}}     \ContinuedFloat  \caption{\normalsize (Continued)} \end{figure} \clearpage
\begin{figure} \centerline{\includegraphics[width=6.5in,angle=0]{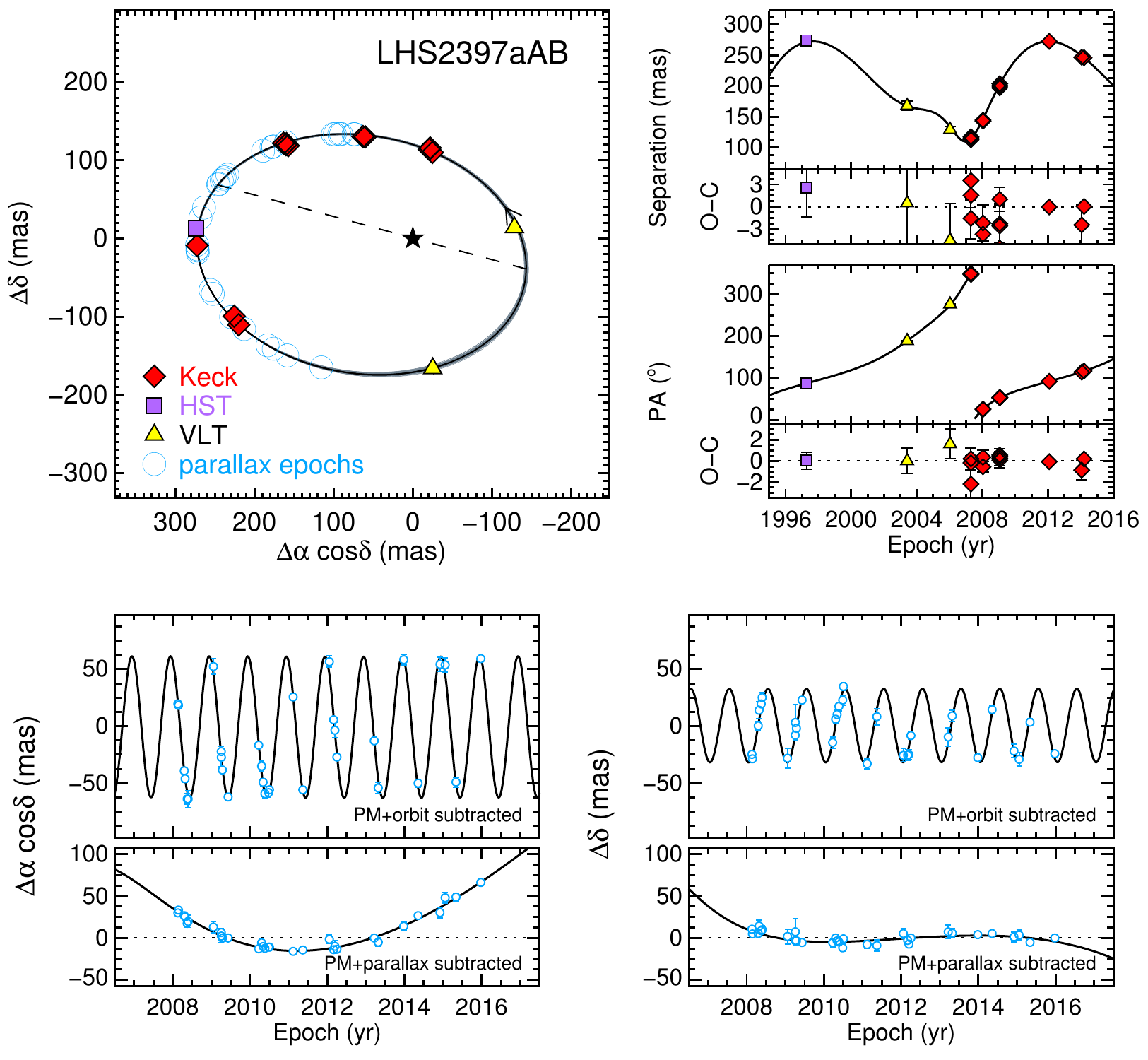}}   \ContinuedFloat  \caption{\normalsize (Continued)} \end{figure} \clearpage
\begin{figure} \centerline{\includegraphics[width=6.5in,angle=0]{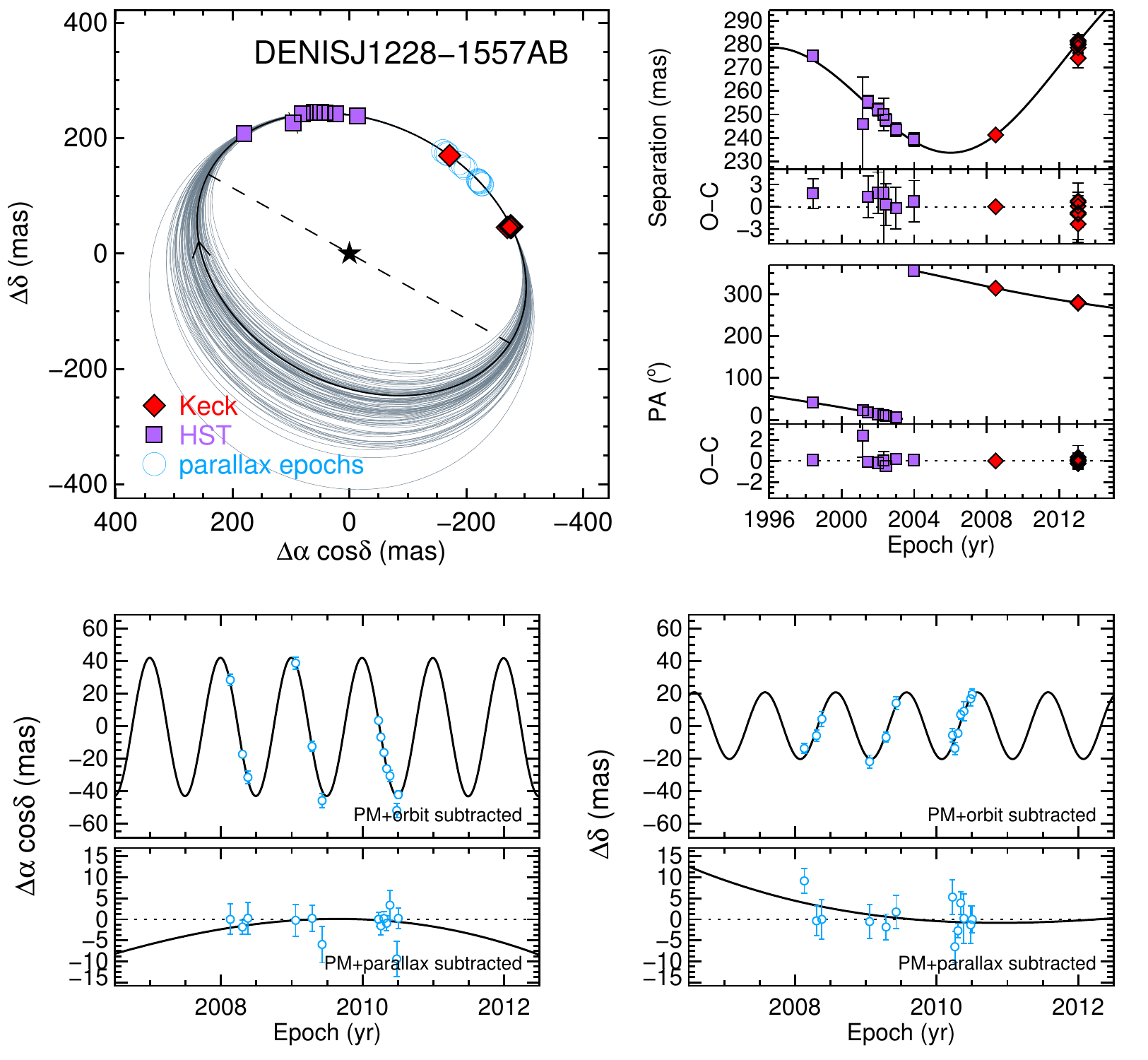}}  \ContinuedFloat  \caption{\normalsize (Continued)} \end{figure} \clearpage
\begin{figure} \centerline{\includegraphics[width=6.5in,angle=0]{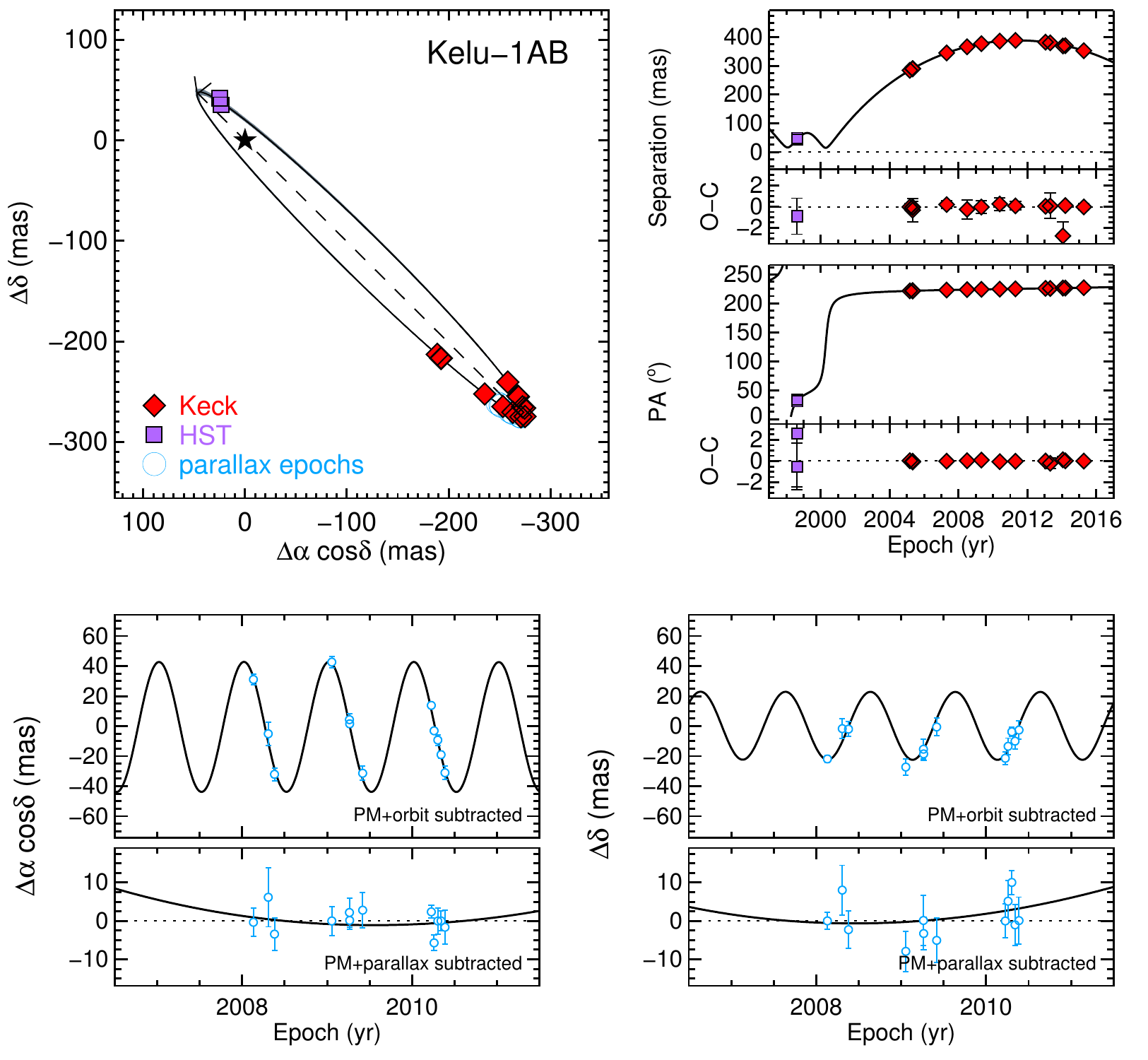}}     \ContinuedFloat  \caption{\normalsize (Continued)} \end{figure} \clearpage
\begin{figure} \centerline{\includegraphics[width=6.5in,angle=0]{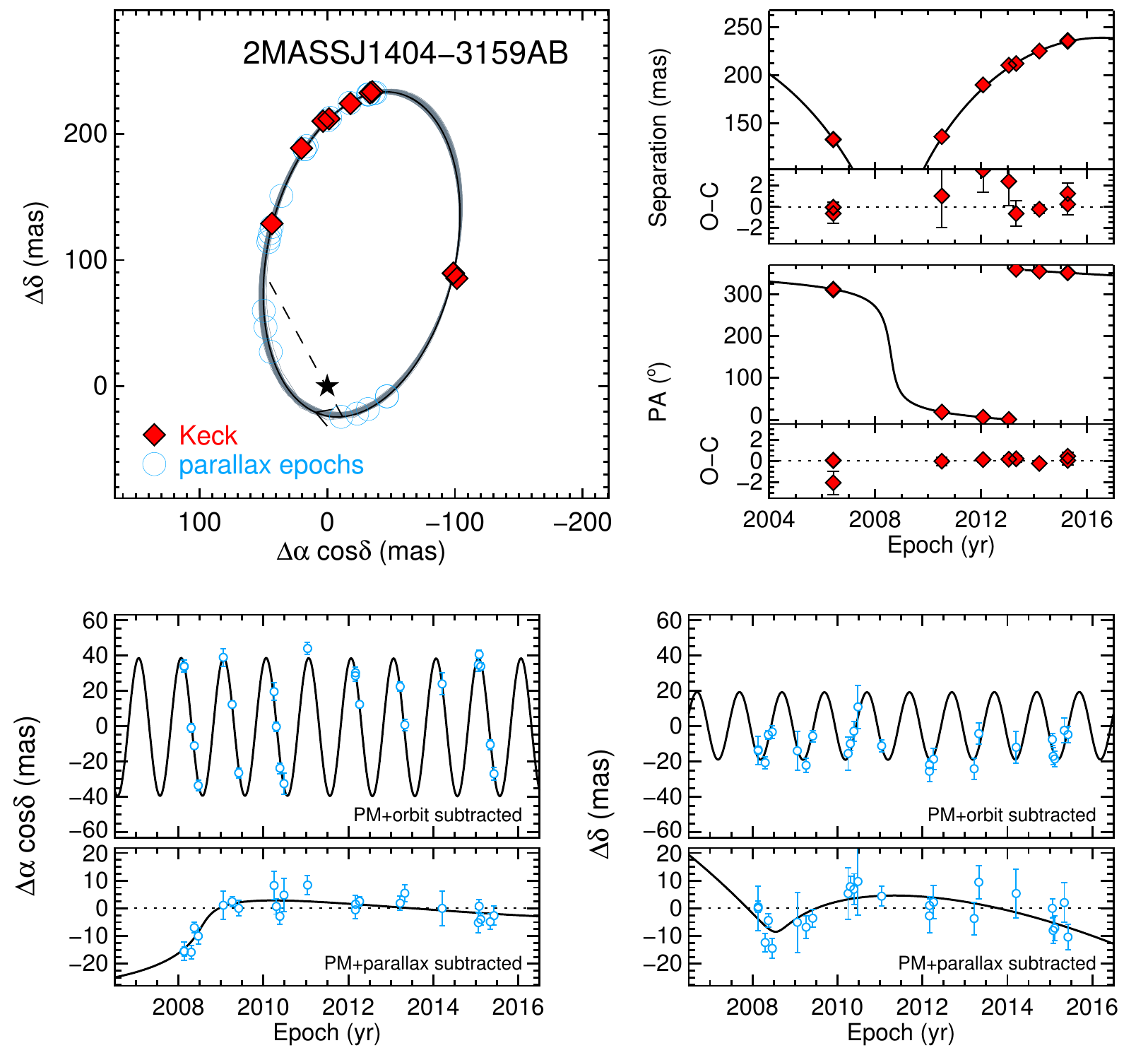}}  \ContinuedFloat  \caption{\normalsize (Continued)} \end{figure} \clearpage
\begin{figure} \centerline{\includegraphics[width=6.5in,angle=0]{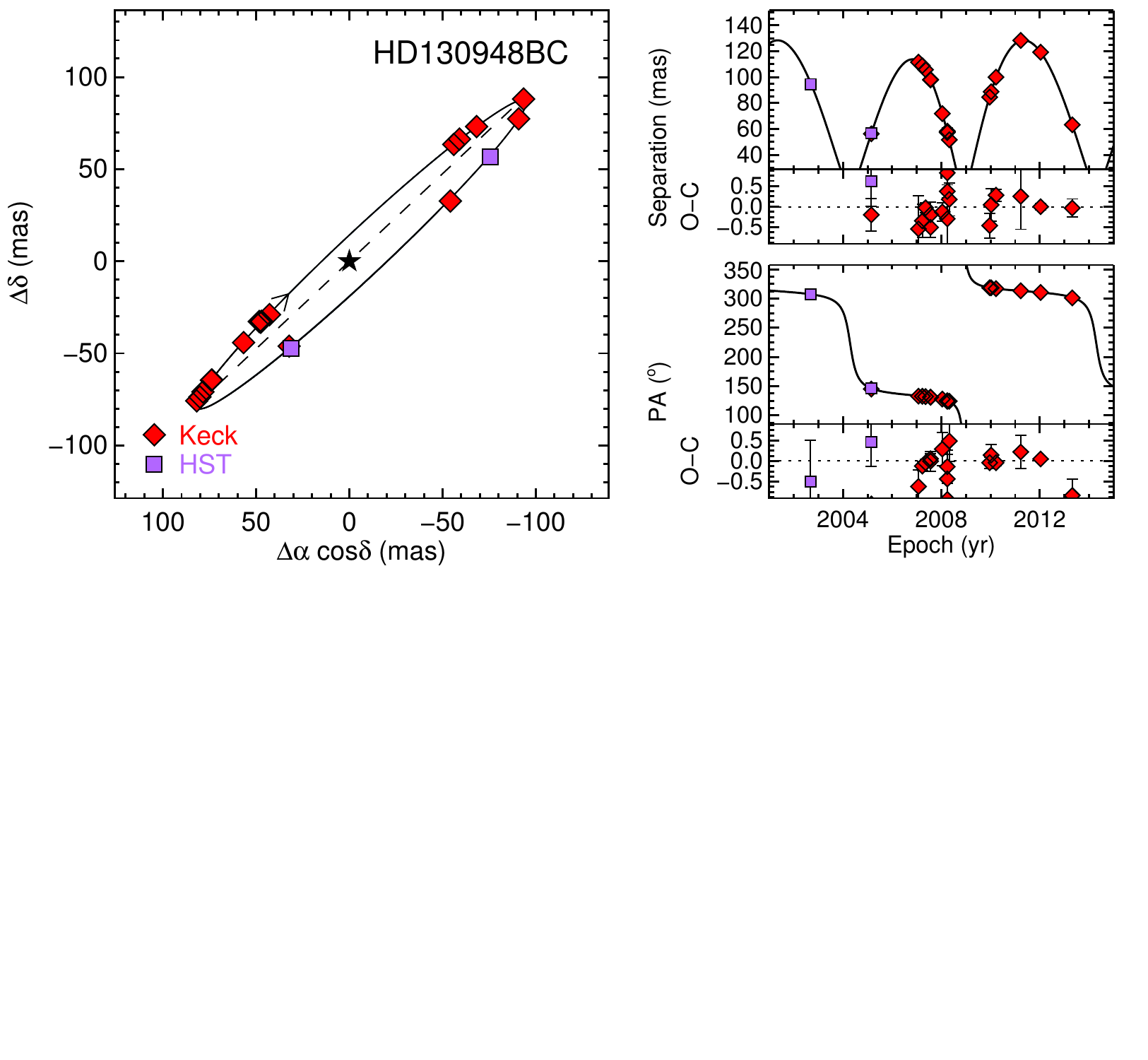}}  \ContinuedFloat  \caption{\normalsize (Continued)} \end{figure} \clearpage
\begin{figure} \centerline{\includegraphics[width=6.5in,angle=0]{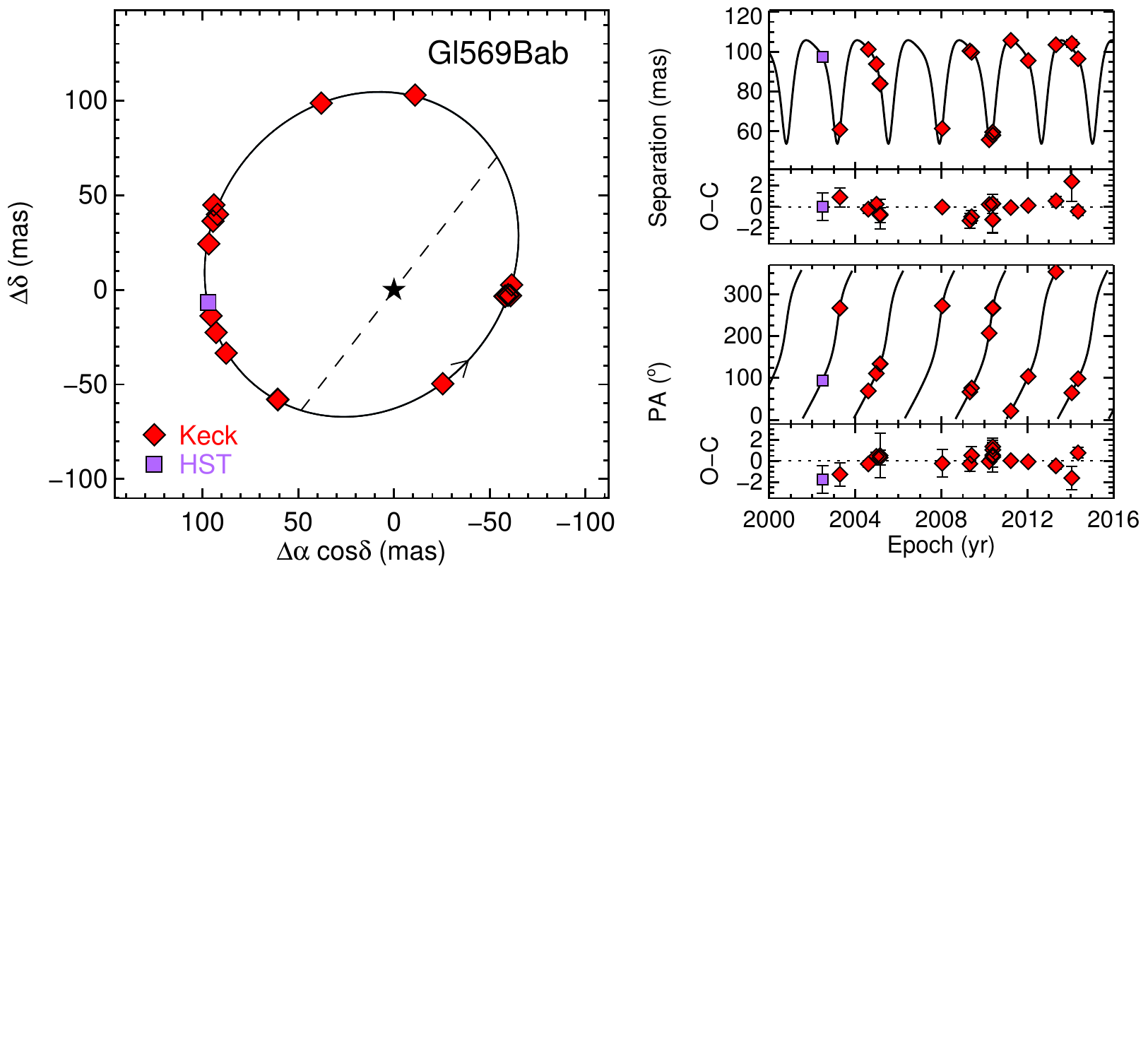}}     \ContinuedFloat  \caption{\normalsize (Continued)} \end{figure} \clearpage
\begin{figure} \centerline{\includegraphics[width=6.5in,angle=0]{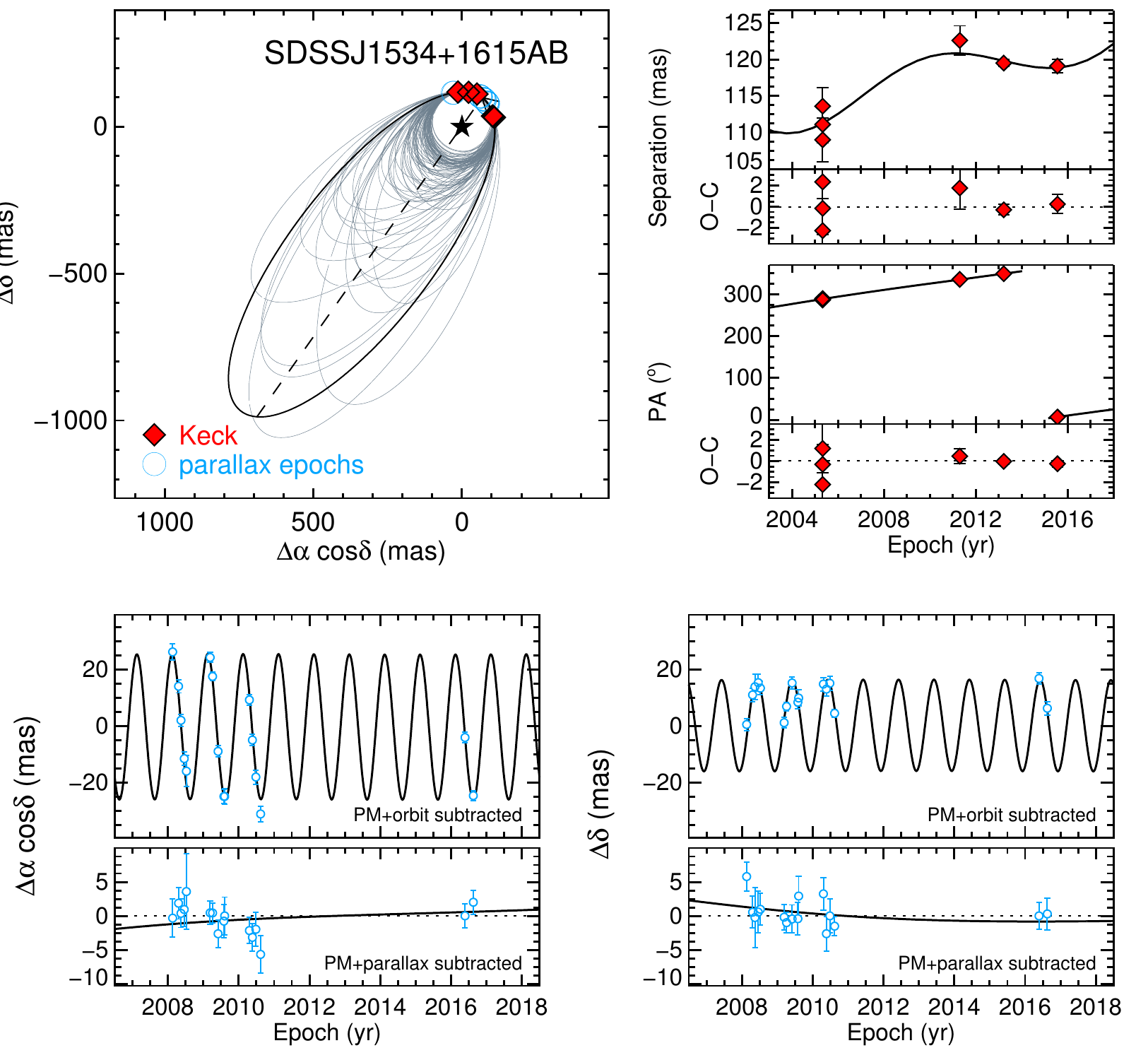}}  \ContinuedFloat  \caption{\normalsize (Continued)} \end{figure} \clearpage
\begin{figure} \centerline{\includegraphics[width=6.5in,angle=0]{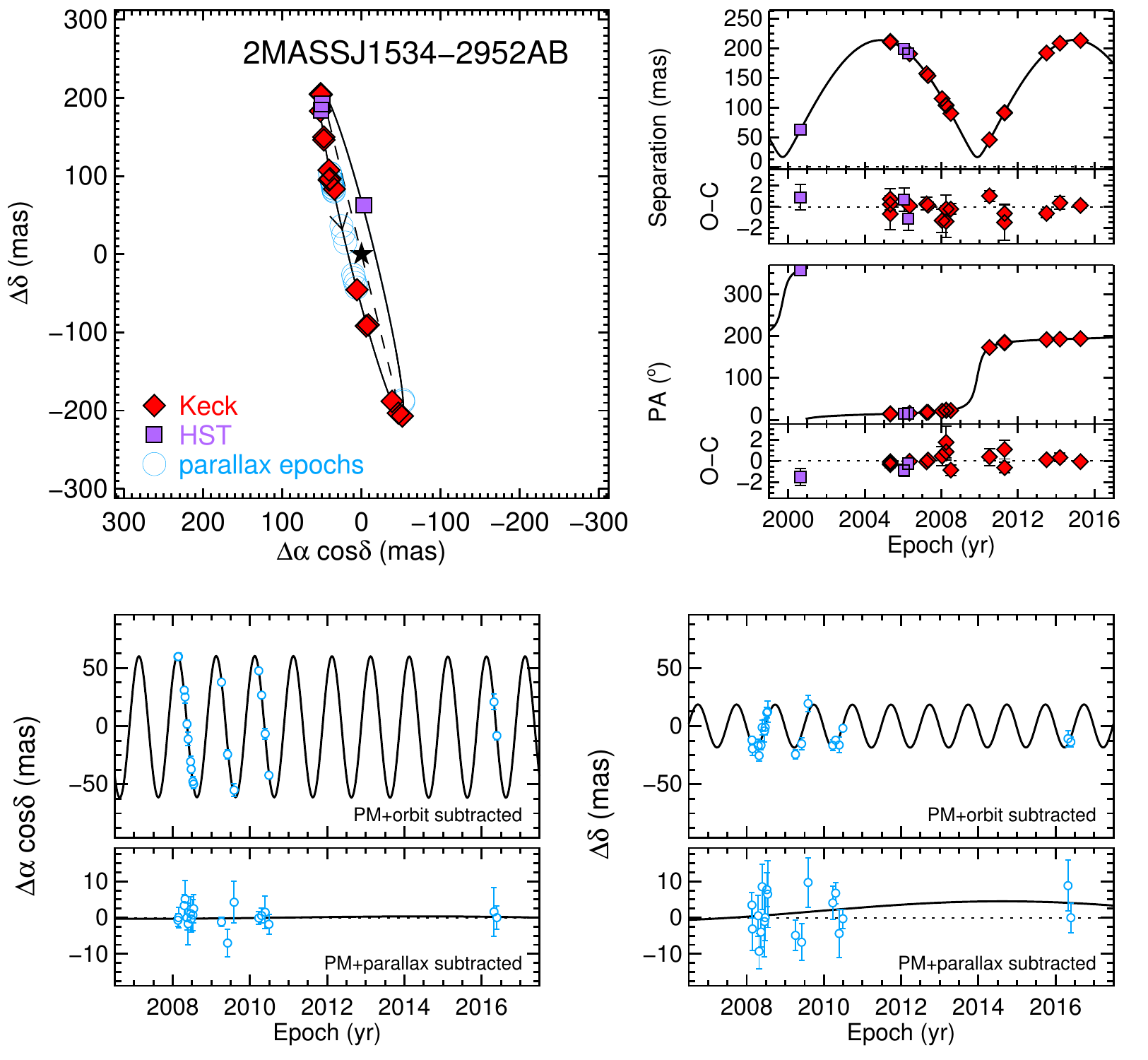}}  \ContinuedFloat  \caption{\normalsize (Continued)} \end{figure} \clearpage
\begin{figure} \centerline{\includegraphics[width=6.5in,angle=0]{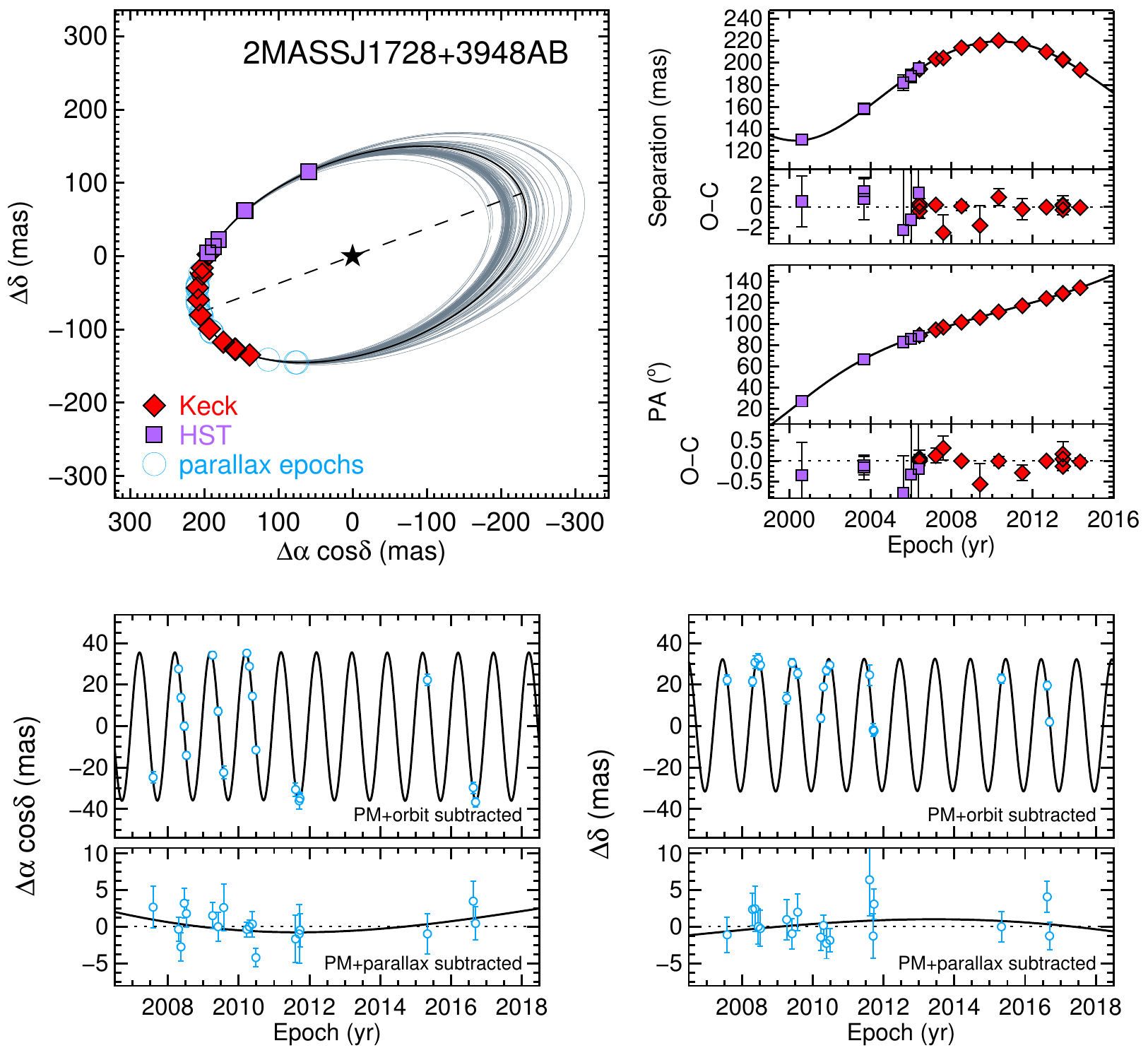}}  \ContinuedFloat  \caption{\normalsize (Continued)} \end{figure} \clearpage
\begin{figure} \centerline{\includegraphics[width=6.5in,angle=0]{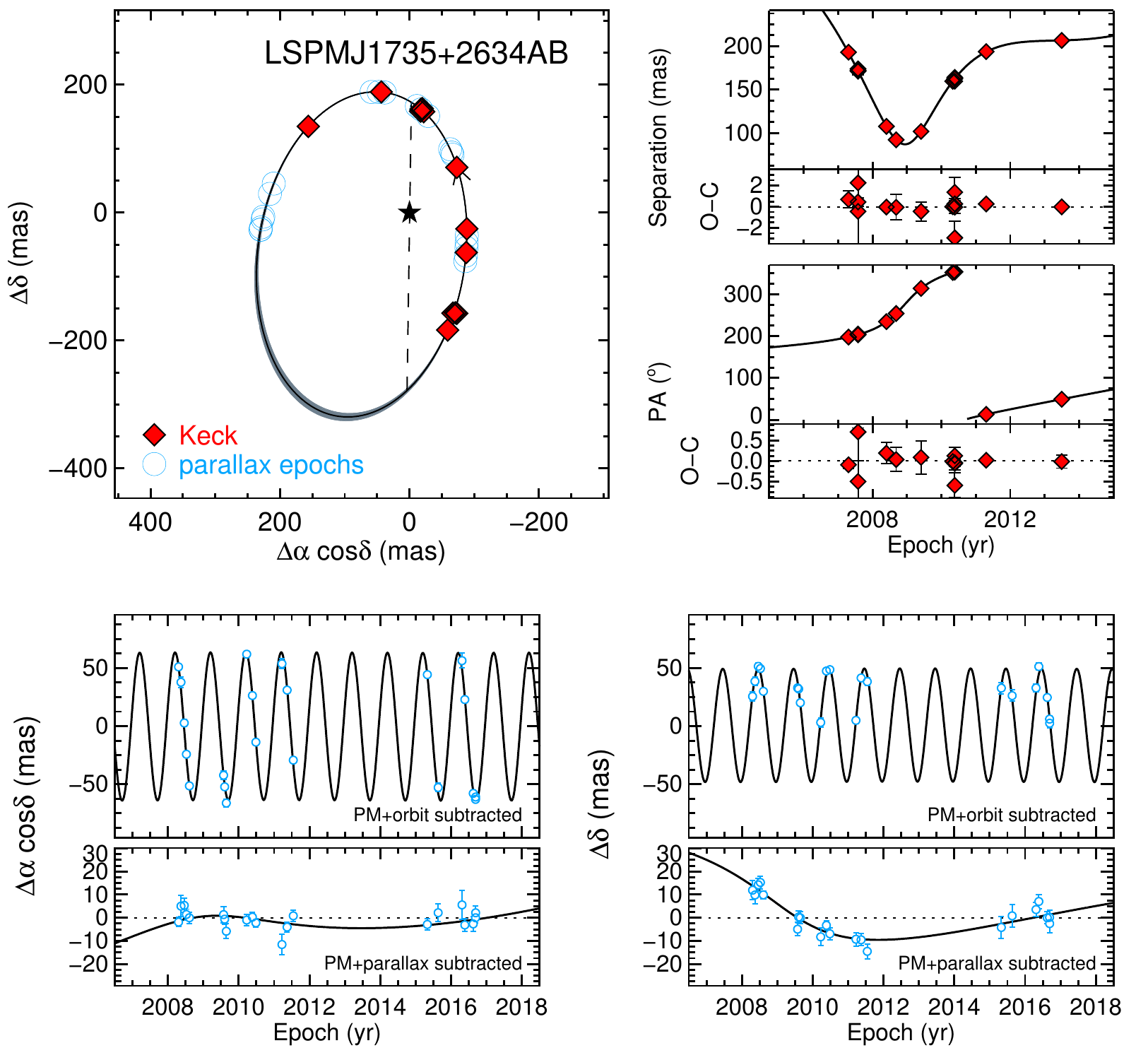}}  \ContinuedFloat  \caption{\normalsize (Continued)} \end{figure} \clearpage
\begin{figure} \centerline{\includegraphics[width=6.5in,angle=0]{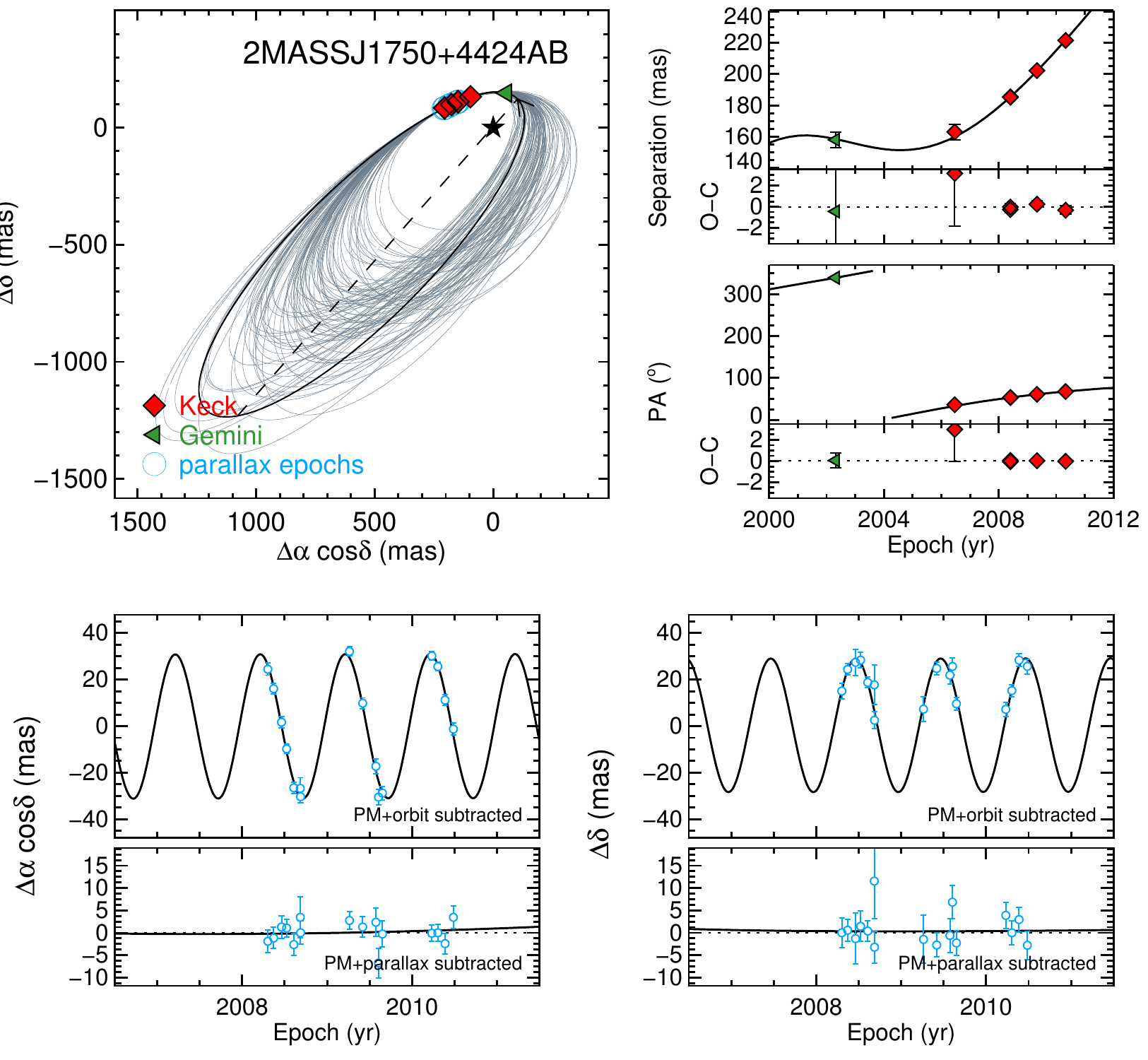}}  \ContinuedFloat  \caption{\normalsize (Continued)} \end{figure} \clearpage
\begin{figure} \centerline{\includegraphics[width=6.5in,angle=0]{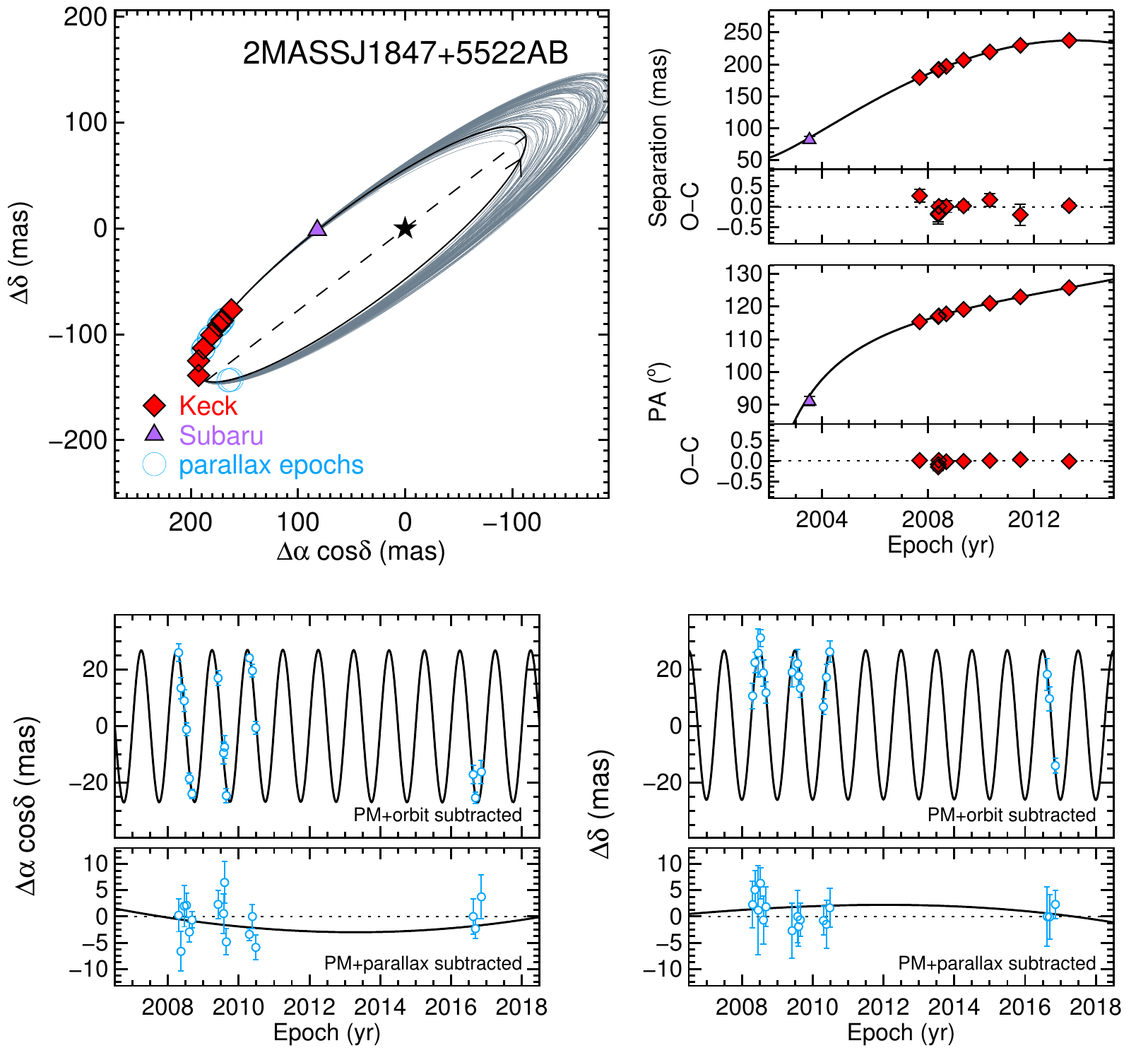}}  \ContinuedFloat  \caption{\normalsize (Continued)} \end{figure} \clearpage
\begin{figure} \centerline{\includegraphics[width=6.5in,angle=0]{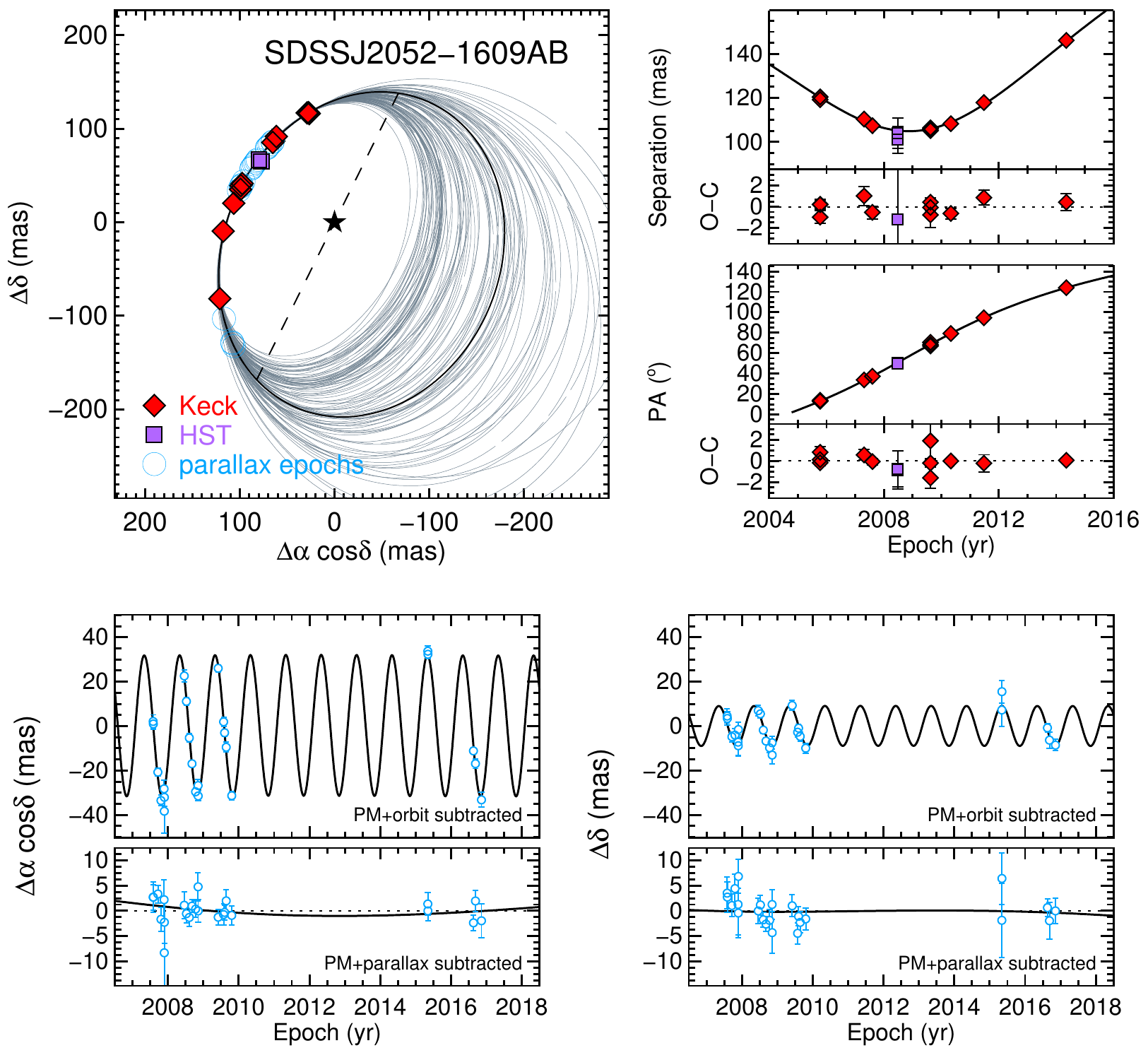}}  \ContinuedFloat  \caption{\normalsize (Continued)} \end{figure} \clearpage
\begin{figure} \centerline{\includegraphics[width=6.5in,angle=0]{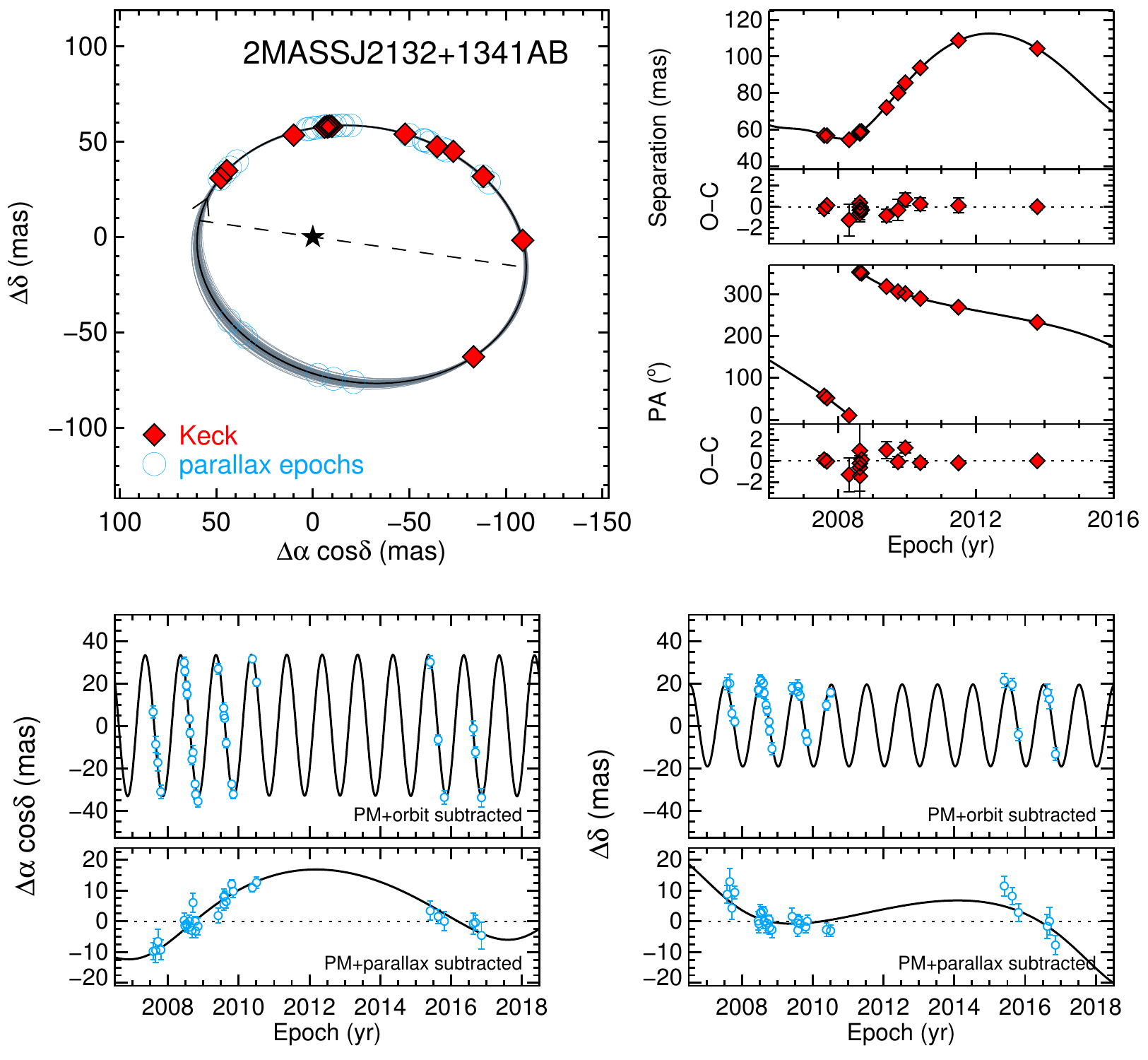}}  \ContinuedFloat  \caption{\normalsize (Continued)} \end{figure} \clearpage
\begin{figure} \centerline{\includegraphics[width=6.5in,angle=0]{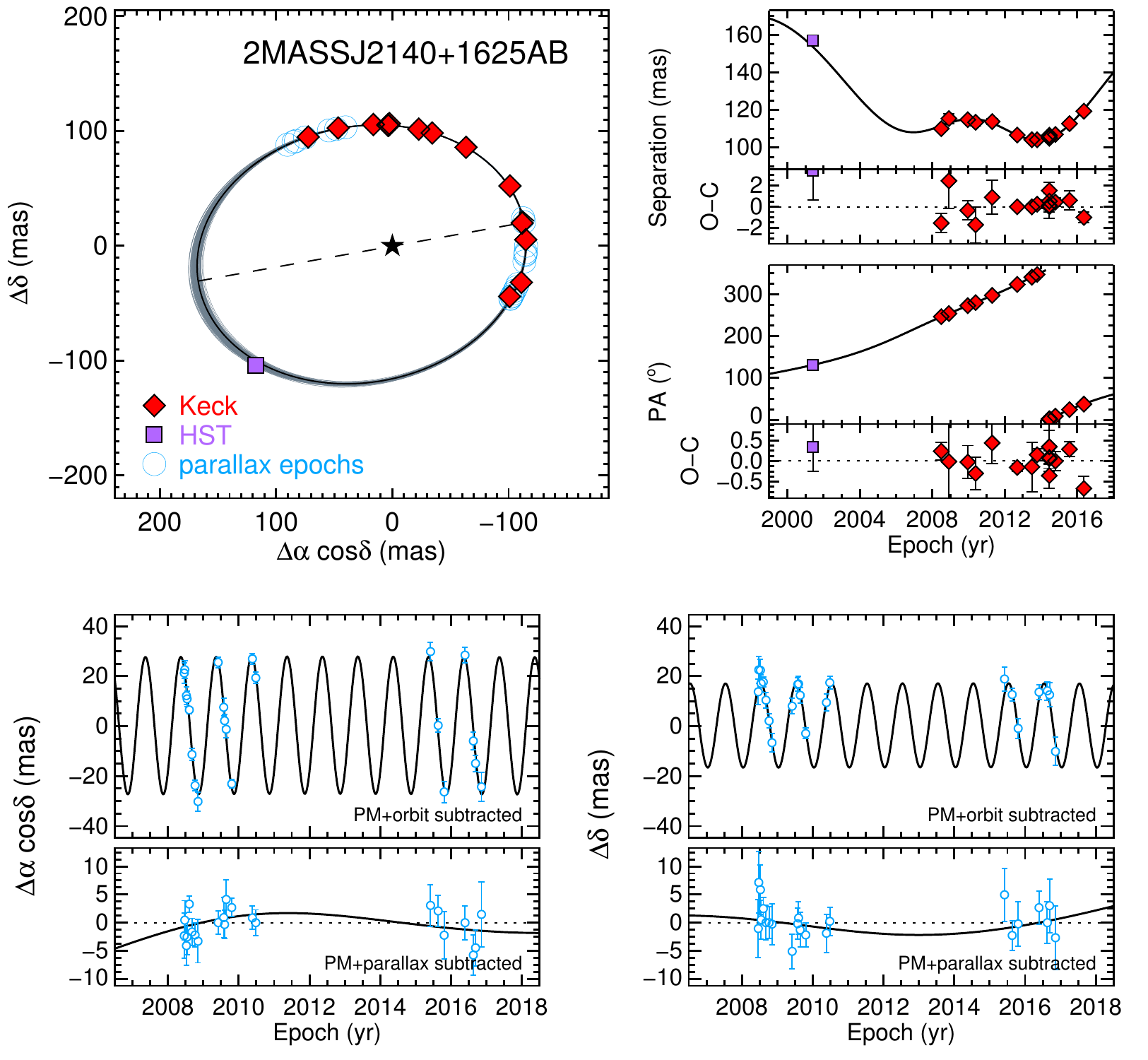}}  \ContinuedFloat  \caption{\normalsize (Continued)} \end{figure} \clearpage
\begin{figure} \centerline{\includegraphics[width=6.5in,angle=0]{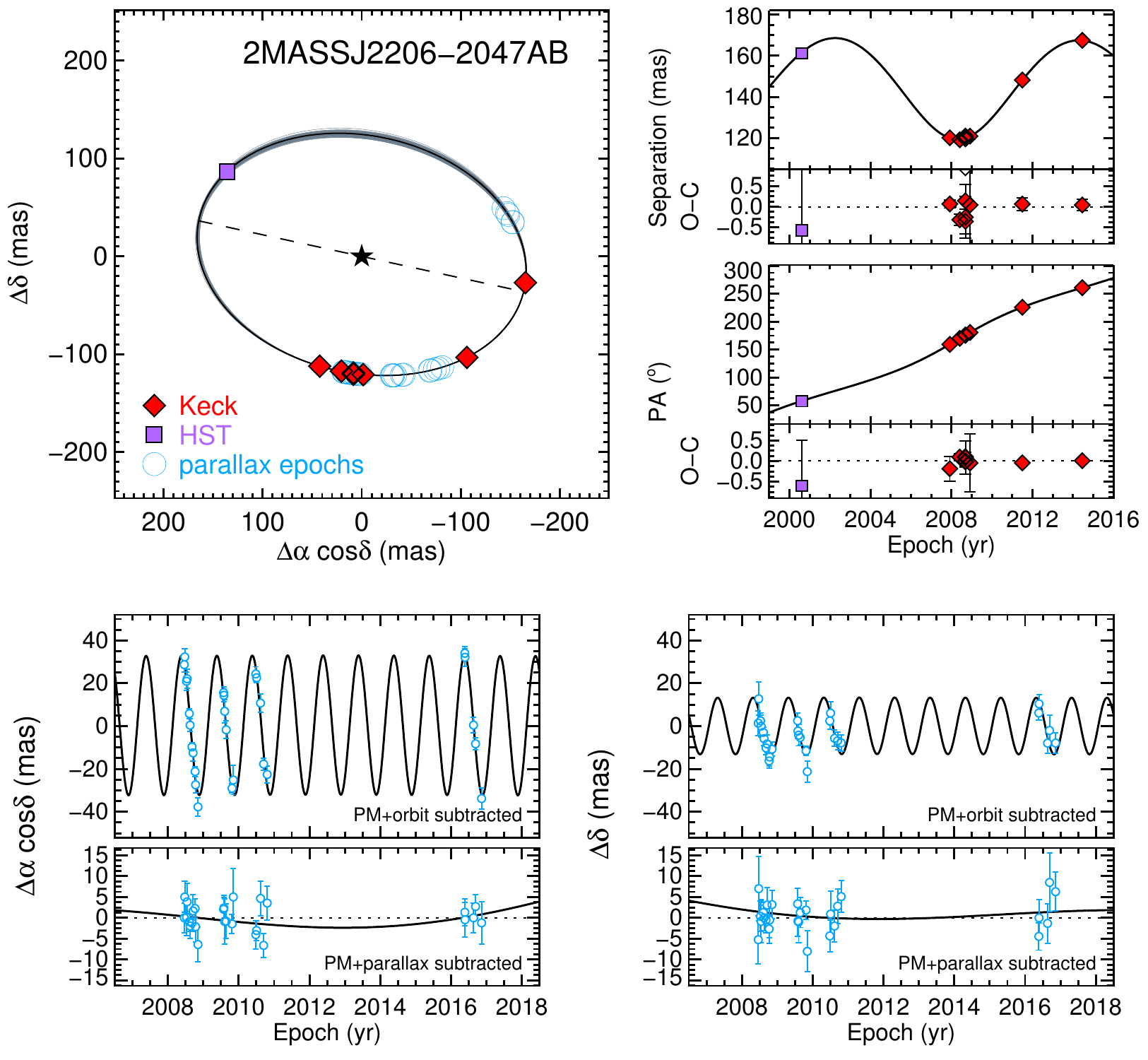}}  \ContinuedFloat  \caption{\normalsize (Continued)} \end{figure} \clearpage
\begin{figure} \centerline{\includegraphics[width=6.5in,angle=0]{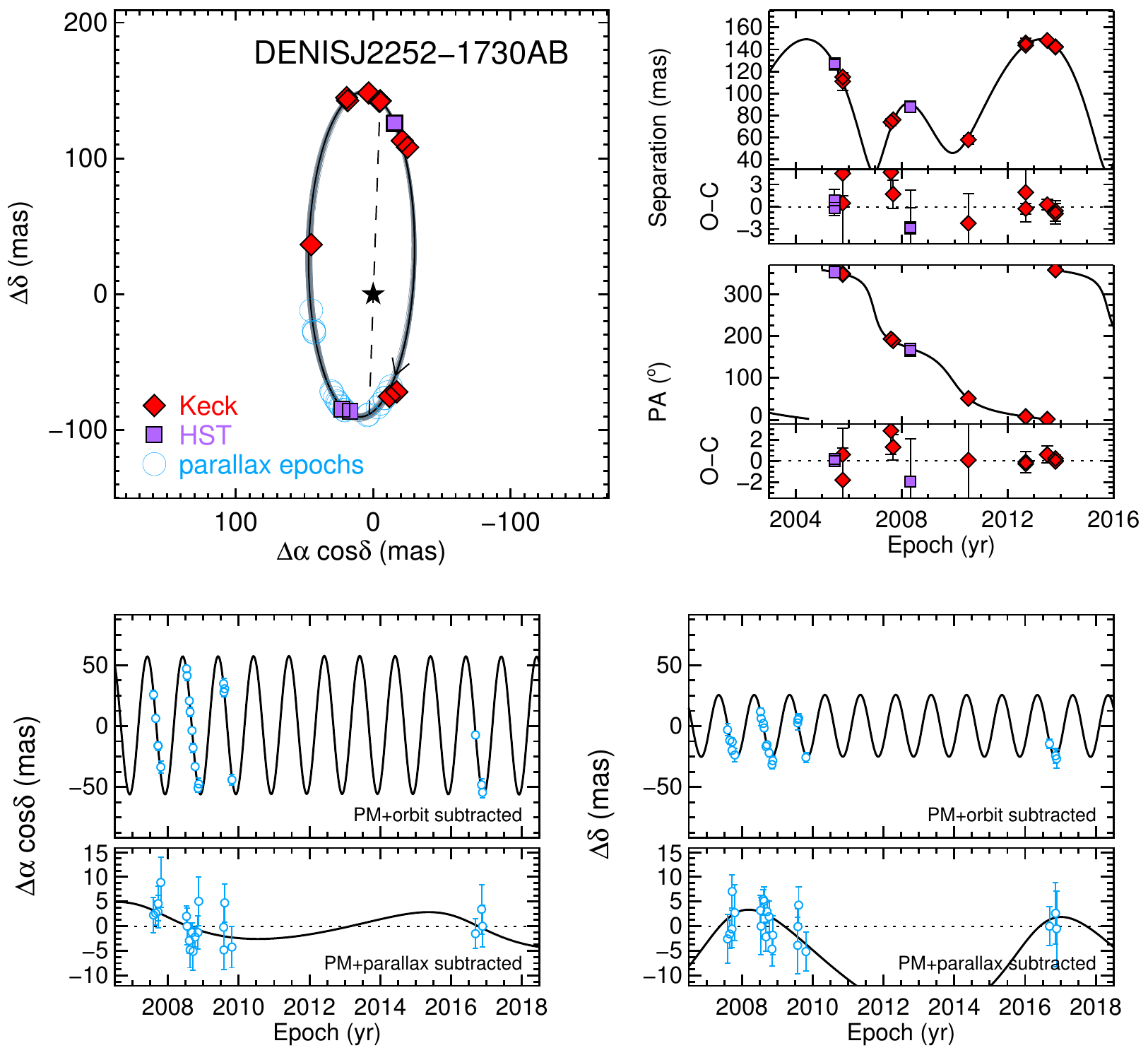}}  \ContinuedFloat  \caption{\normalsize (Continued)} \end{figure} \clearpage

\begin{figure} 
  \vskip -1.2in
  \centerline{\includegraphics[width=6.5in,angle=0]{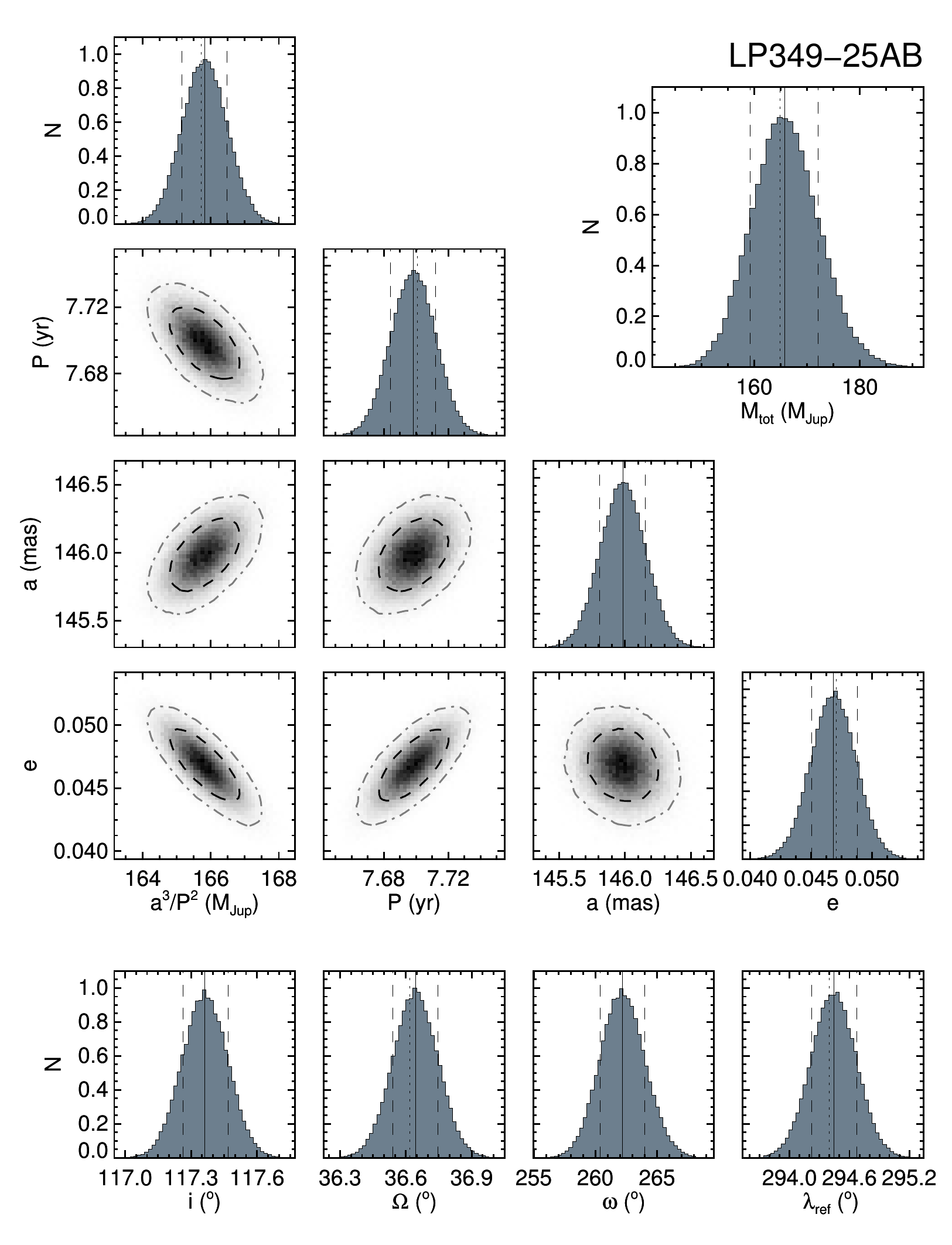}}
  \caption{\normalsize Posterior distributions of orbital parameters
    from our MCMC analysis.  In histograms, solid lines show the
    medians, dotted lines show the best-fit values, and dashed lines
    show the 68.3\% (1$\sigma$) credible intervals.  In contour plots,
    regions containing 68.3\% and 95.4\% of the posterior are
    indicated by black dashed lines and gray dash-dotted lines,
    respectively. \emph{Top right:} the final posterior in total
    system mass after including both the orbit and parallax
    uncertainties.  \emph{Middle triangle plot:} Histograms and
    correlations for the astrophysical parameters of mass, period,
    semimajor axis, and eccentricity.  In the triangle plot neither
    mass nor semimajor axis includes the error in distance from the
    parallax, thus showing the uncertainties and correlations in
    parameters from the orbit fit alone.  \emph{Bottom:} Histograms of
    the various viewing angles that are part of the orbit
    fit. \label{fig:orb-hist}}
\end{figure}
\clearpage
\begin{figure} \vskip -0.4in \centerline{\includegraphics[width=6.5in,angle=0]{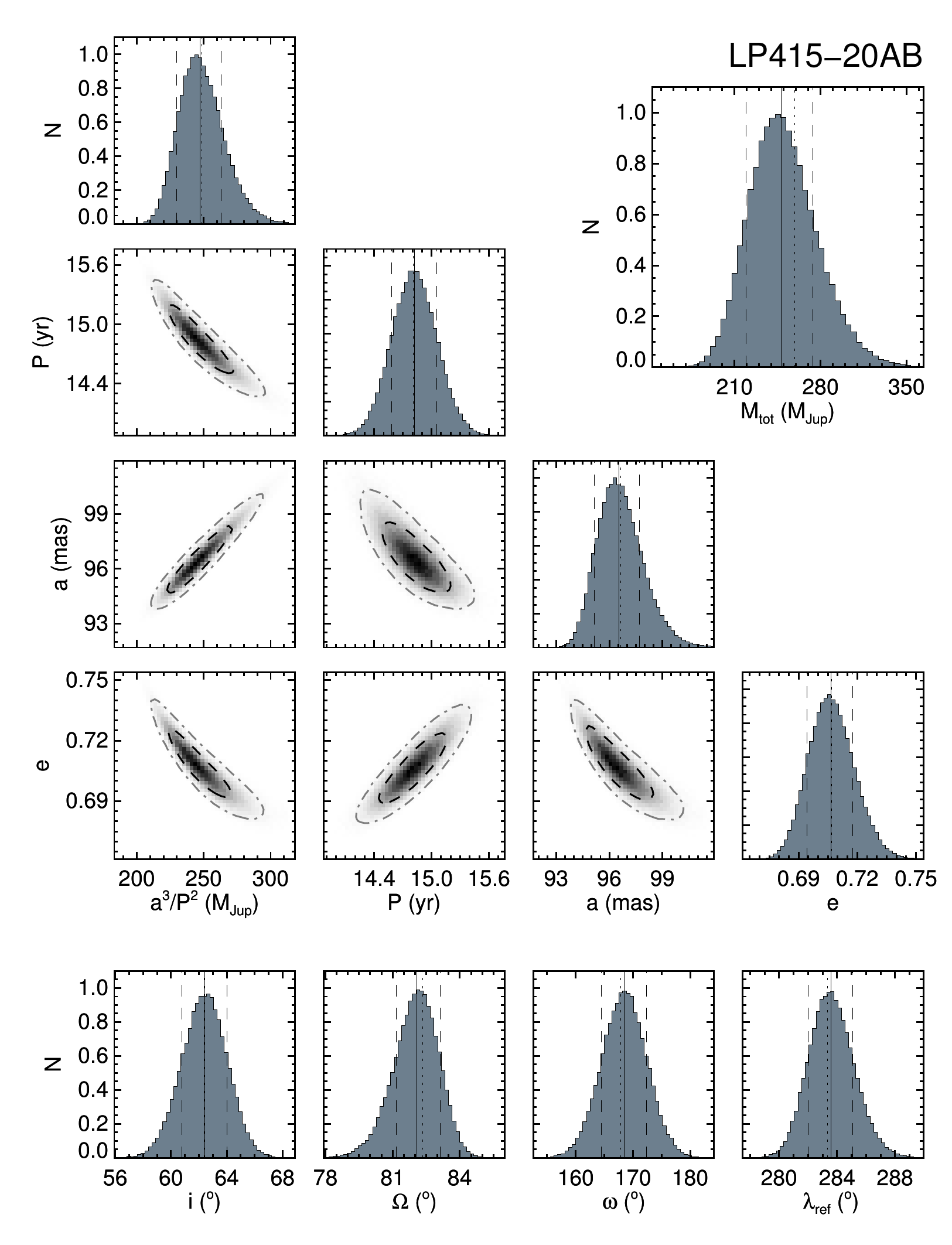}}   \ContinuedFloat  \caption{\normalsize (Continued)} \end{figure} \clearpage
\begin{figure} \vskip -0.4in \centerline{\includegraphics[width=6.5in,angle=0]{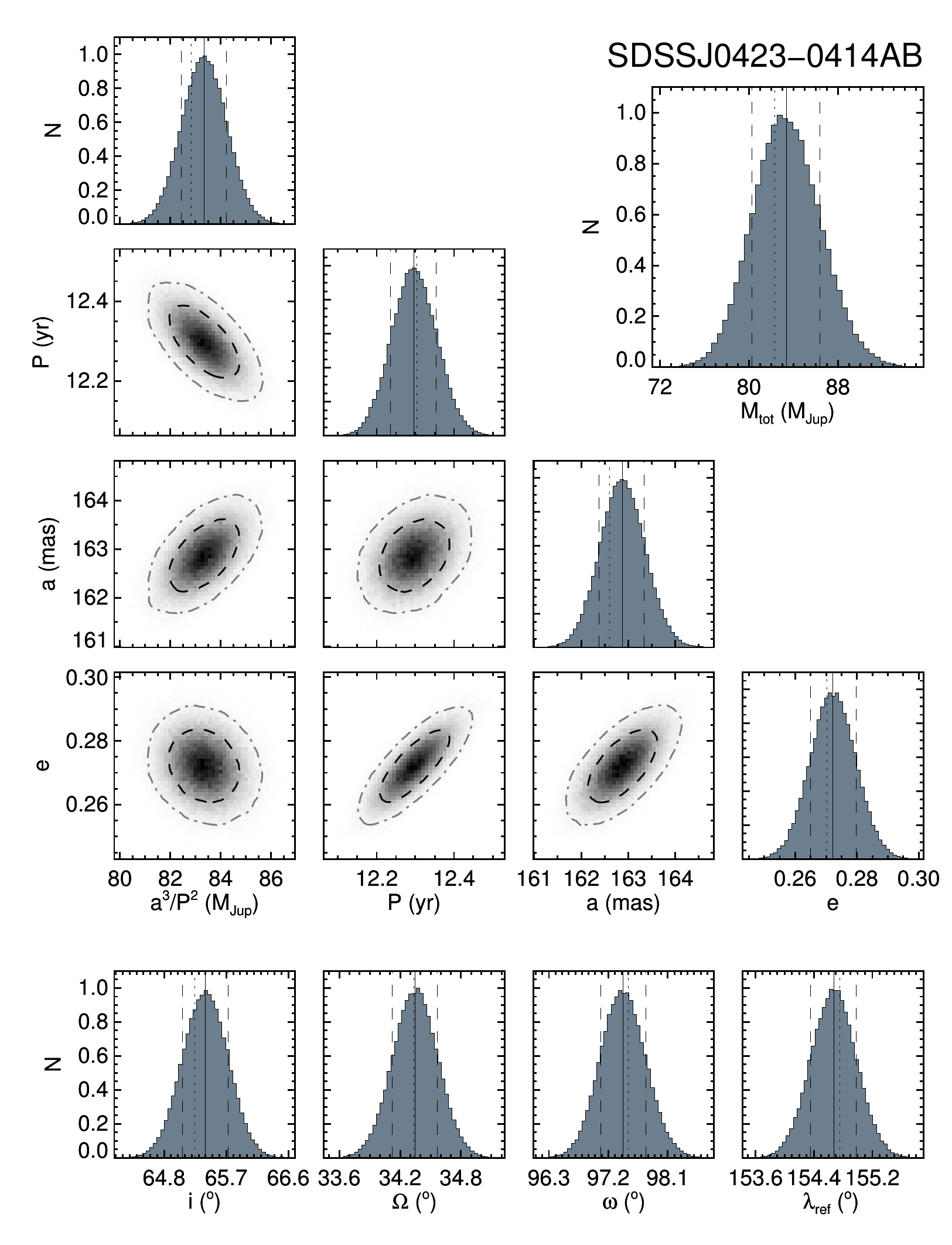}}  \ContinuedFloat  \caption{\normalsize (Continued)} \end{figure} \clearpage
\begin{figure} \vskip -0.4in \centerline{\includegraphics[width=6.5in,angle=0]{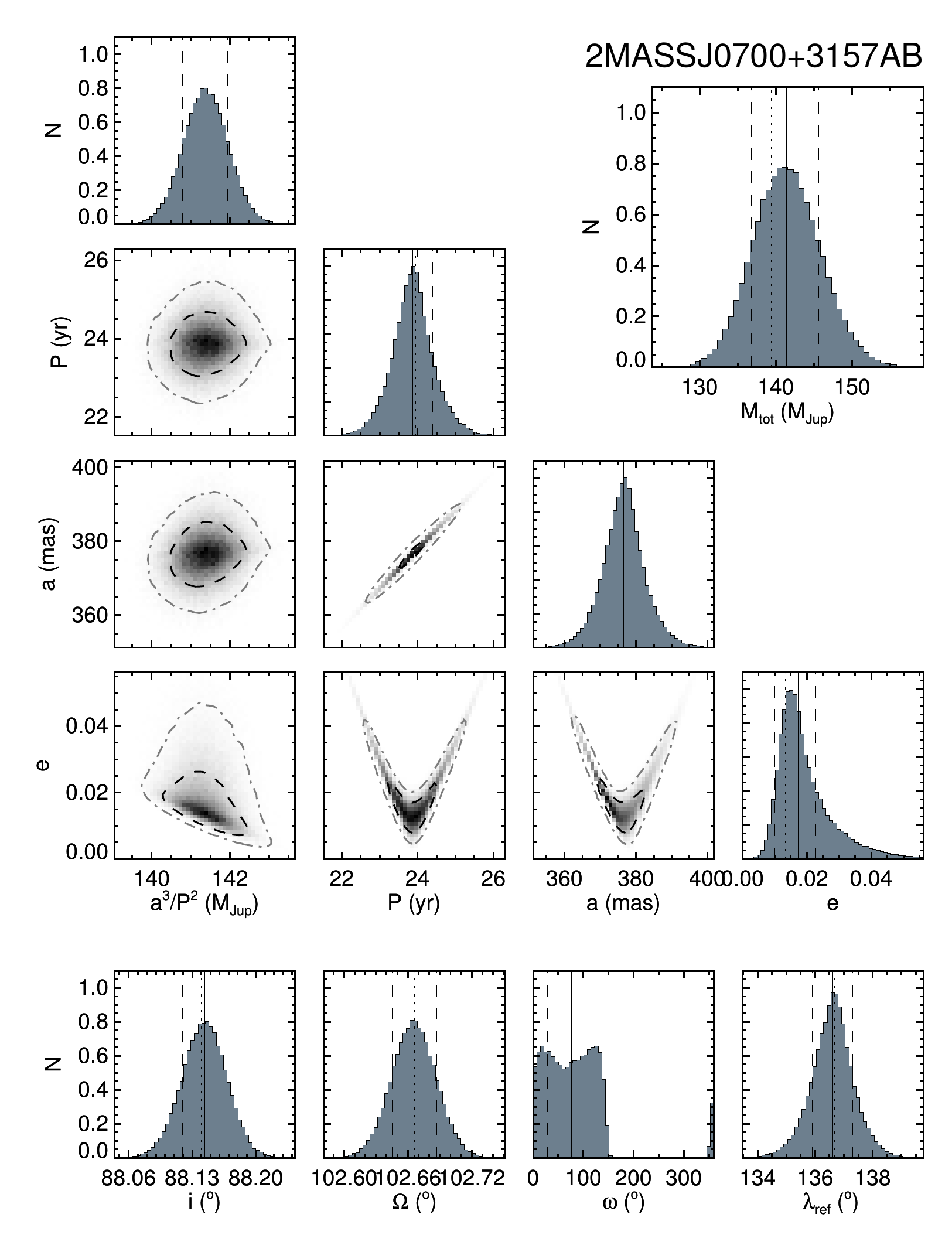}}  \ContinuedFloat  \caption{\normalsize (Continued)} \end{figure} \clearpage
\begin{figure} \vskip -0.4in \centerline{\includegraphics[width=6.5in,angle=0]{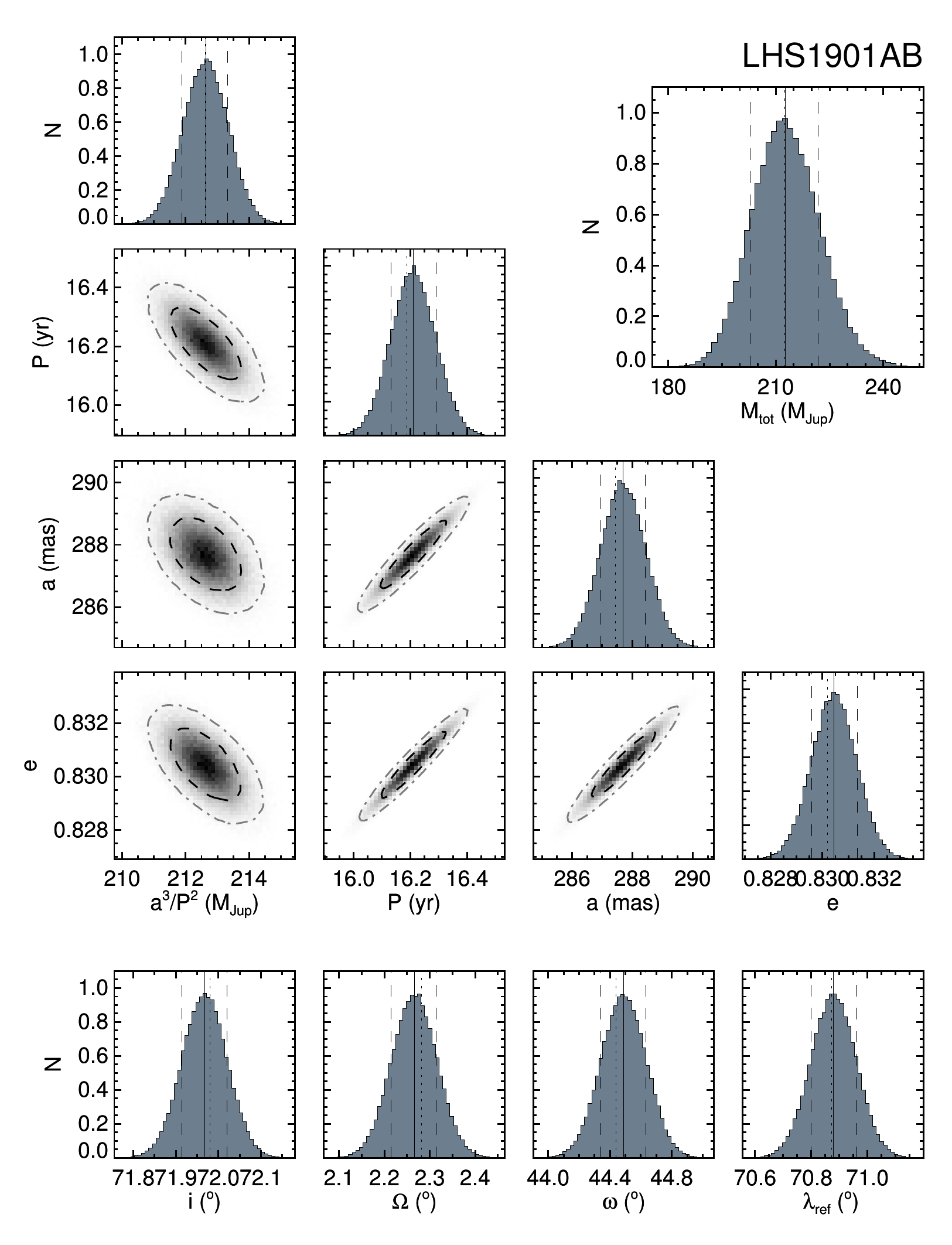}}    \ContinuedFloat  \caption{\normalsize (Continued)} \end{figure} \clearpage
\begin{figure} \vskip -0.4in \centerline{\includegraphics[width=6.5in,angle=0]{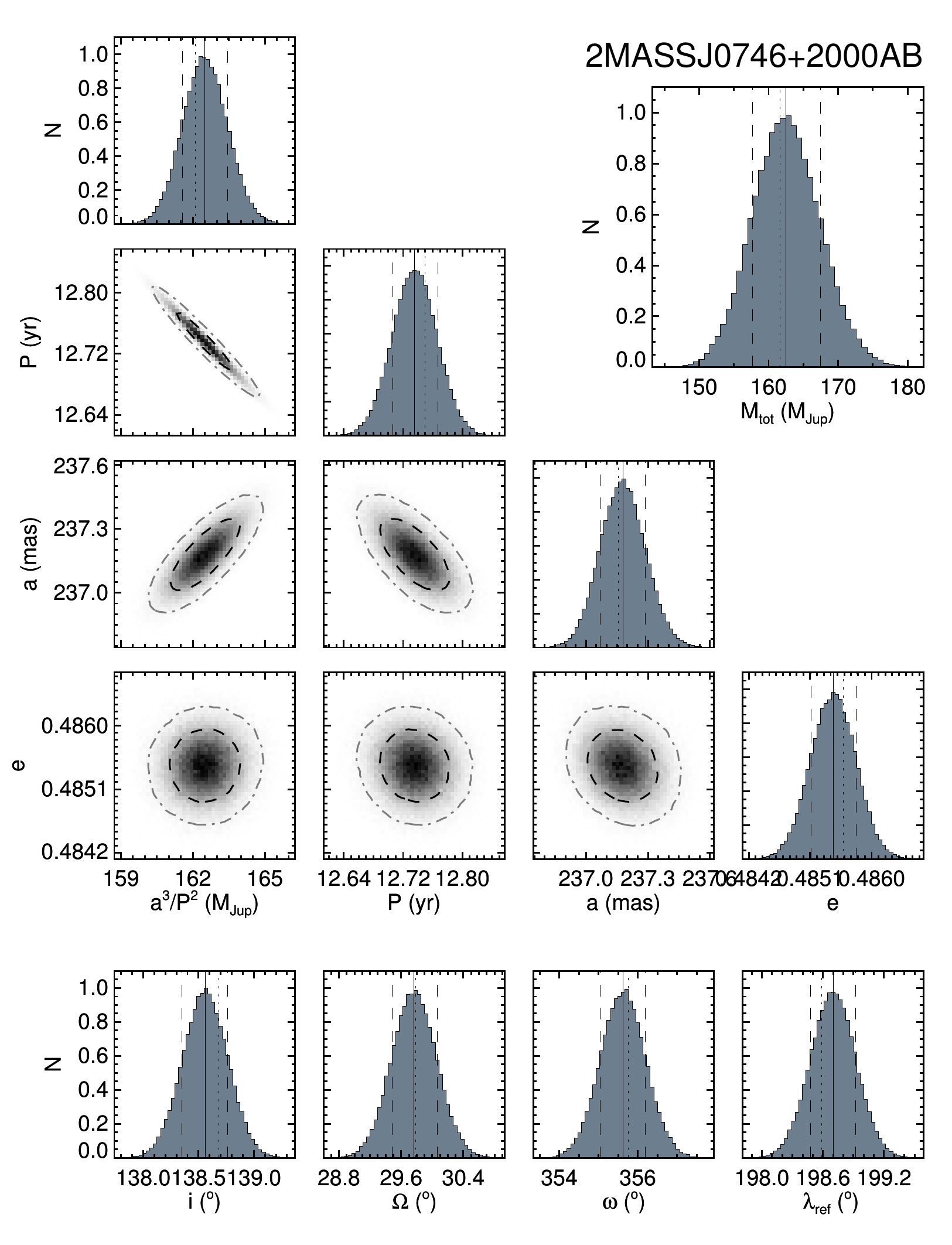}}  \ContinuedFloat  \caption{\normalsize (Continued)} \end{figure} \clearpage
\begin{figure} \vskip -0.4in \centerline{\includegraphics[width=6.5in,angle=0]{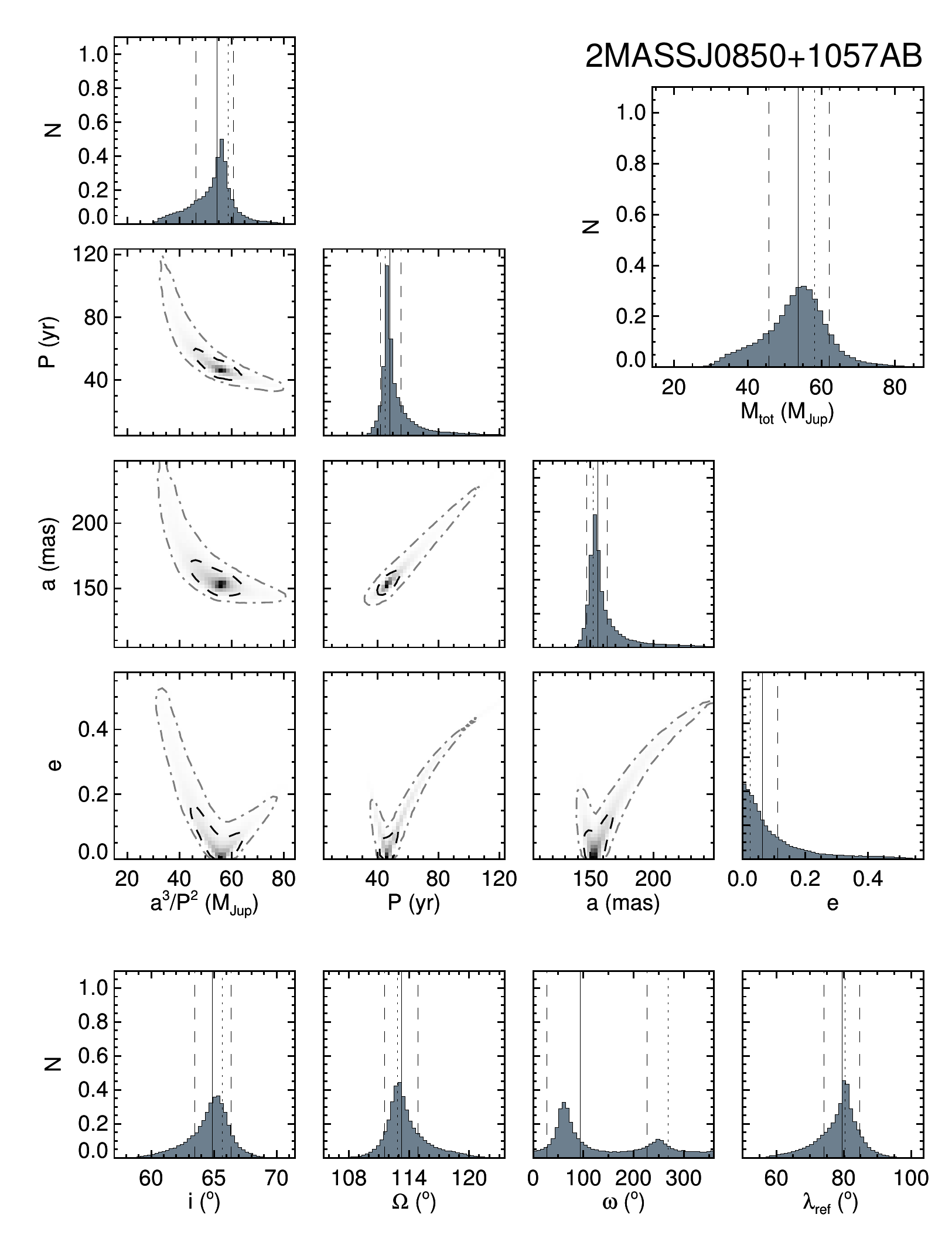}}  \ContinuedFloat  \caption{\normalsize (Continued)} \end{figure} \clearpage
\begin{figure} \vskip -0.4in \centerline{\includegraphics[width=6.5in,angle=0]{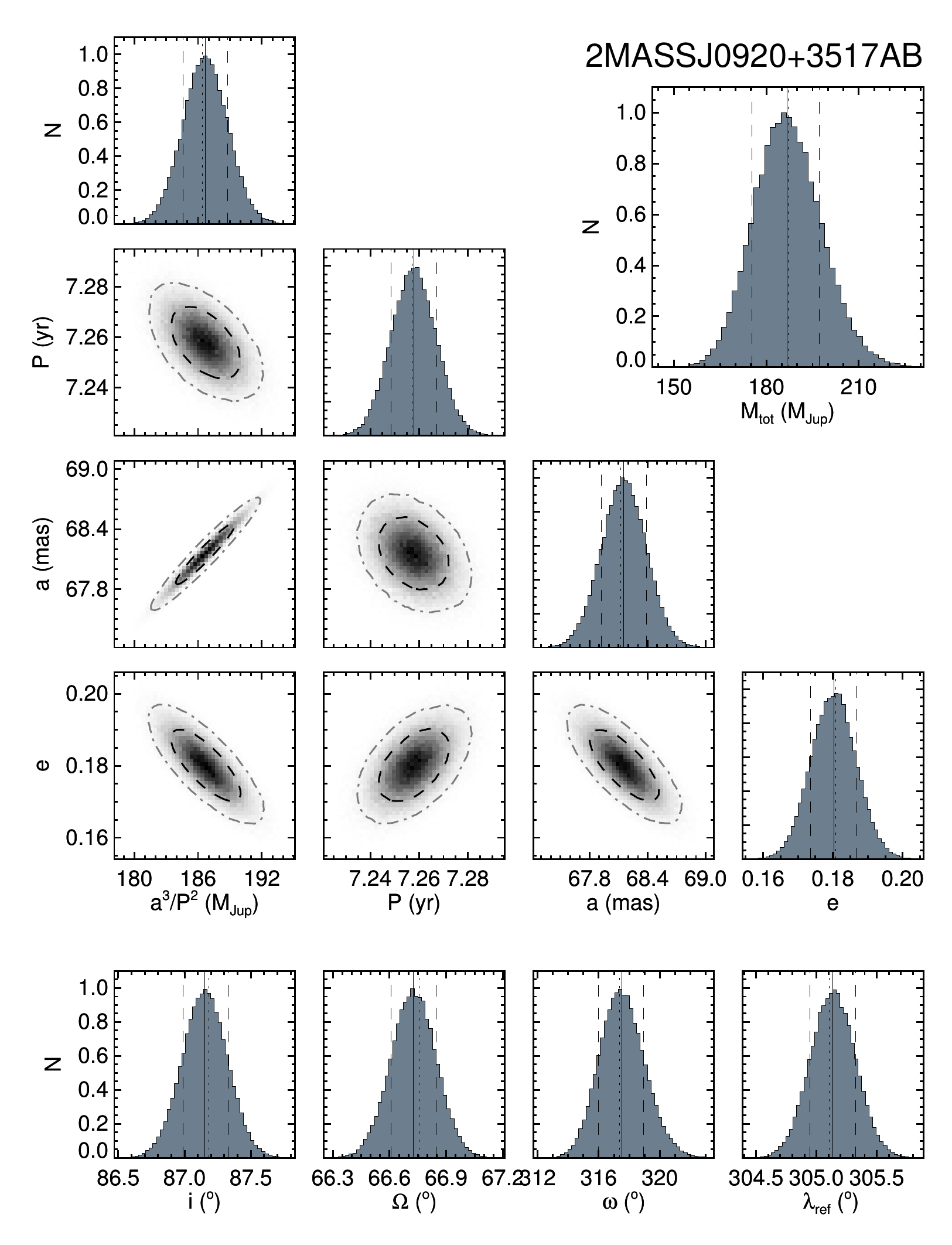}}  \ContinuedFloat  \caption{\normalsize (Continued)} \end{figure} \clearpage
\begin{figure} \vskip -0.4in \centerline{\includegraphics[width=6.5in,angle=0]{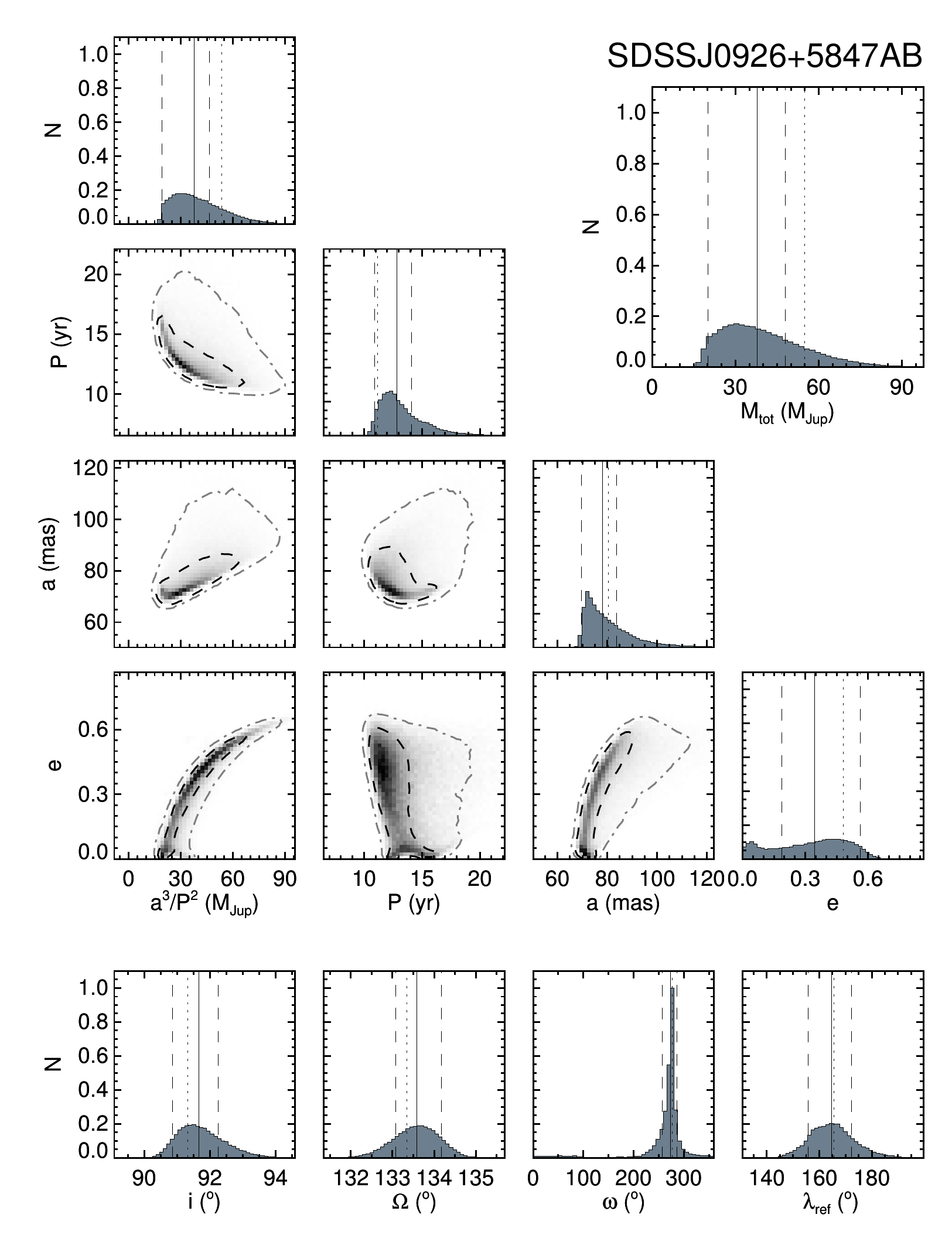}}  \ContinuedFloat  \caption{\normalsize (Continued)} \end{figure} \clearpage
\begin{figure} \vskip -0.4in \centerline{\includegraphics[width=6.5in,angle=0]{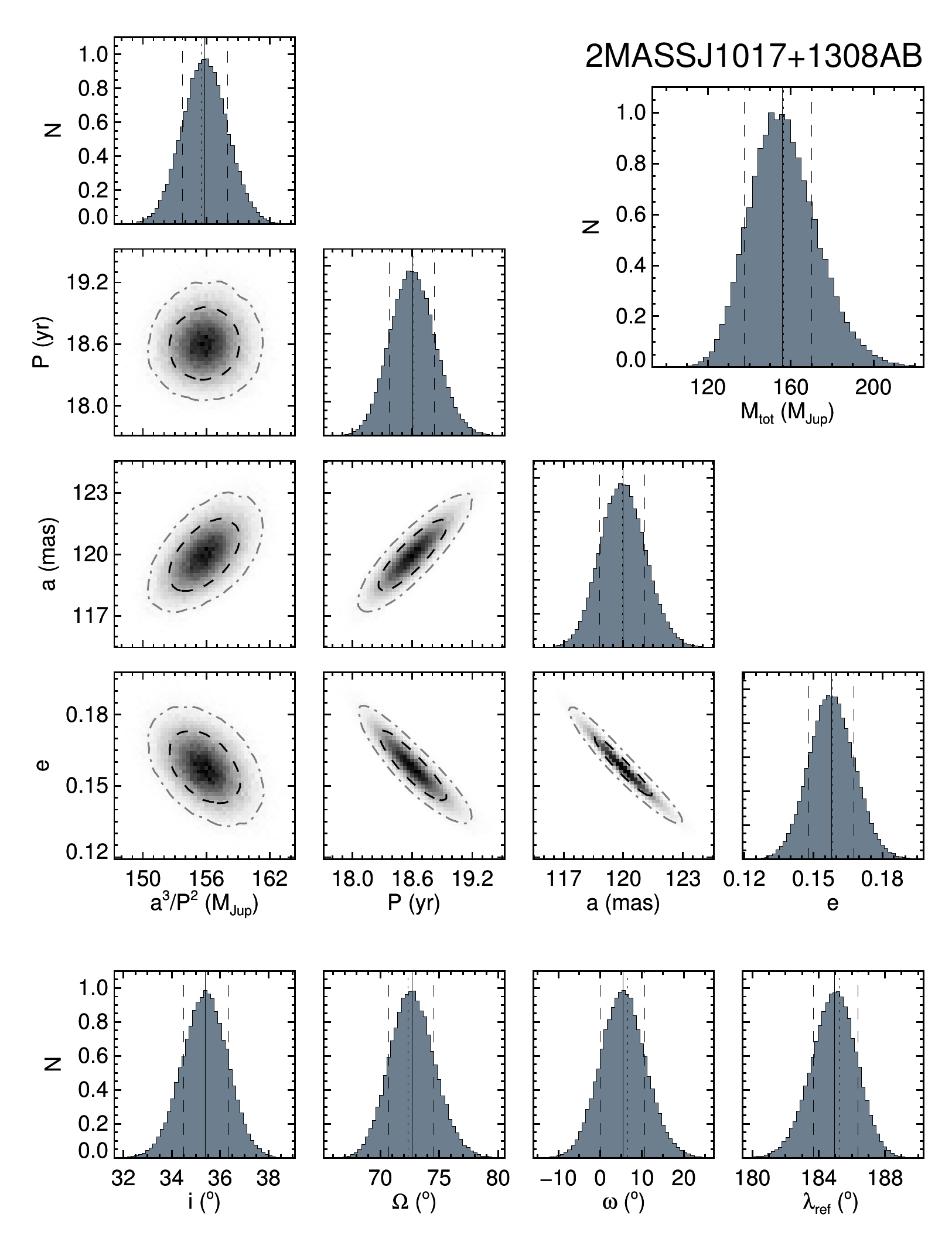}}  \ContinuedFloat  \caption{\normalsize (Continued)} \end{figure} \clearpage
\begin{figure} \vskip -0.4in \centerline{\includegraphics[width=6.5in,angle=0]{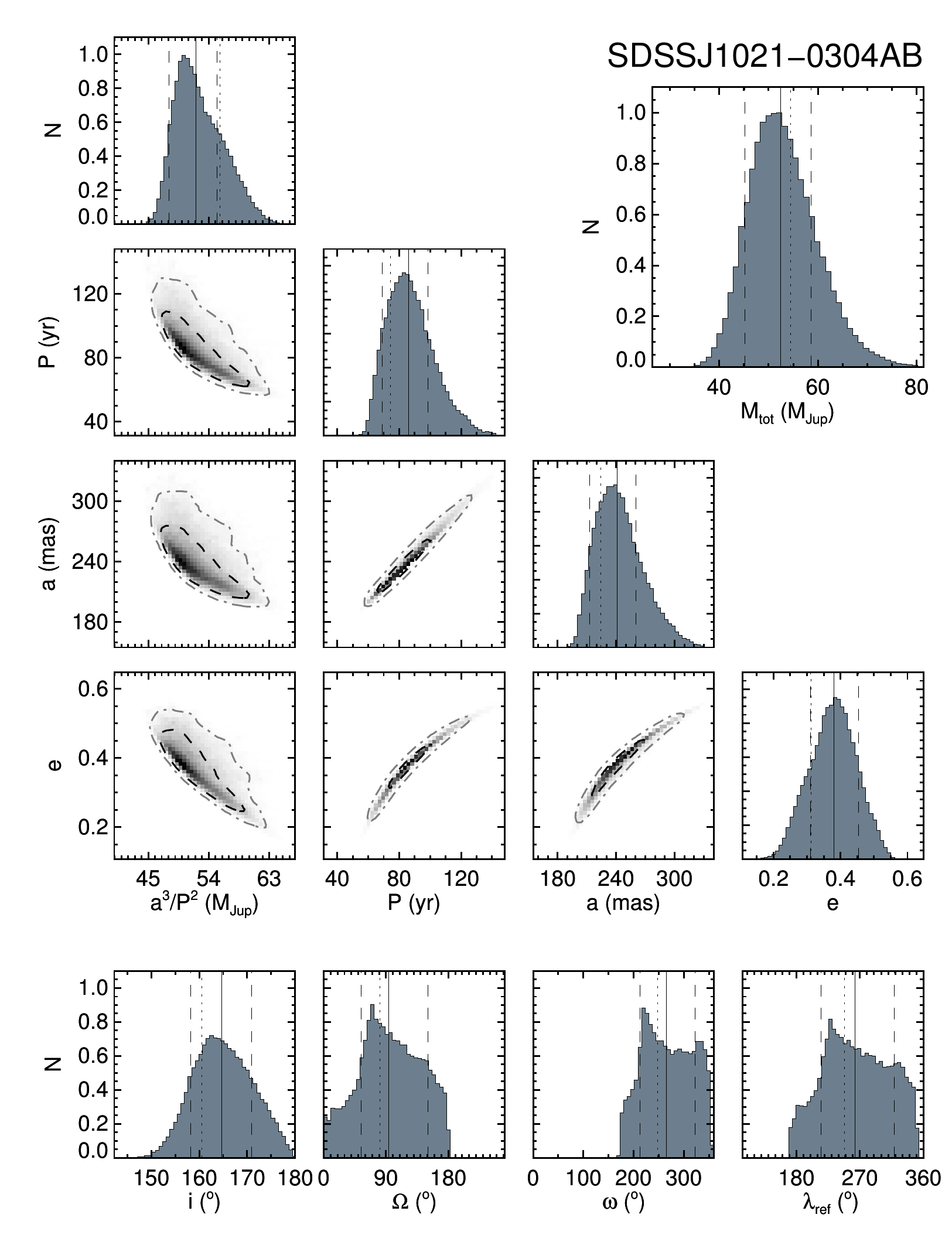}}  \ContinuedFloat  \caption{\normalsize (Continued)} \end{figure} \clearpage
\begin{figure} \vskip -0.4in \centerline{\includegraphics[width=6.5in,angle=0]{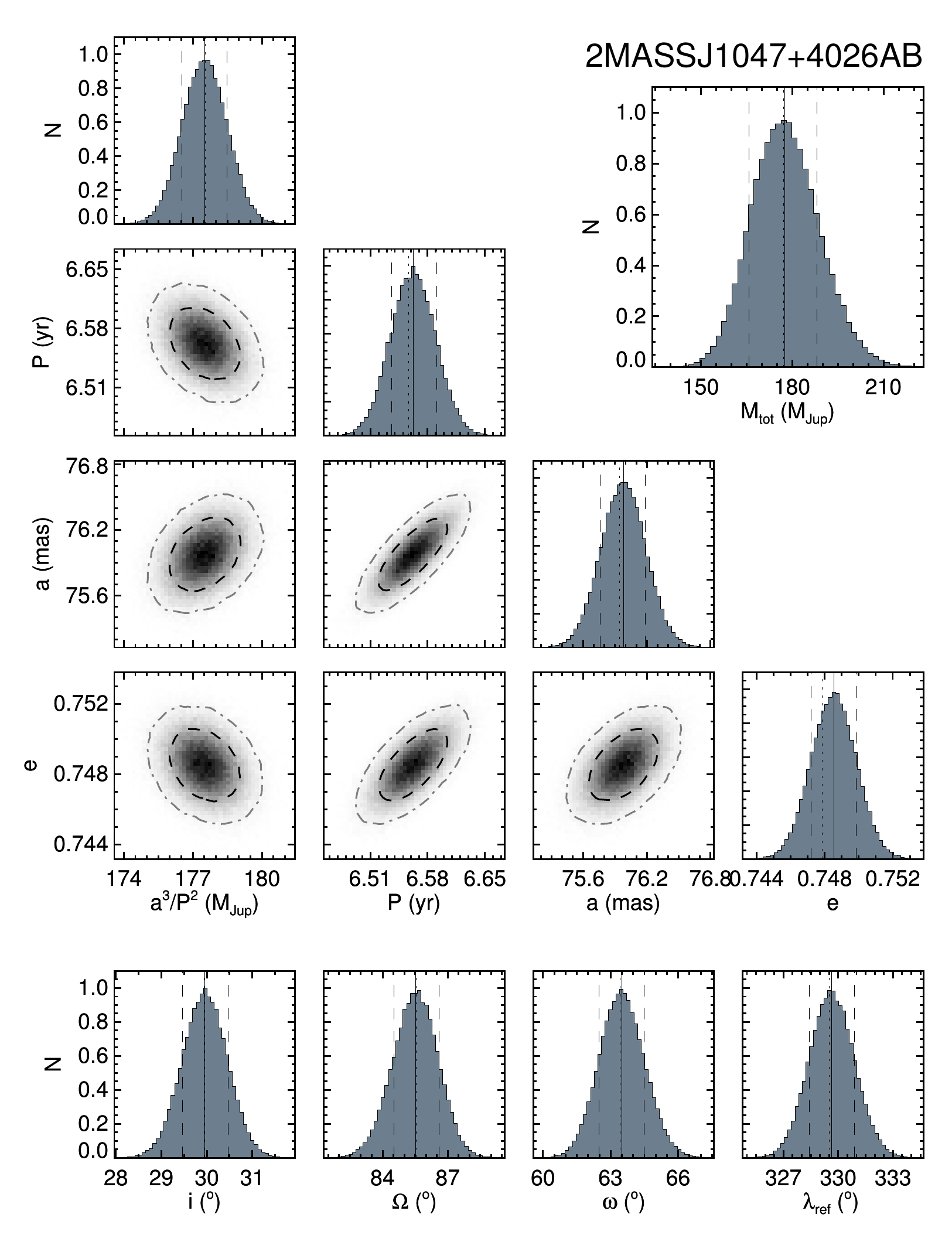}}  \ContinuedFloat  \caption{\normalsize (Continued)} \end{figure} \clearpage
\begin{figure} \vskip -0.4in \centerline{\includegraphics[width=6.5in,angle=0]{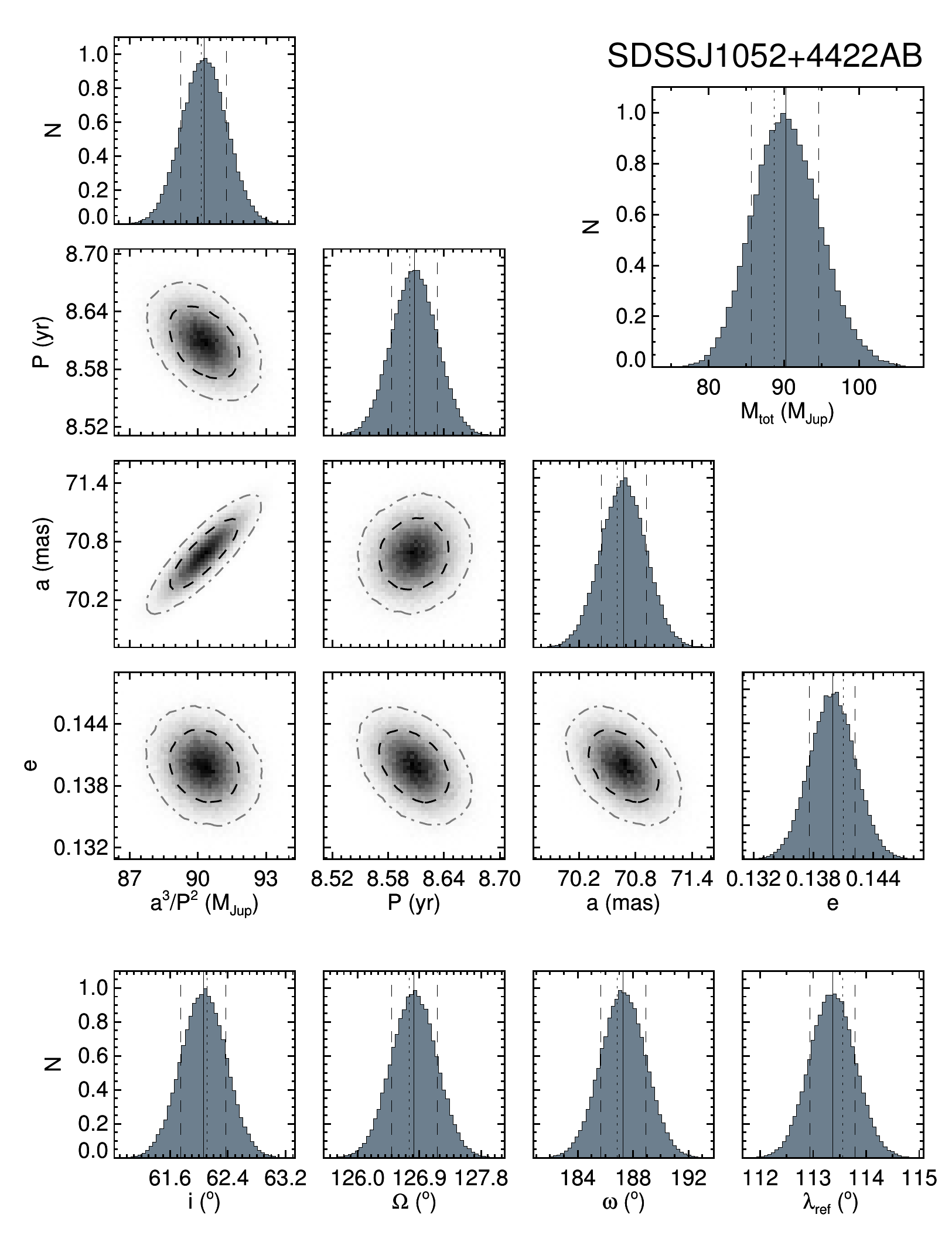}}  \ContinuedFloat  \caption{\normalsize (Continued)} \end{figure} \clearpage
\begin{figure} \vskip -0.4in \centerline{\includegraphics[width=6.5in,angle=0]{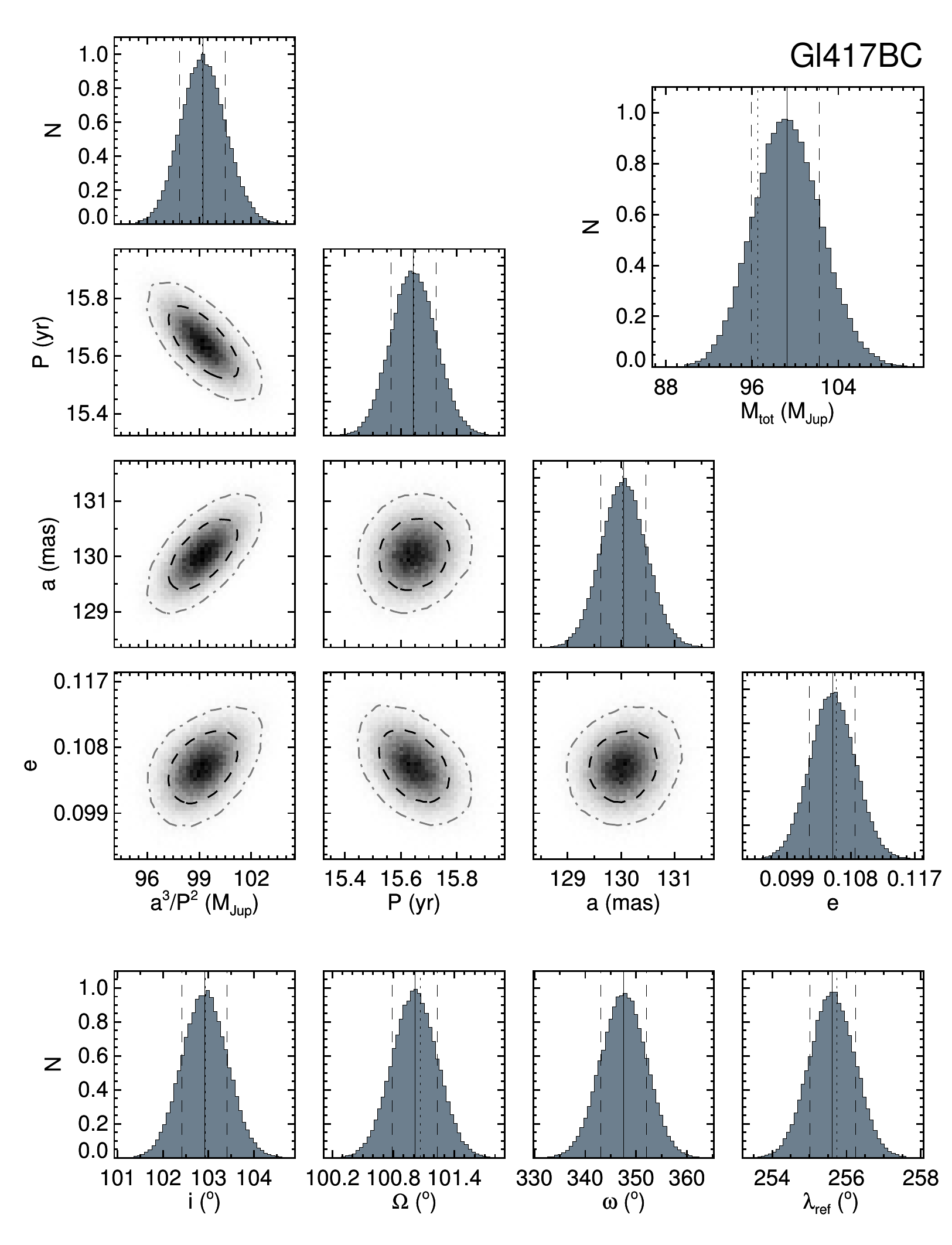}}     \ContinuedFloat  \caption{\normalsize (Continued)} \end{figure} \clearpage
\begin{figure} \vskip -0.4in \centerline{\includegraphics[width=6.5in,angle=0]{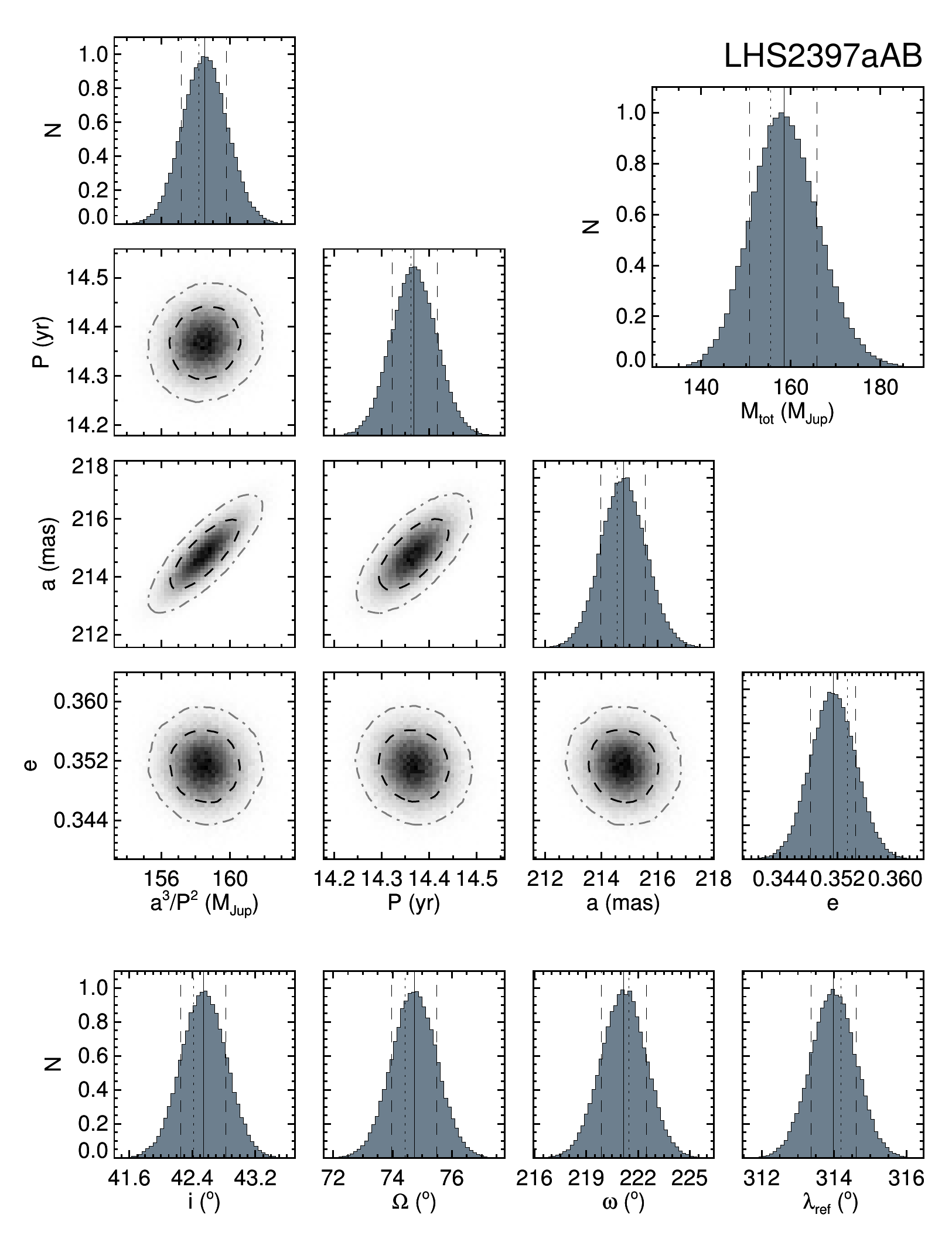}}   \ContinuedFloat  \caption{\normalsize (Continued)} \end{figure} \clearpage
\begin{figure} \vskip -0.4in \centerline{\includegraphics[width=6.5in,angle=0]{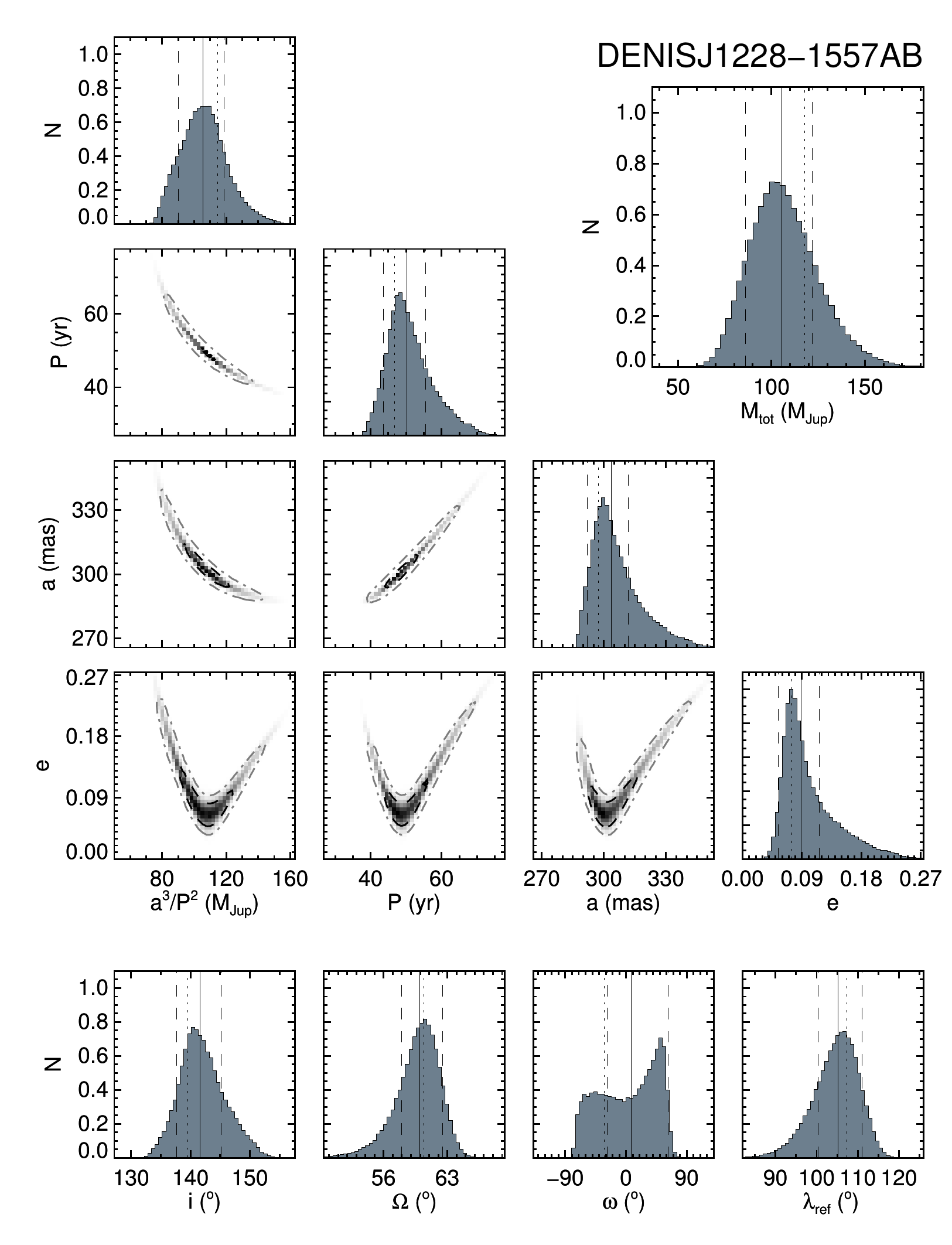}}  \ContinuedFloat  \caption{\normalsize (Continued)} \end{figure} \clearpage
\begin{figure} \vskip -0.4in \centerline{\includegraphics[width=6.5in,angle=0]{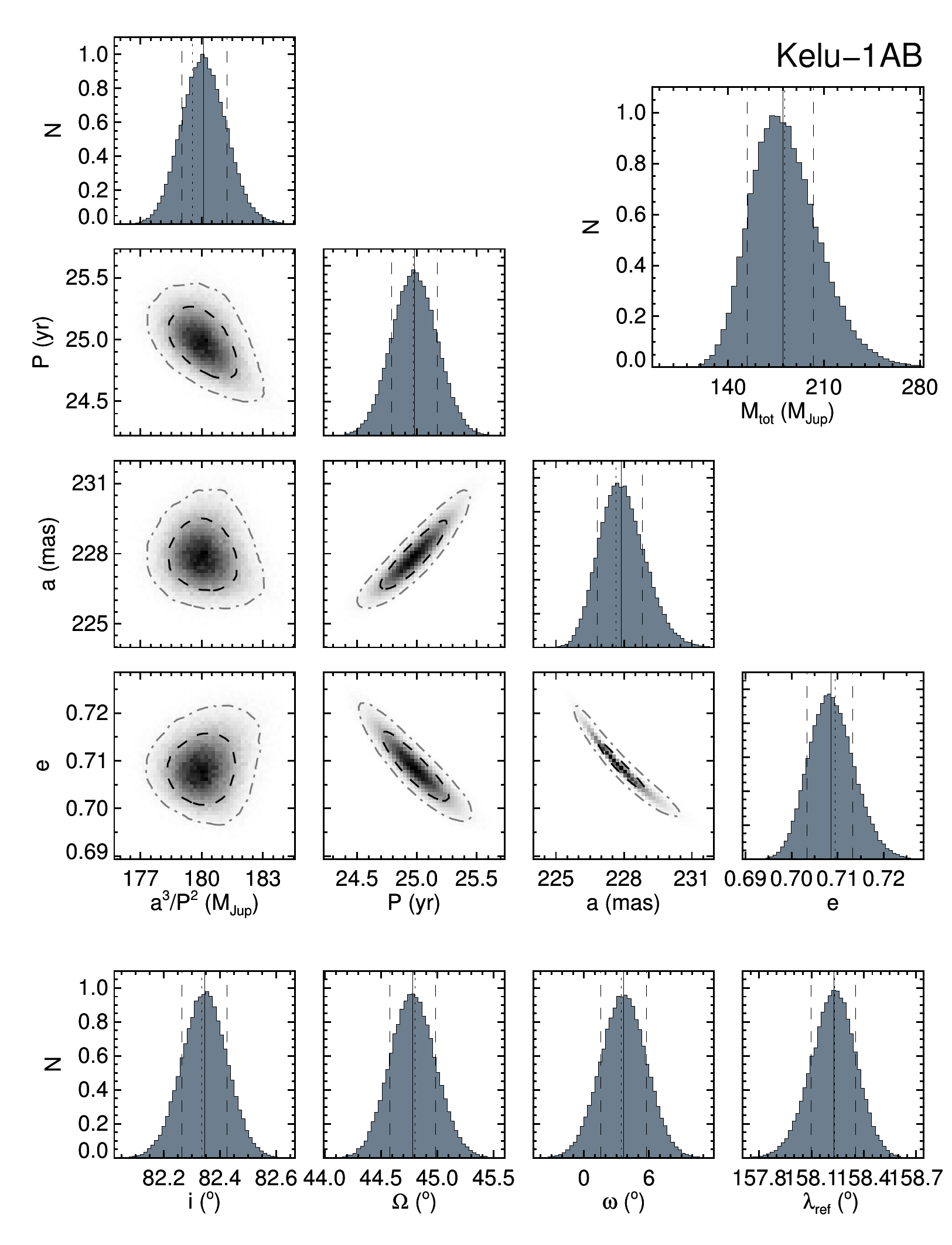}}     \ContinuedFloat  \caption{\normalsize (Continued)} \end{figure} \clearpage
\begin{figure} \vskip -0.4in \centerline{\includegraphics[width=6.5in,angle=0]{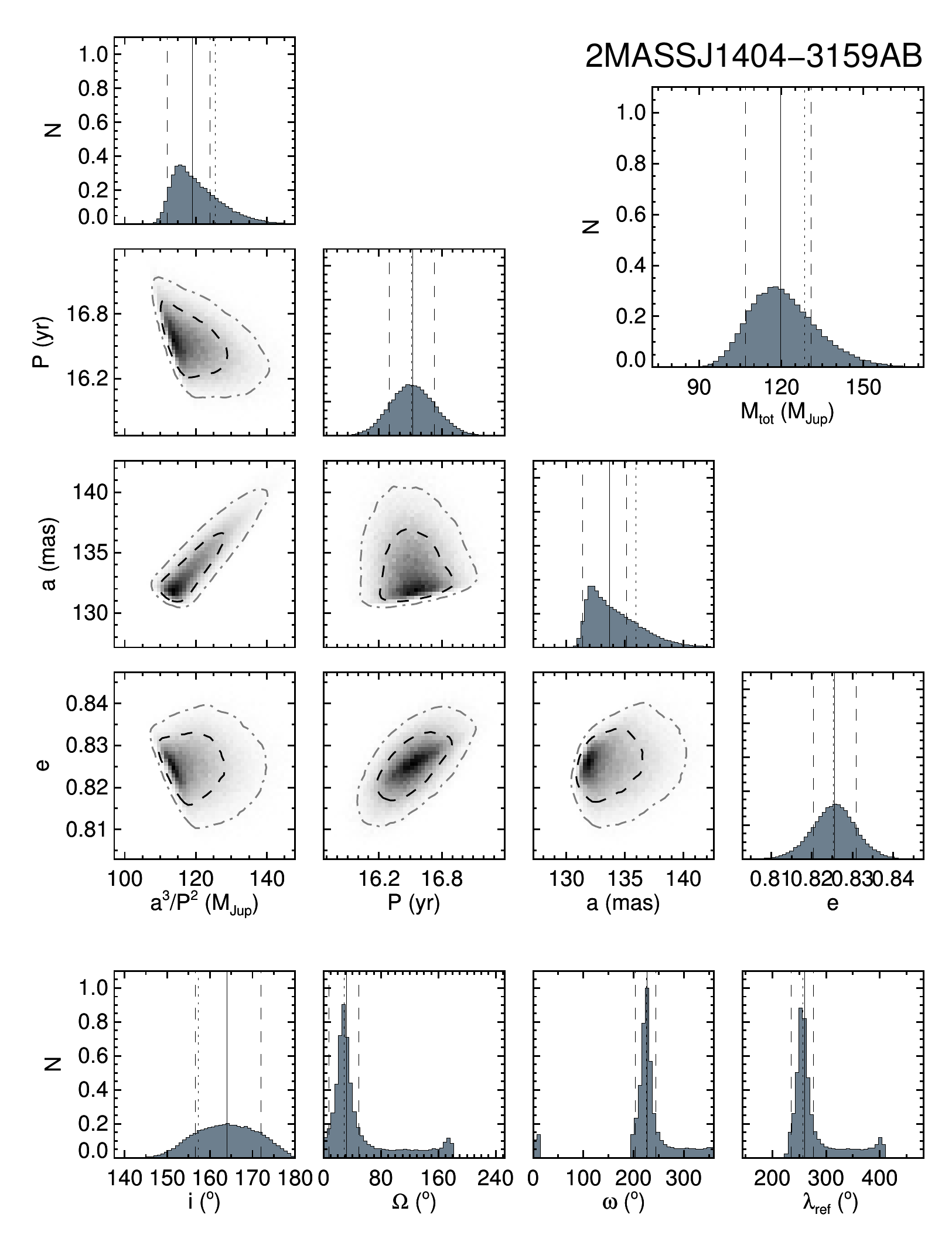}}  \ContinuedFloat  \caption{\normalsize (Continued)} \end{figure} \clearpage
\begin{figure} \vskip -0.4in \centerline{\includegraphics[width=6.5in,angle=0]{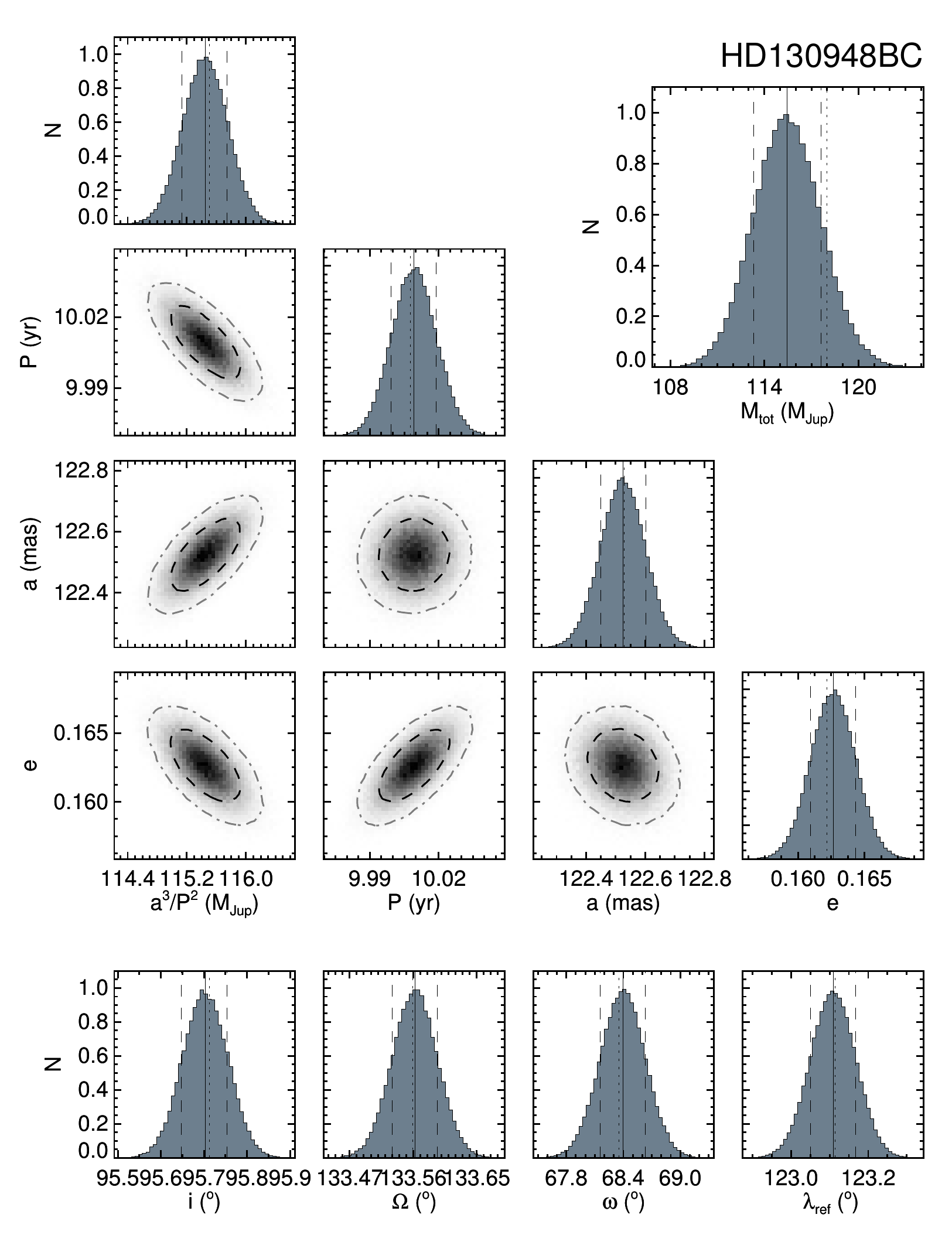}}  \ContinuedFloat  \caption{\normalsize (Continued)} \end{figure} \clearpage
\begin{figure} \vskip -0.4in \centerline{\includegraphics[width=6.5in,angle=0]{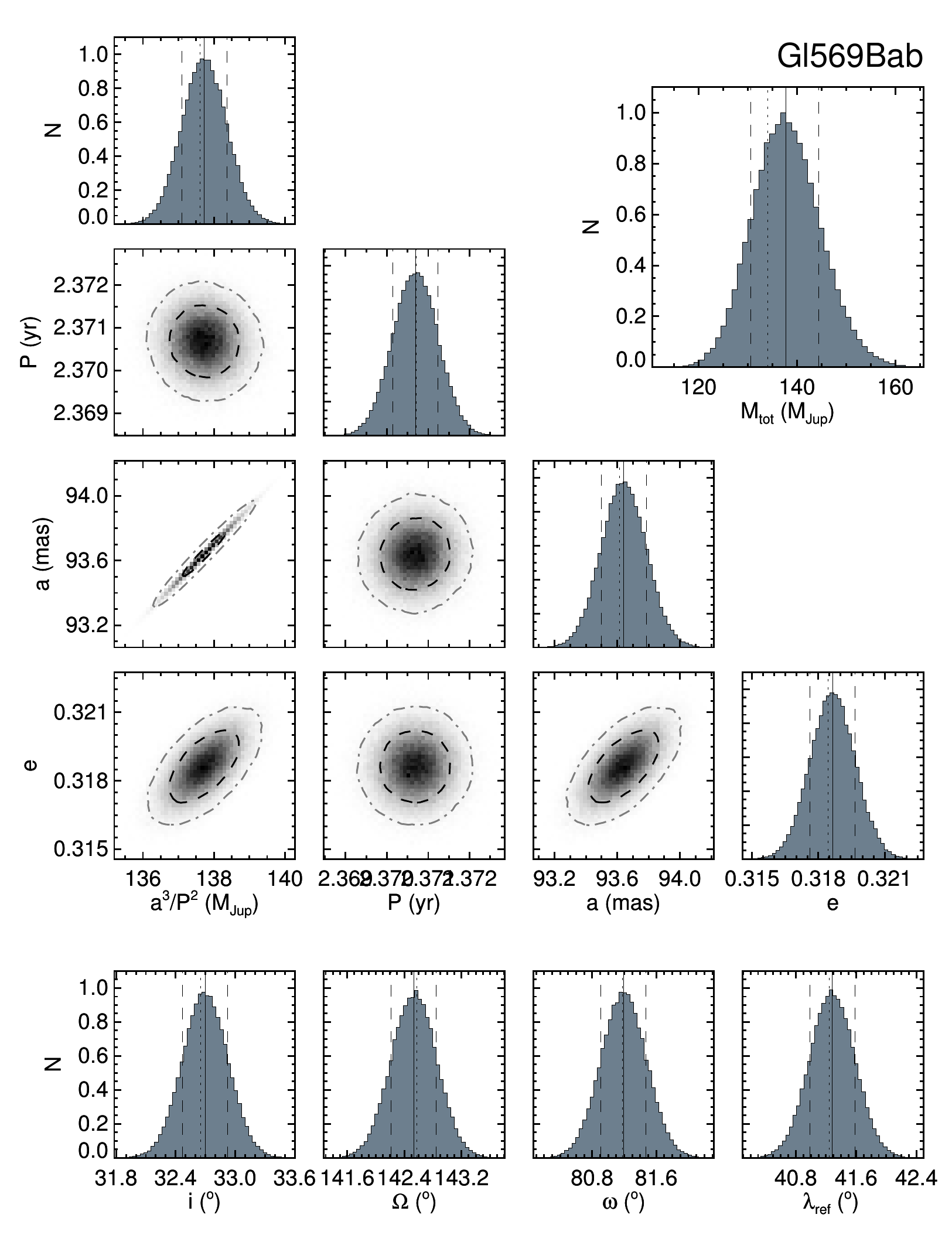}}     \ContinuedFloat  \caption{\normalsize (Continued)} \end{figure} \clearpage
\begin{figure} \vskip -0.4in \centerline{\includegraphics[width=6.5in,angle=0]{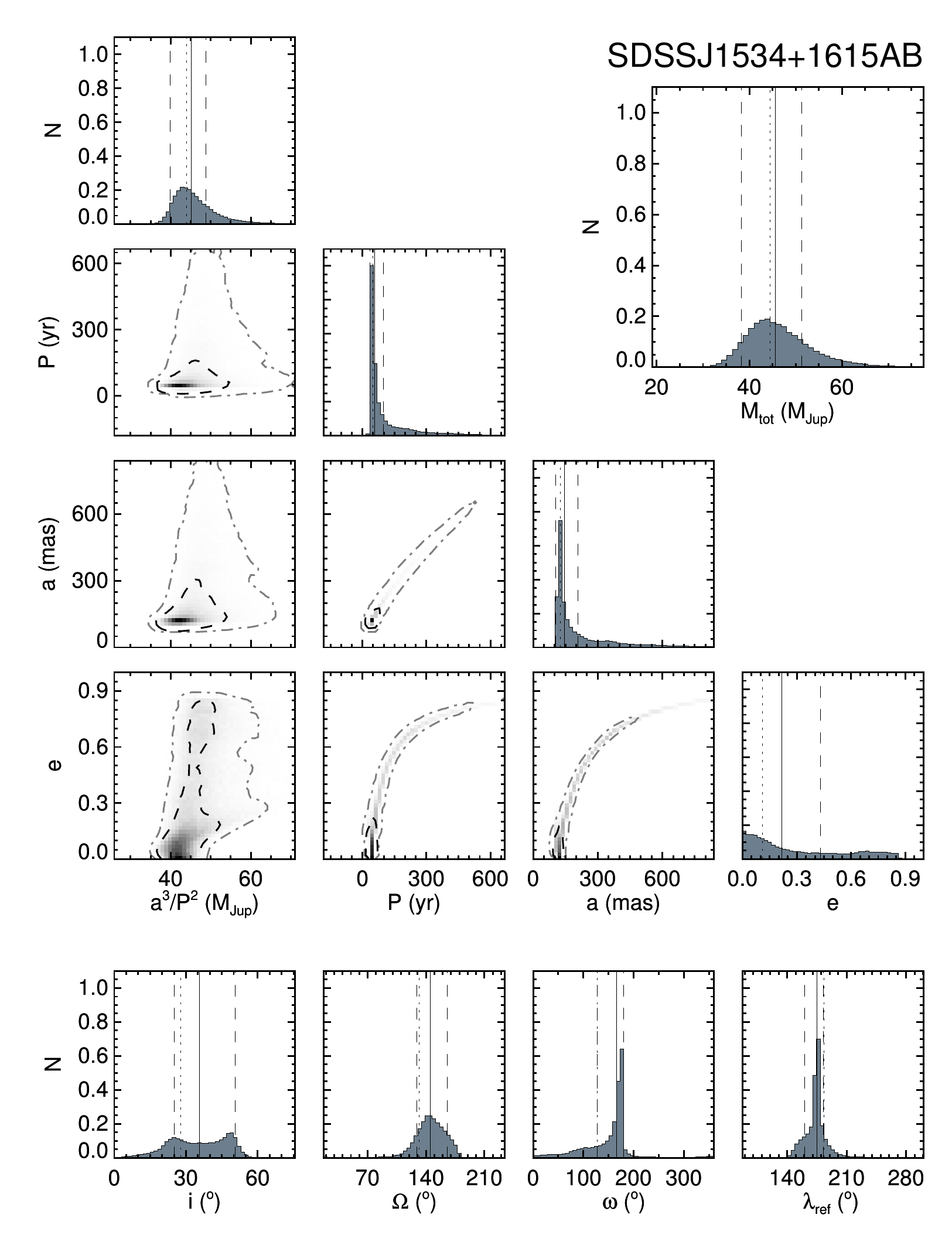}}  \ContinuedFloat  \caption{\normalsize (Continued)} \end{figure} \clearpage
\begin{figure} \vskip -0.4in \centerline{\includegraphics[width=6.5in,angle=0]{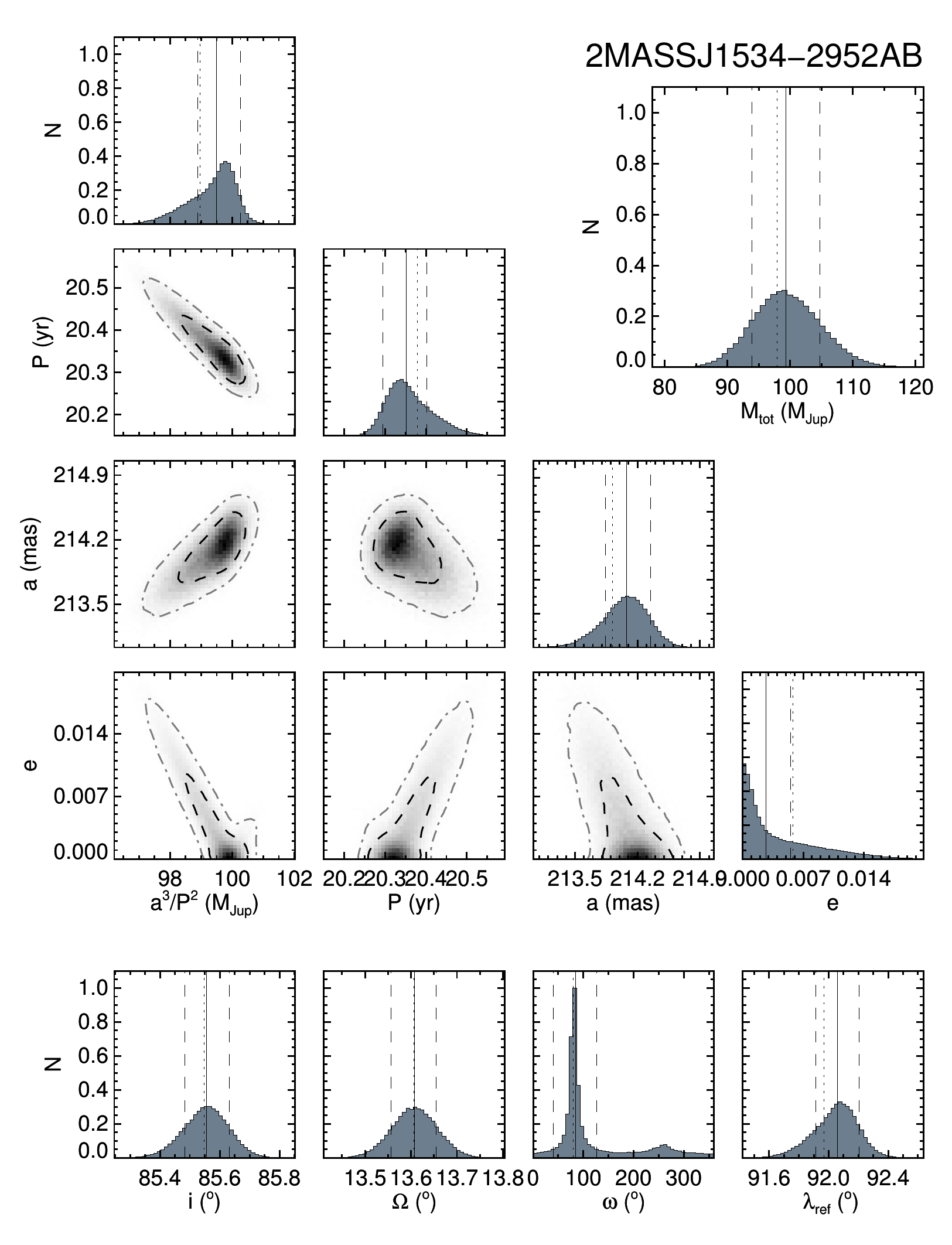}}  \ContinuedFloat  \caption{\normalsize (Continued)} \end{figure} \clearpage
\begin{figure} \vskip -0.4in \centerline{\includegraphics[width=6.5in,angle=0]{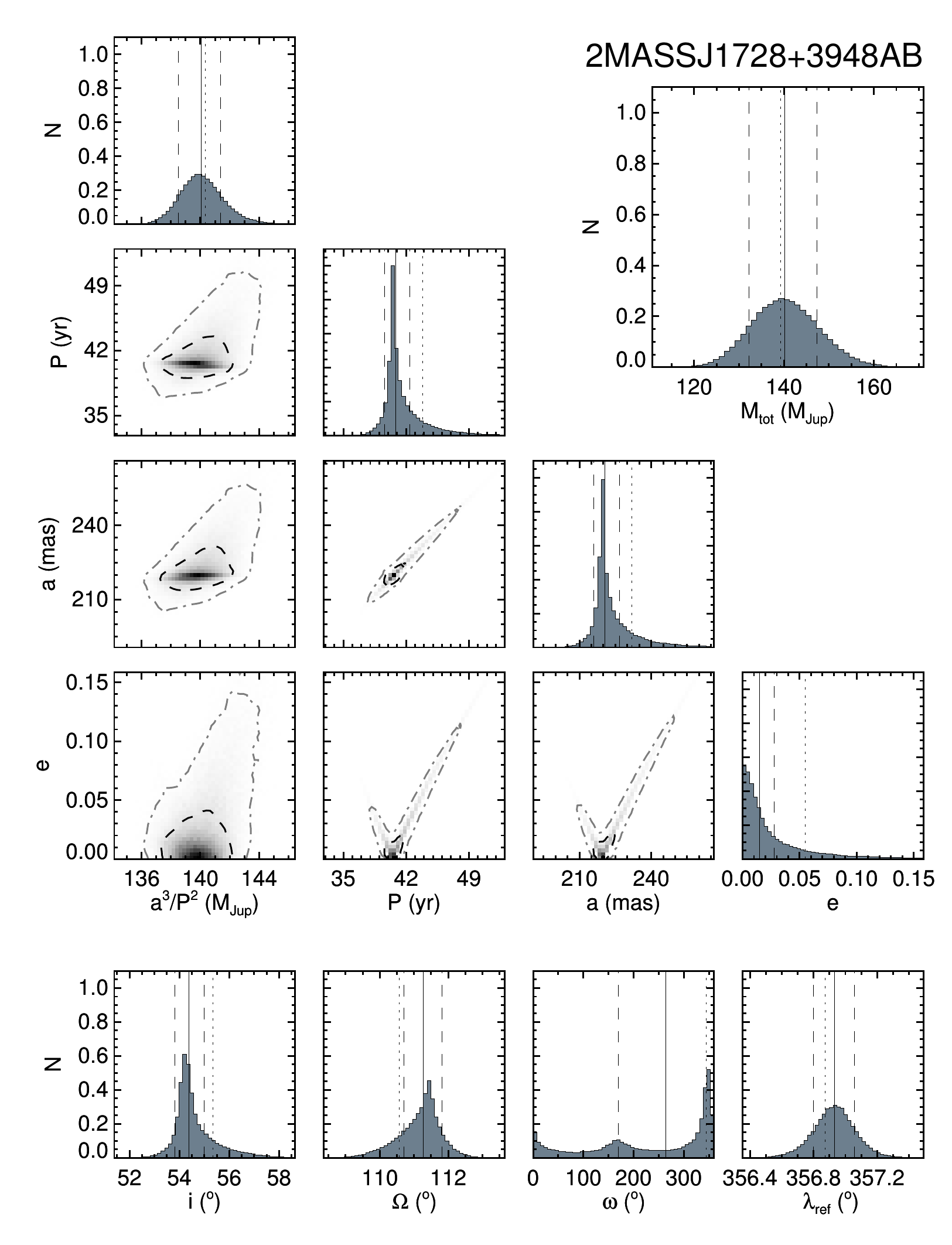}}  \ContinuedFloat  \caption{\normalsize (Continued)} \end{figure} \clearpage
\begin{figure} \vskip -0.4in \centerline{\includegraphics[width=6.5in,angle=0]{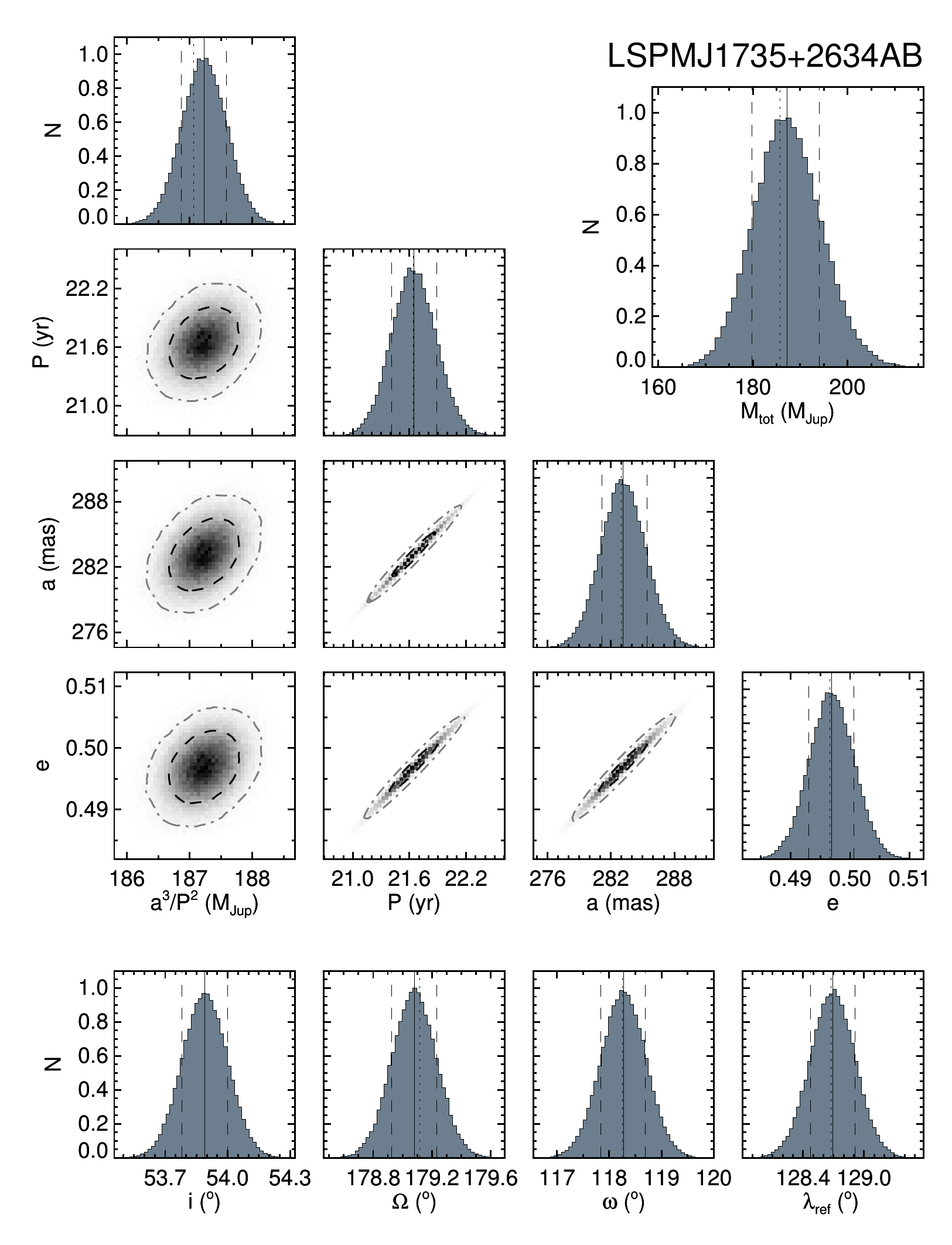}}  \ContinuedFloat  \caption{\normalsize (Continued)} \end{figure} \clearpage
\begin{figure} \vskip -0.4in \centerline{\includegraphics[width=6.5in,angle=0]{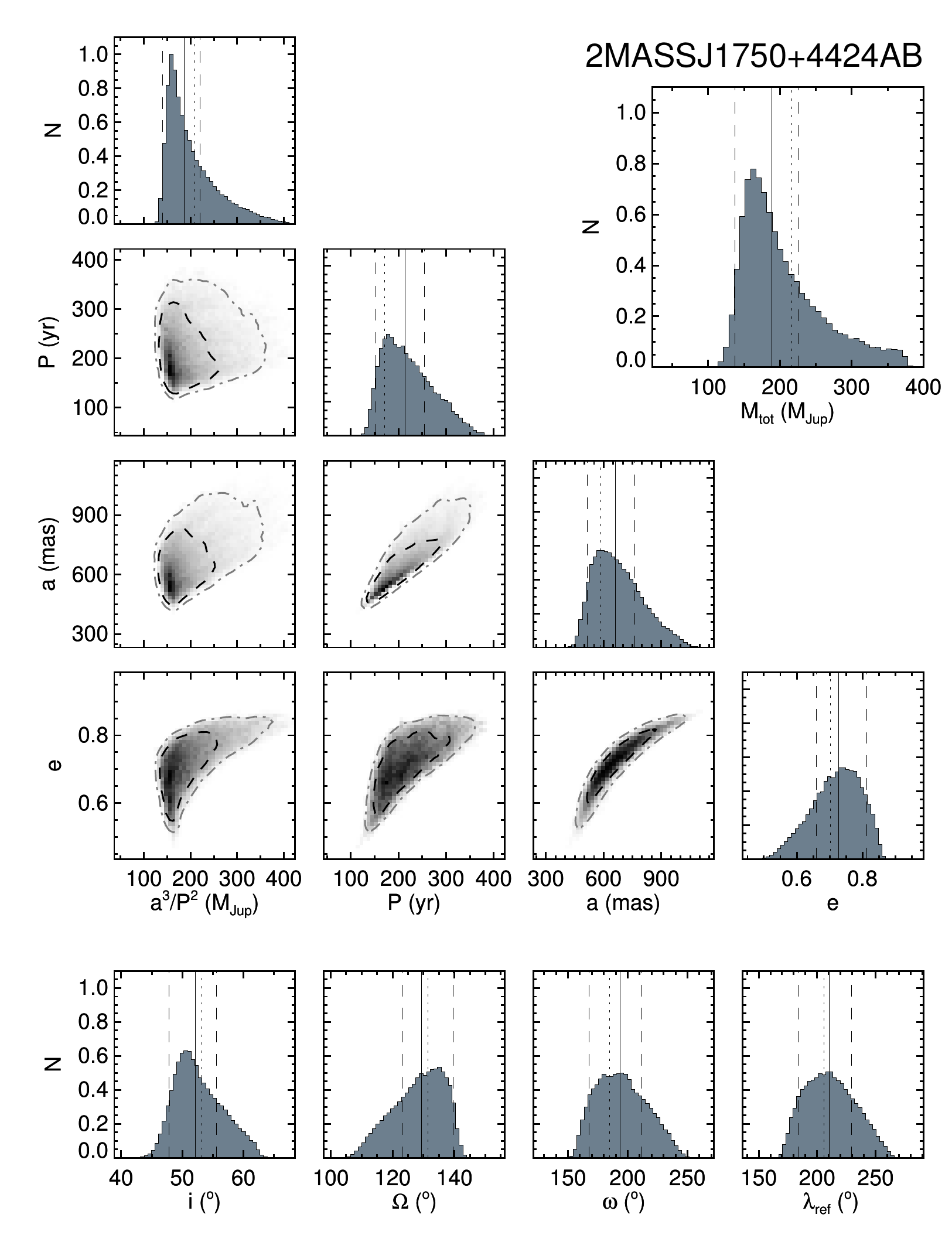}}  \ContinuedFloat  \caption{\normalsize (Continued)} \end{figure} \clearpage
\begin{figure} \vskip -0.4in \centerline{\includegraphics[width=6.5in,angle=0]{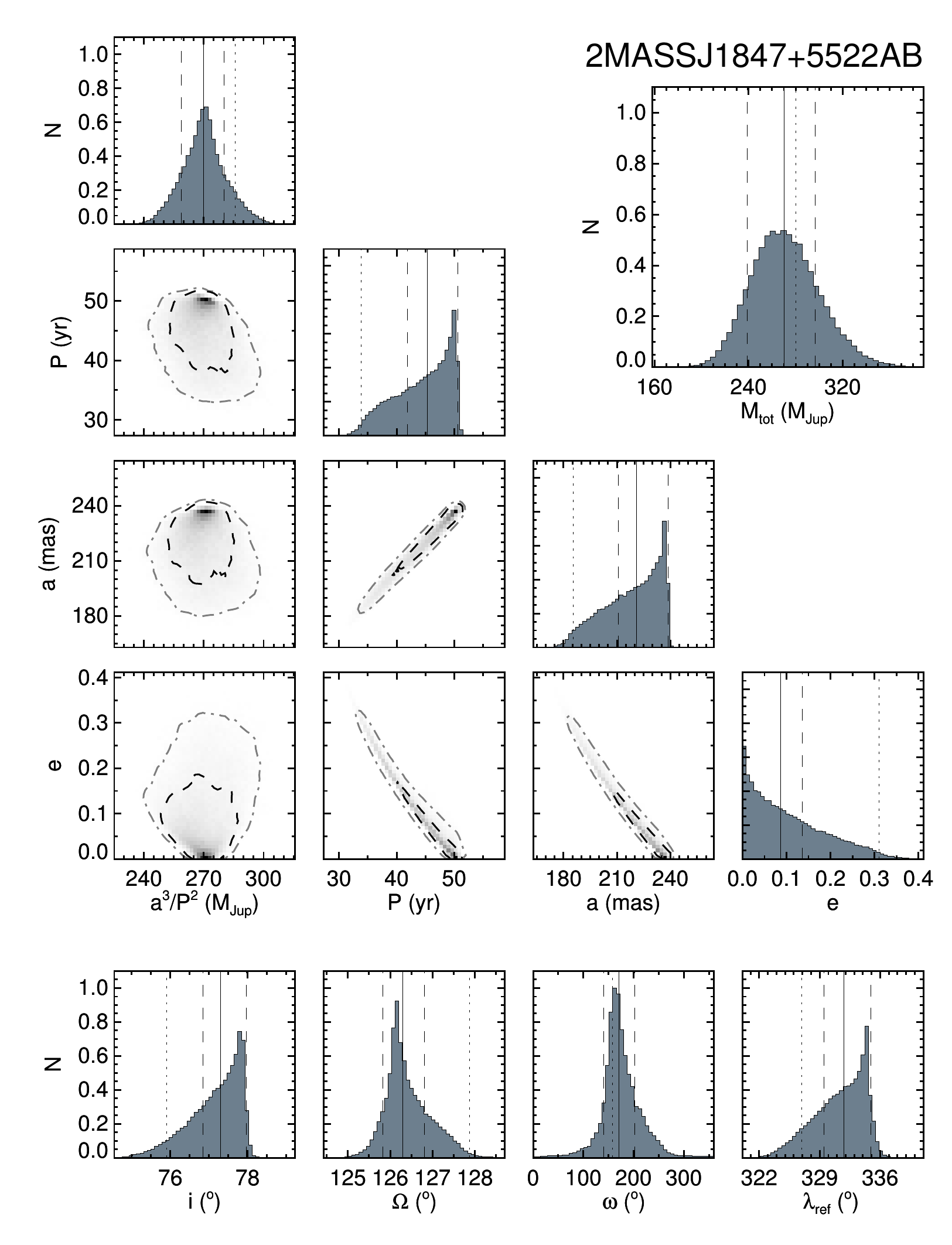}}  \ContinuedFloat  \caption{\normalsize (Continued)} \end{figure} \clearpage
\begin{figure} \vskip -0.4in \centerline{\includegraphics[width=6.5in,angle=0]{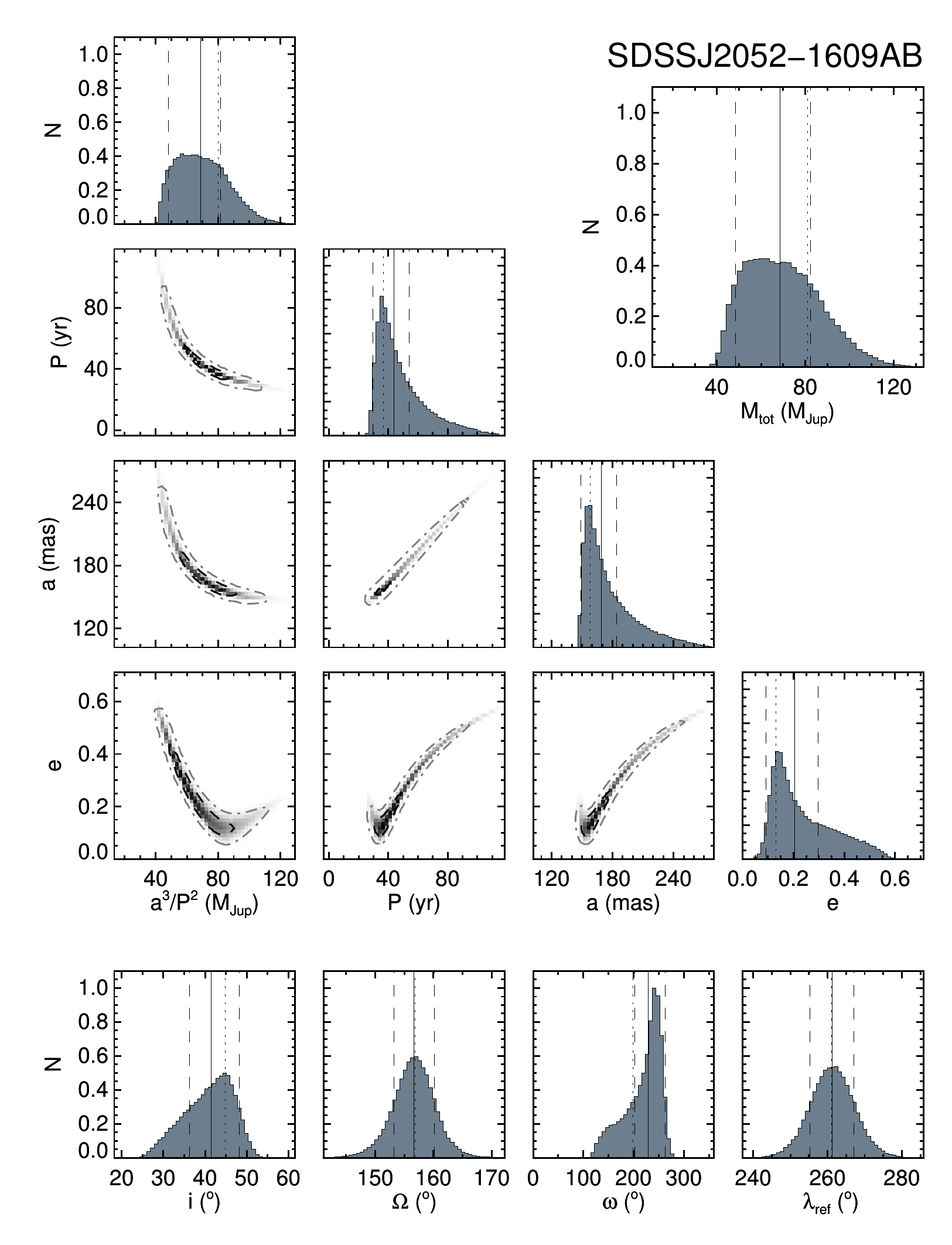}}  \ContinuedFloat  \caption{\normalsize (Continued)} \end{figure} \clearpage
\begin{figure} \vskip -0.4in \centerline{\includegraphics[width=6.5in,angle=0]{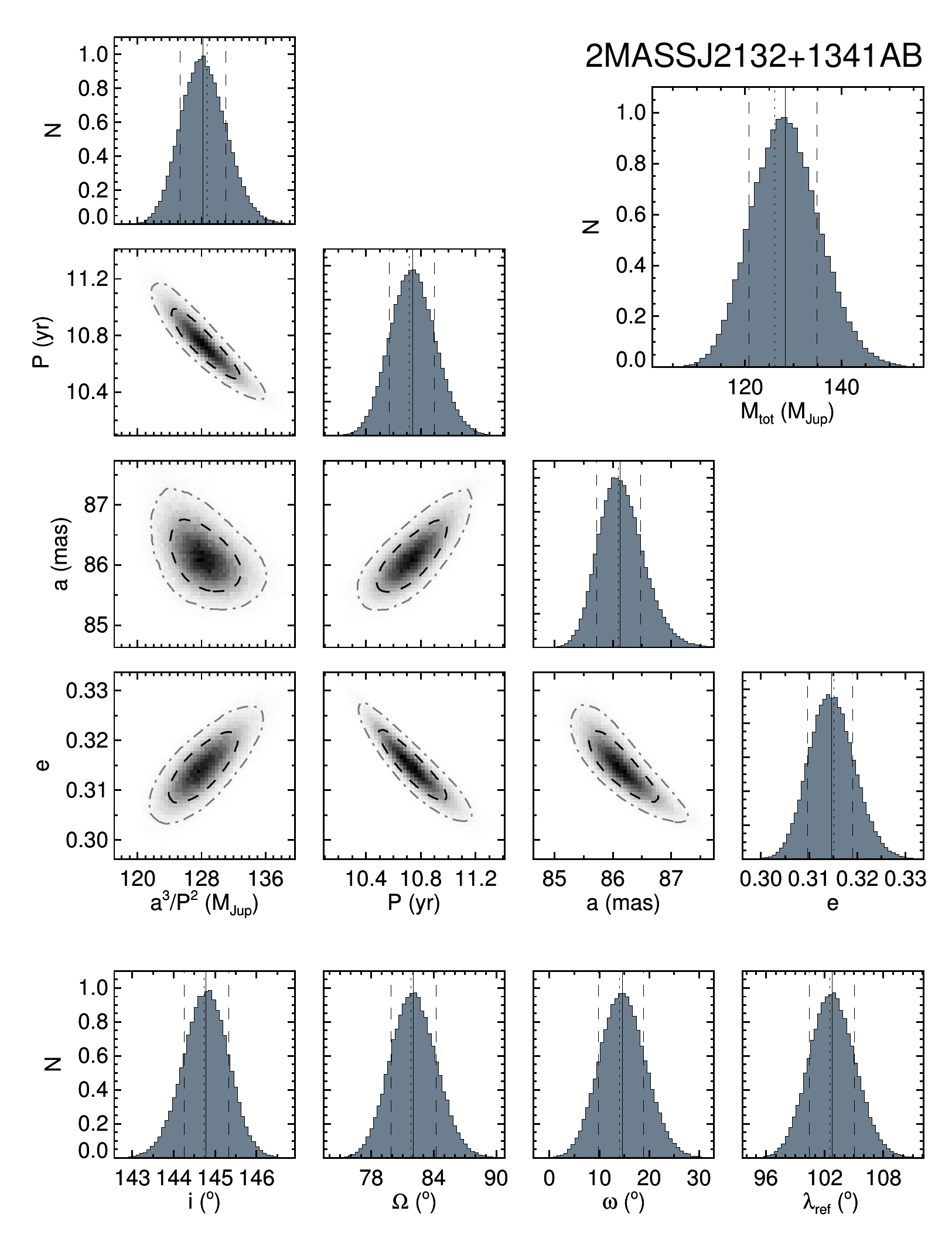}}  \ContinuedFloat  \caption{\normalsize (Continued)} \end{figure} \clearpage
\begin{figure} \vskip -0.4in \centerline{\includegraphics[width=6.5in,angle=0]{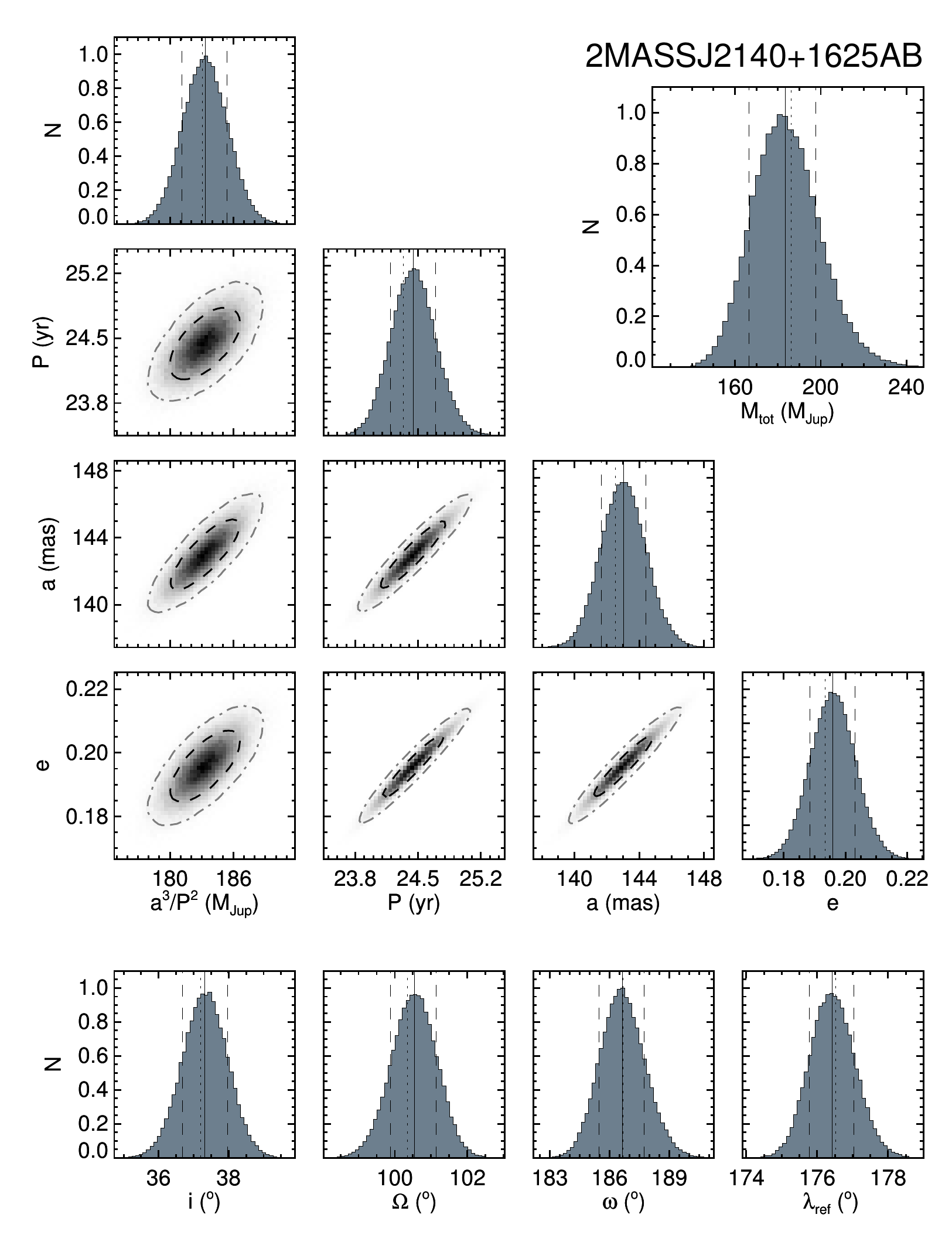}}  \ContinuedFloat  \caption{\normalsize (Continued)} \end{figure} \clearpage
\begin{figure} \vskip -0.4in \centerline{\includegraphics[width=6.5in,angle=0]{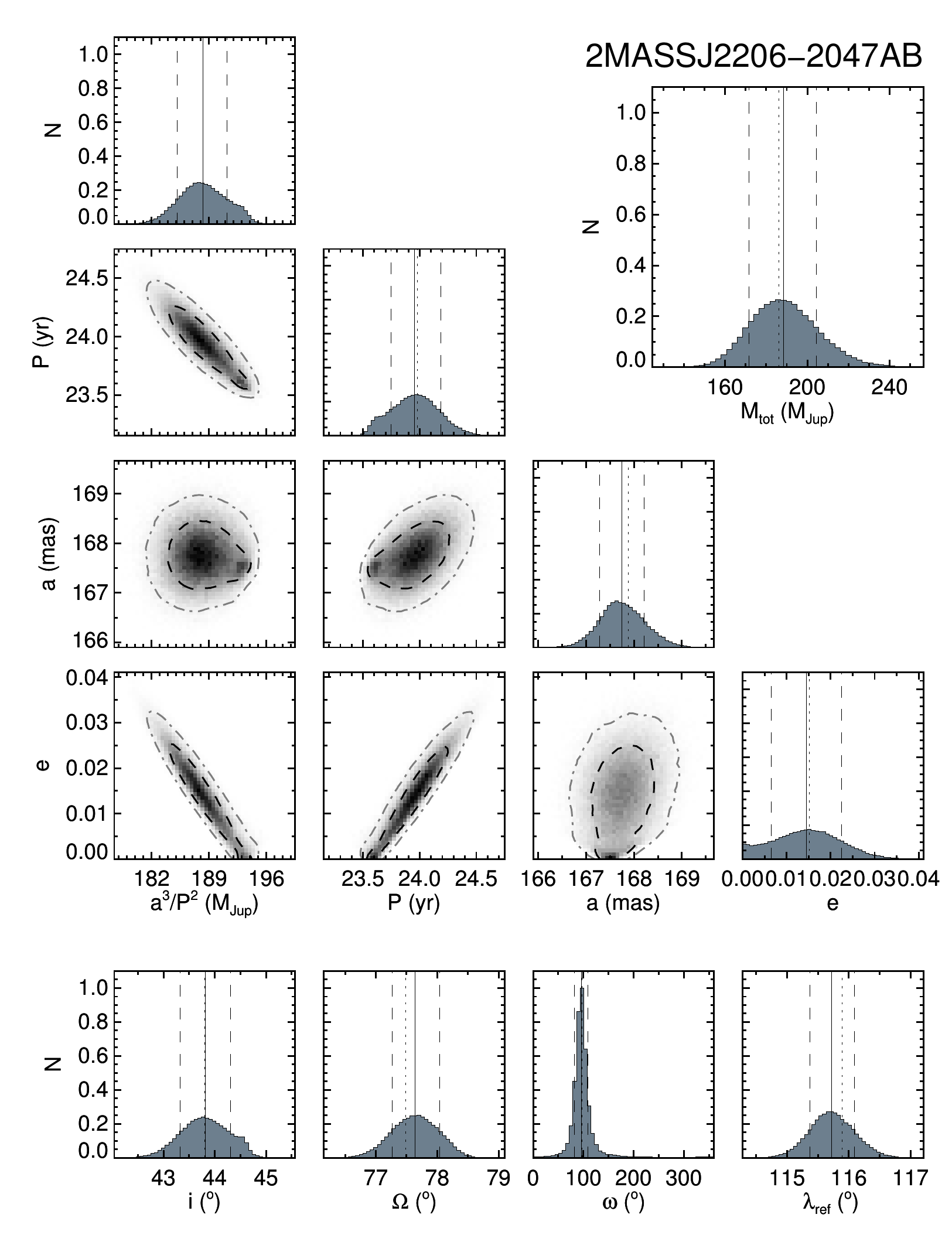}}  \ContinuedFloat  \caption{\normalsize (Continued)} \end{figure} \clearpage
\begin{figure} \vskip -0.4in \centerline{\includegraphics[width=6.5in,angle=0]{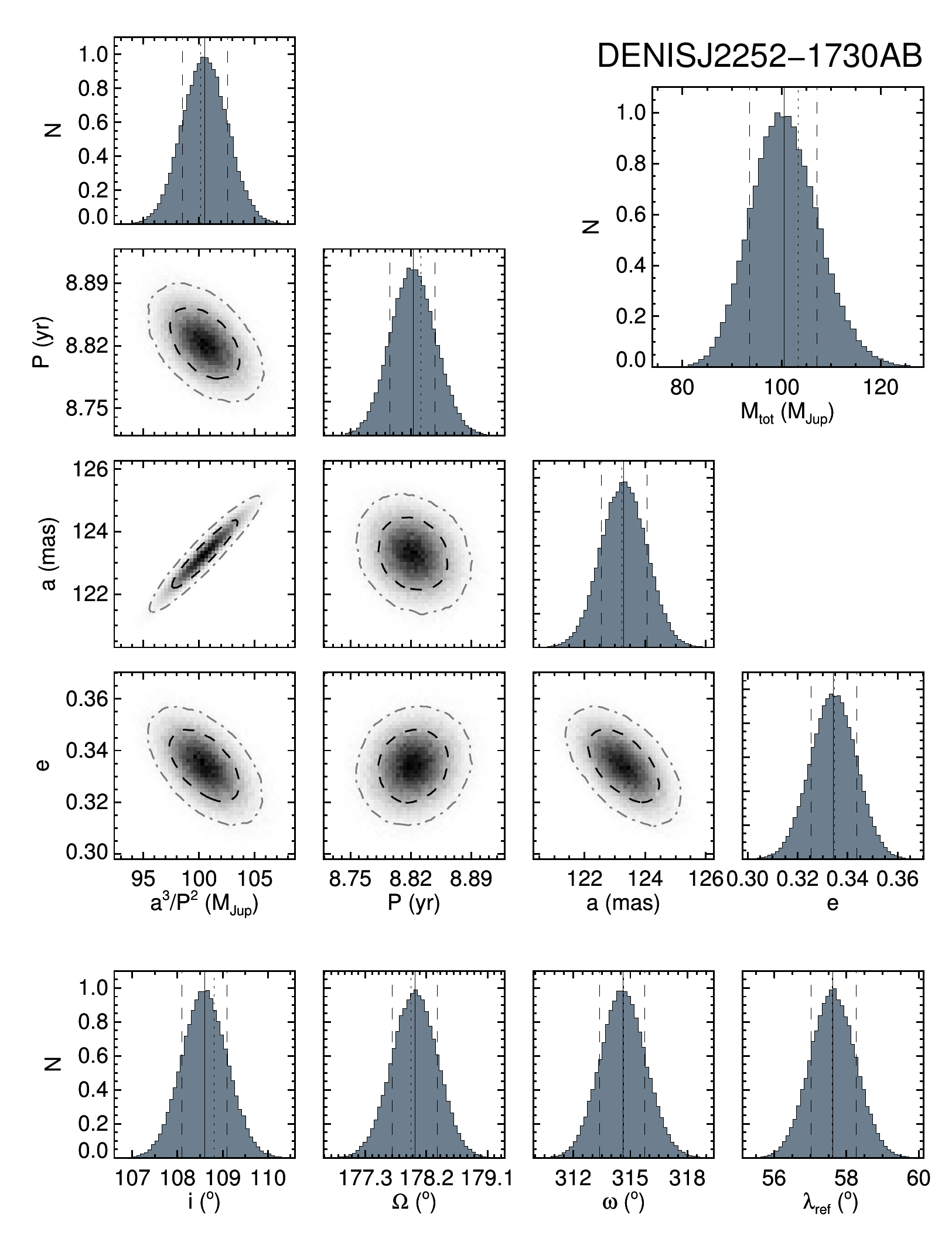}}  \ContinuedFloat  \caption{\normalsize (Continued)} \end{figure} \clearpage

\begin{figure} 
  \centerline{\includegraphics[width=6.5in,angle=0]{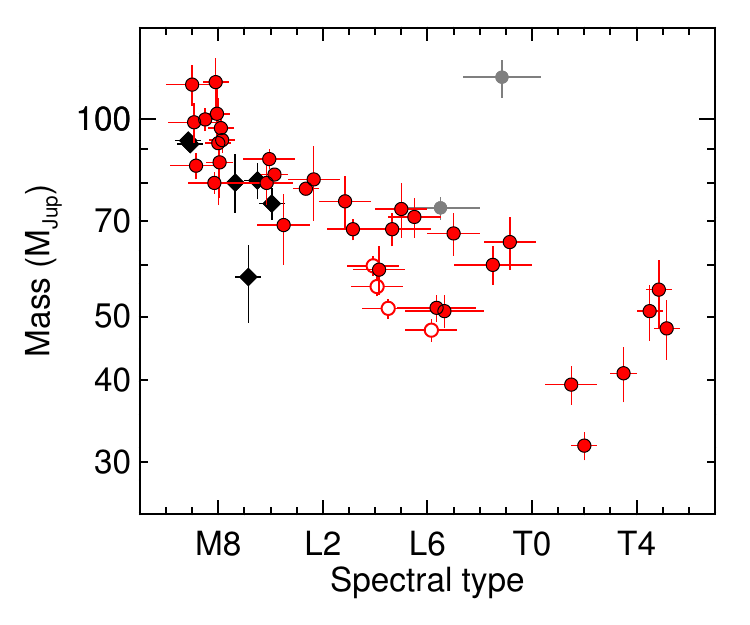}} 
  \caption{\normalsize Mass as function of spectral type for objects
    with directly measured individual masses in our sample (filled red
    circles) and from the literature (filled black diamonds).  Open
    red circles show objects from our sample with individual masses
    computed from our measured total mass and SM08 model-derived mass
    ratios.  The likely unresolved multiple systems (gray circles)
    have high masses for their spectral type, as expected.  Otherwise,
    no objects later than L4 have masses above 70\,\Mjup, implying
    that this is roughly the spectral type and mass corresponding to
    the hydrogen-fusion limit for the field population.  The lowest
    mass object in our sample (SDSS~J0423$-$0414B) is not the latest
    spectral type, illustrating that our field sample spans a range of
    ages.  (For clarity of viewing many data points at the same
    spectral type, small offsets to the spectral type values have been
    added for plotting here.) \label{fig:spt-m}}
\end{figure} 
\clearpage

\begin{figure} 
  \centerline{\includegraphics[width=6.5in,angle=0]{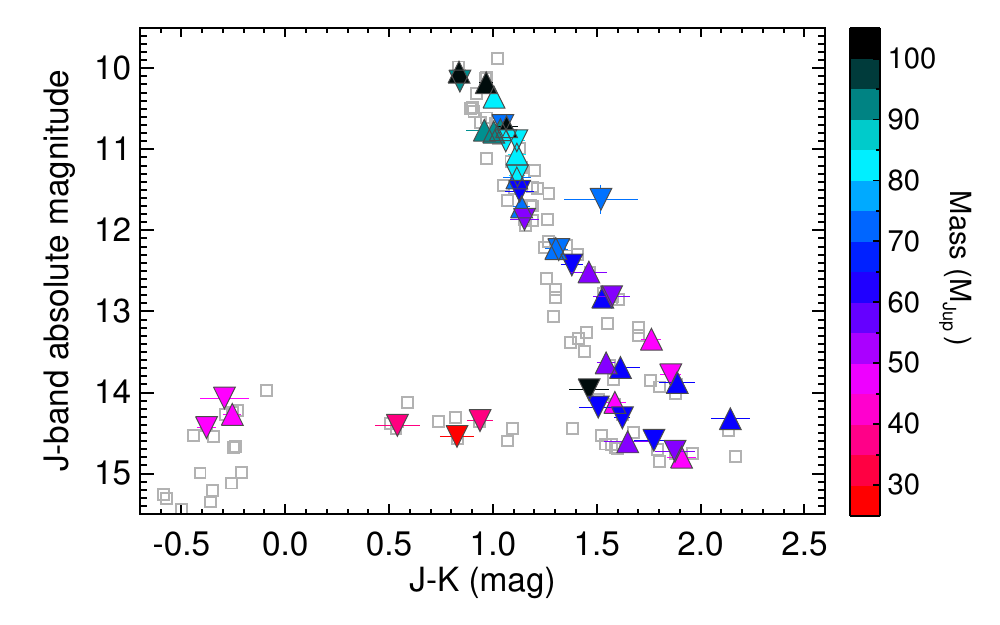}} 
  \caption{\normalsize Color--magnitude diagram (CMD) showing our
    dynamical mass sample with symbols colored according to their
    directly measured individual masses.  (The color bar is the same
    as used in all other plots with this color scheme.)  Up-pointing
    triangles indicate primary components, down-pointing triangles
    indicate secondary components, and the field sequence is shown for
    reference as gray squares using the latest compilation of
    ultracool dwarf parallaxes at
    \protect\url{http://www.as.utexas.edu/~tdupuy/plx}.  (We plot only
    normal, single field objects that have ${\rm S/N}> 10$ parallaxes
    and $J-K$ uncertainties $<$0.10\,mag.)  While mass generally
    decreases through the CMD sequence, there is not a one-to-one
    correspondence with mass and CMD location given that our sample is
    drawn from the field population of ultracool dwarfs spanning a
    range of ages.  For example, the latest-type T dwarfs
    ($J-K = -0.5$ to 0.0\,mag) happen to not be the least massive
    objects in our sample.  By chance, the least massive objects are
    all located roughly in the middle of the L/T transition
    ($J-K = 0.5$--1.0\,mag).  The massive object in the bottom right
    part of the CMD (black down-pointing triangle near
    $M_J = 14$\,mag) is 2MASS~J0920+3517B, an unresolved pair of brown
    dwarfs in a triple system. \label{fig:cmd-mass}}
\end{figure} 
\clearpage

\begin{figure} 
  \centerline{\includegraphics[height=3.0in,angle=0]{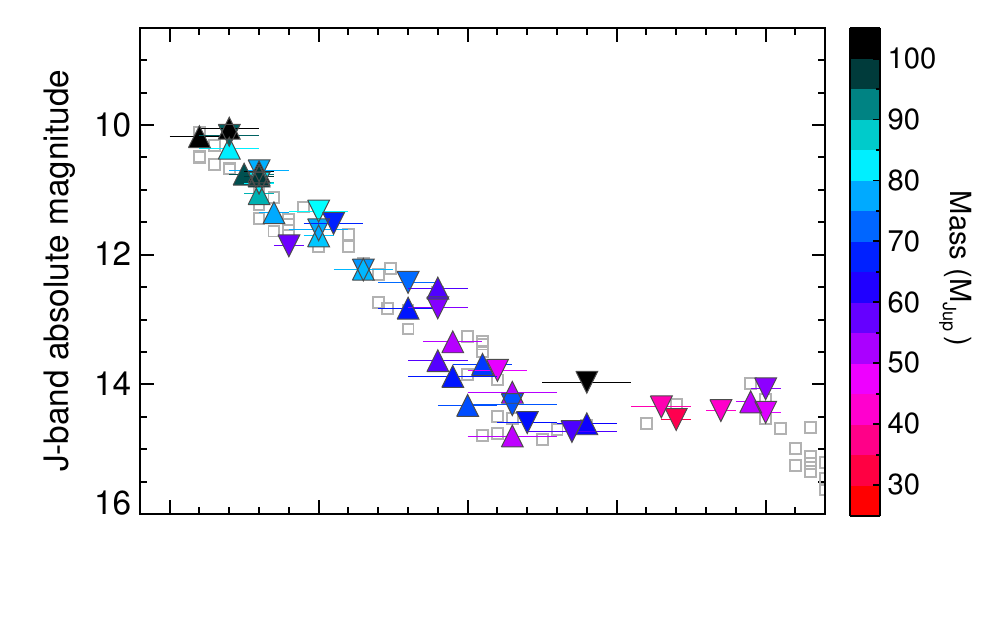}} 
  \vskip -0.6in
  \centerline{\includegraphics[height=3.0in,angle=0]{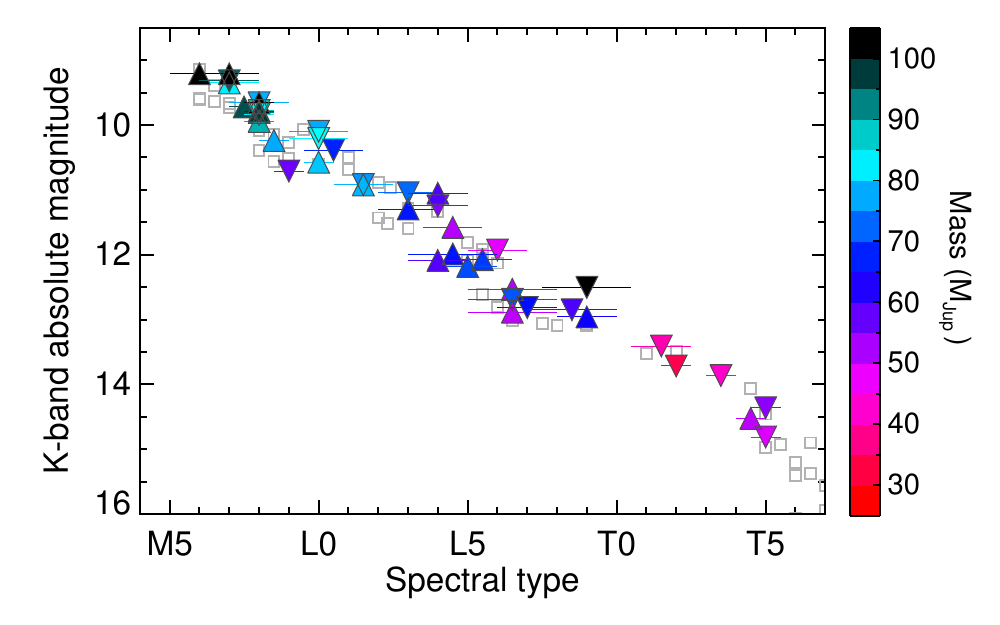}} 
  \caption{\normalsize Absolute magnitude plotted as a function of
    spectral type for our dynamical mass sample with symbols colored
    according to their directly measured individual masses.  (The
    color bar is the same as used in all other plots with this color
    scheme.)  Up-pointing triangles indicate primary components,
    down-pointing triangles indicate secondary components, and the
    field sequence is shown for reference as gray squares using the
    latest compilation of ultracool dwarf parallaxes at
    \protect\url{http://www.as.utexas.edu/~tdupuy/plx}.  (We plot only
    normal, single field objects that have ${\rm S/N}> 10$ parallaxes
    and $J-K$ uncertainties $<$0.10\,mag.)  Our sample is broadly
    consistent with the field sequence, indicating the accuracy of our
    spectral types that we derived by decomposing each binary's
    integrated-light spectrum using our resolved photometry and a
    library of spectral templates.  The largest outlier is
    2MASS~J0920+3517B (L$9.0\pm1.5$ sitting well above the L/T
    transition sequence on the $J$-band plot), which is consistent
    with our dynamical masses showing that this is an unresolved pair
    of brown dwarfs in a triple system.  2MASS~J0700+3157B, the other
    such unresolved pair, is not an outlier on these plots because our
    derived spectral type (L$6.5\pm1.5$) and absolute magnitudes
    happen to be consistent with the field
    sequence. \label{fig:spt-abs}}
\end{figure} 
\clearpage

\begin{figure} 
  \centerline{\includegraphics[width=6.5in,angle=0]{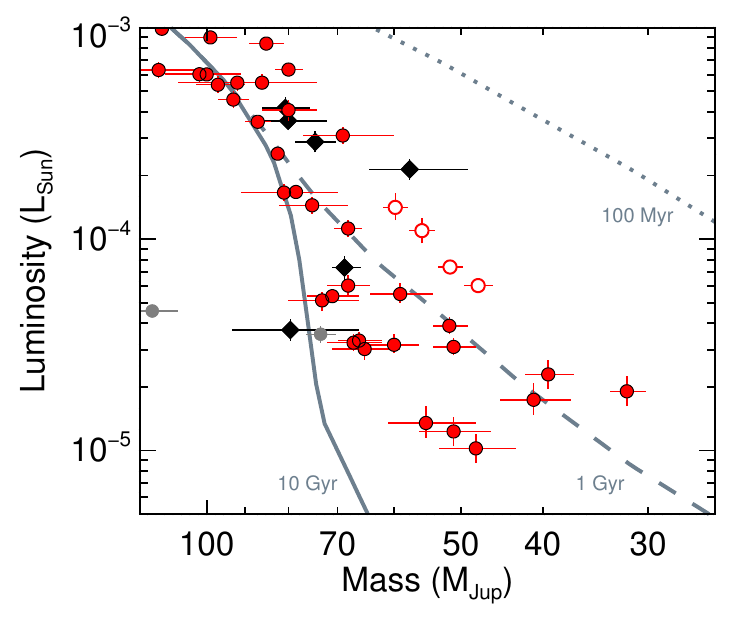}} 
  \caption{\normalsize Luminosity as function of mass for objects with
    directly measured individual masses in our sample (filled red
    circles) and from the literature (filled black diamonds).  Open
    red circles show objects from our sample with individual masses
    computed from our measured total mass and SM08 model-derived mass
    ratios.  Filled gray circles indicate the likely unresolved
    multiples in our sample 2MASS~J0920+3517B and 2MASS~J0700+3157B.
    The other possible unresolved multiple LP~415-20A
    ($156^{+17}_{-18}$\,\Mjup) does not appear on this plot because it
    is too massive. Cond model isochrones are shown for reference,
    indicating that most objects in our sample are consistent with
    having ages of $\sim$1--10\,Gyr.  For Gl~569Bab we use black
    diamonds to indicate that the literature mass ratio is used to
    compute individual masses, even though it is in our sample and we
    use our dynamical total mass here.  \label{fig:m-l}}
\end{figure} 
\clearpage

\begin{figure} 
  \centerline{\includegraphics[width=6.5in,angle=0]{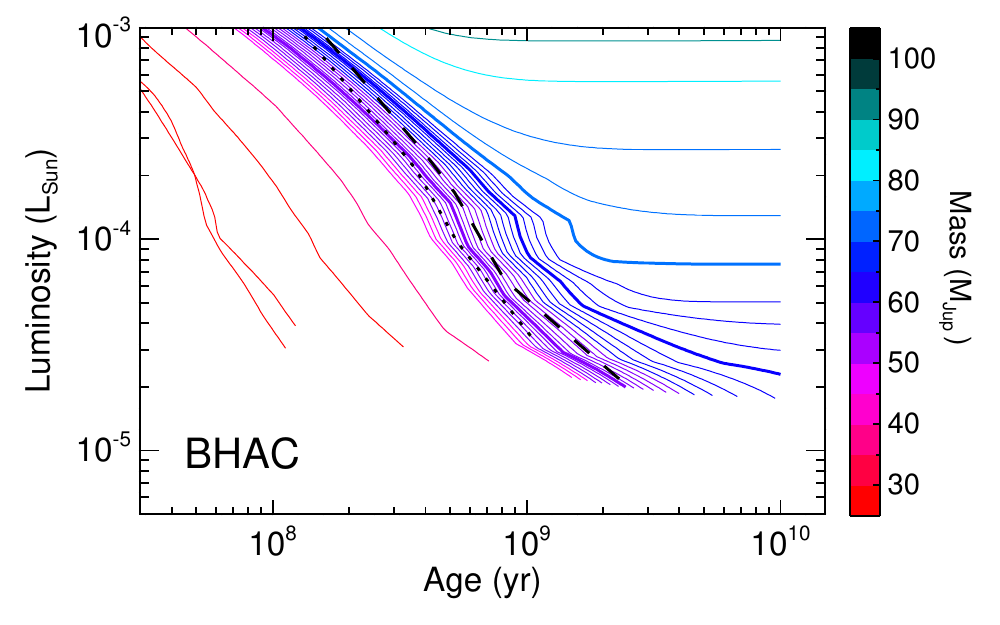}} 
  \caption{\normalsize Evolutionary models used in our analysis.  Each
    colored line shows the predicted luminosity as a function of age
    for an object of given mass.  The 75\,\Mjup, 70\,\Mjup, and
    60\,\Mjup\ models are plotted with thicker lines as a visual
    aid. All model masses are integer multiples of 0.001\,\Msun\
    (1.048\,\Mjup), and only masses of 0.015--0.100\,\Msun\
    (15--105\,\Mjup) are plotted here.  BHAC models are uniformly
    sampled at steps of 0.001\,\Msun\ over the range
    0.050--0.070\,\Msun\ (52--73\,\Mjup) and at coarser steps at other
    masses.  Dotted and dashed and lines indicate boundaries in
    lithium depletion of 50\% and 99.9\% depleted, respectively.  BHAC
    models are only available down to $\Teff \approx 1300$\,K, while
    SM08 models are only available below $\approx$2400\,K. (The color
    bar is the same as used in all other plots with this color
    scheme.) \label{fig:models-t-l}}
\end{figure} 
\clearpage
\begin{figure} 
  \centerline{\includegraphics[width=6.5in,angle=0]{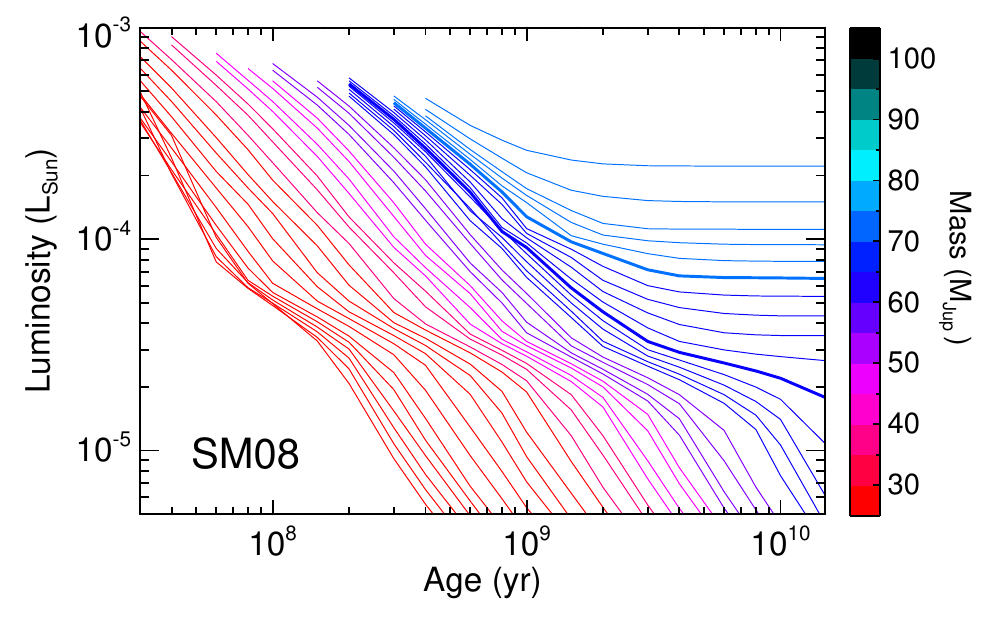}} 
  \ContinuedFloat  \caption{\normalsize (Continued)}
\end{figure} 
\clearpage
\begin{figure} 
  \centerline{\includegraphics[width=6.5in,angle=0]{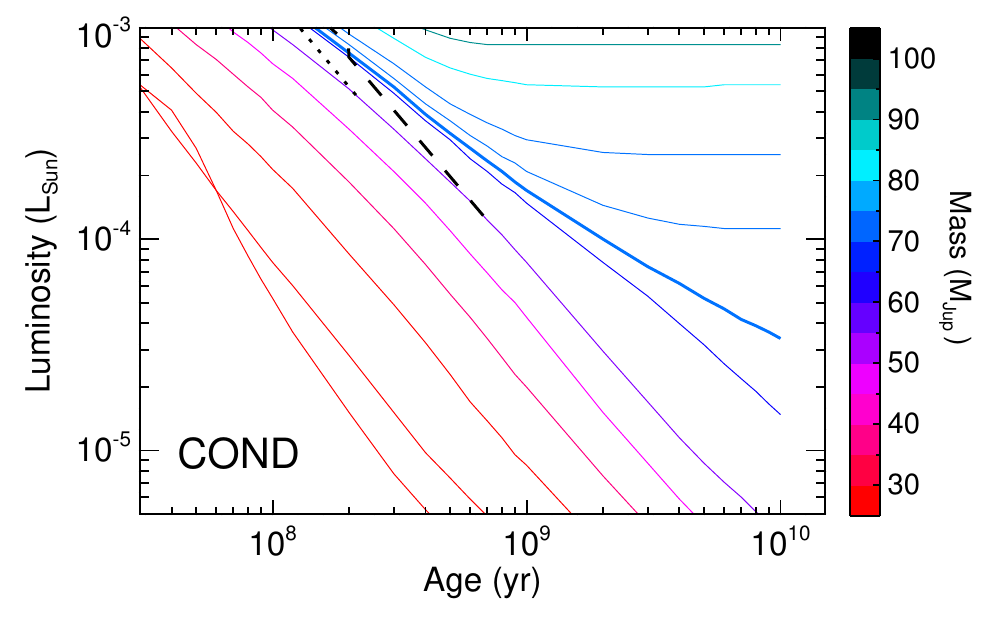}} 
  \ContinuedFloat  \caption{\normalsize (Continued)}
\end{figure} 
\clearpage

\begin{landscape}
\begin{figure} 
  \centerline{\includegraphics[width=3.0in,angle=0]{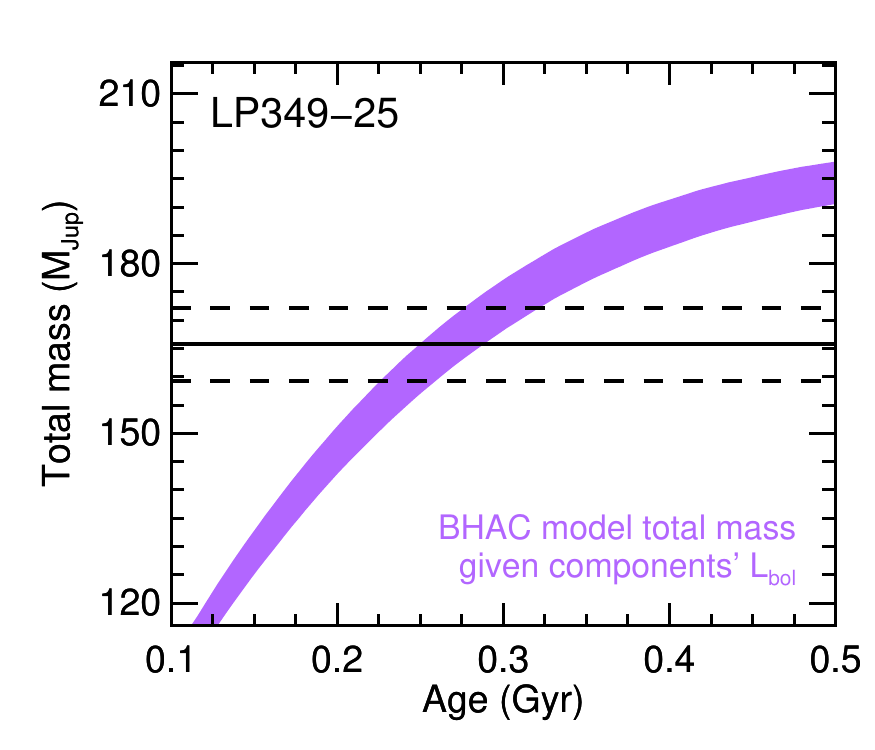} \includegraphics[width=3.0in,angle=0]{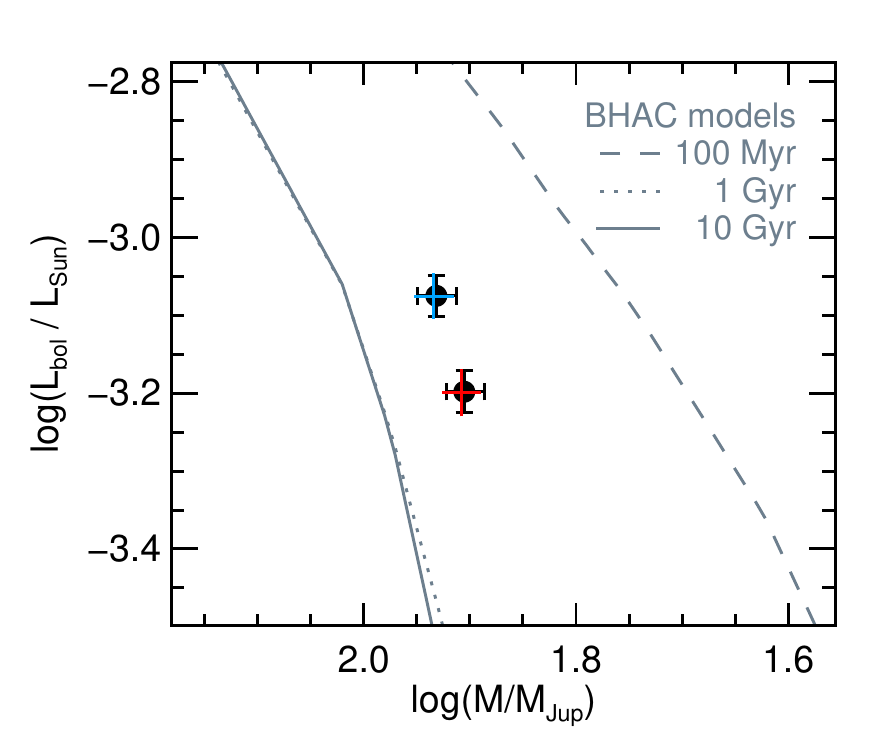} \includegraphics[width=3.0in,angle=0]{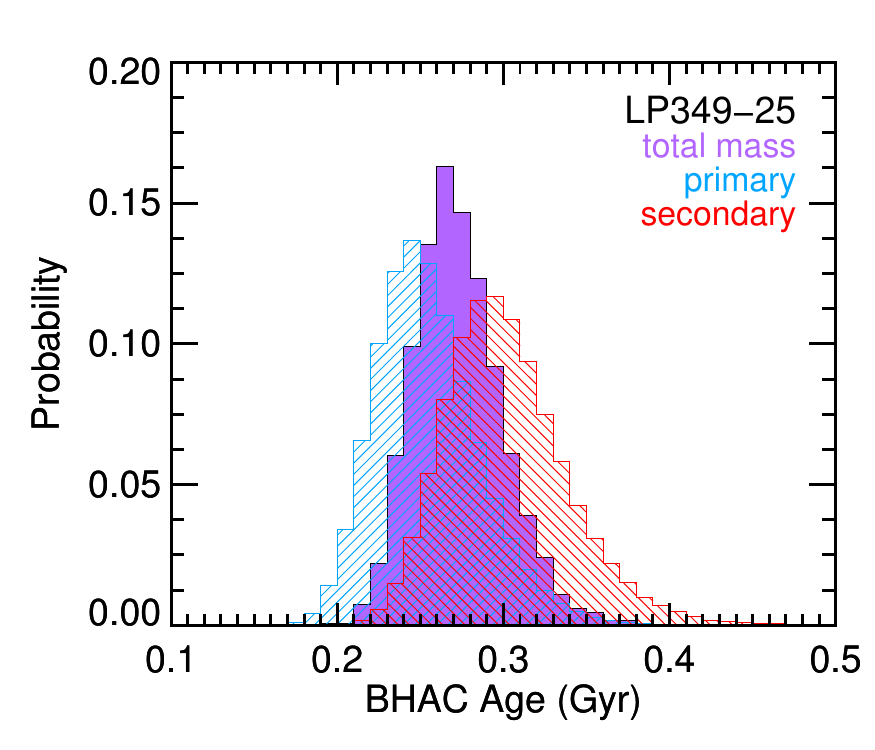}} 
  \caption{{\bf Left panel:} Total mass derived from models using the
    component luminosities over a range of ages (purple), where the
    vertical extent of the shaded region corresponds to the 1$\sigma$
    uncertainties in the luminosities.  Our measured total mass and
    1$\sigma$ confidence interval are shown by a solid line and dashed
    lines, respectively.  (This panel is for display purposes only as
    our actual analysis is based on rejection sampling as described in
    Section~\ref{sec:model}.)  {\bf Middle panel:} Our directly
    measured individual masses and luminosities are plotted as black
    filled circles with 1$\sigma$ error bars.  Model isochrones are
    shown as gray lines.  The 1$\sigma$ interval in mass and
    luminosity after performing our rejection sampling analysis is
    shown by the colored error bars for the primary (blue) and
    secondary (red).  When there are offsets between these colored
    points and the measurements, or differences in the error bars, it
    is caused by rejecting input Monte Carlo samples that do not agree
    with models and/or by imposing our uniform age prior.  {\bf Right
      panel:} Final age distribution derived from our rejection
    sampling analysis using the total mass and individual luminosities
    (solid purple), the primary mass and luminosity (blue hatched),
    and the secondary mass and luminosity (red hatched).  Our uniform
    prior in age results in many stellar binaries having nearly flat
    age distributions over the range of main sequence ages covered by
    models, whereas brown dwarfs and pre--main-seuqence stars like
    LP~349-25AB have well constrained model-derived
    ages. \label{fig:model-derive}}
\end{figure}
\clearpage
\begin{figure} 
  \centerline{\includegraphics[width=3.0in,angle=0]{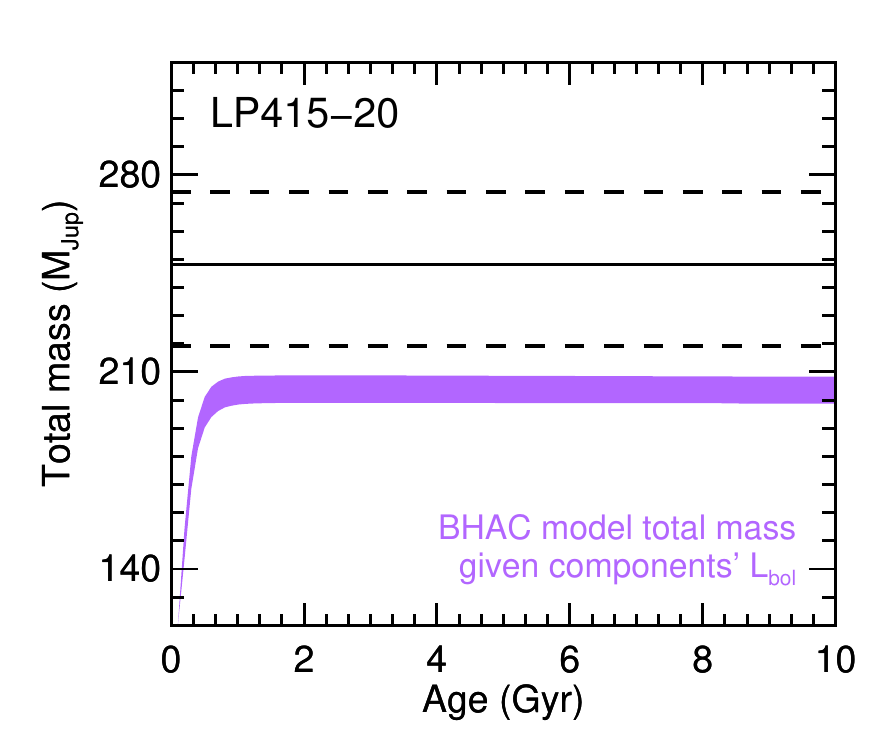} \includegraphics[width=3.0in,angle=0]{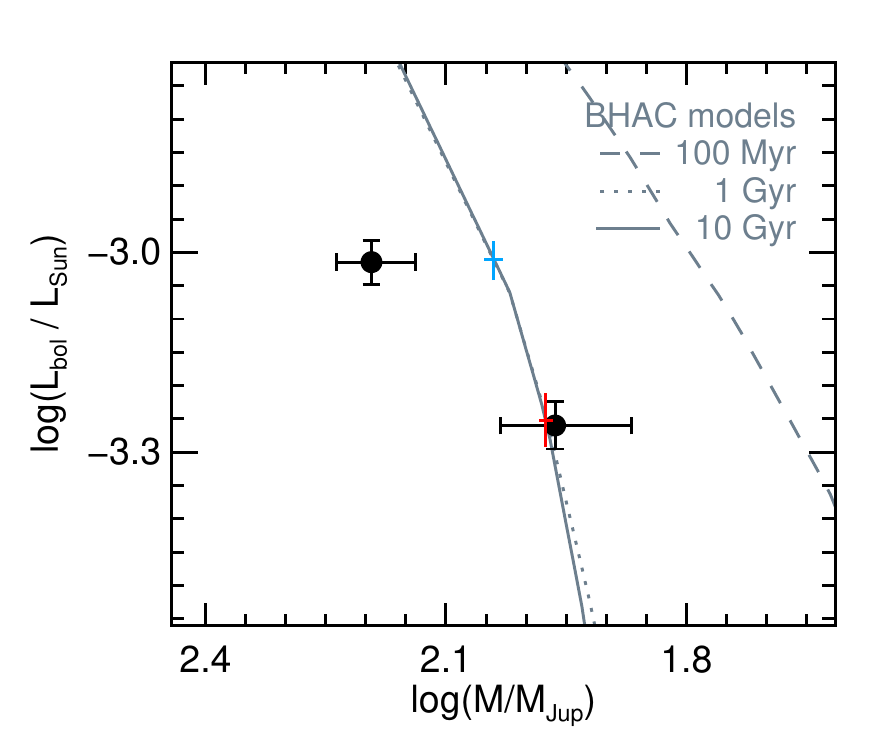} \includegraphics[width=3.0in,angle=0]{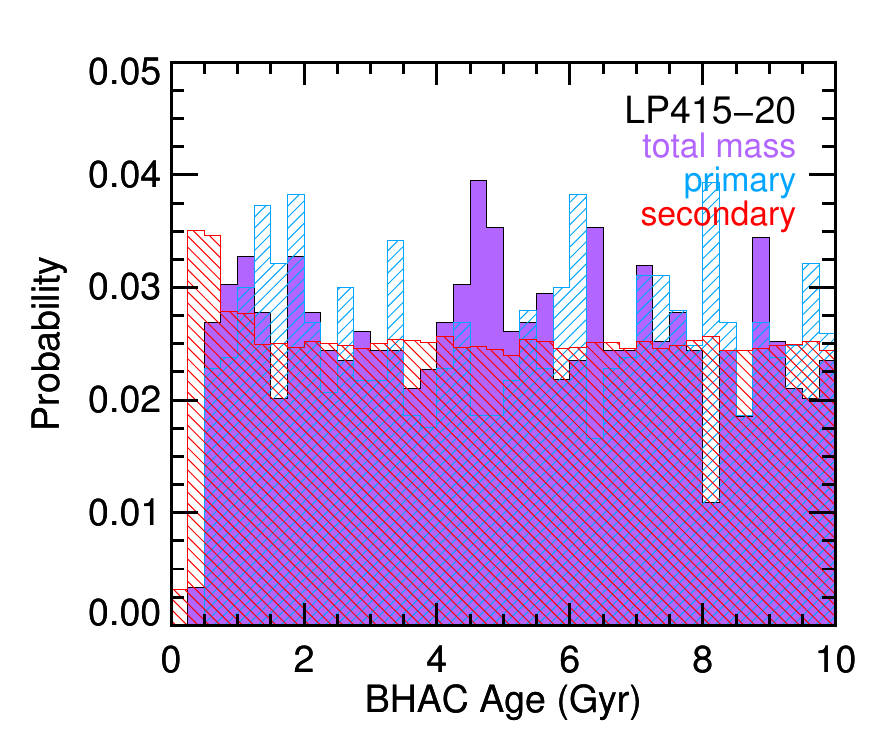}} 
\ContinuedFloat \caption{\normalsize (Continued)} \end{figure}
\clearpage
\begin{figure} 
  \centerline{\includegraphics[width=3.0in,angle=0]{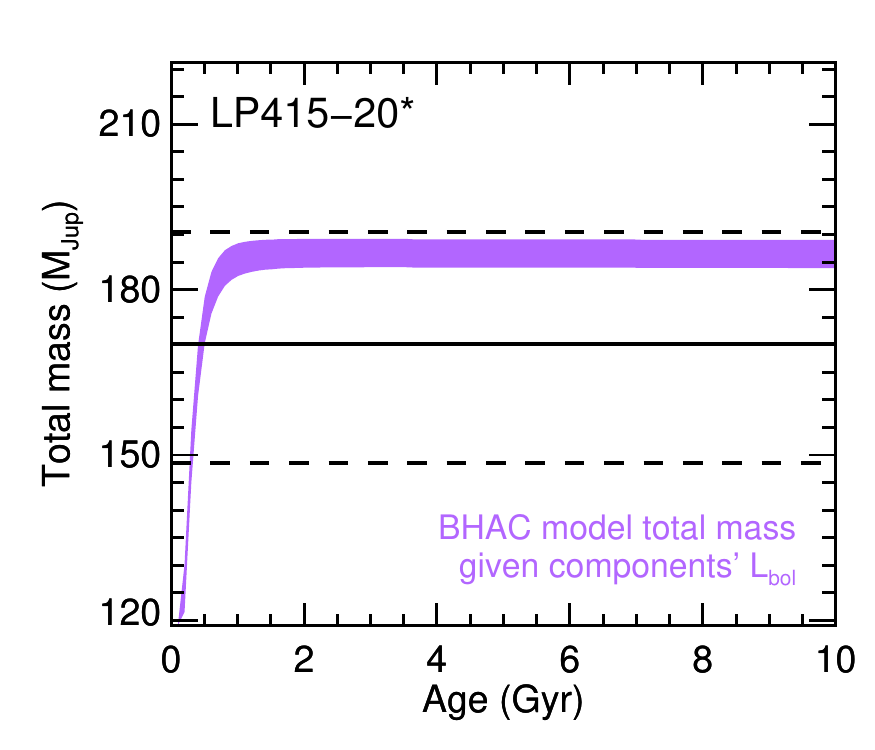} \includegraphics[width=3.0in,angle=0]{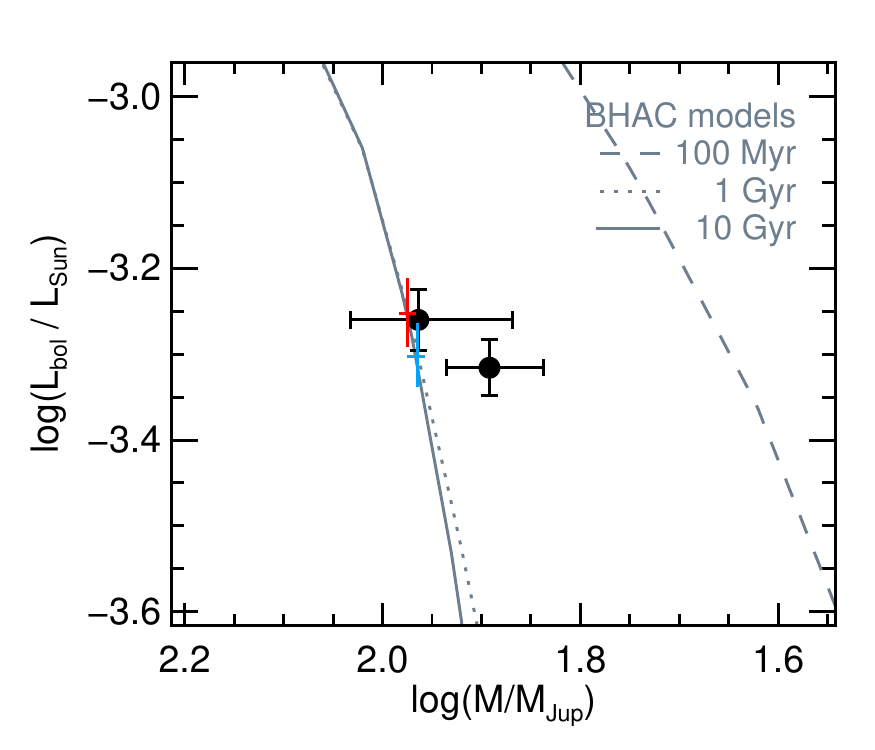} \includegraphics[width=3.0in,angle=0]{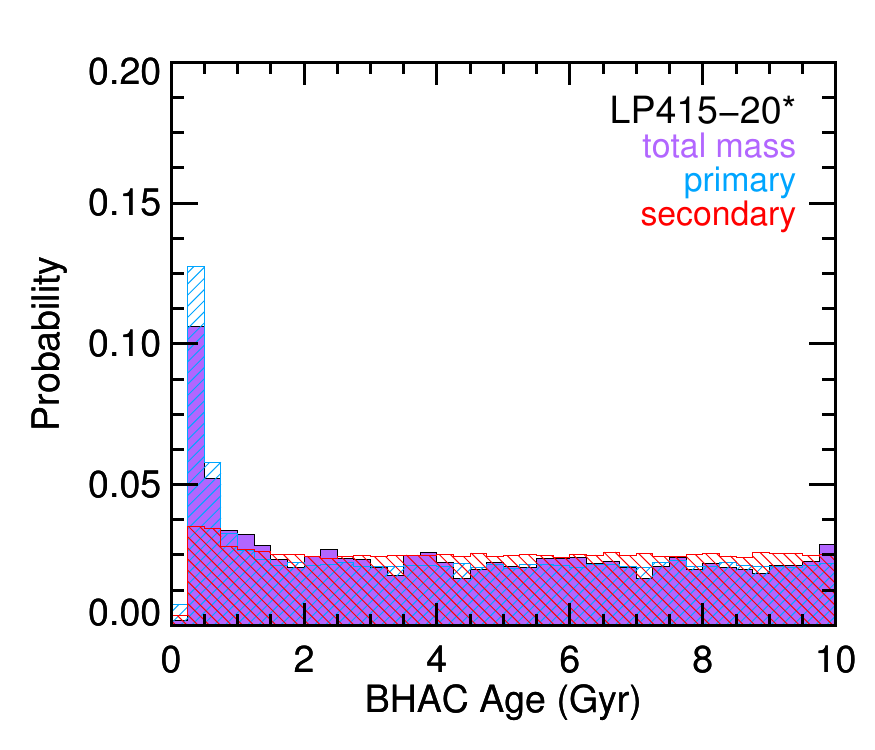}} 
\ContinuedFloat \caption{\normalsize (Continued)} \end{figure}
\clearpage
\begin{figure} 
  \centerline{\includegraphics[width=3.0in,angle=0]{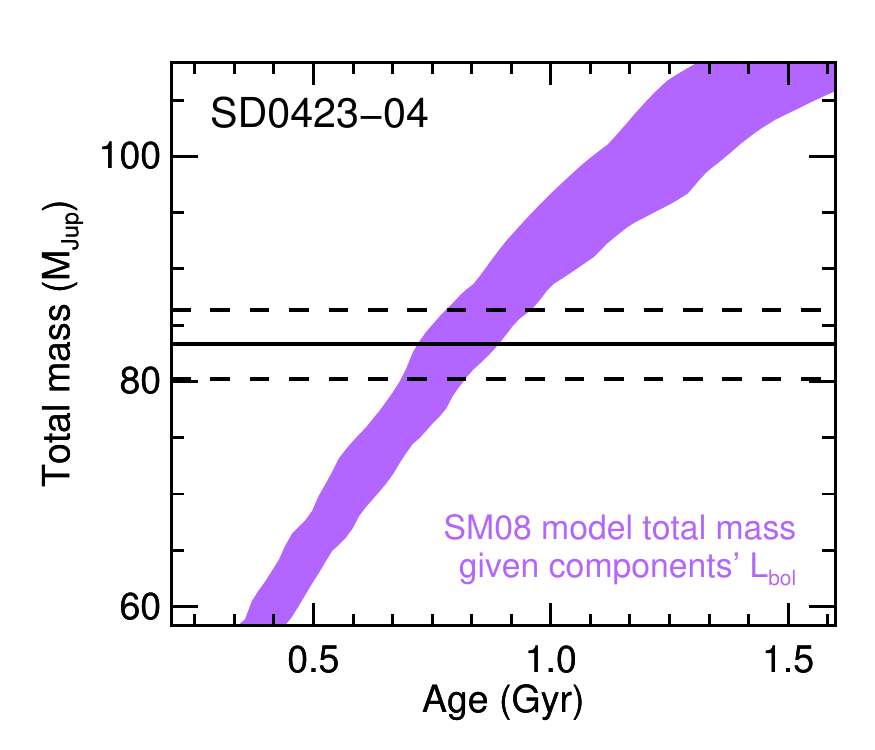} \includegraphics[width=3.0in,angle=0]{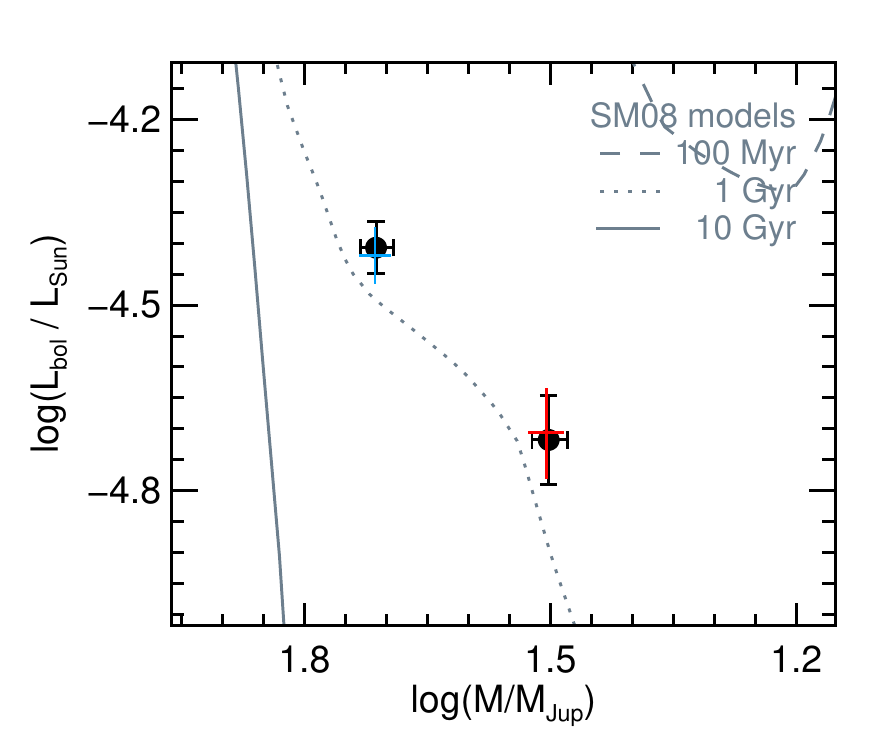} \includegraphics[width=3.0in,angle=0]{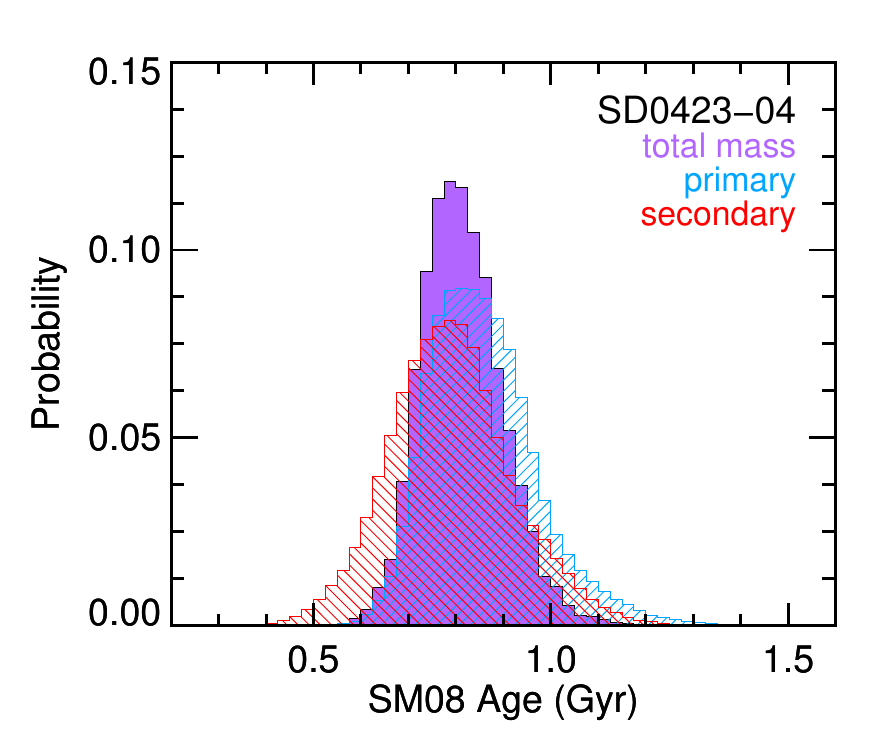}} 
  \centerline{\includegraphics[width=3.0in,angle=0]{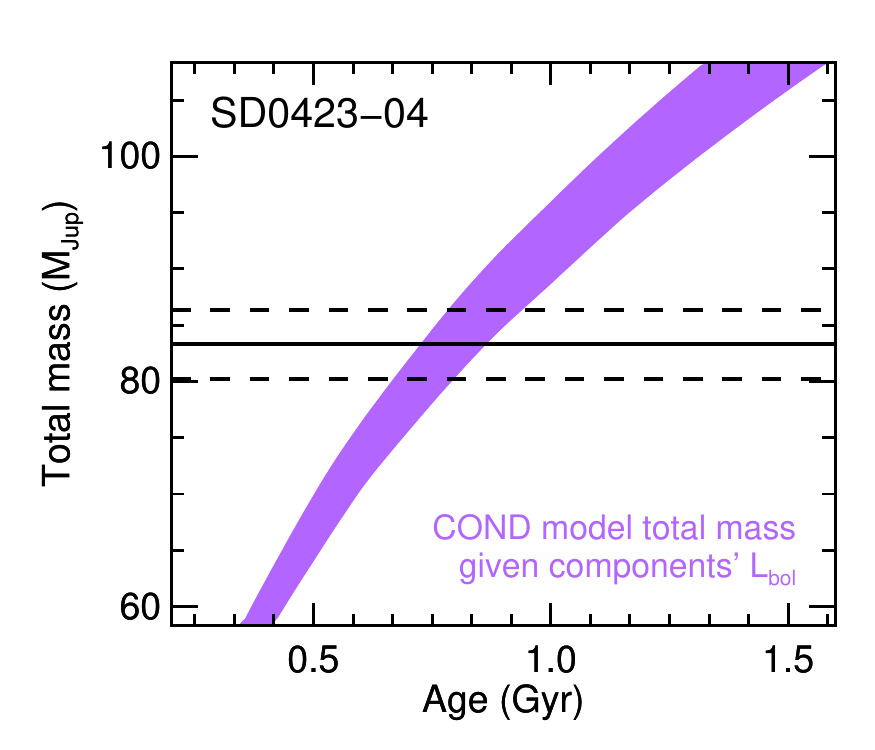} \includegraphics[width=3.0in,angle=0]{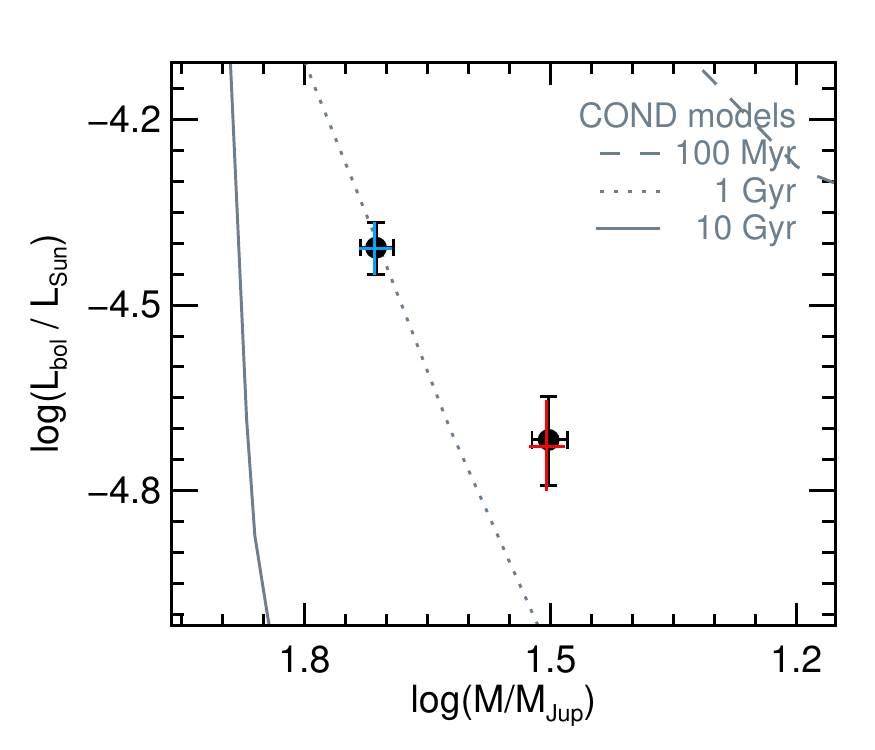} \includegraphics[width=3.0in,angle=0]{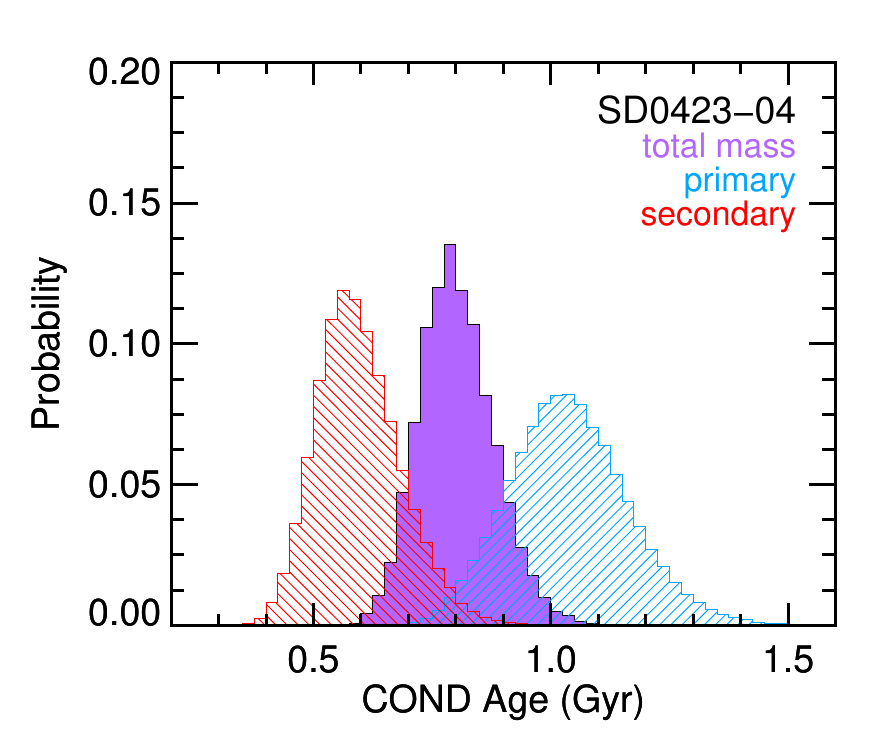}} 
\ContinuedFloat \caption{\normalsize (Continued)} \end{figure} 
\clearpage
\begin{figure} 
  \centerline{\includegraphics[width=3.0in,angle=0]{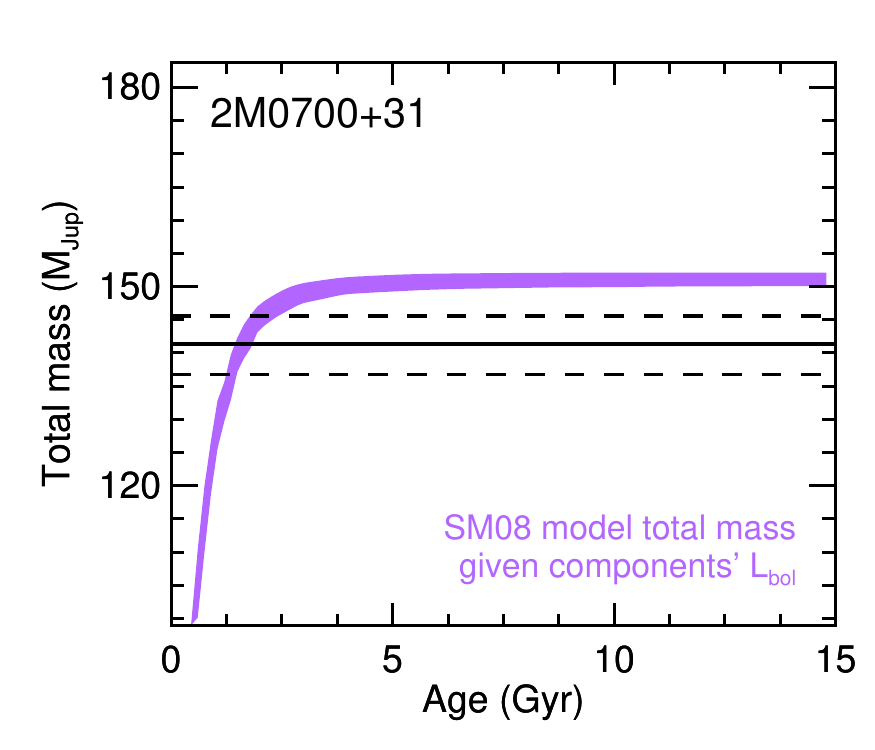} \includegraphics[width=3.0in,angle=0]{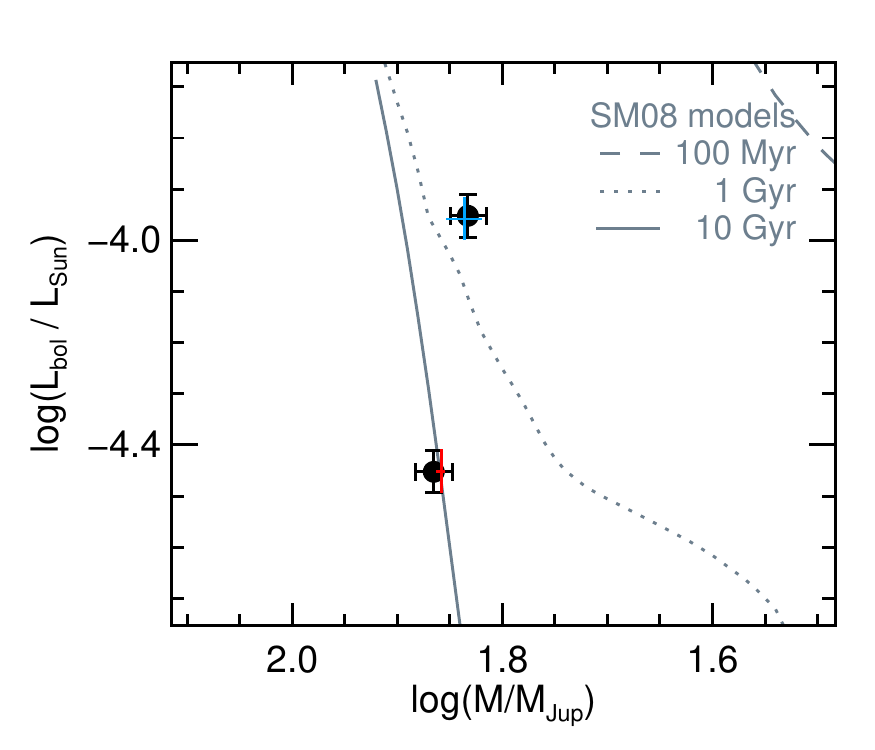} \includegraphics[width=3.0in,angle=0]{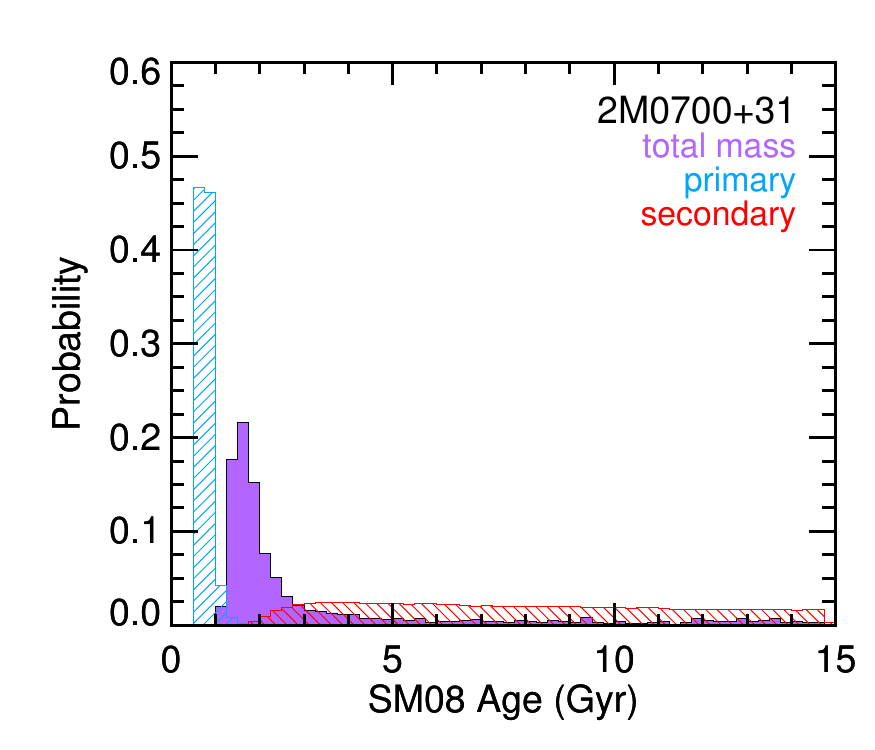}} 
  \centerline{\includegraphics[width=3.0in,angle=0]{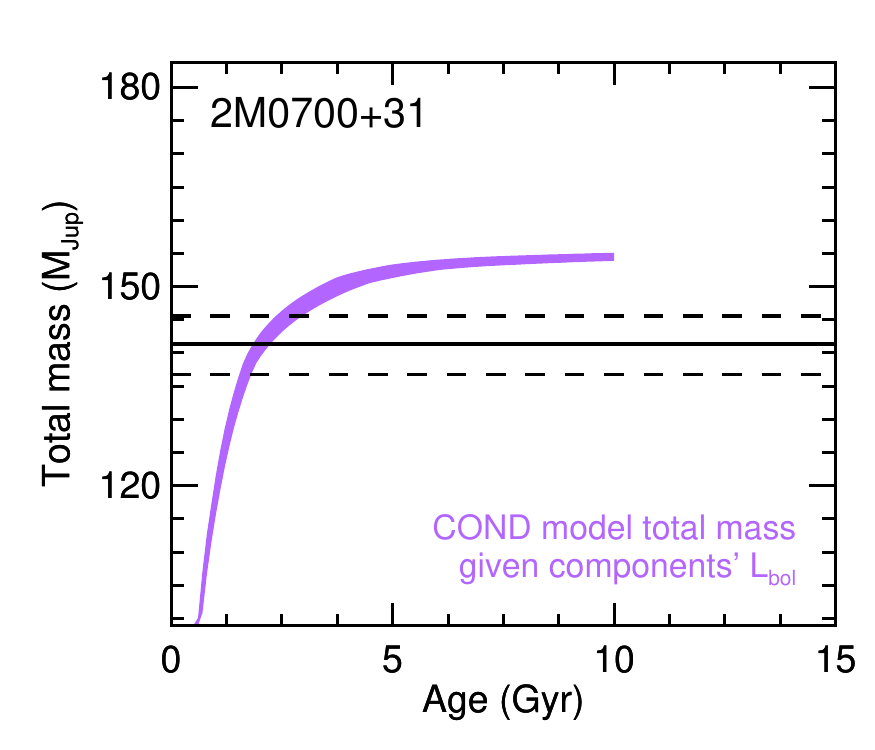} \includegraphics[width=3.0in,angle=0]{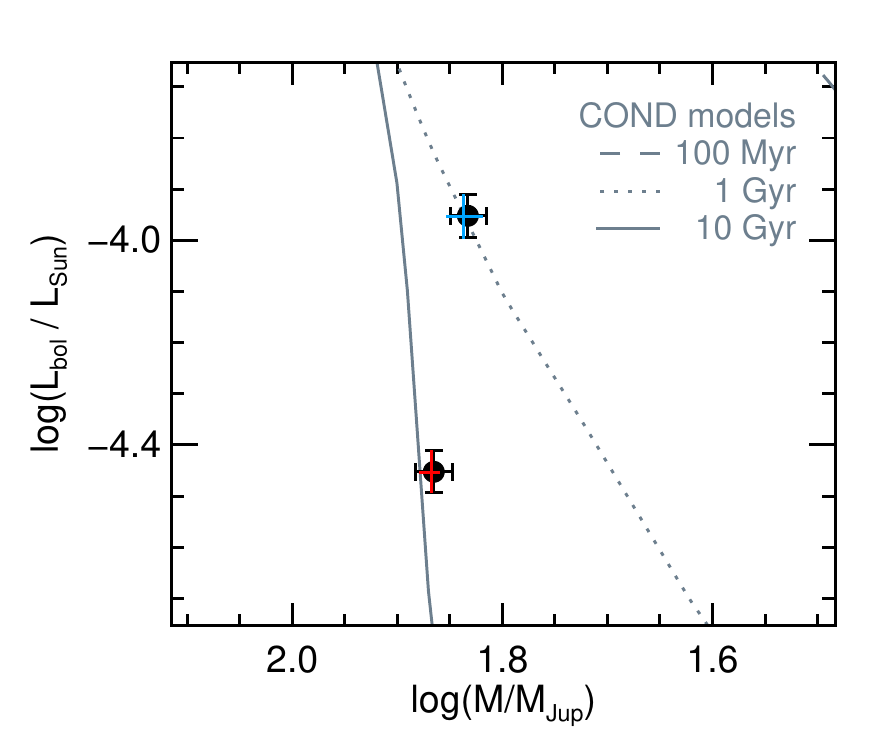} \includegraphics[width=3.0in,angle=0]{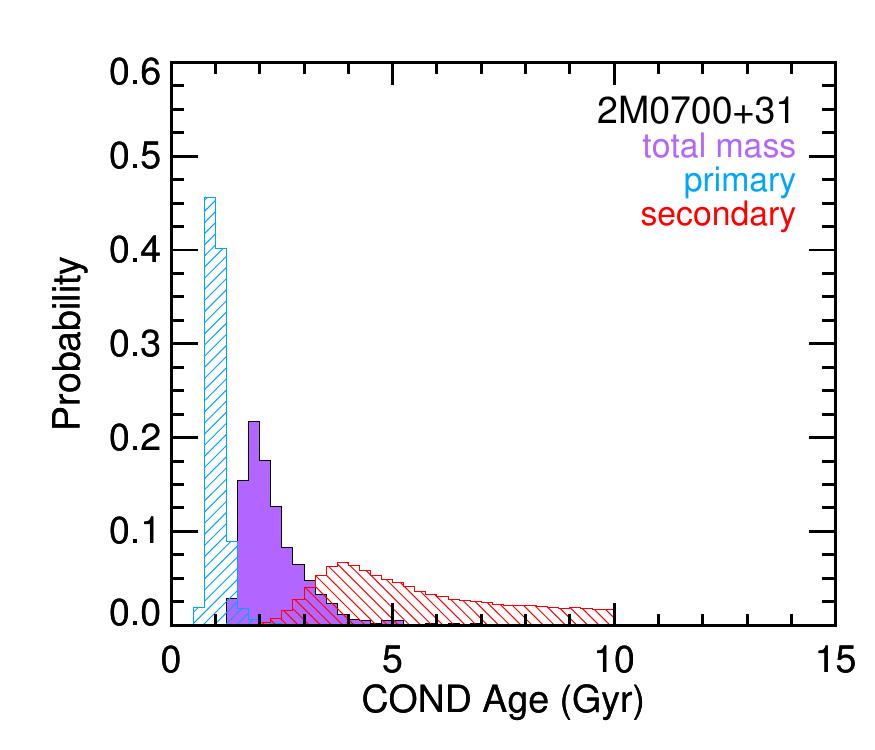}} 
\ContinuedFloat \caption{\normalsize (Continued)} \end{figure}
\clearpage
\begin{figure} 
  \centerline{\includegraphics[width=3.0in,angle=0]{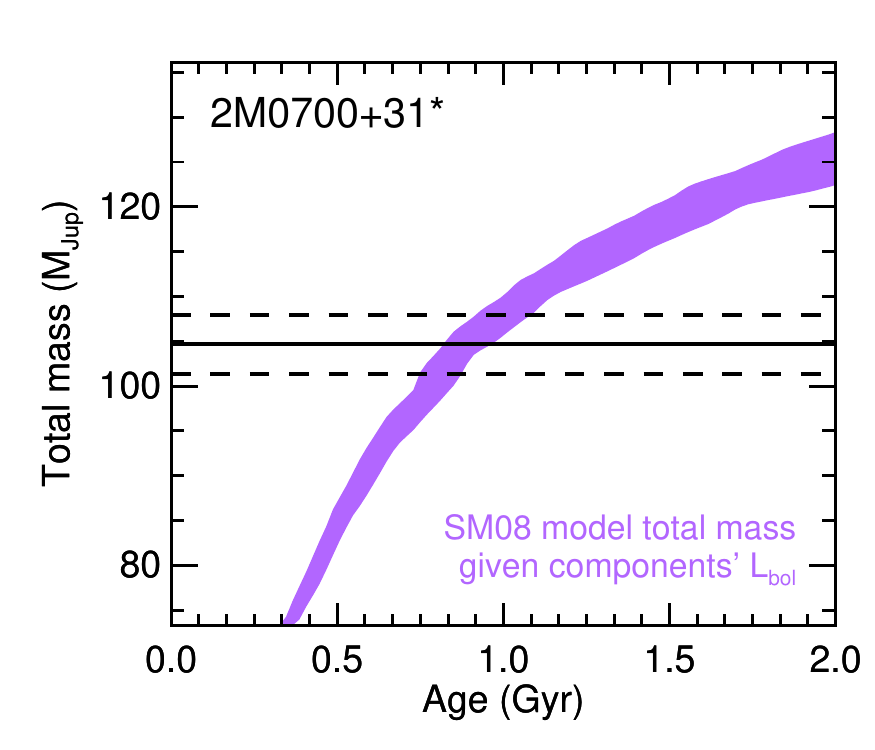} \includegraphics[width=3.0in,angle=0]{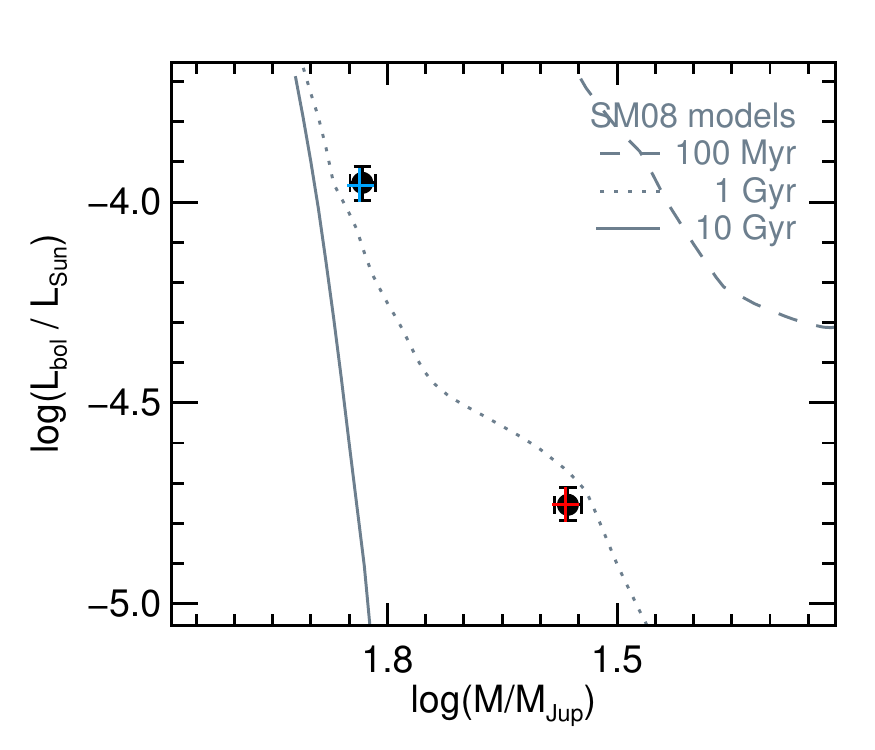} \includegraphics[width=3.0in,angle=0]{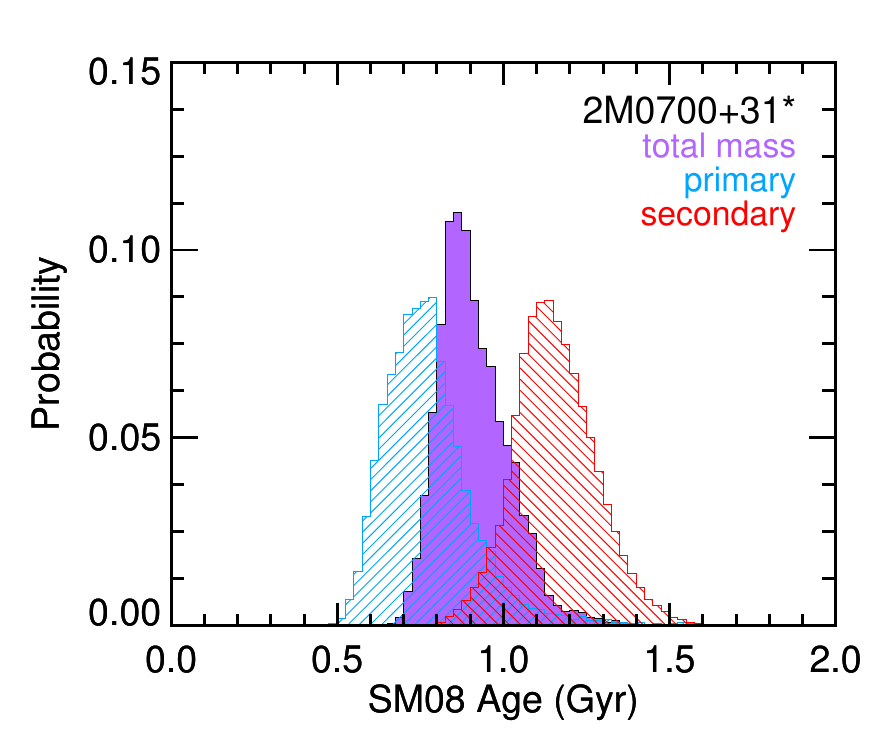}} 
  \centerline{\includegraphics[width=3.0in,angle=0]{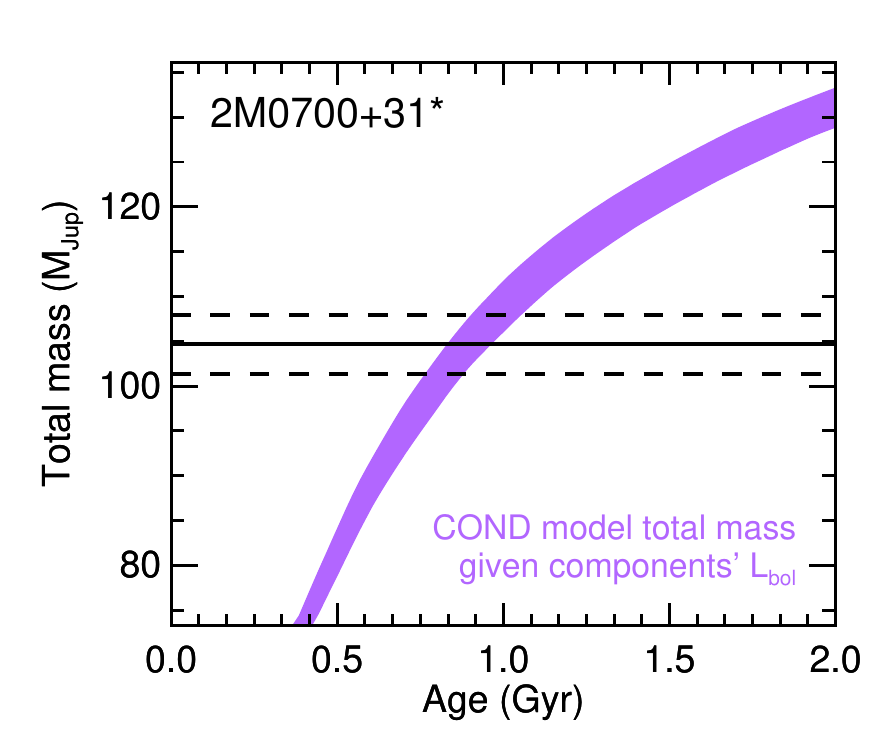} \includegraphics[width=3.0in,angle=0]{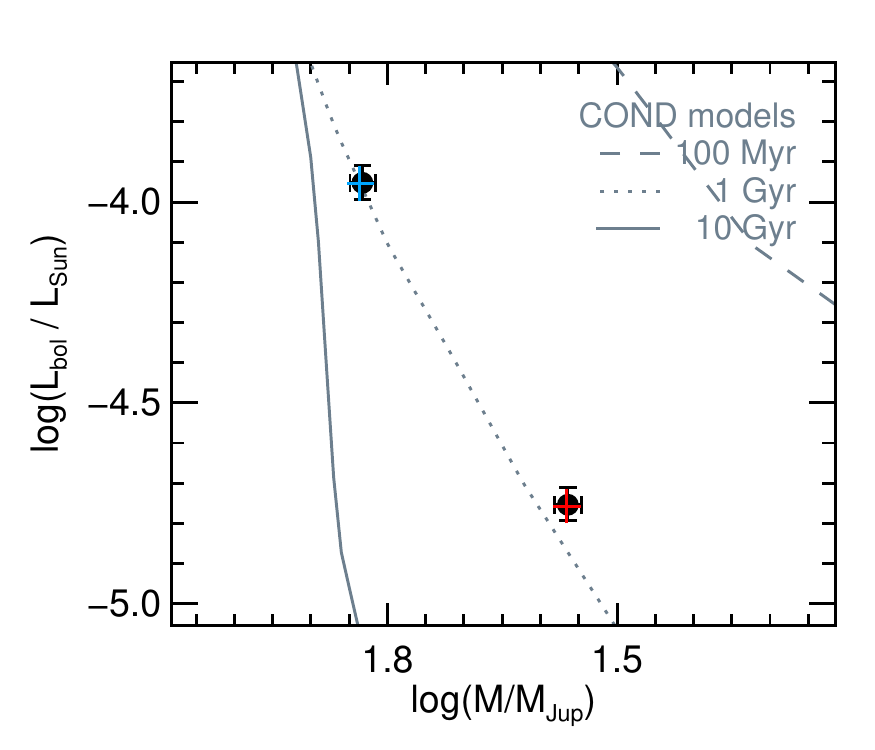} \includegraphics[width=3.0in,angle=0]{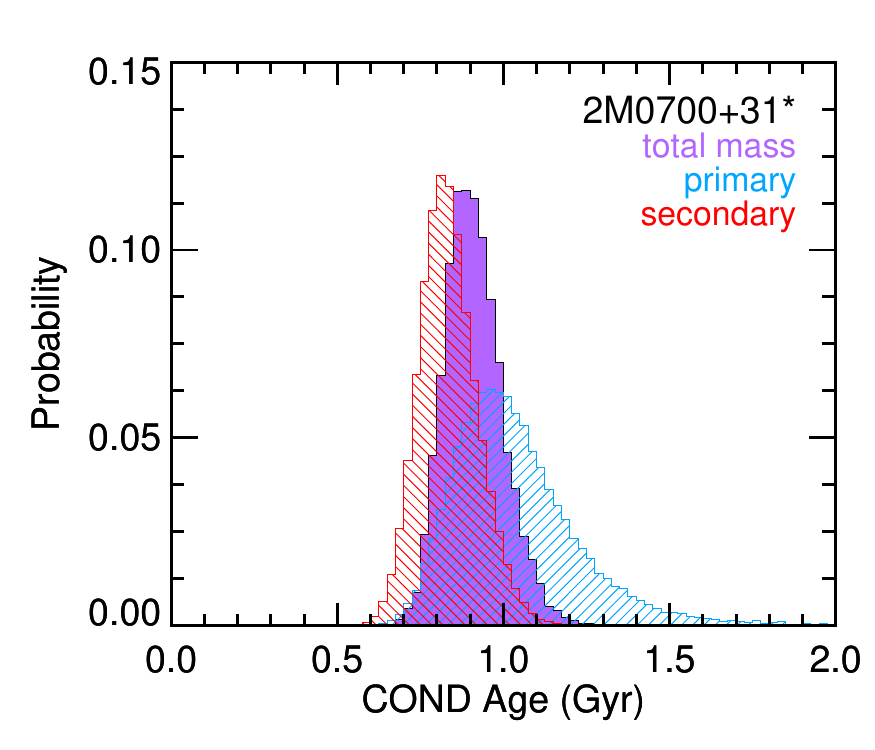}} 
\ContinuedFloat \caption{\normalsize (Continued)} \end{figure}
\clearpage
\begin{figure} 
  \centerline{\includegraphics[width=3.0in,angle=0]{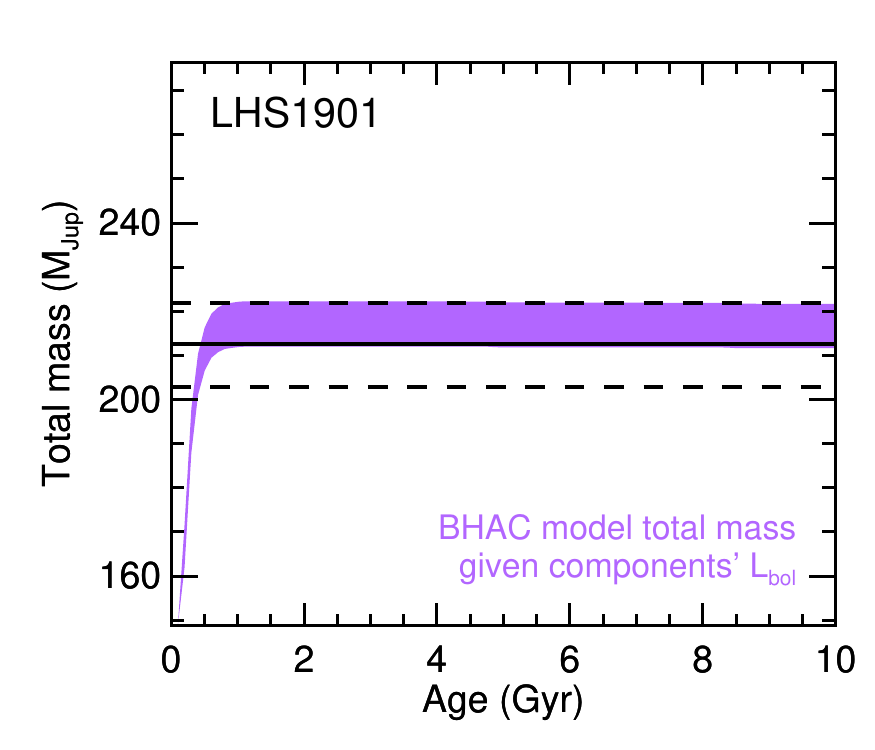} \includegraphics[width=3.0in,angle=0]{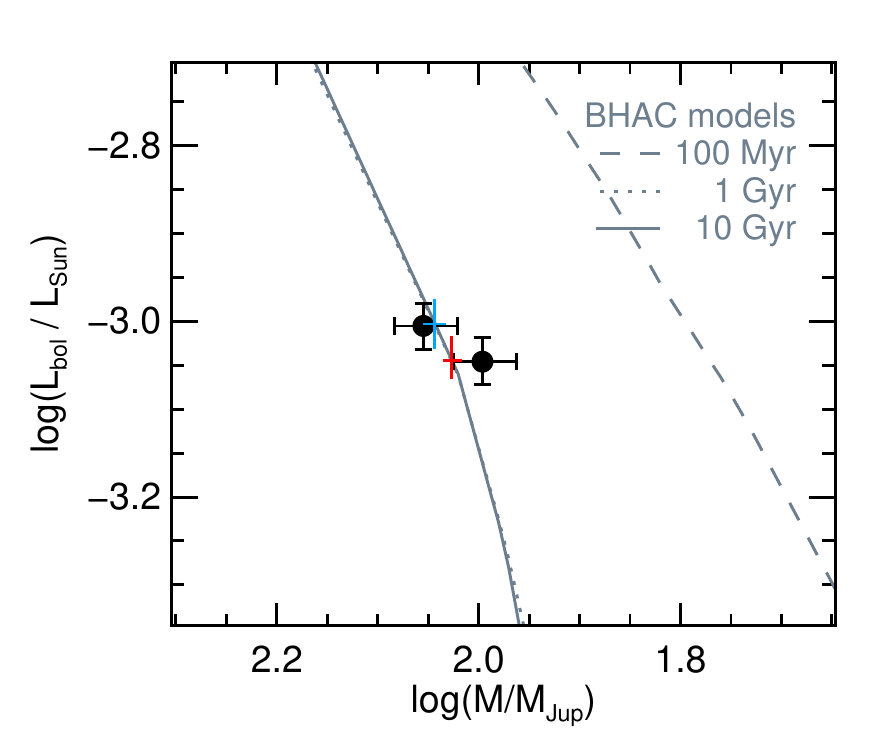} \includegraphics[width=3.0in,angle=0]{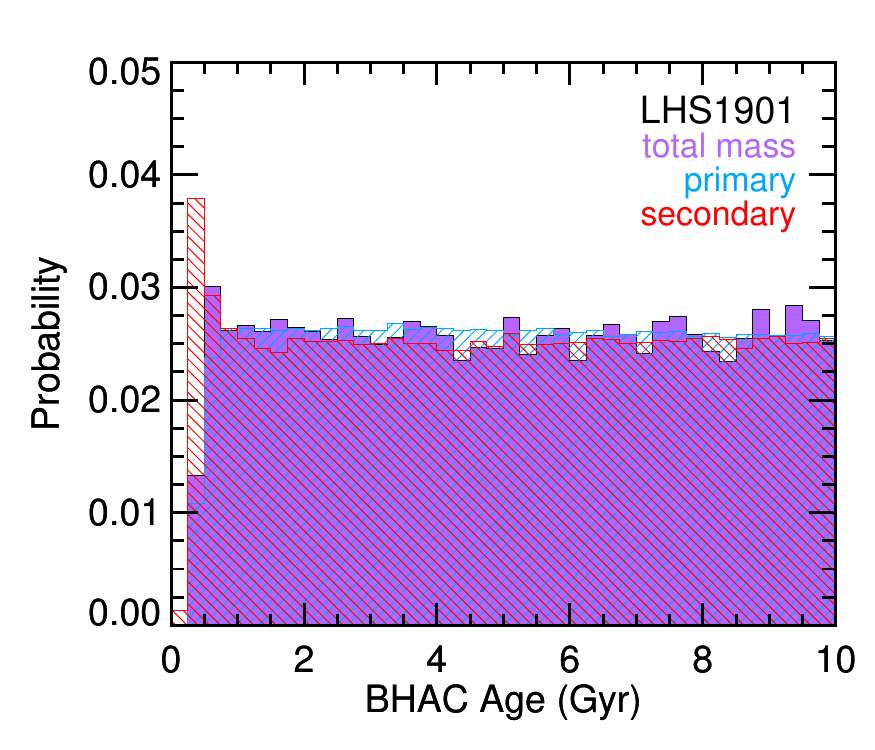}} 
\ContinuedFloat \caption{\normalsize (Continued)} \end{figure} 
\clearpage
\begin{figure} 
  \centerline{\includegraphics[width=3.0in,angle=0]{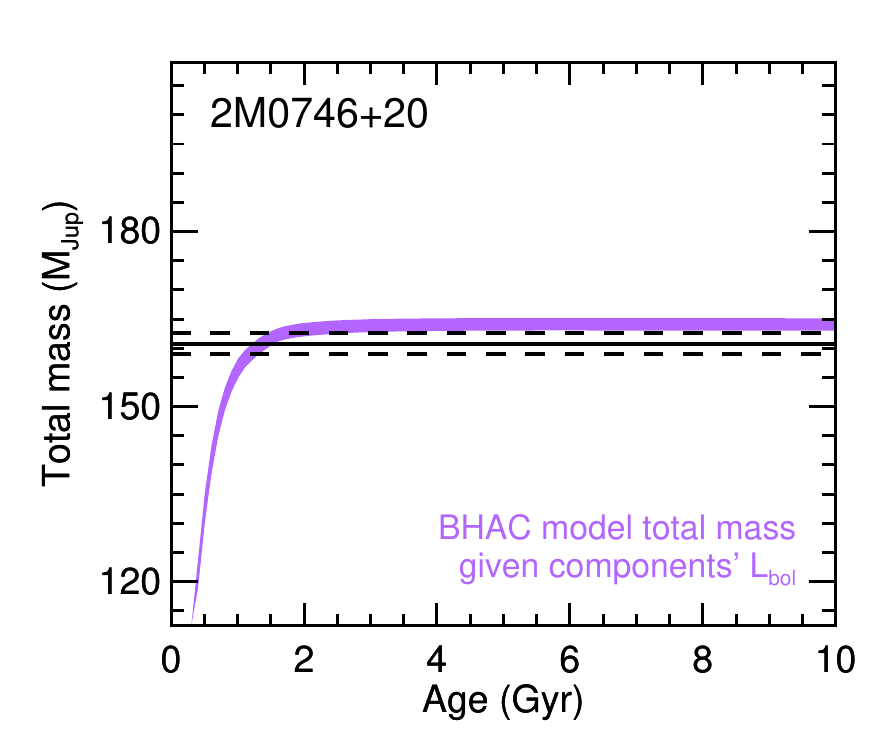} \includegraphics[width=3.0in,angle=0]{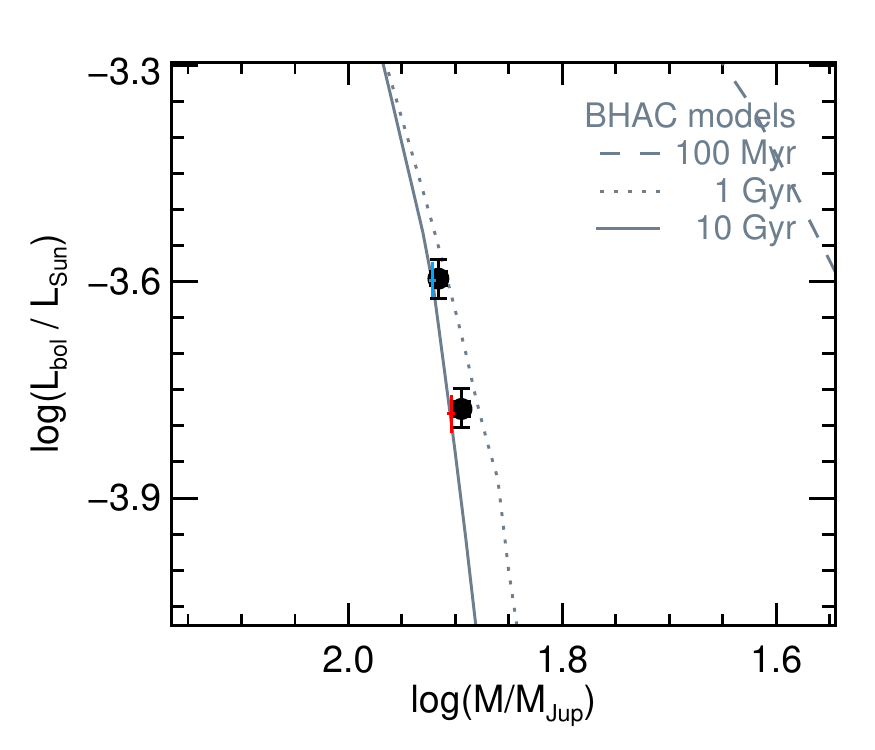} \includegraphics[width=3.0in,angle=0]{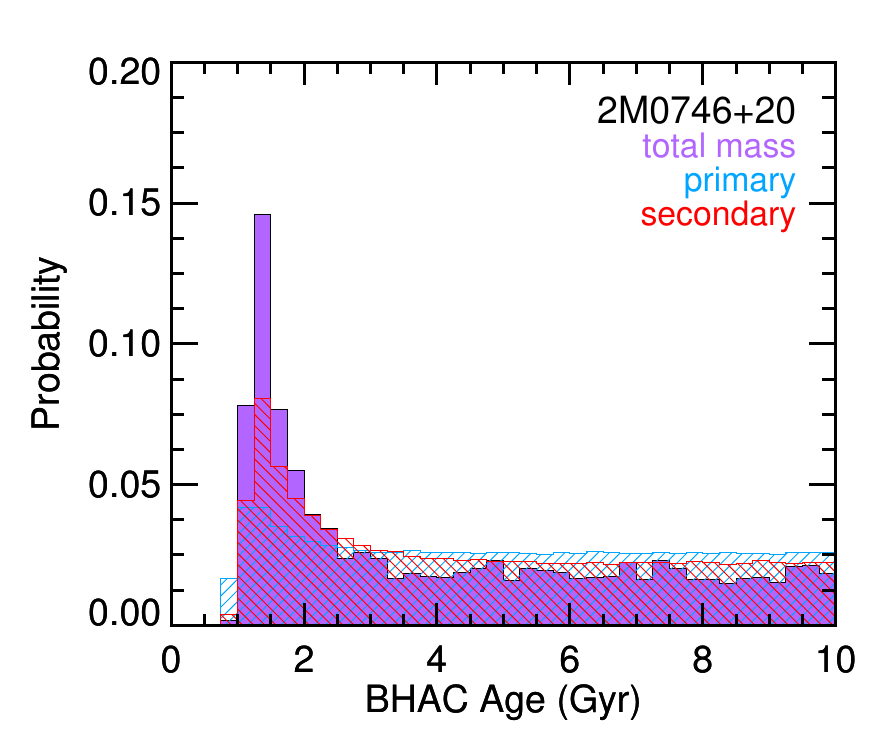}} 
\ContinuedFloat \caption{\normalsize (Continued)} \end{figure} 
\clearpage
\begin{figure} 
  \centerline{\includegraphics[width=3.0in,angle=0]{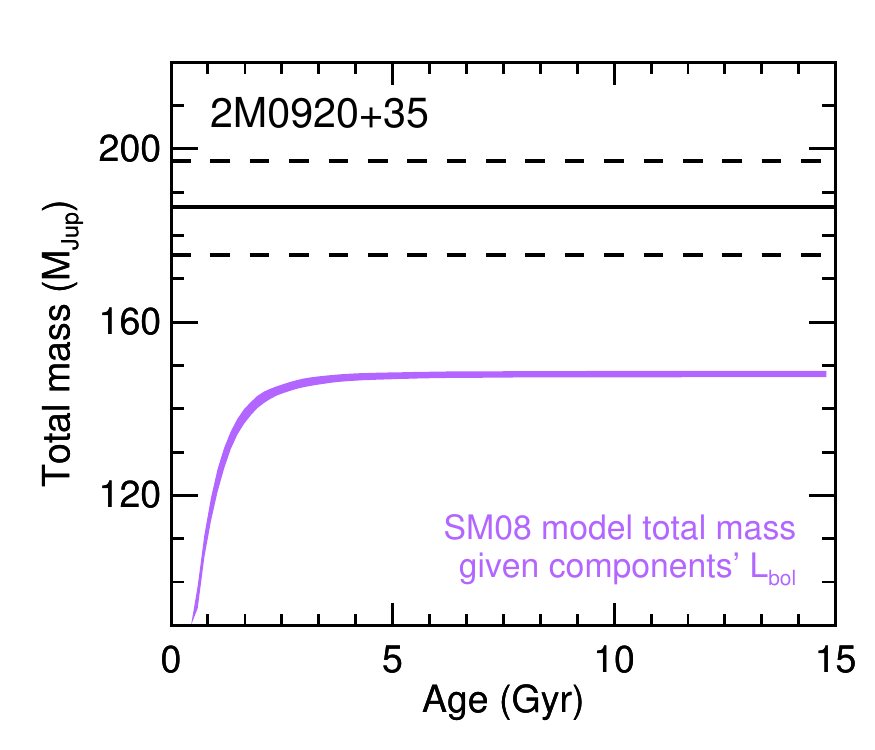} \includegraphics[width=3.0in,angle=0]{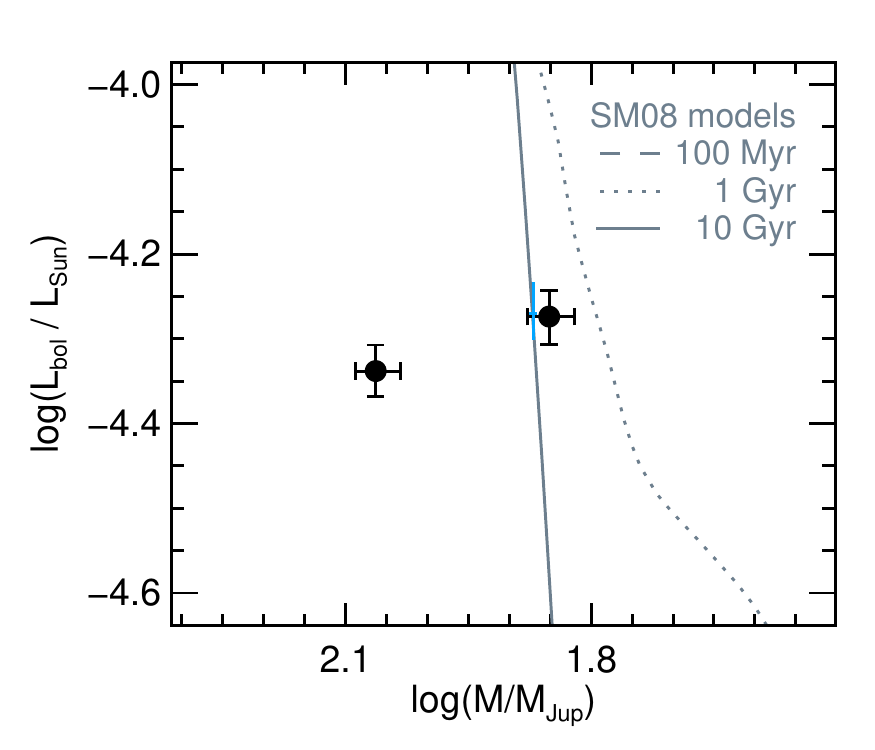} \includegraphics[width=3.0in,angle=0]{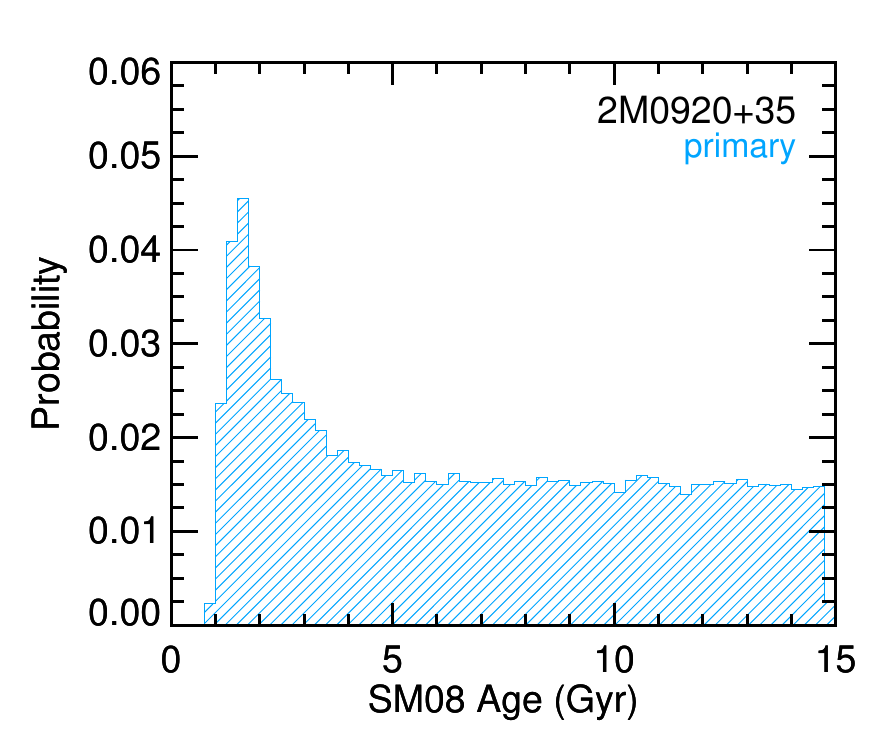}} 
  \centerline{\includegraphics[width=3.0in,angle=0]{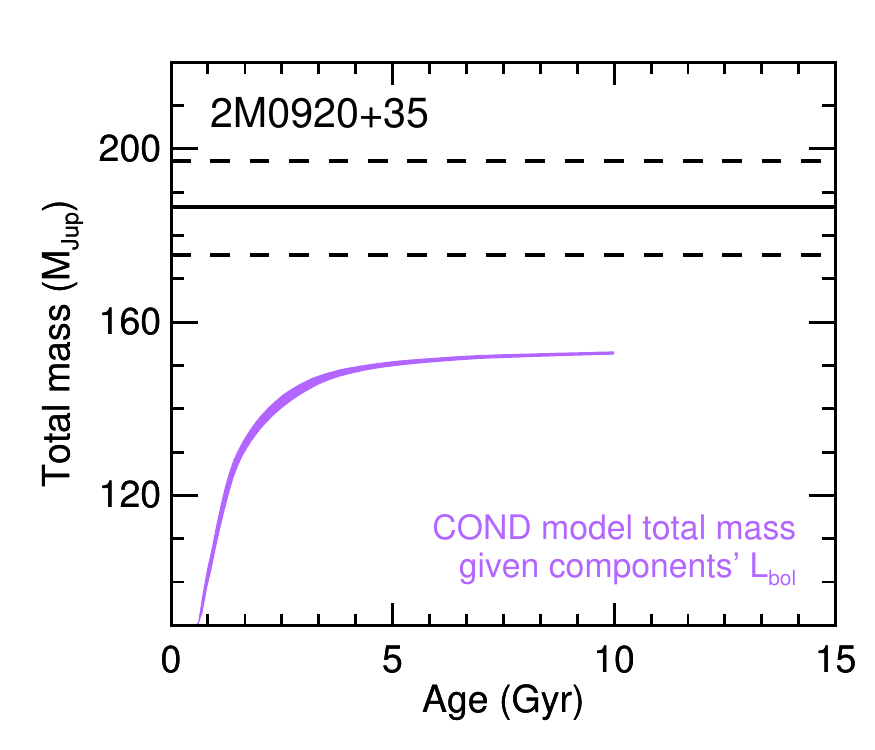} \includegraphics[width=3.0in,angle=0]{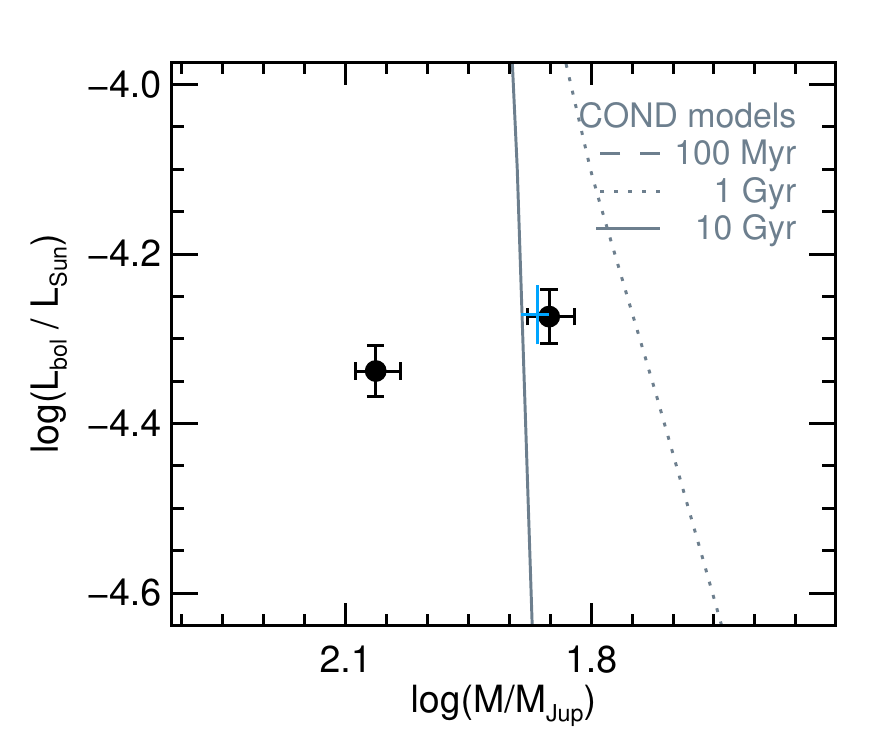} \includegraphics[width=3.0in,angle=0]{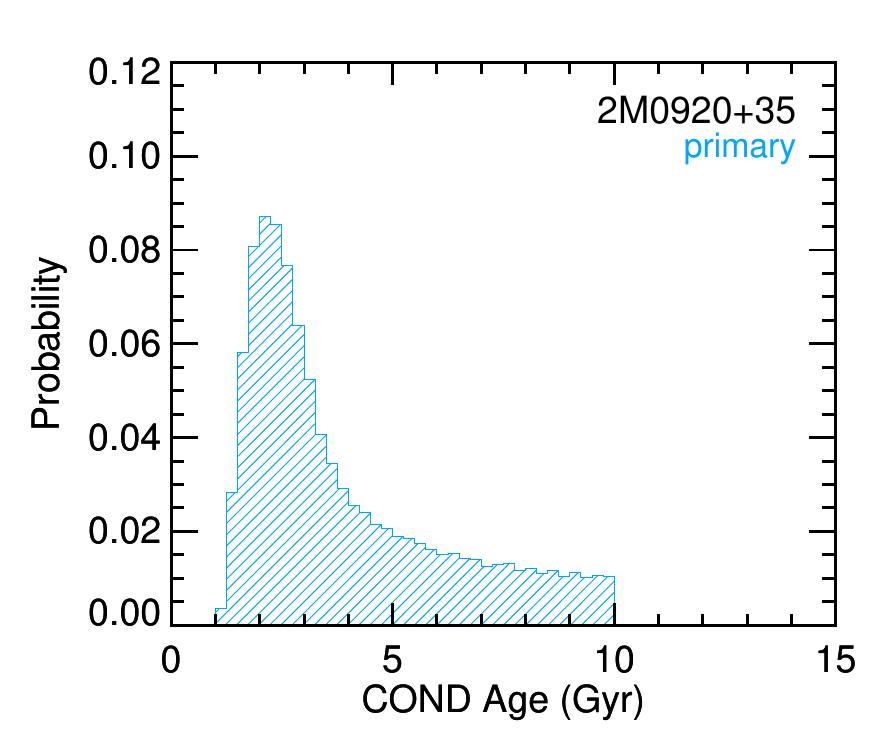}} 
\ContinuedFloat \caption{\normalsize (Continued)} \end{figure} 
\begin{figure} 
  \centerline{\includegraphics[width=3.0in,angle=0]{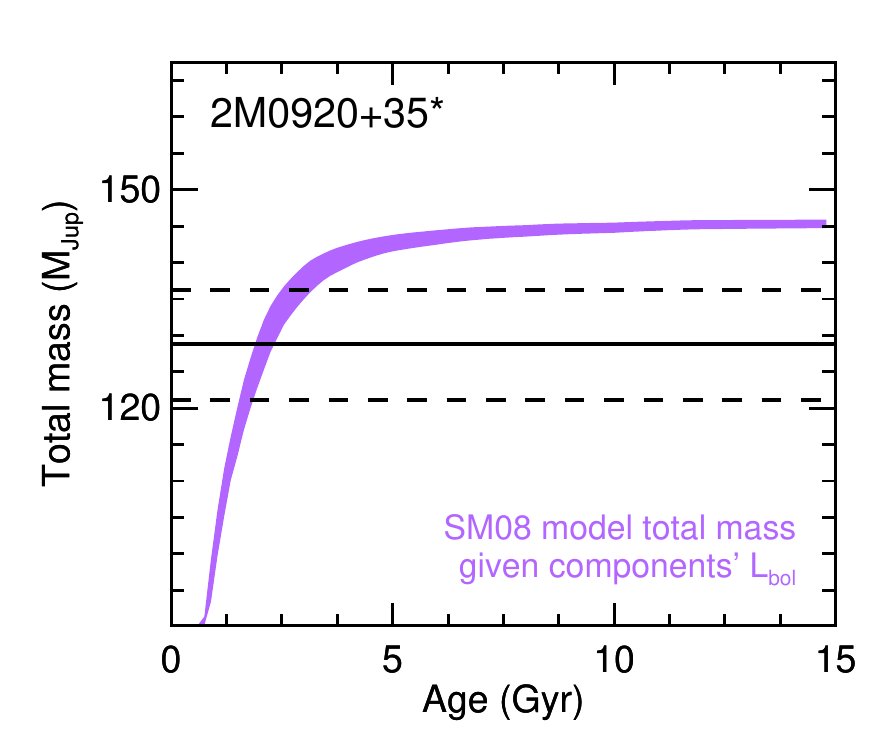} \includegraphics[width=3.0in,angle=0]{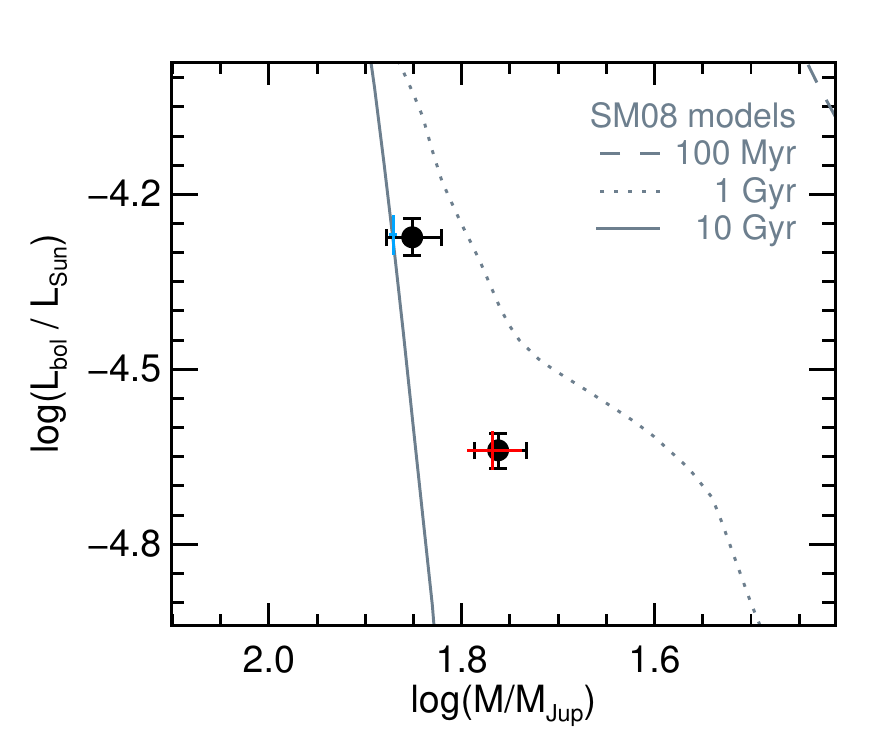} \includegraphics[width=3.0in,angle=0]{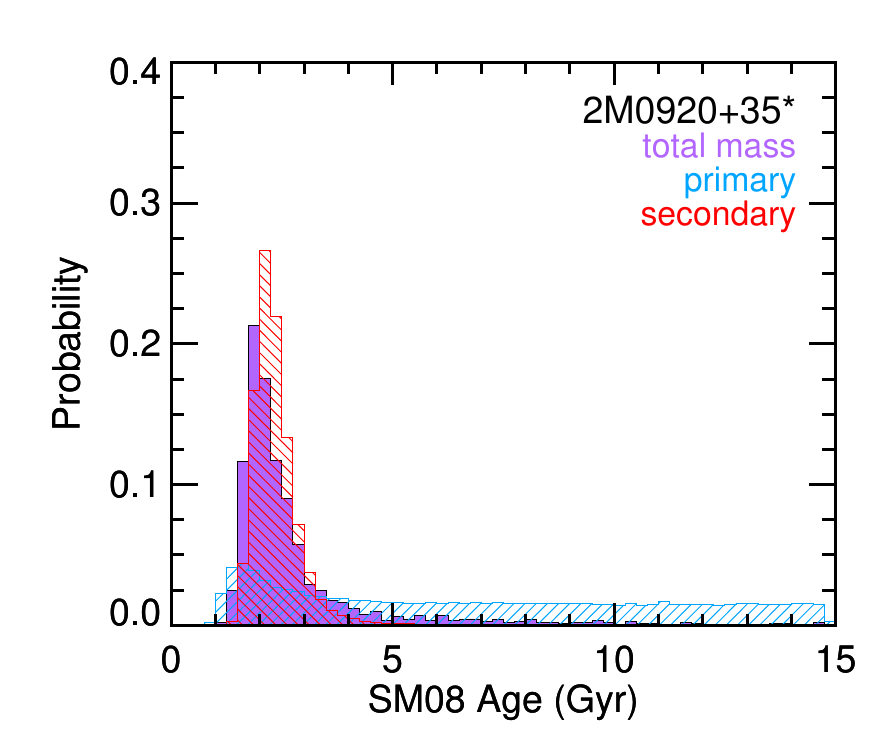}} 
  \centerline{\includegraphics[width=3.0in,angle=0]{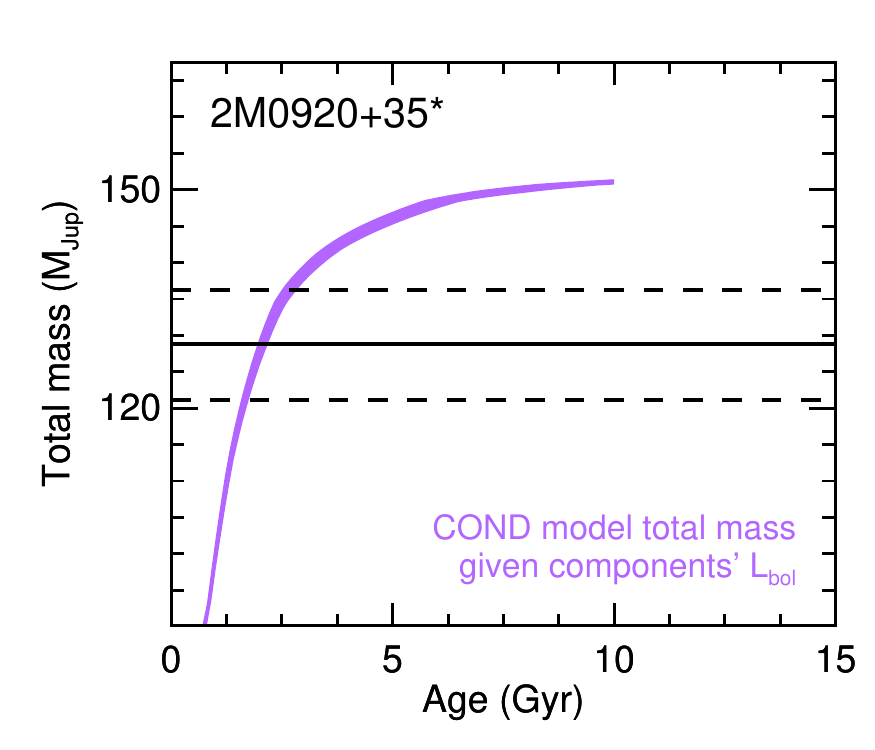} \includegraphics[width=3.0in,angle=0]{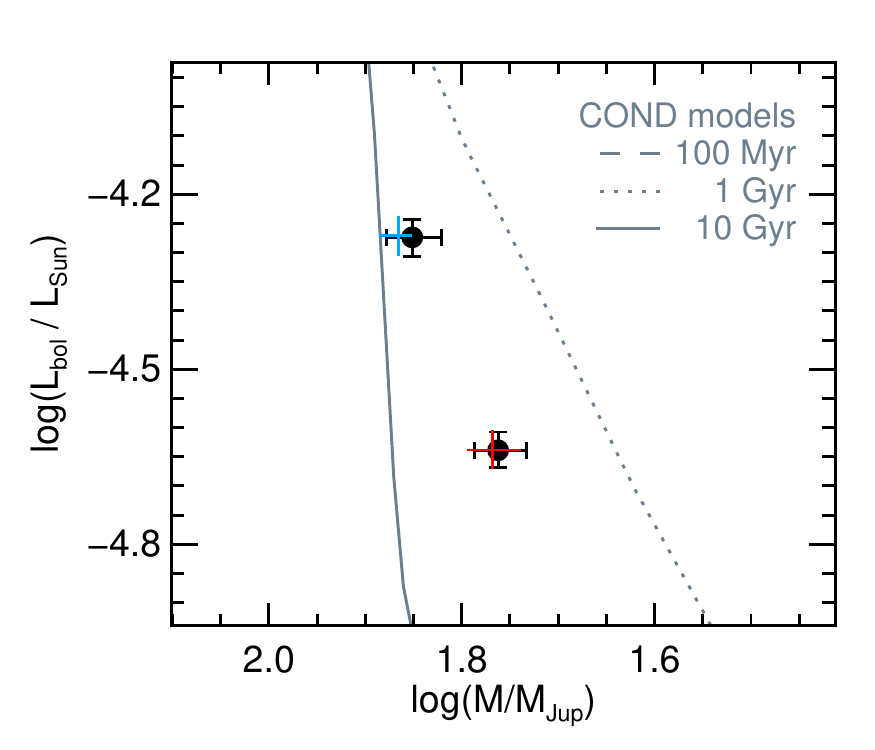} \includegraphics[width=3.0in,angle=0]{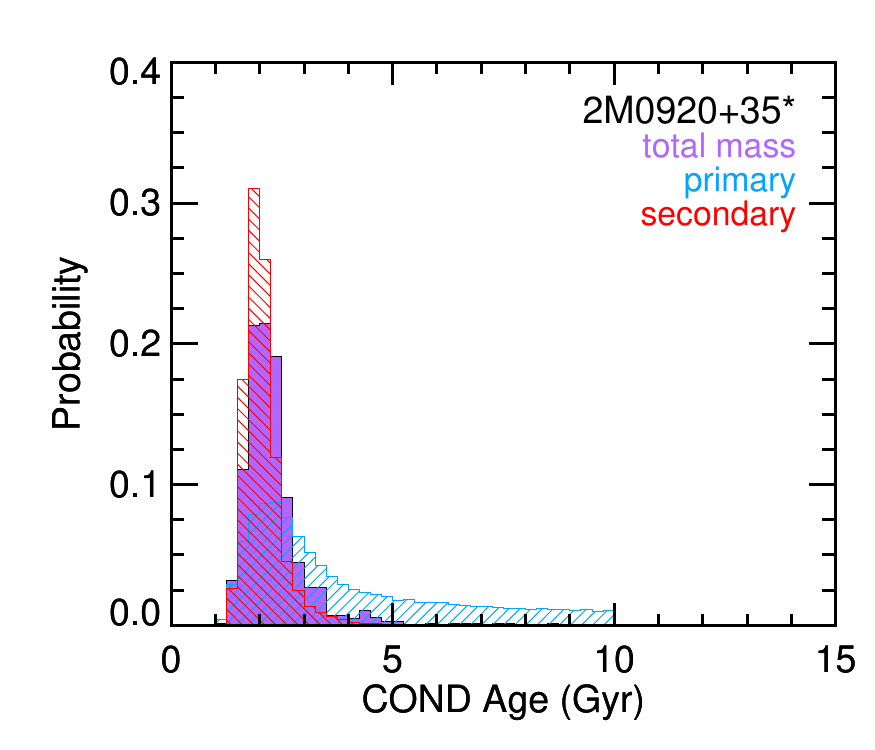}} 
\ContinuedFloat \caption{\normalsize (Continued)} \end{figure} 
\clearpage
\begin{figure} 
  \centerline{\includegraphics[width=3.0in,angle=0]{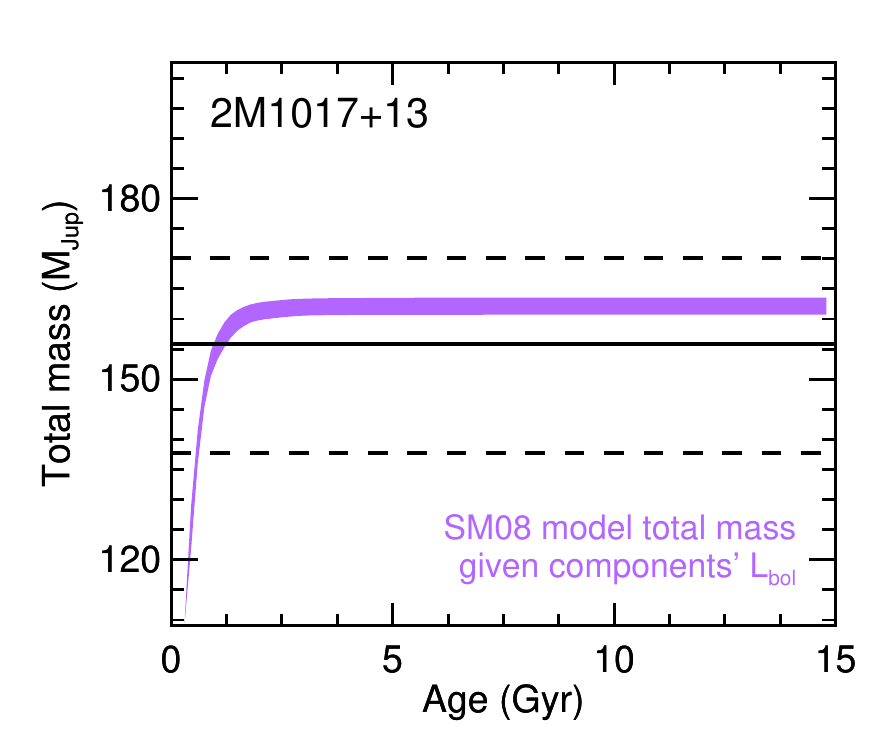} \includegraphics[width=3.0in,angle=0]{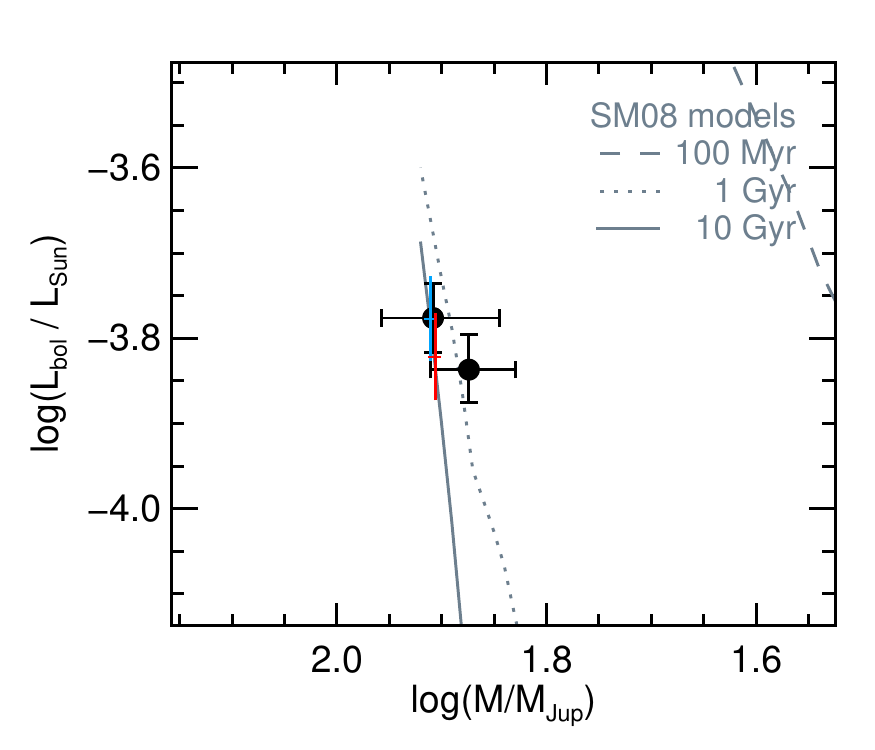} \includegraphics[width=3.0in,angle=0]{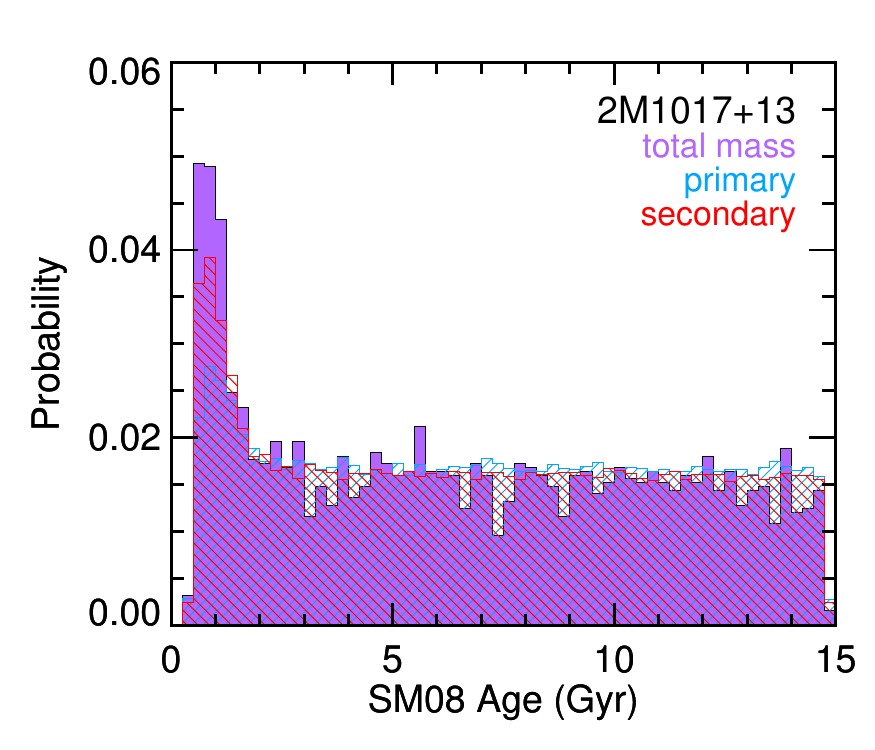}} 
  \centerline{\includegraphics[width=3.0in,angle=0]{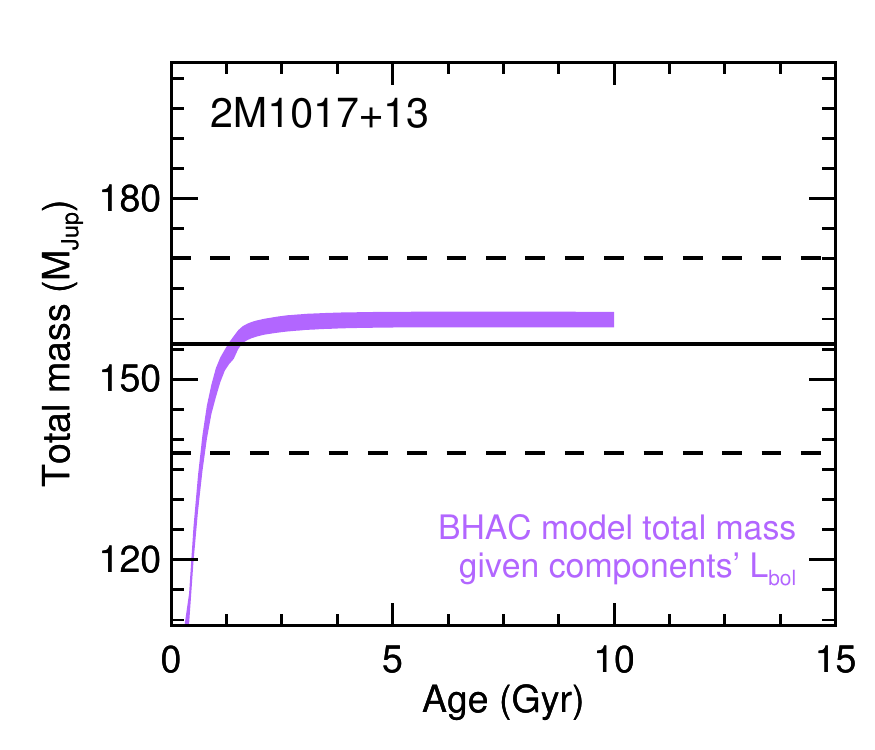} \includegraphics[width=3.0in,angle=0]{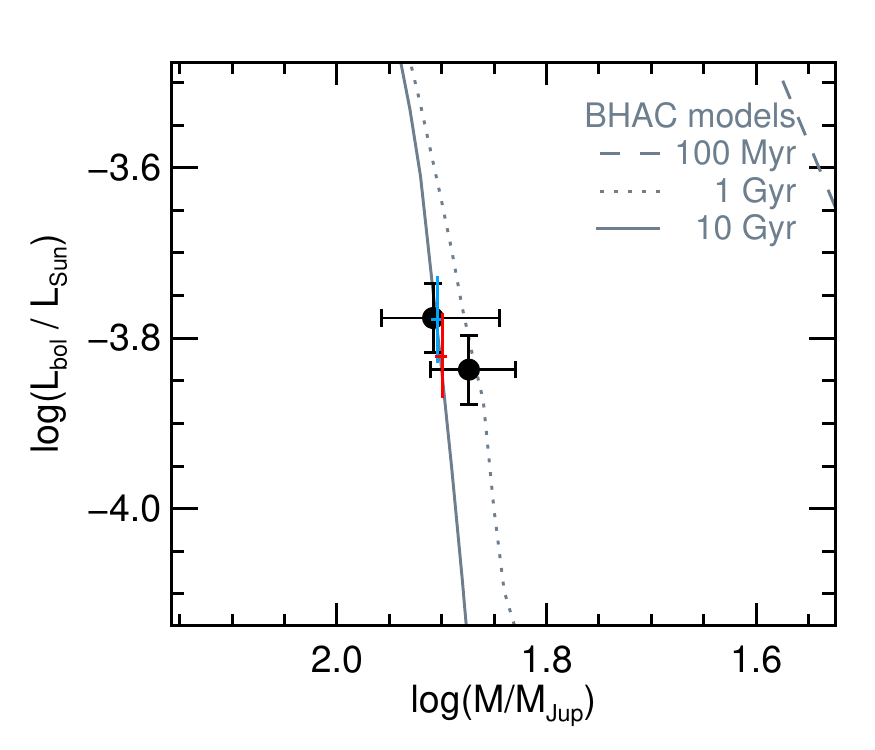} \includegraphics[width=3.0in,angle=0]{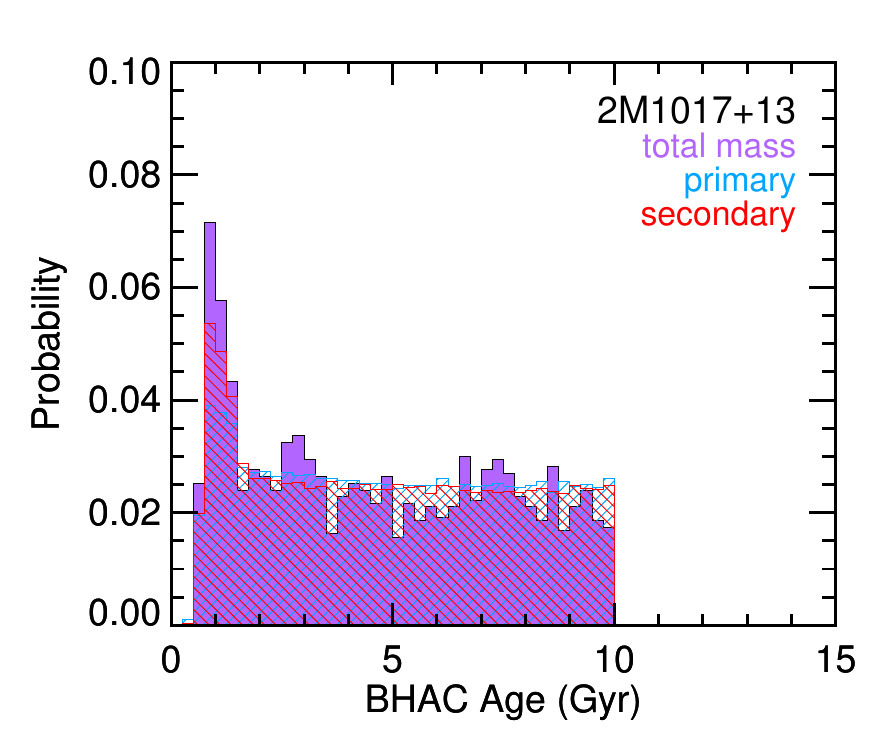}} 
\ContinuedFloat \caption{\normalsize (Continued)} \end{figure}
\clearpage
\begin{figure} 
  \centerline{\includegraphics[width=3.0in,angle=0]{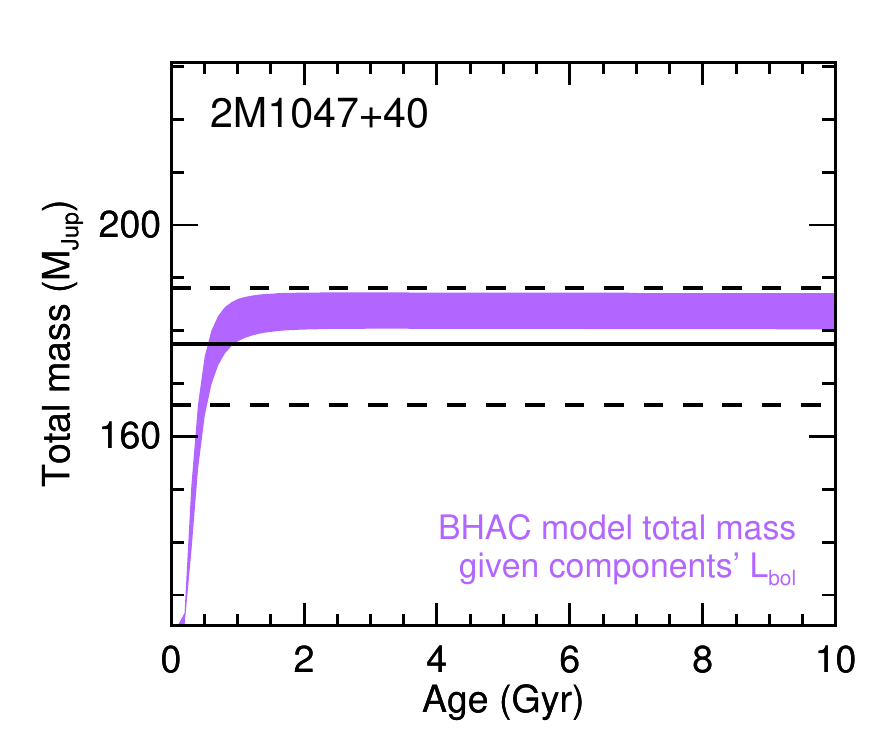} \includegraphics[width=3.0in,angle=0]{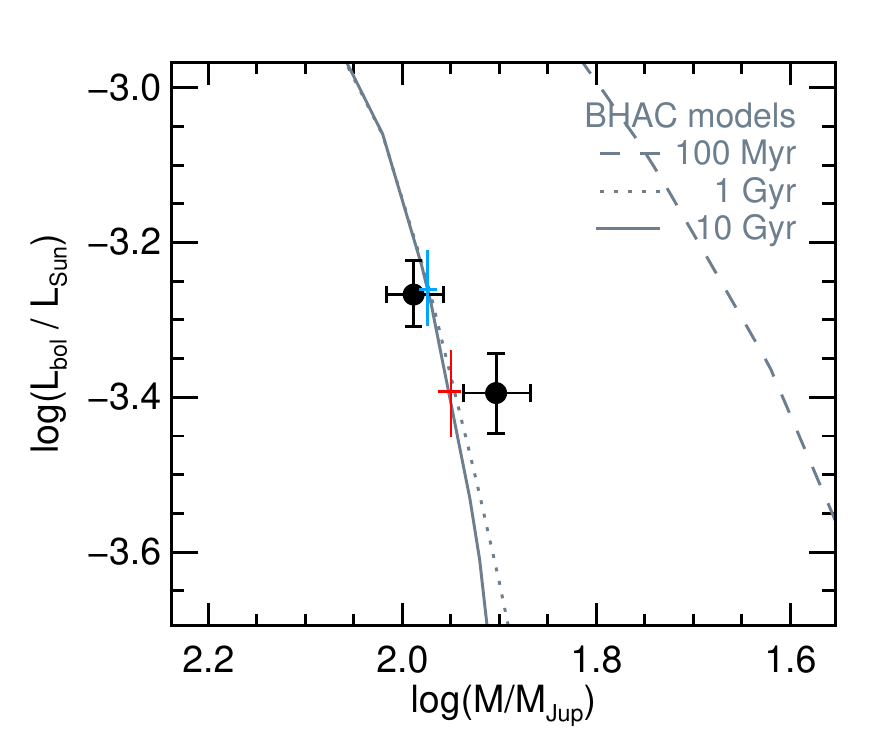} \includegraphics[width=3.0in,angle=0]{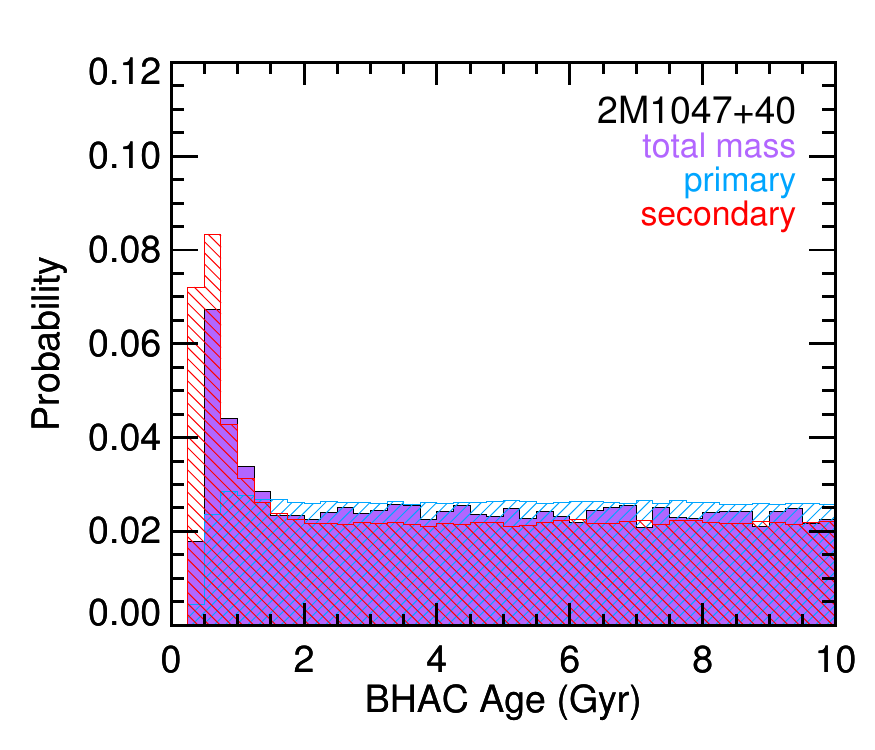}} 
\ContinuedFloat \caption{\normalsize (Continued)} \end{figure} 
\clearpage
\begin{figure} 
  \centerline{\includegraphics[width=3.0in,angle=0]{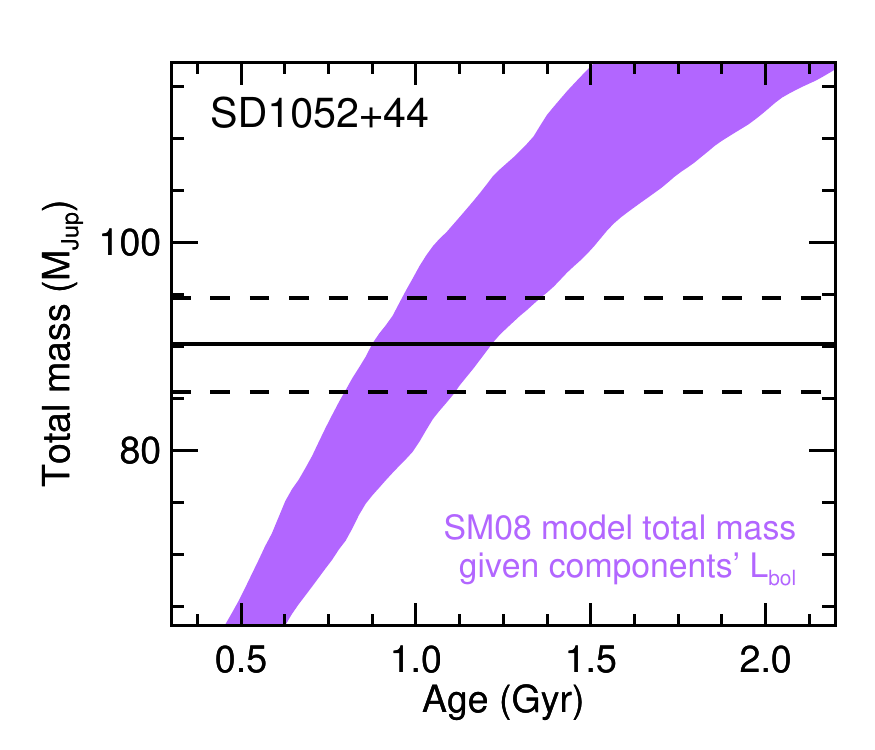} \includegraphics[width=3.0in,angle=0]{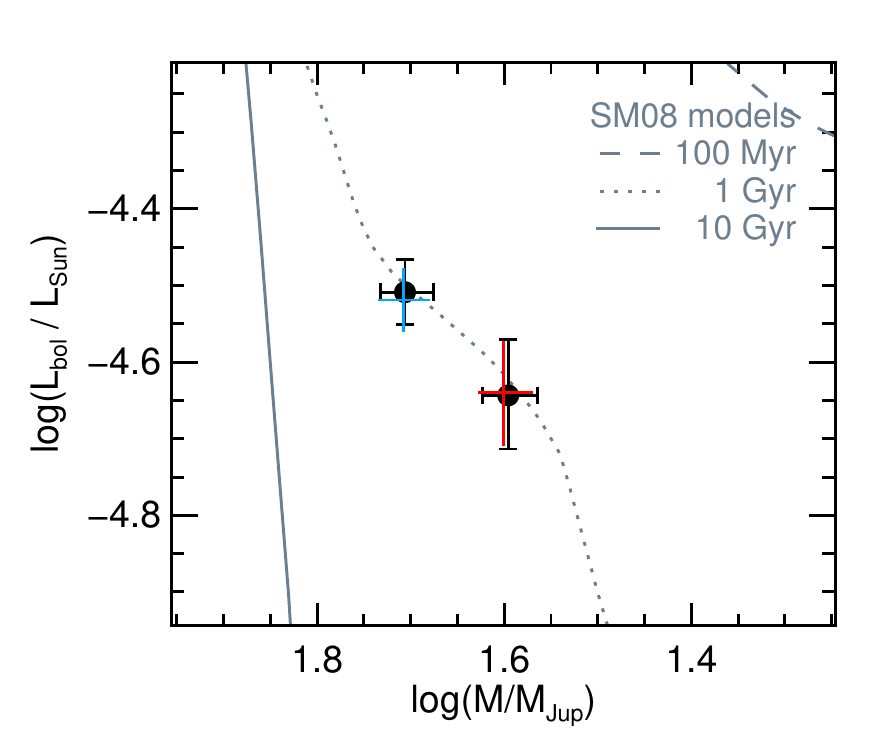} \includegraphics[width=3.0in,angle=0]{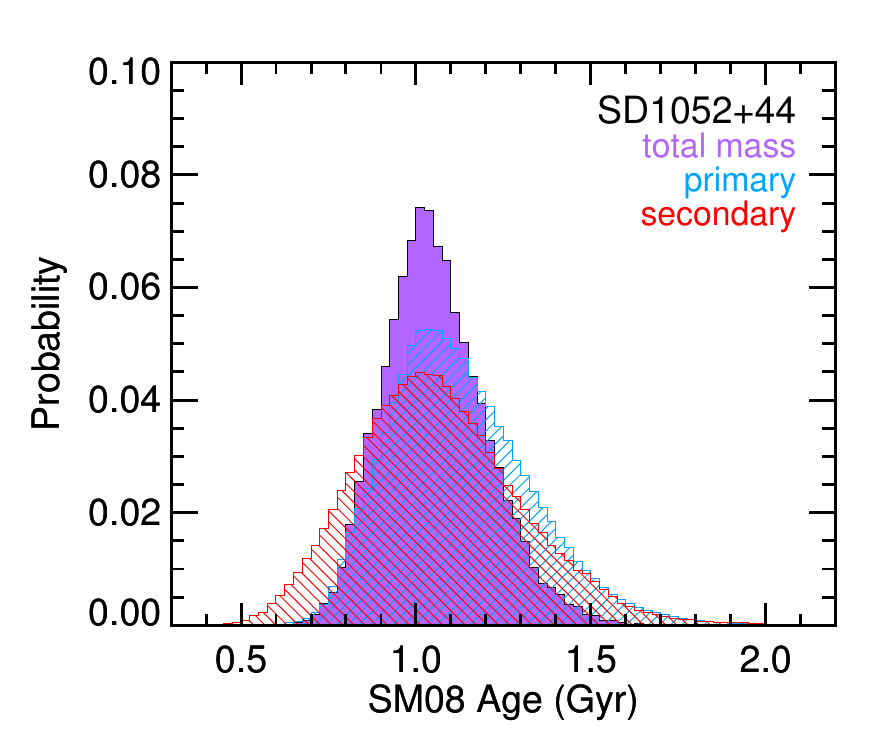}} 
  \centerline{\includegraphics[width=3.0in,angle=0]{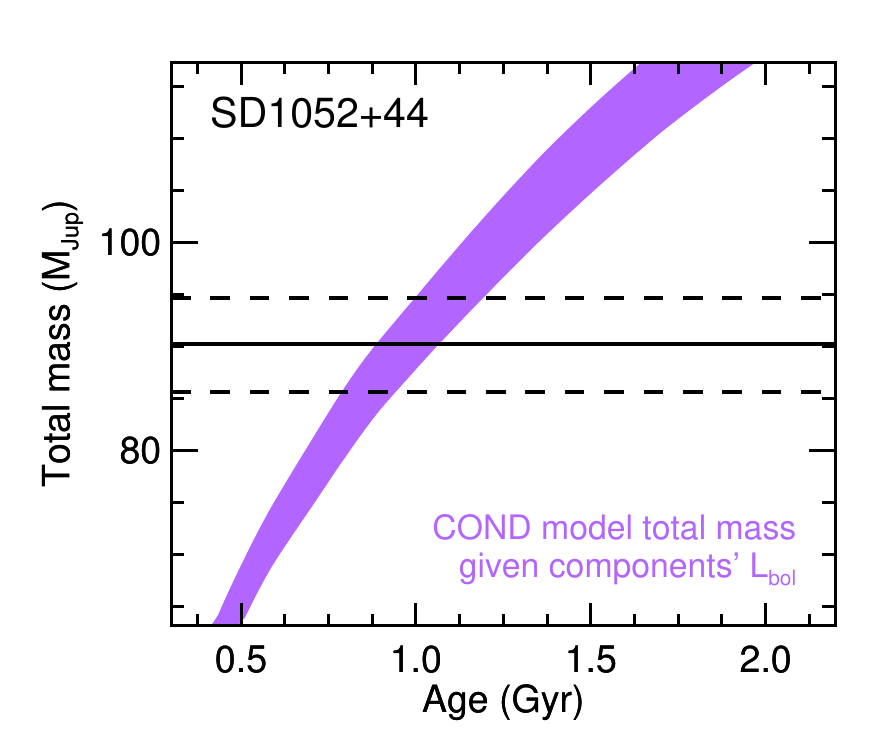} \includegraphics[width=3.0in,angle=0]{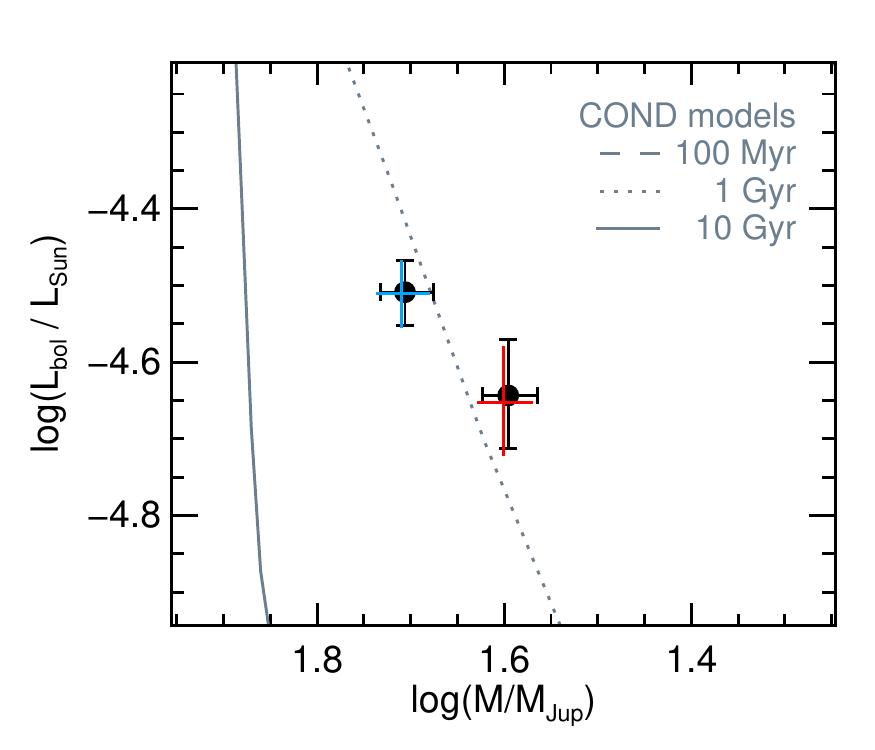} \includegraphics[width=3.0in,angle=0]{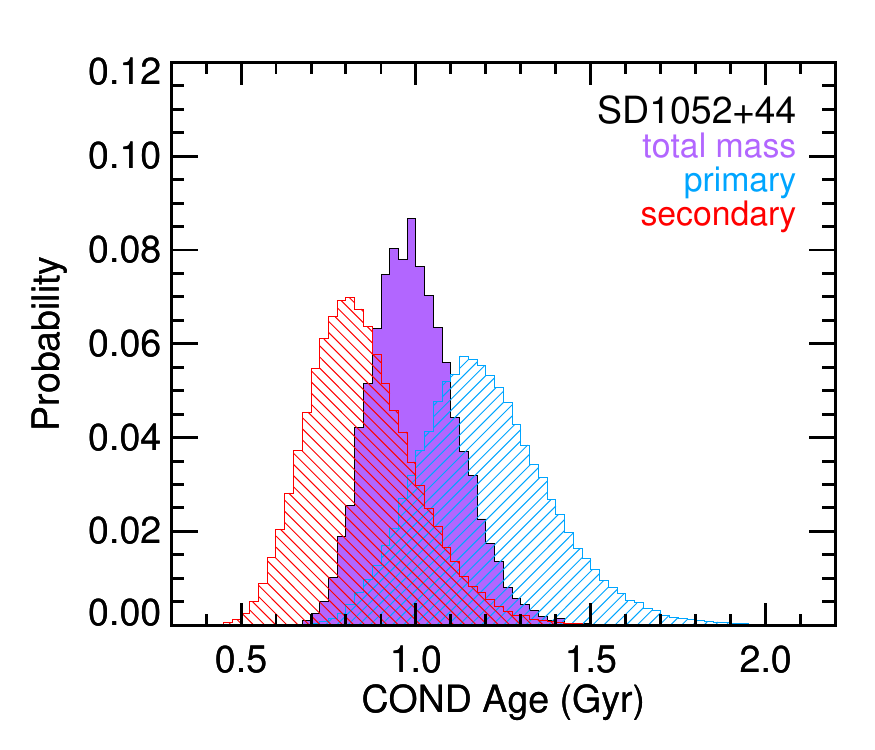}} 
\ContinuedFloat \caption{\normalsize (Continued)} \end{figure} 
\clearpage
\begin{figure} 
  \centerline{\includegraphics[width=3.0in,angle=0]{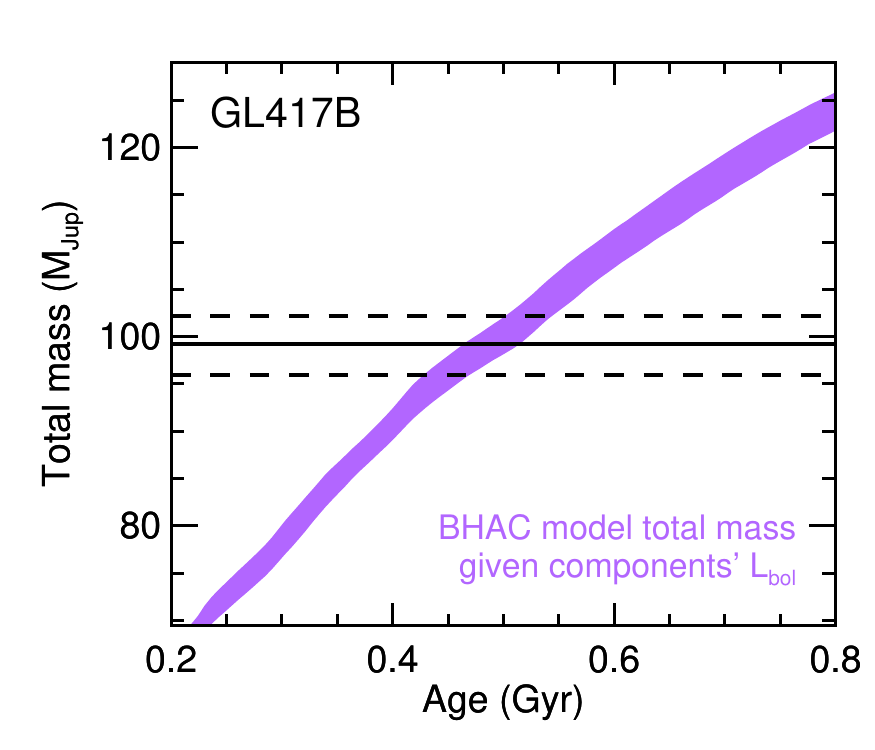} \includegraphics[width=3.0in,angle=0]{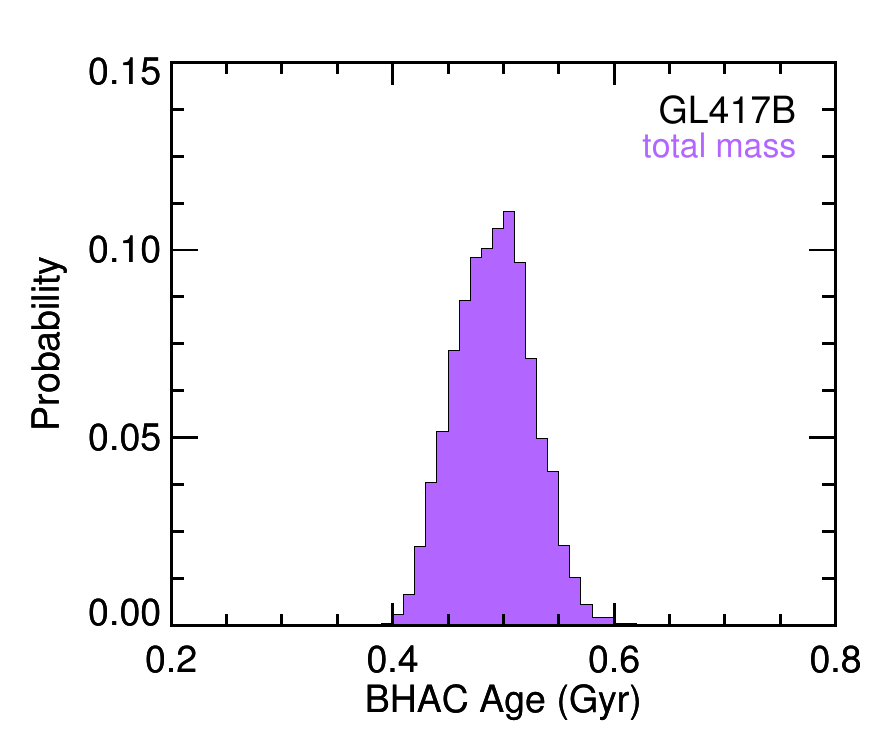}} 
  \centerline{\includegraphics[width=3.0in,angle=0]{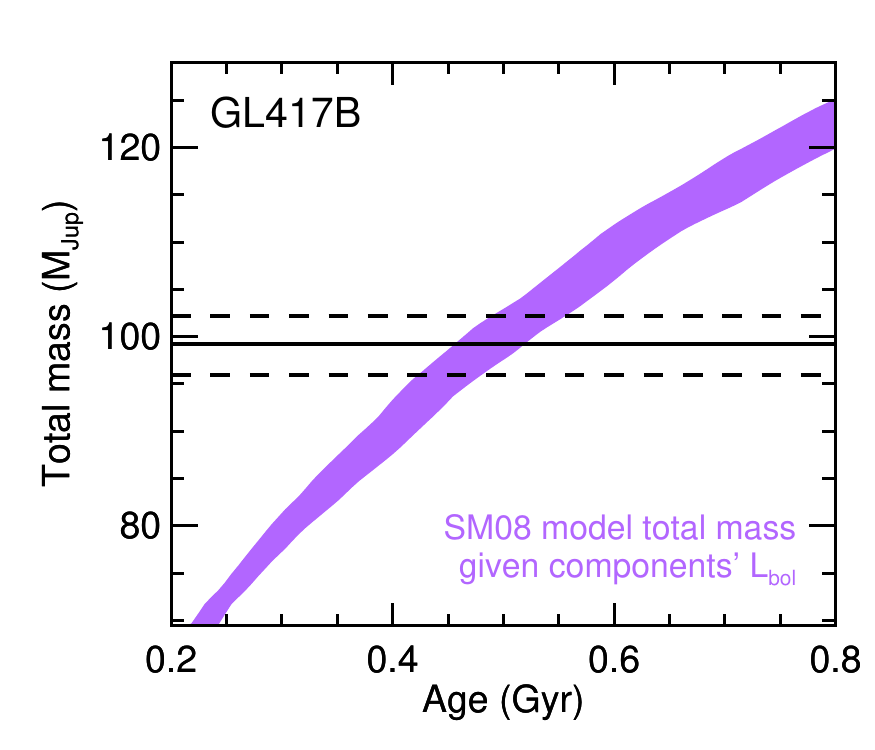} \includegraphics[width=3.0in,angle=0]{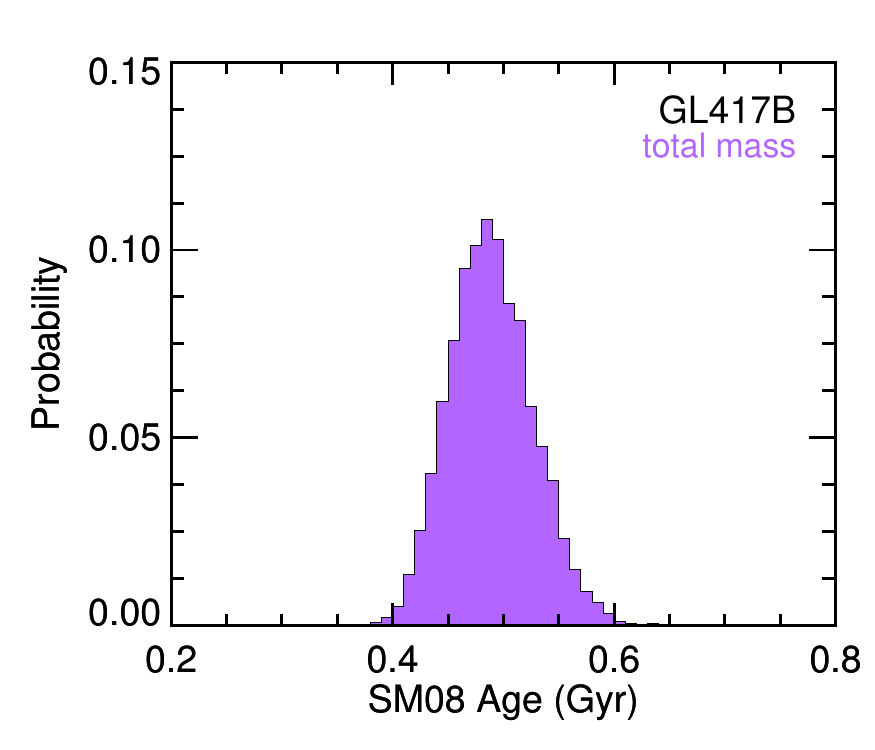}} 
\ContinuedFloat \caption{\normalsize (Continued)} \end{figure} 
\clearpage
\begin{figure} 
  \centerline{\includegraphics[width=3.0in,angle=0]{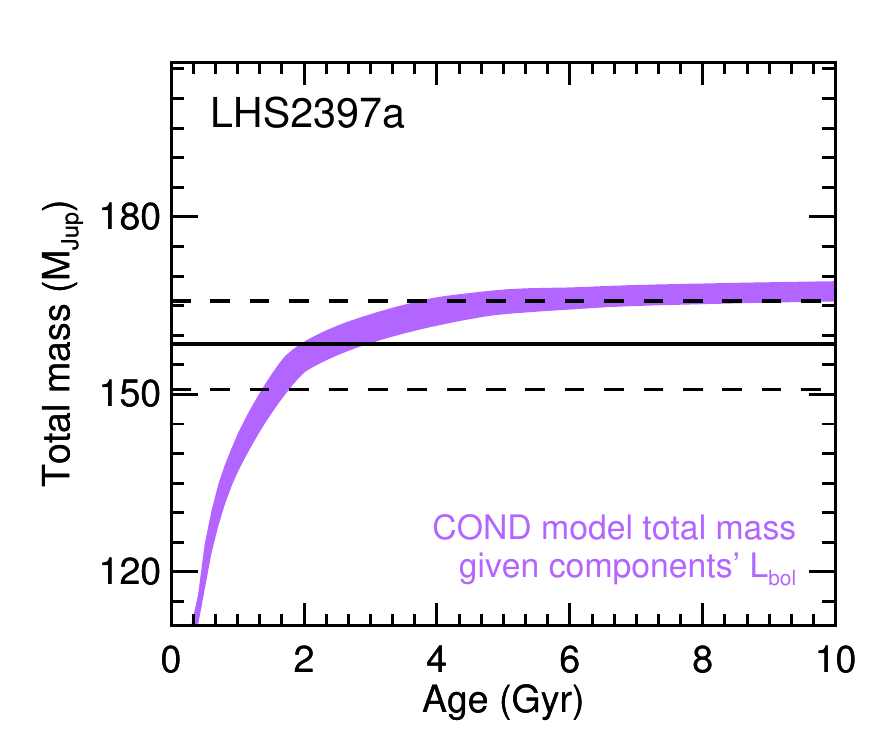} \includegraphics[width=3.0in,angle=0]{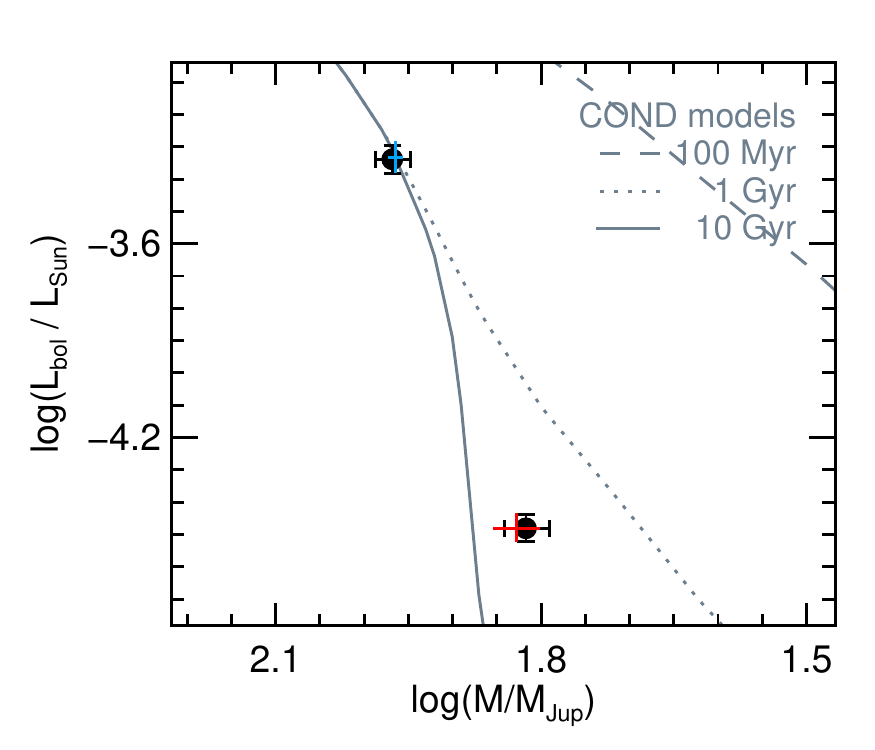} \includegraphics[width=3.0in,angle=0]{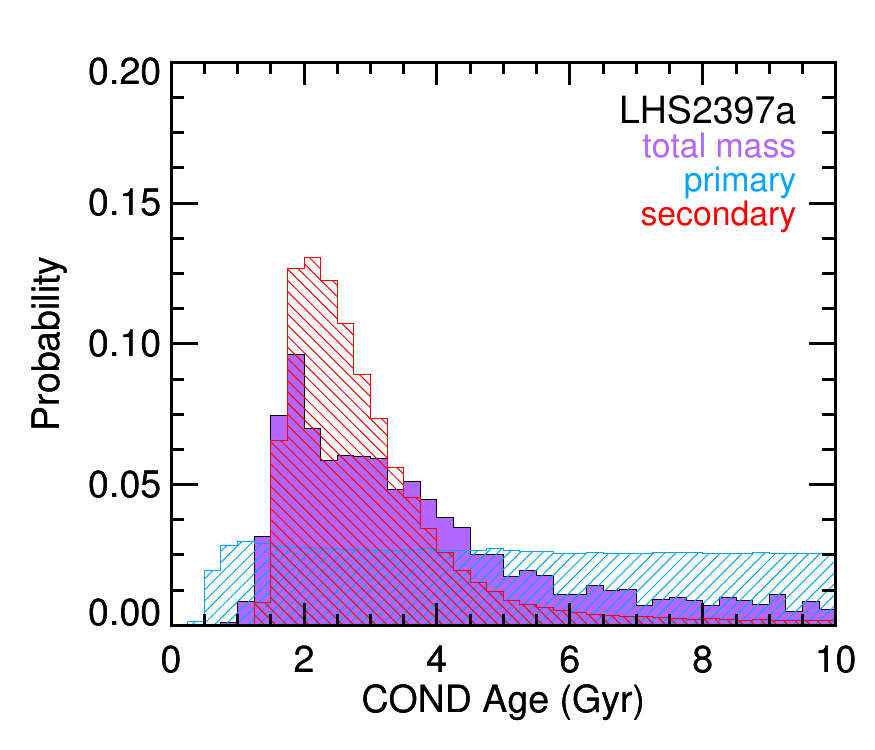}} 
\ContinuedFloat \caption{\normalsize (Continued)} \end{figure}
\clearpage
\begin{figure}
  \centerline{\includegraphics[width=3.0in,angle=0]{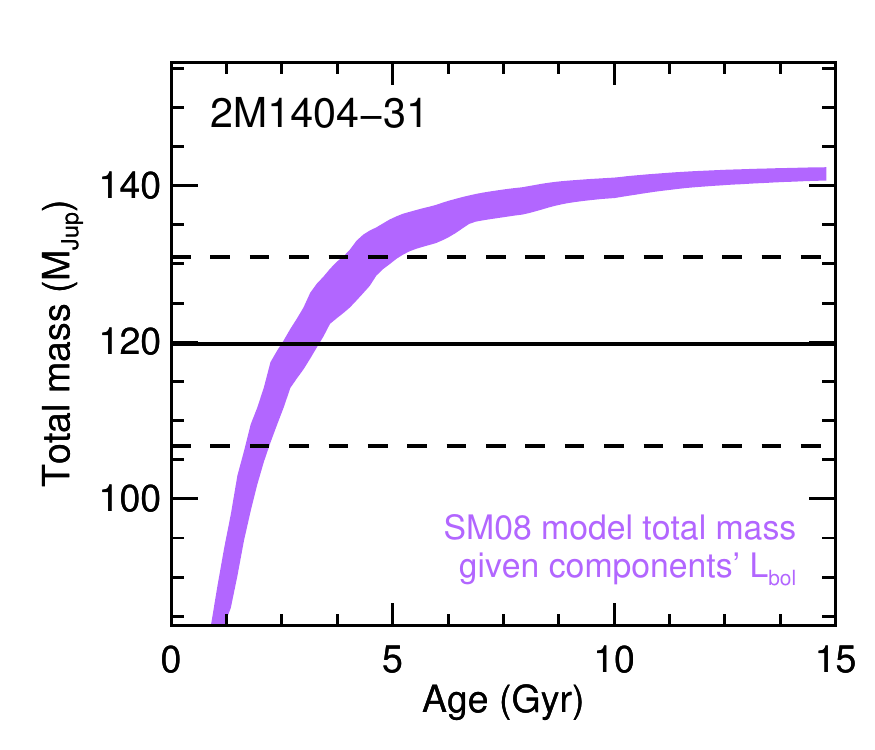} \includegraphics[width=3.0in,angle=0]{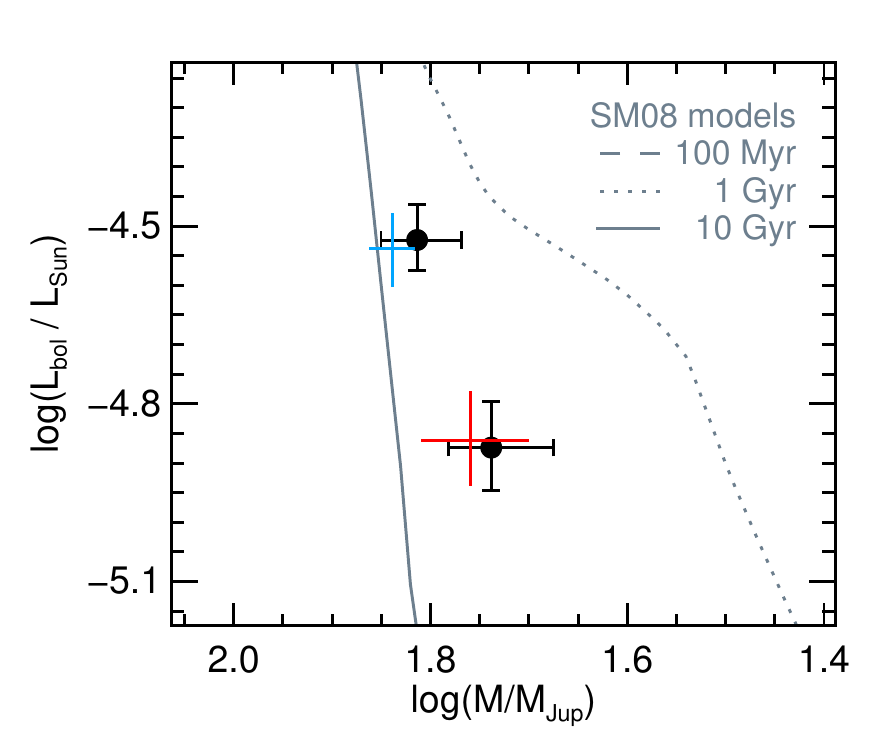} \includegraphics[width=3.0in,angle=0]{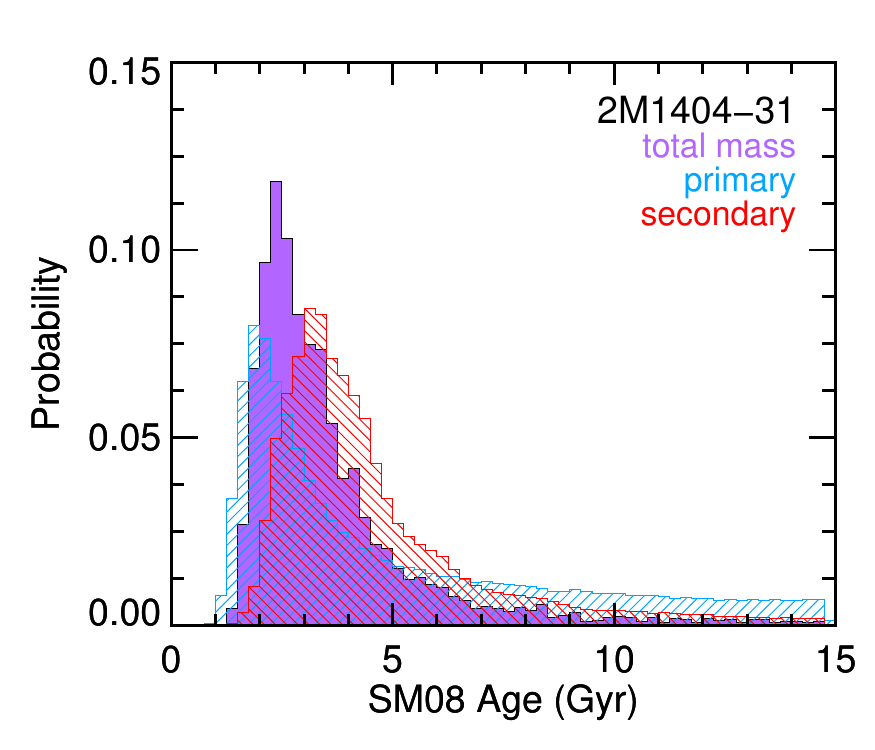}} 
  \centerline{\includegraphics[width=3.0in,angle=0]{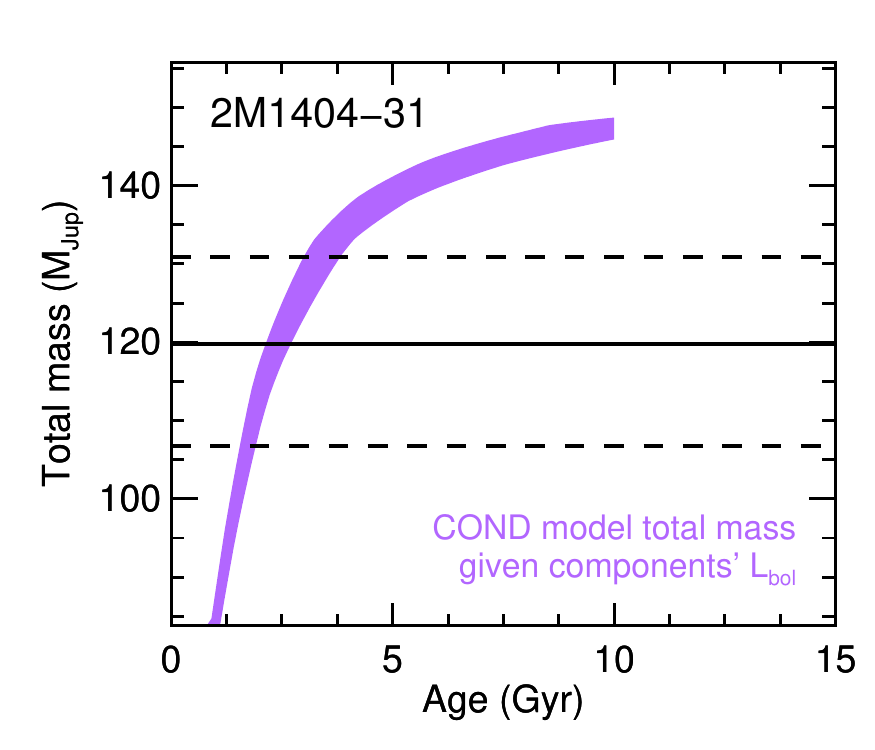} \includegraphics[width=3.0in,angle=0]{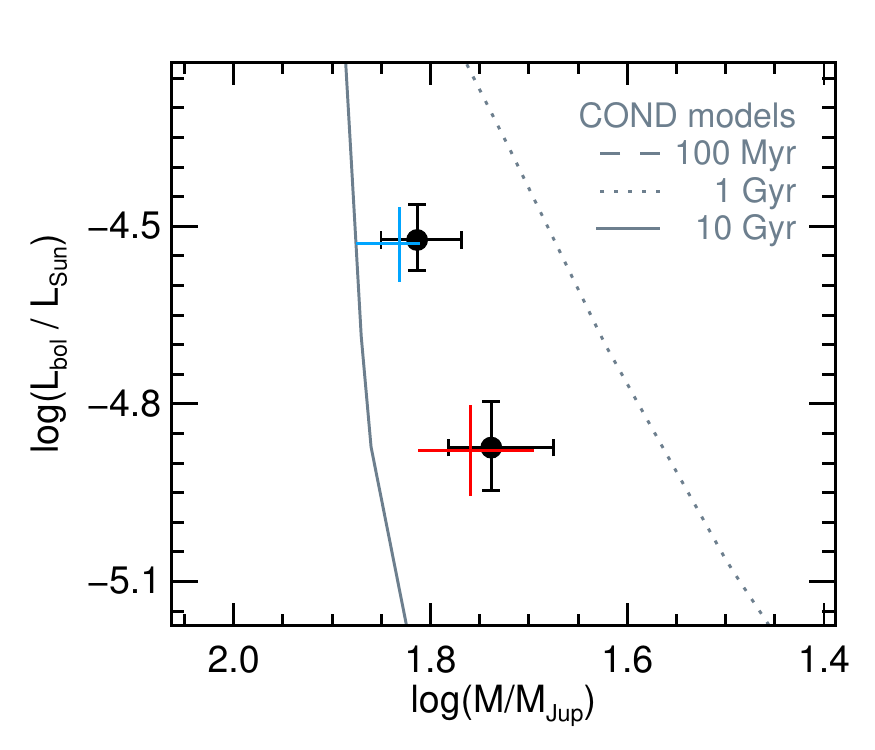} \includegraphics[width=3.0in,angle=0]{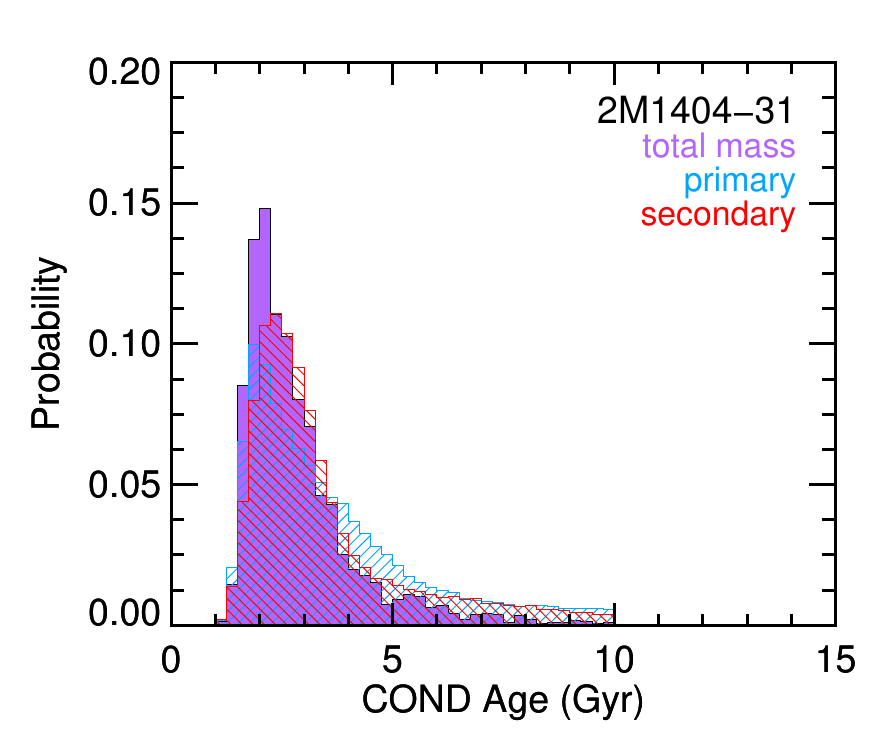}} 
\ContinuedFloat \caption{\normalsize (Continued)} \end{figure} 
\clearpage
\begin{figure} 
  \centerline{\includegraphics[width=3.0in,angle=0]{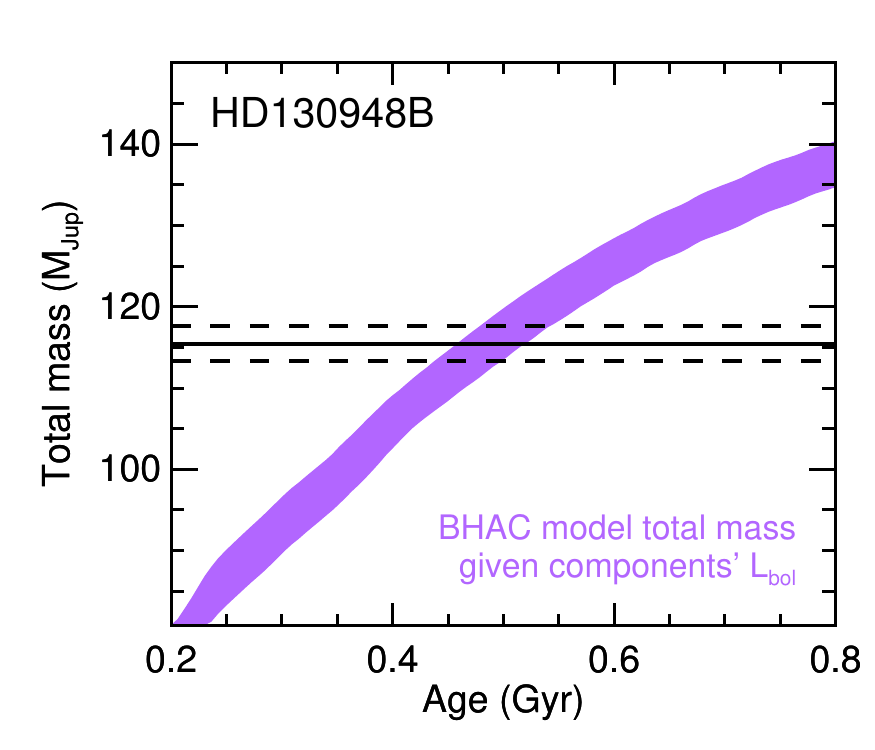} \includegraphics[width=3.0in,angle=0]{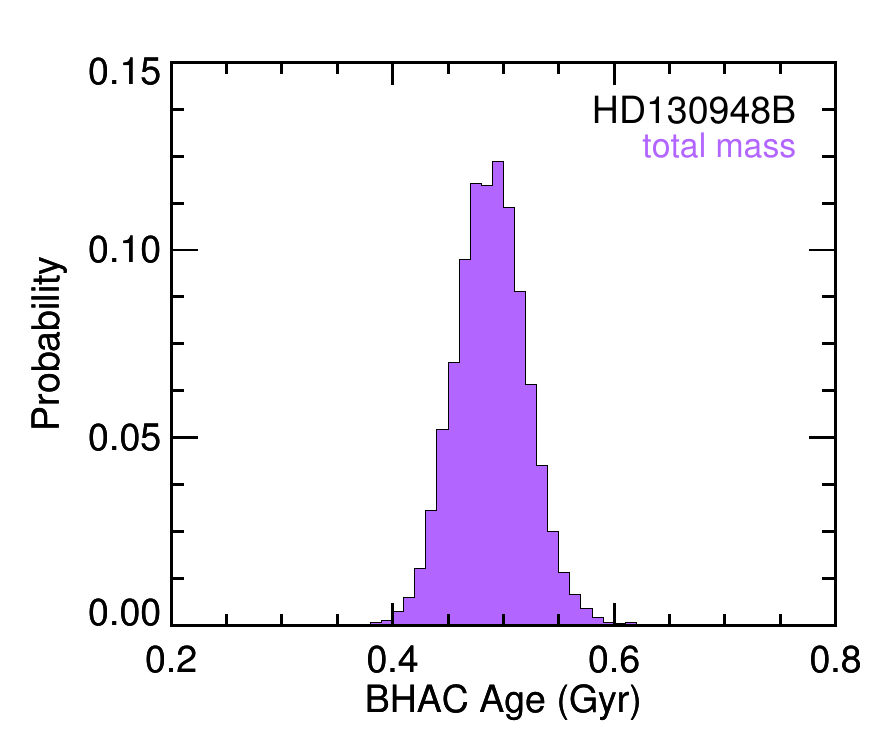}} 
  \centerline{\includegraphics[width=3.0in,angle=0]{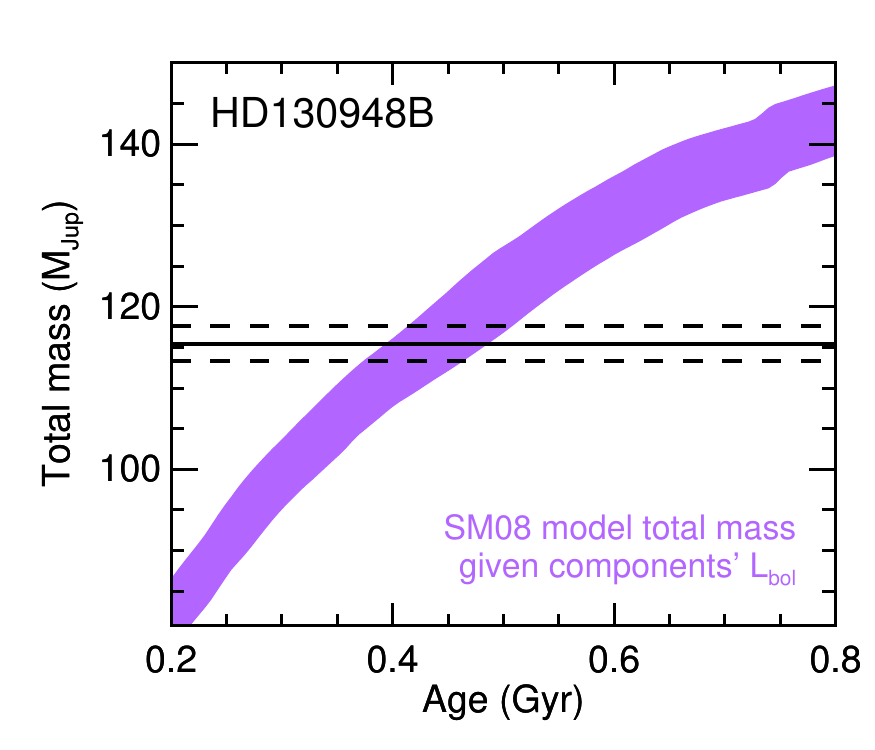} \includegraphics[width=3.0in,angle=0]{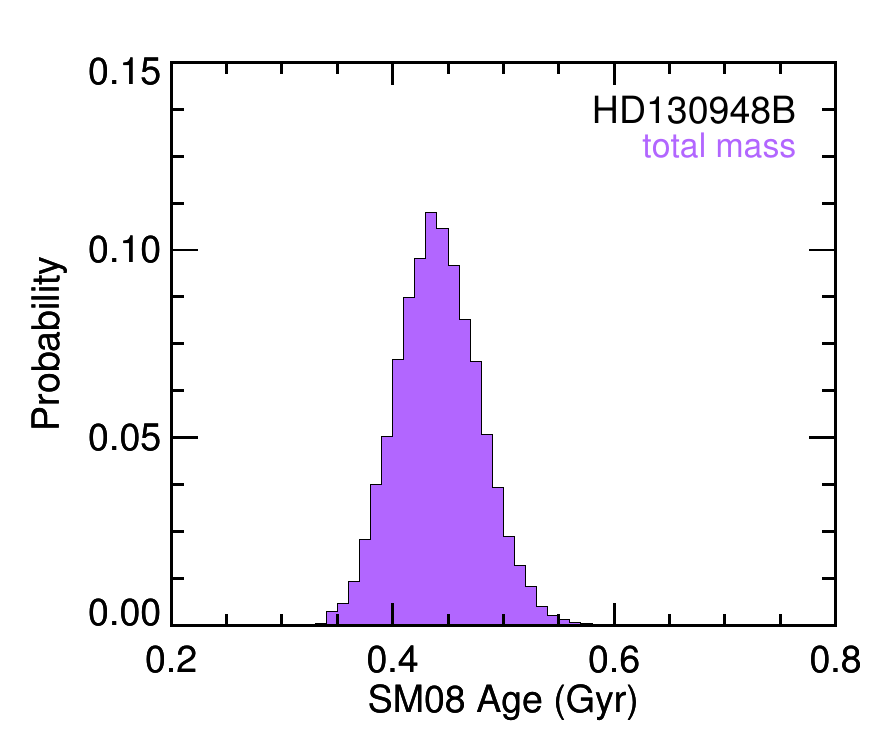}} 
\ContinuedFloat \caption{\normalsize (Continued)} \end{figure} 
\clearpage
\begin{figure} 
  \centerline{\includegraphics[width=3.0in,angle=0]{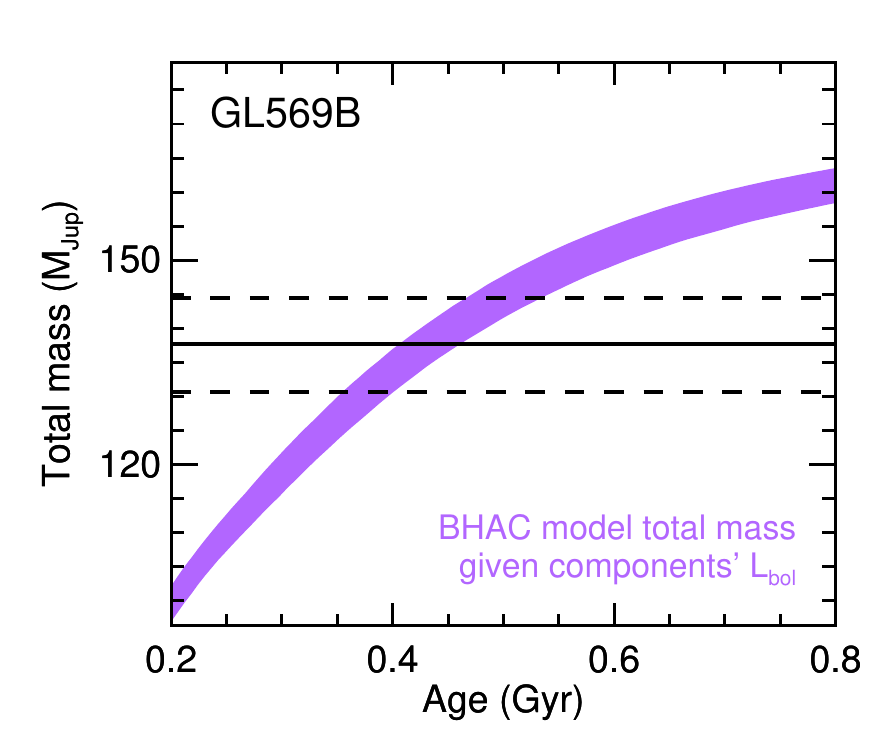} \includegraphics[width=3.0in,angle=0]{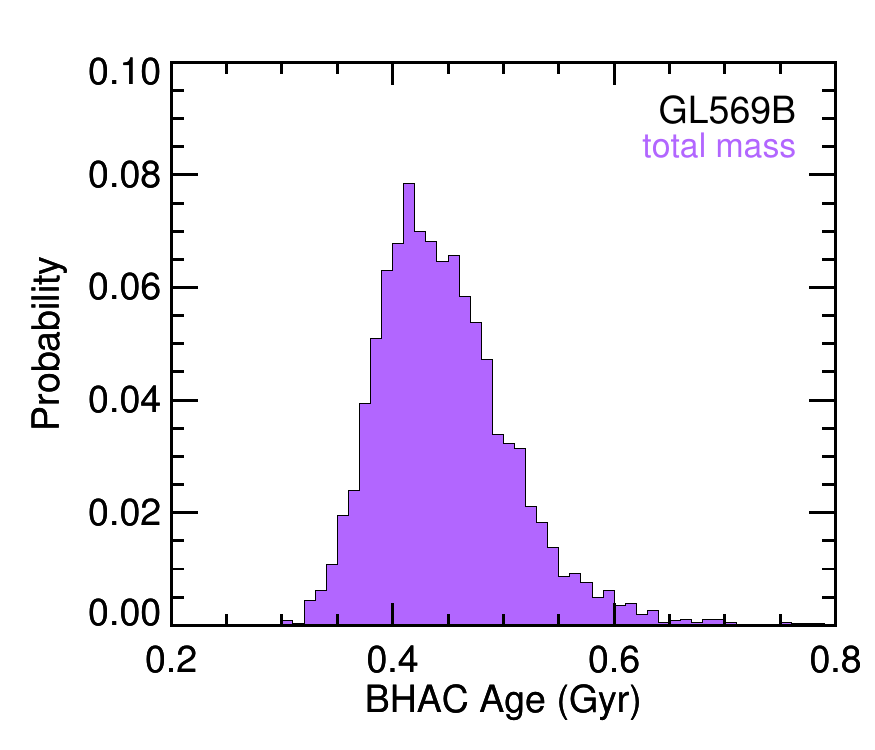}} 
  \centerline{\includegraphics[width=3.0in,angle=0]{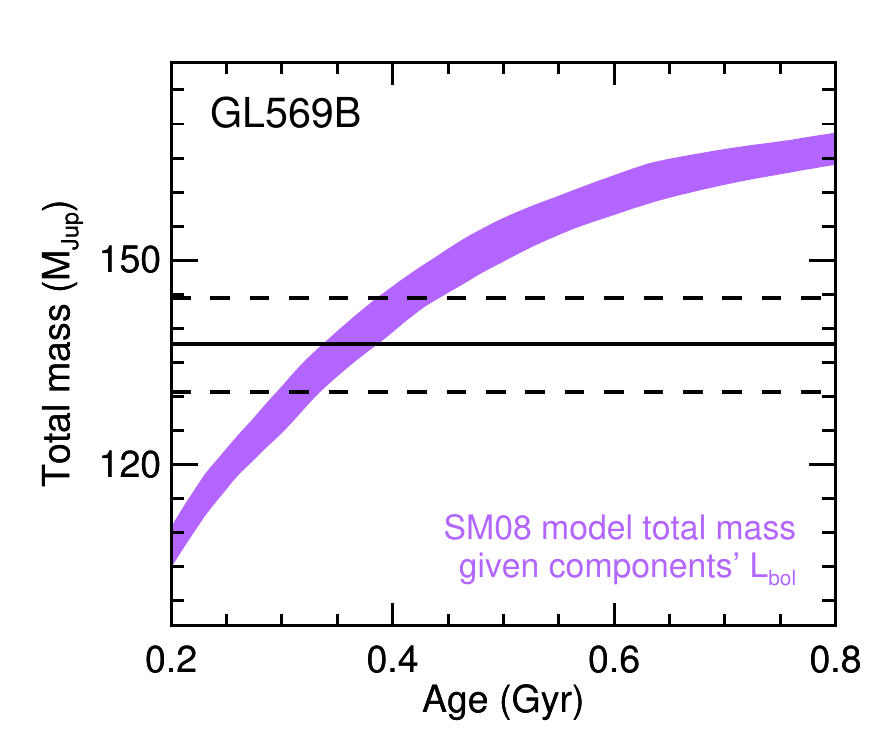} \includegraphics[width=3.0in,angle=0]{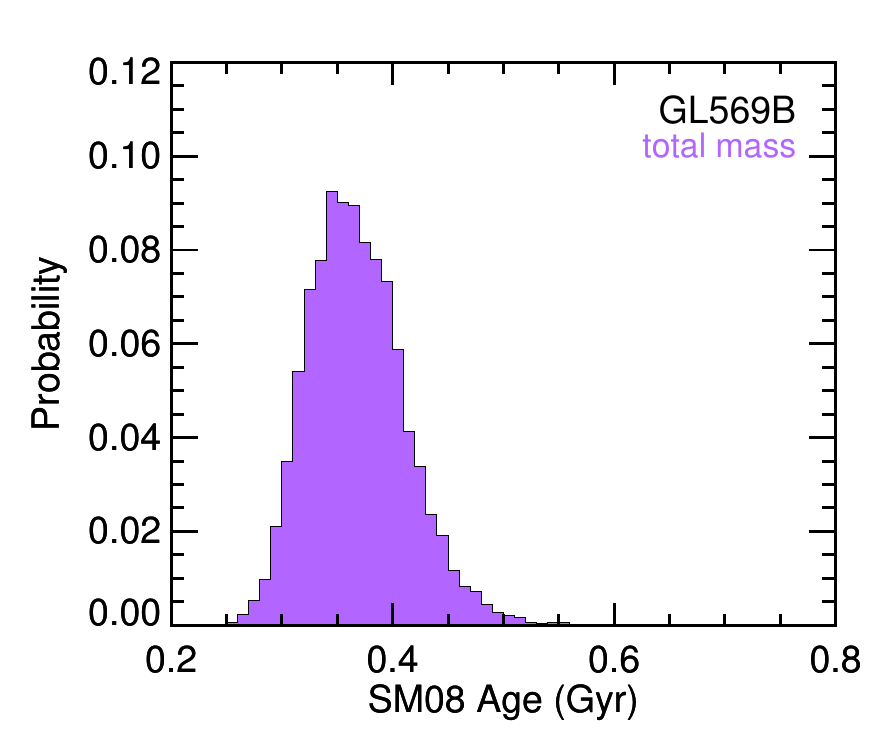}} 
\ContinuedFloat \caption{\normalsize (Continued)} \end{figure} 
\clearpage
\begin{figure}
  \centerline{\includegraphics[width=3.0in,angle=0]{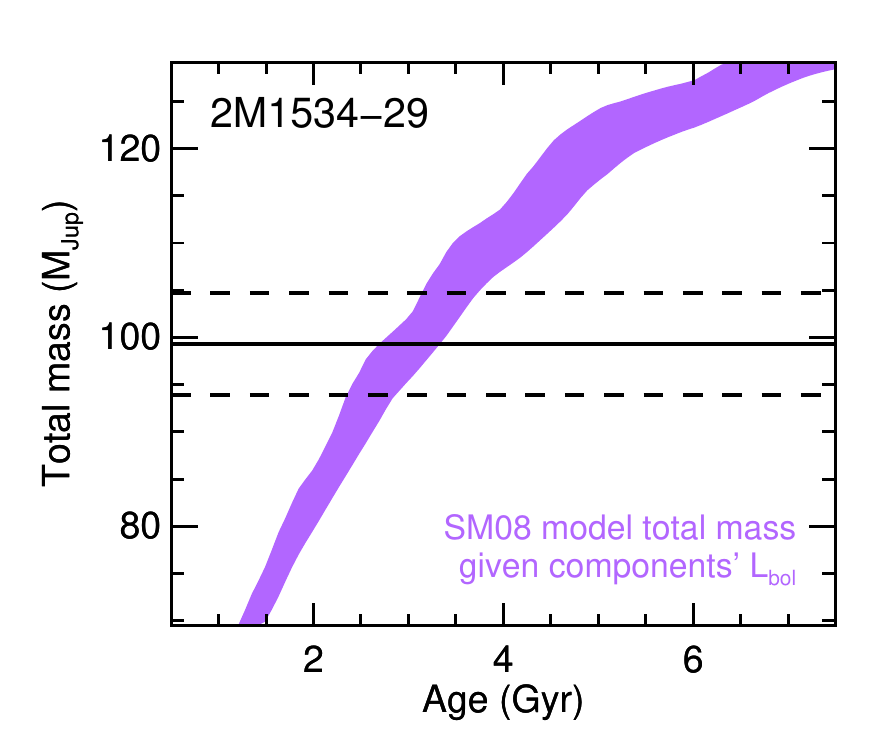} \includegraphics[width=3.0in,angle=0]{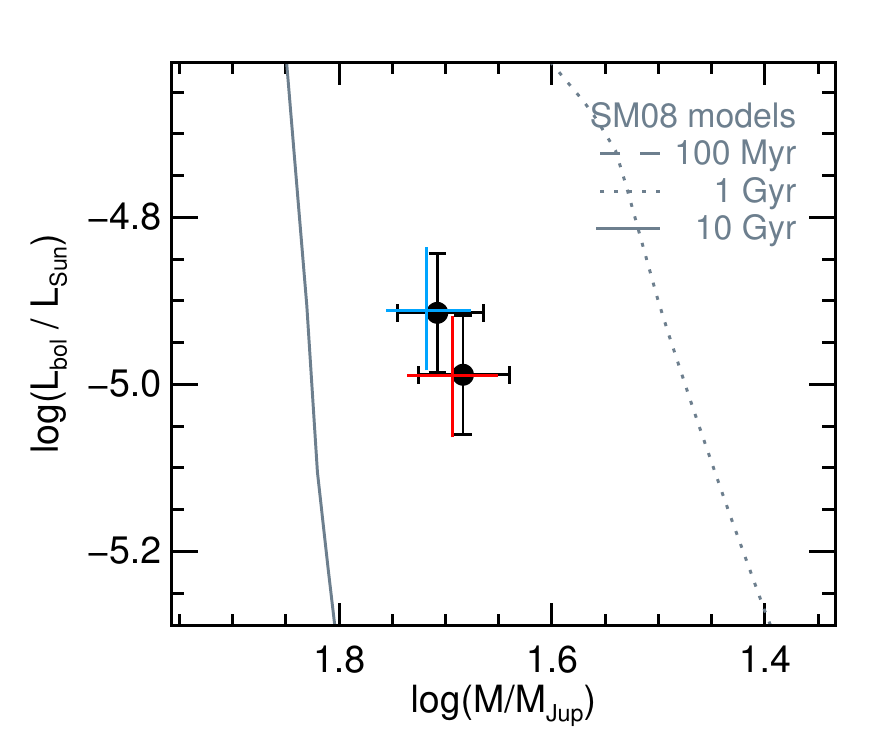} \includegraphics[width=3.0in,angle=0]{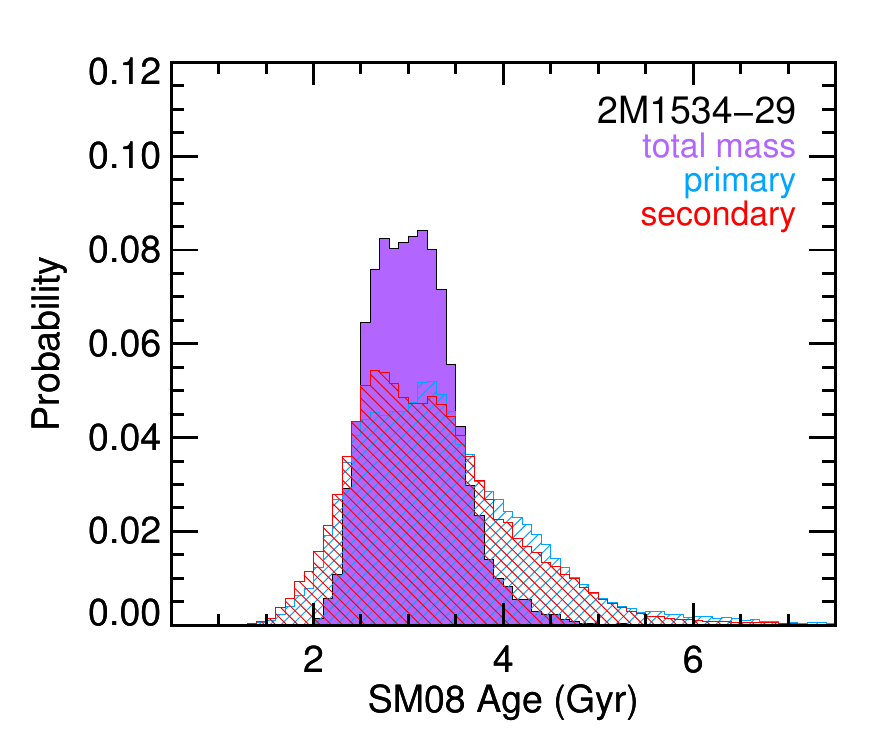}} 
  \centerline{\includegraphics[width=3.0in,angle=0]{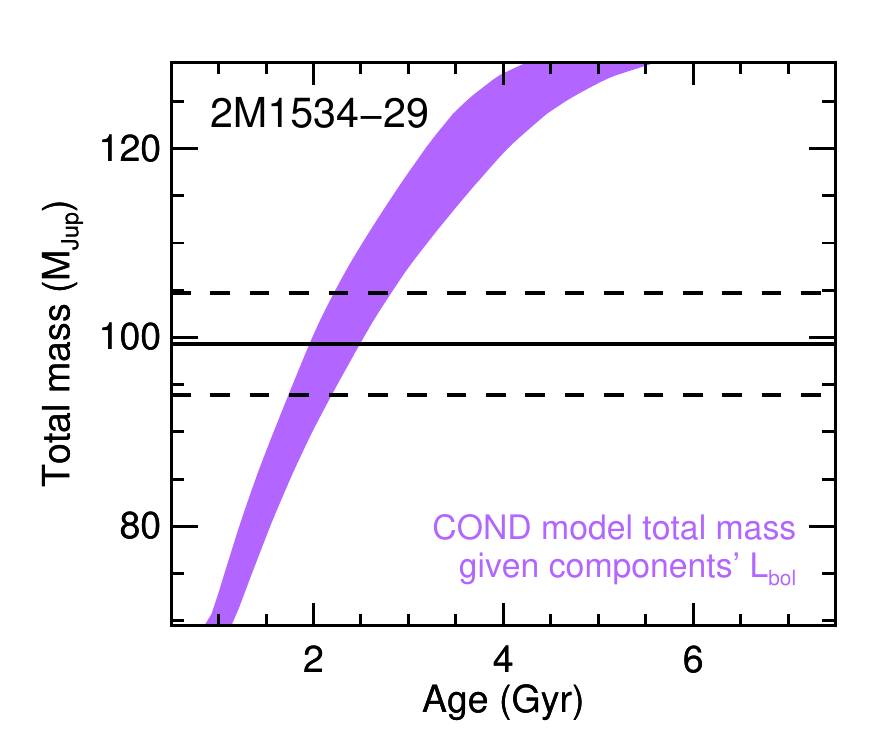} \includegraphics[width=3.0in,angle=0]{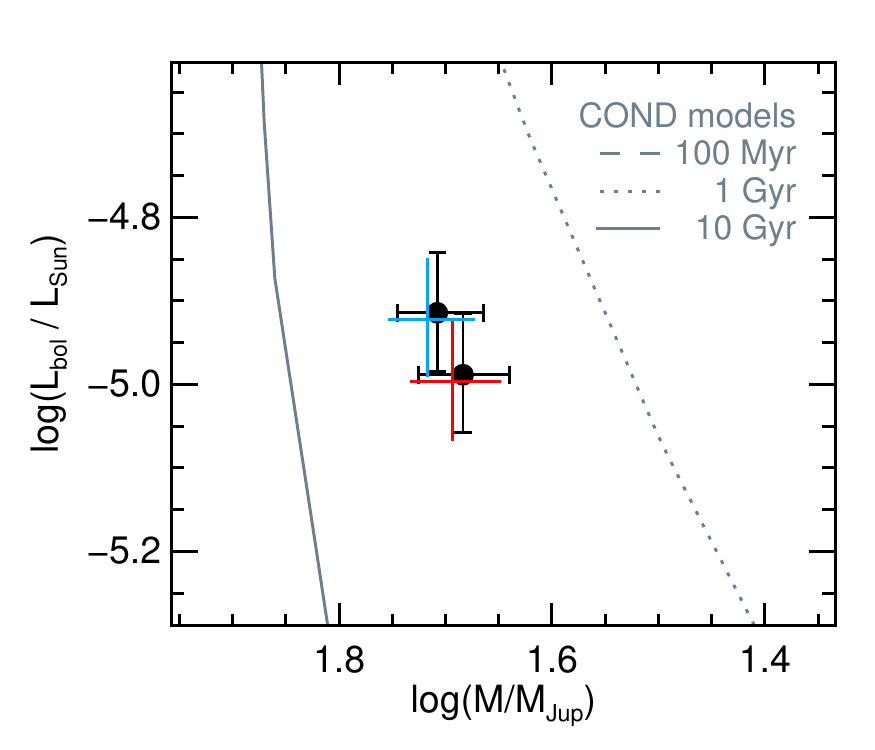} \includegraphics[width=3.0in,angle=0]{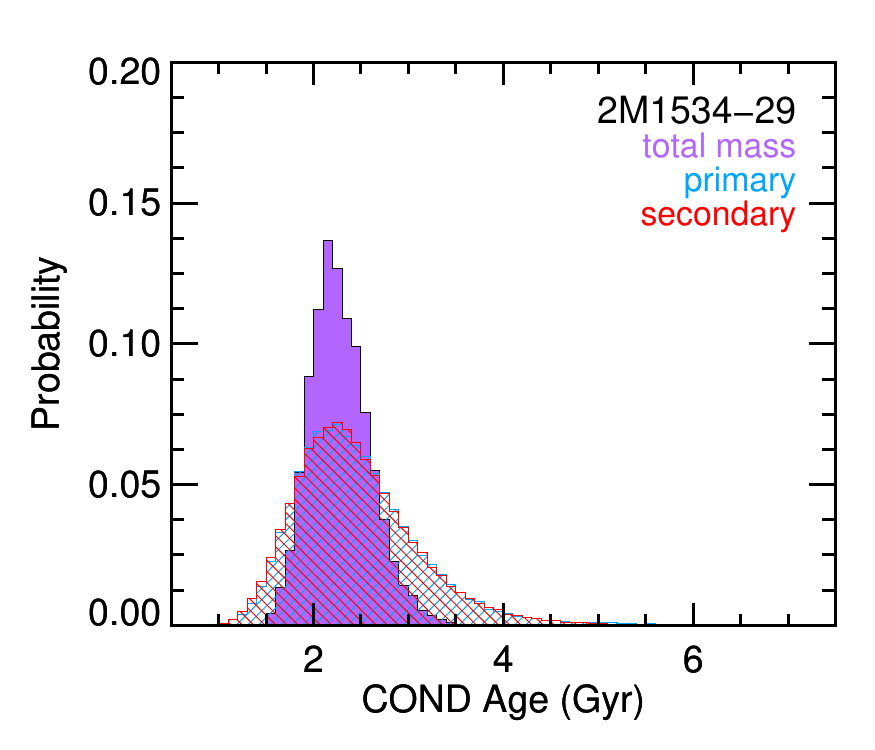}} 
\ContinuedFloat \caption{\normalsize (Continued)} \end{figure} 
\clearpage
\begin{figure}
  \centerline{\includegraphics[width=3.0in,angle=0]{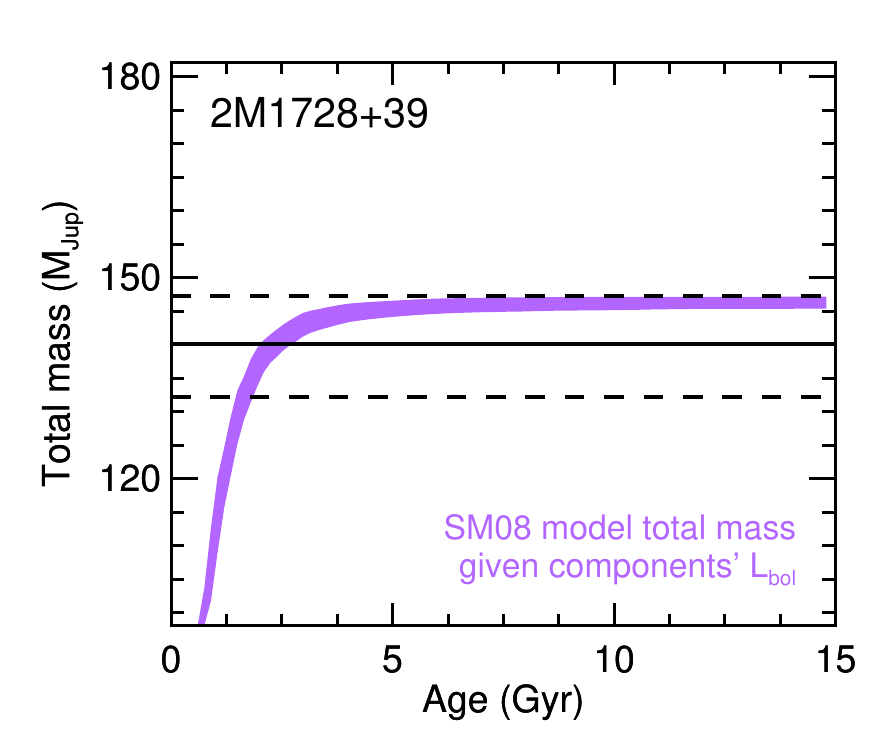} \includegraphics[width=3.0in,angle=0]{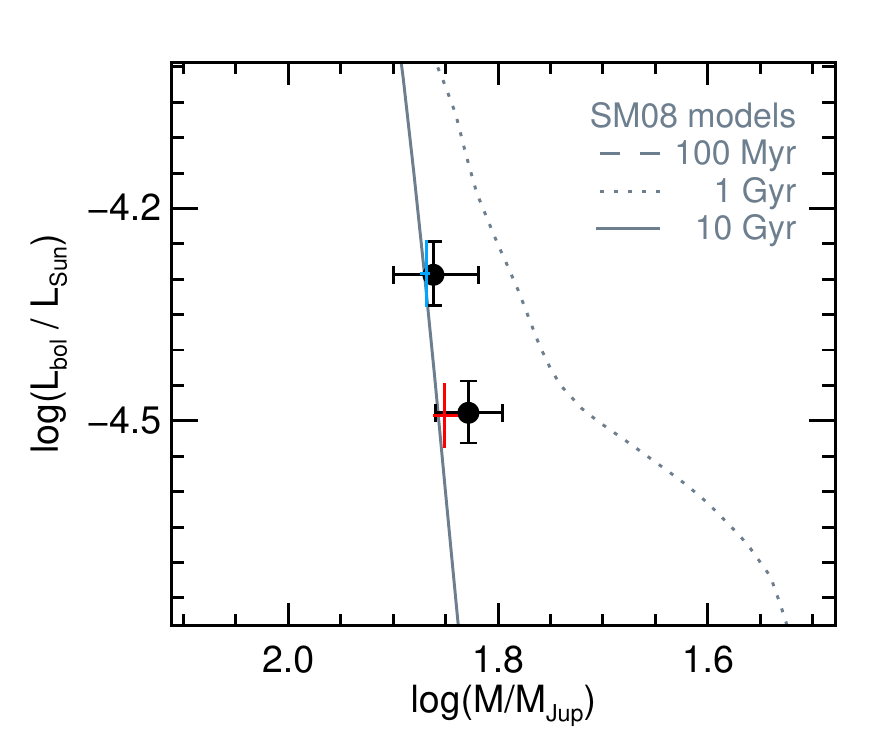} \includegraphics[width=3.0in,angle=0]{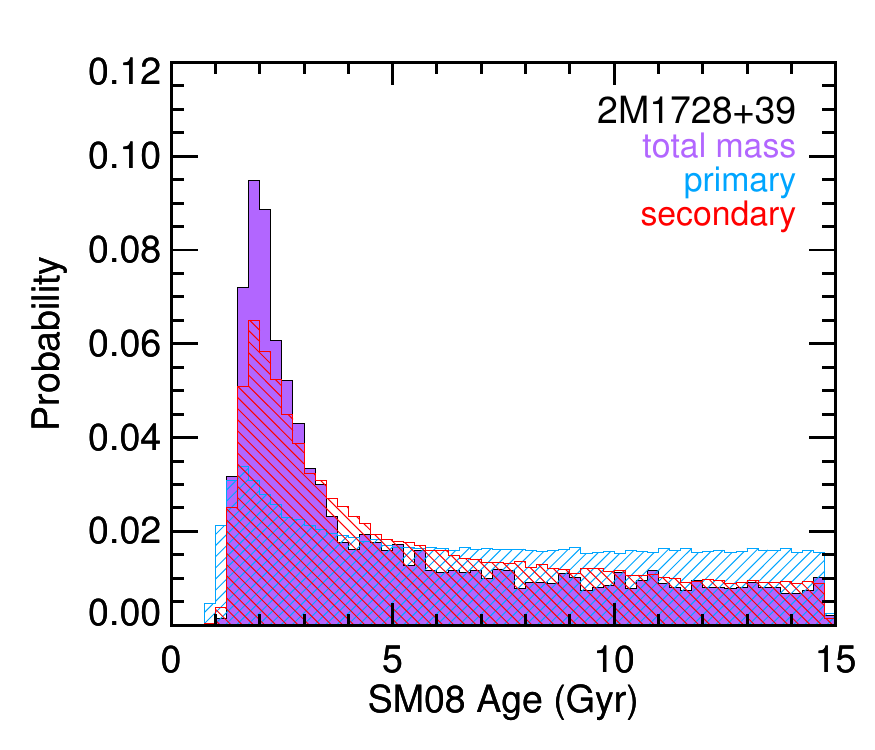}} 
  \centerline{\includegraphics[width=3.0in,angle=0]{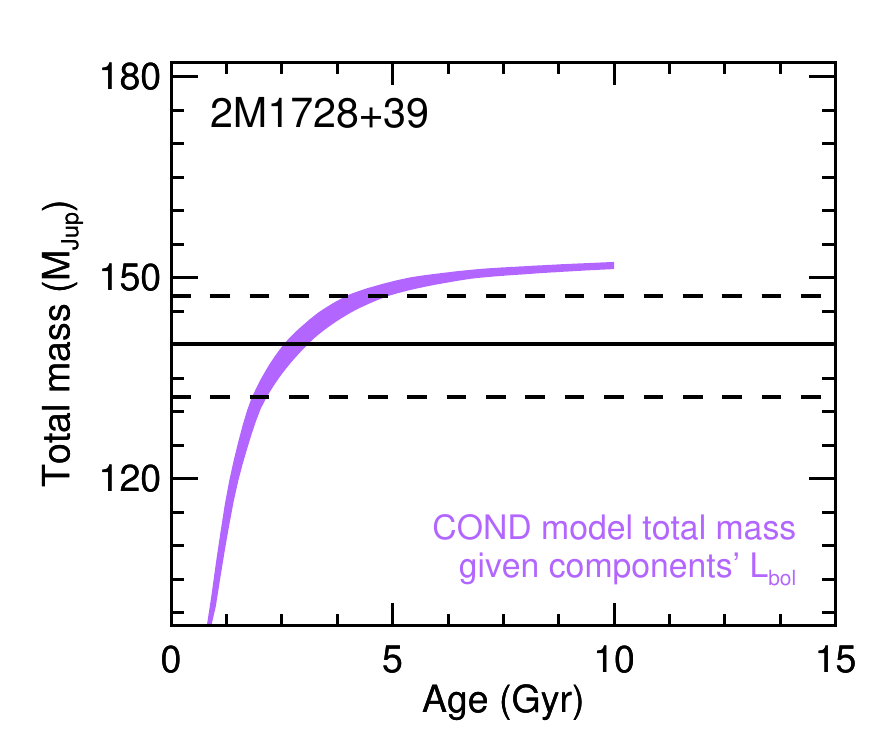} \includegraphics[width=3.0in,angle=0]{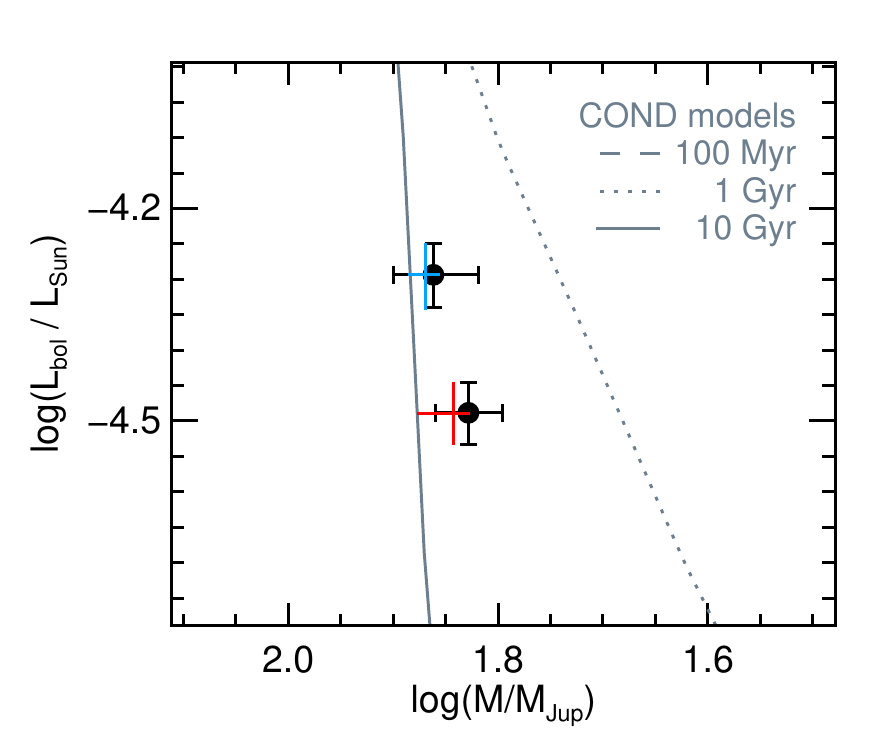} \includegraphics[width=3.0in,angle=0]{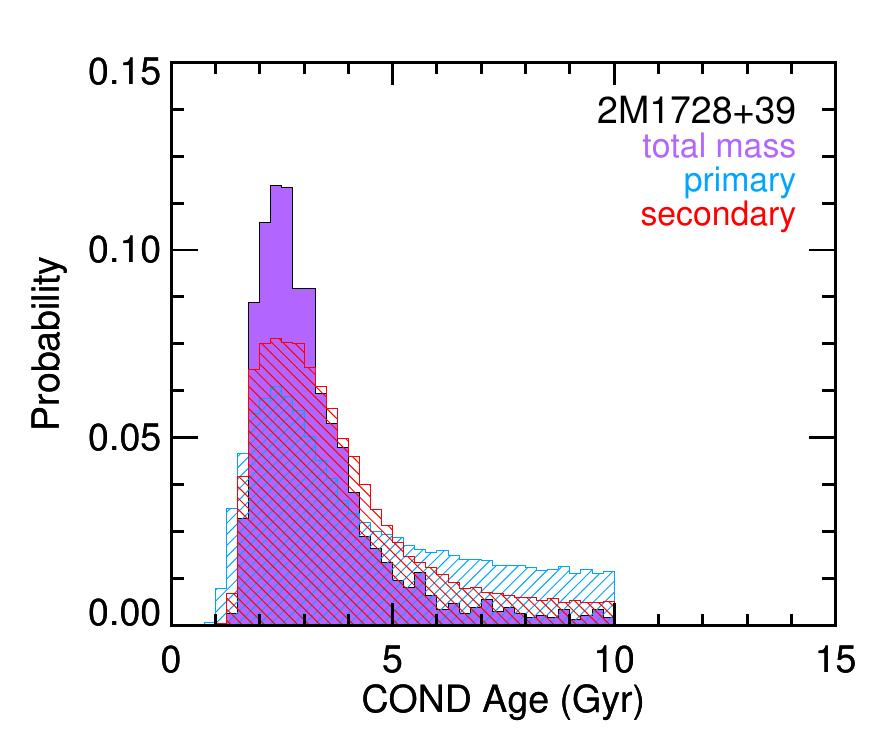}} 
\ContinuedFloat \caption{\normalsize (Continued)} \end{figure} 
\clearpage
\begin{figure} 
  \centerline{\includegraphics[width=3.0in,angle=0]{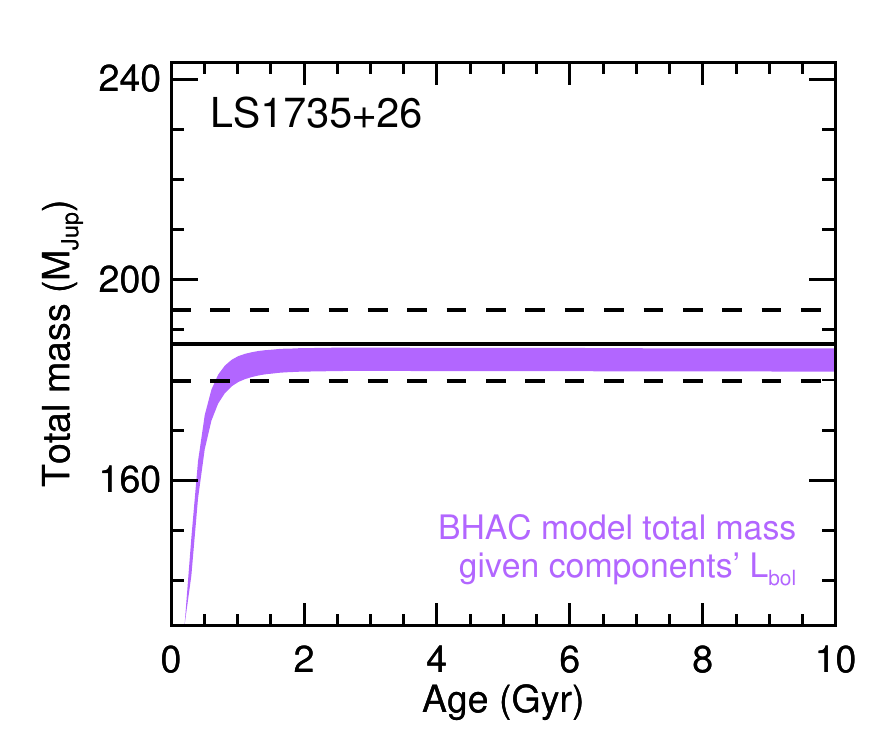} \includegraphics[width=3.0in,angle=0]{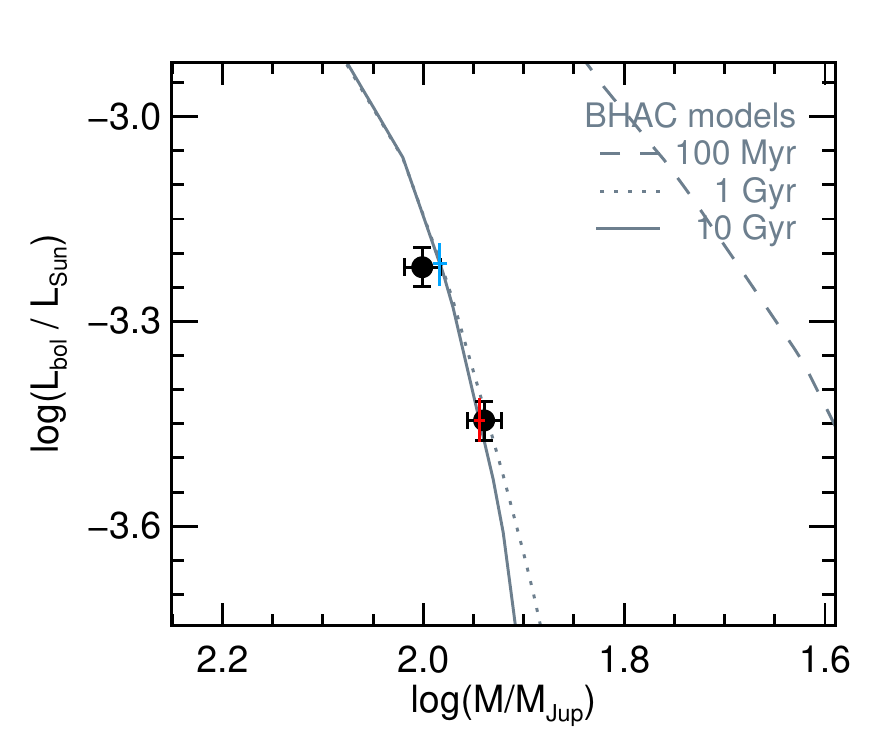} \includegraphics[width=3.0in,angle=0]{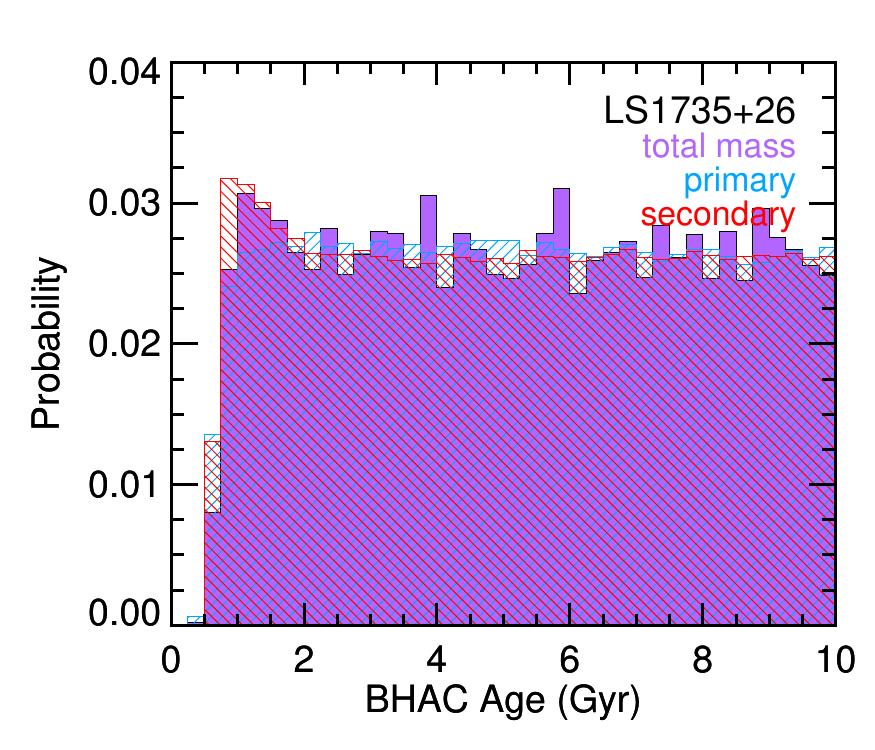}} 
\ContinuedFloat \caption{\normalsize (Continued)} \end{figure} 
\clearpage
\begin{figure} 
  \centerline{\includegraphics[width=3.0in,angle=0]{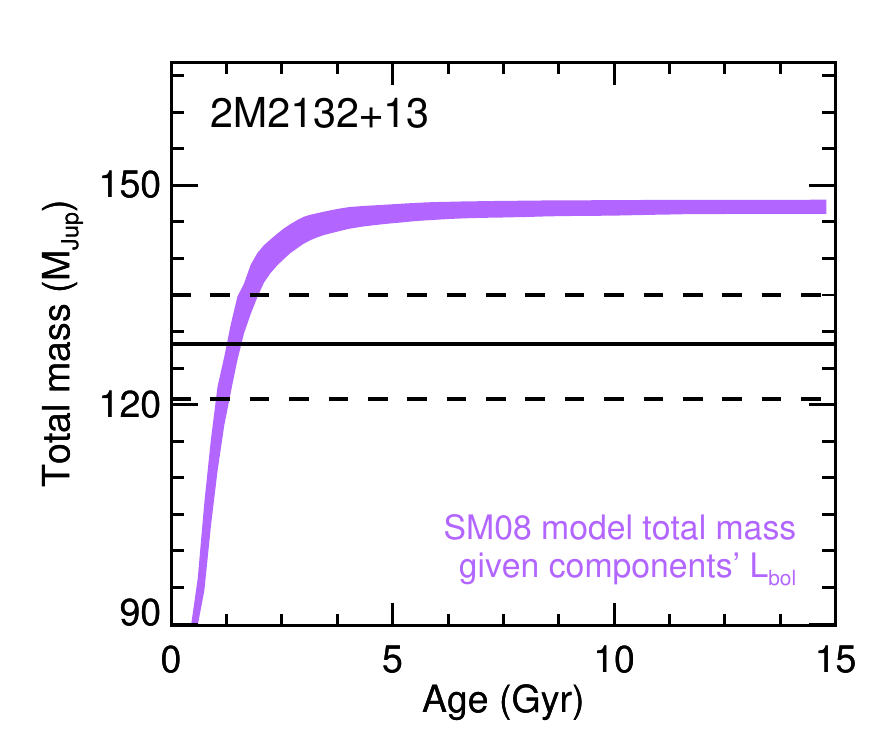} \includegraphics[width=3.0in,angle=0]{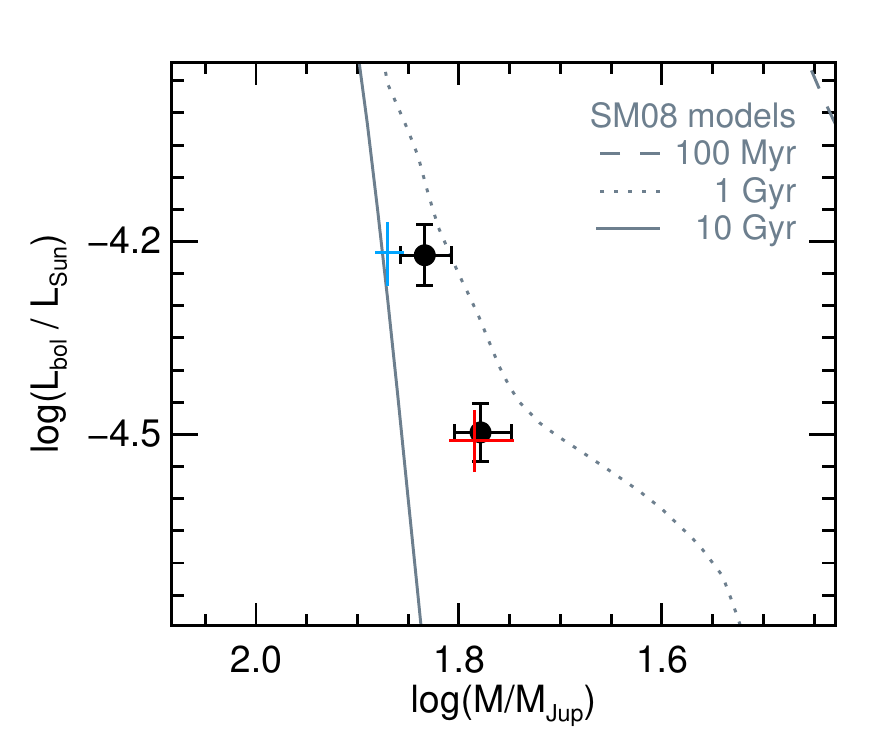} \includegraphics[width=3.0in,angle=0]{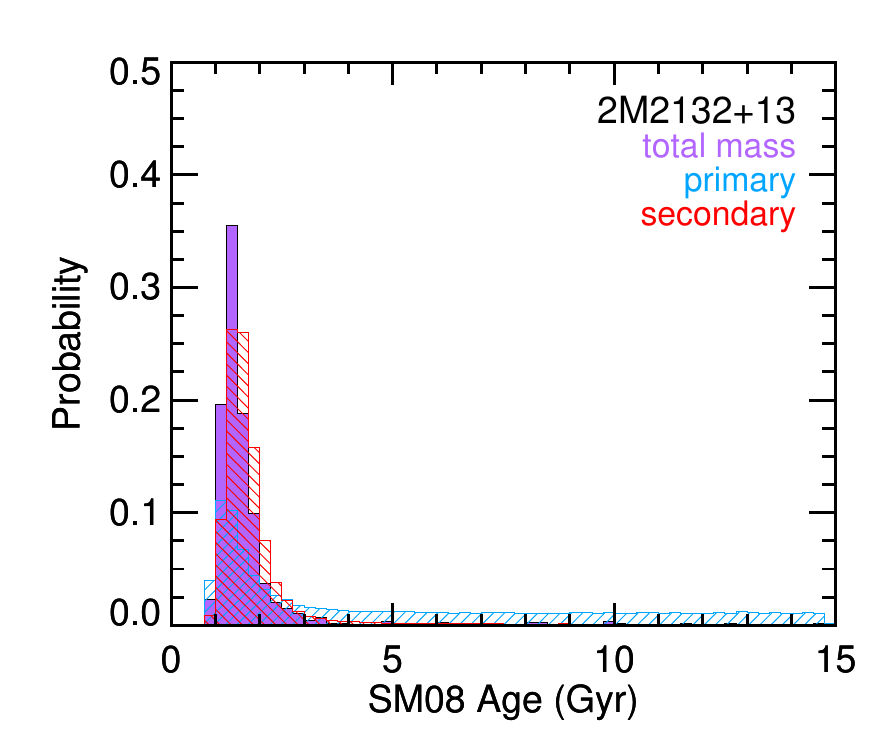}} 
  \centerline{\includegraphics[width=3.0in,angle=0]{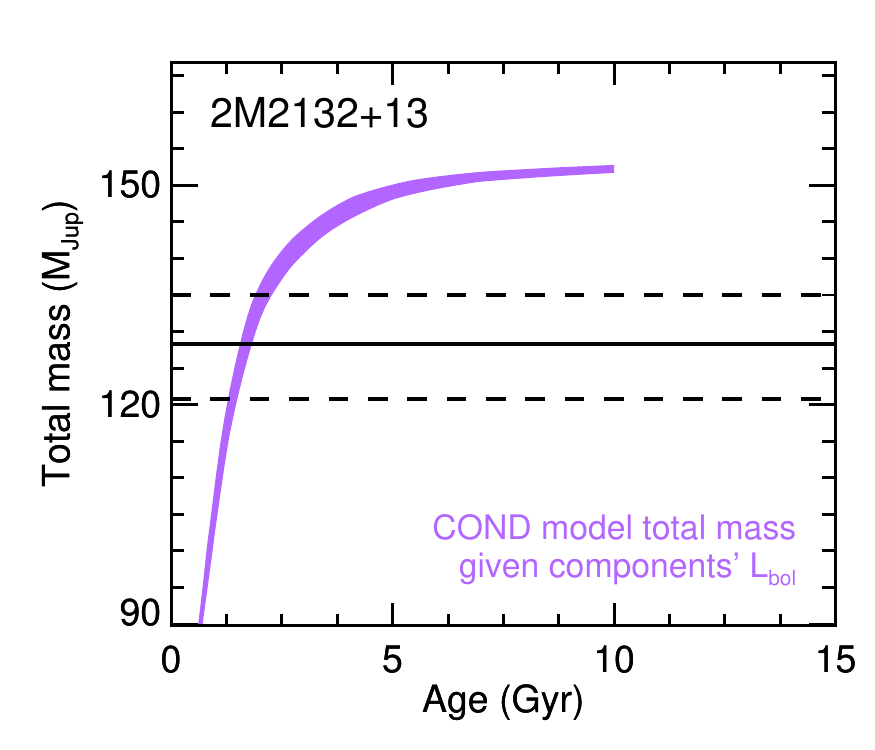} \includegraphics[width=3.0in,angle=0]{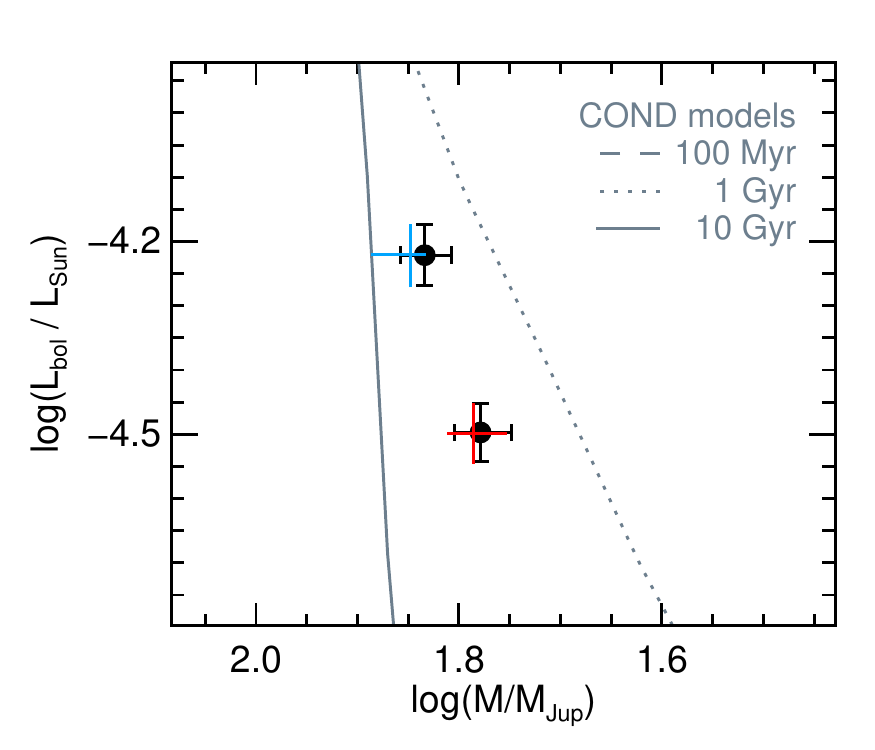} \includegraphics[width=3.0in,angle=0]{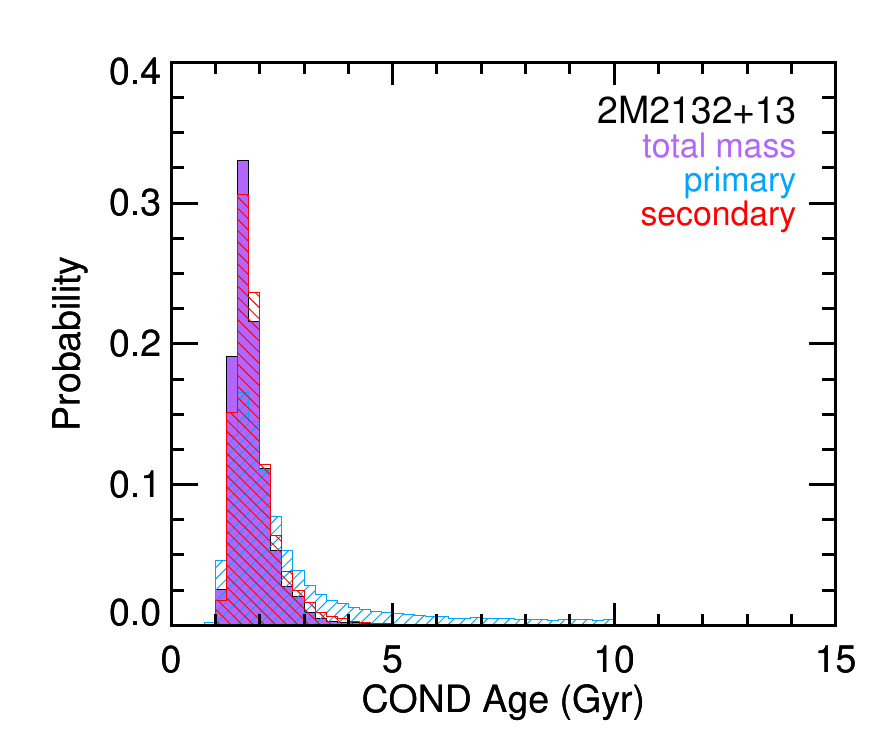}} 
\ContinuedFloat \caption{\normalsize (Continued)} \end{figure} 
\clearpage
\begin{figure} 
  \centerline{\includegraphics[width=3.0in,angle=0]{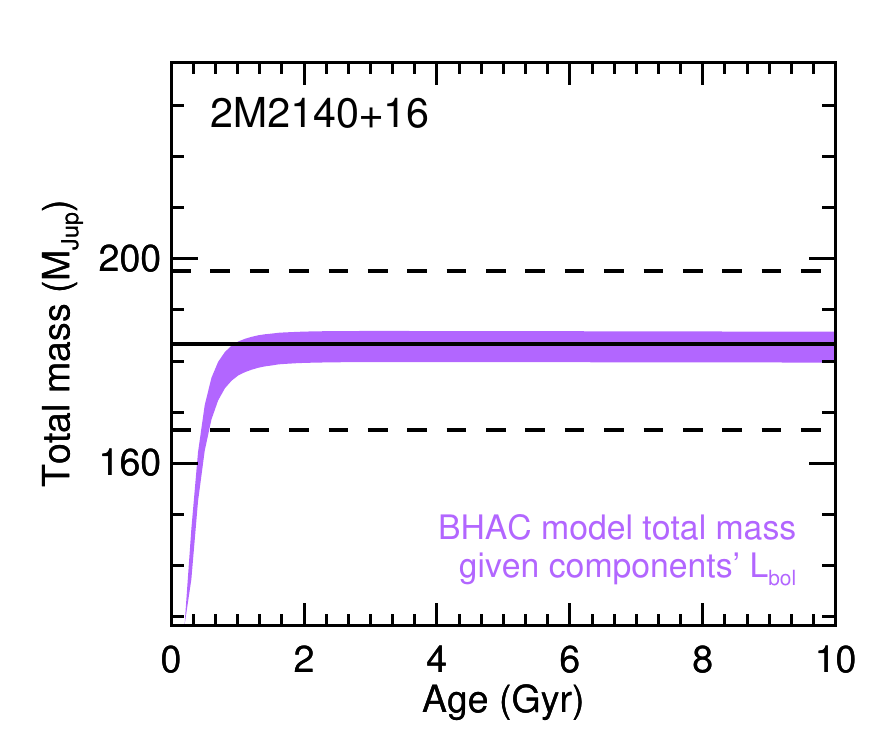} \includegraphics[width=3.0in,angle=0]{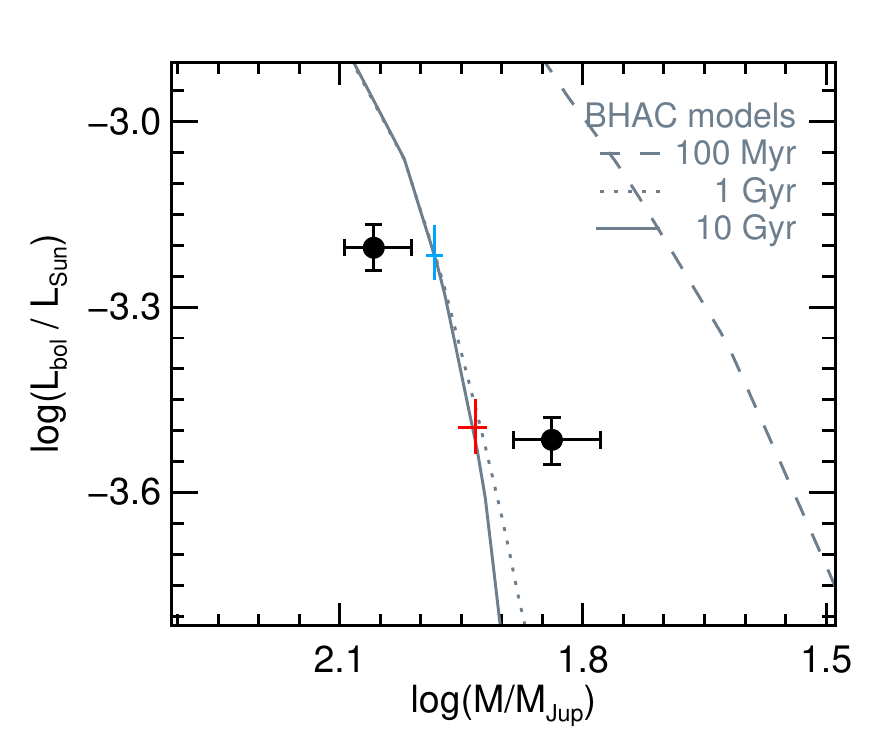} \includegraphics[width=3.0in,angle=0]{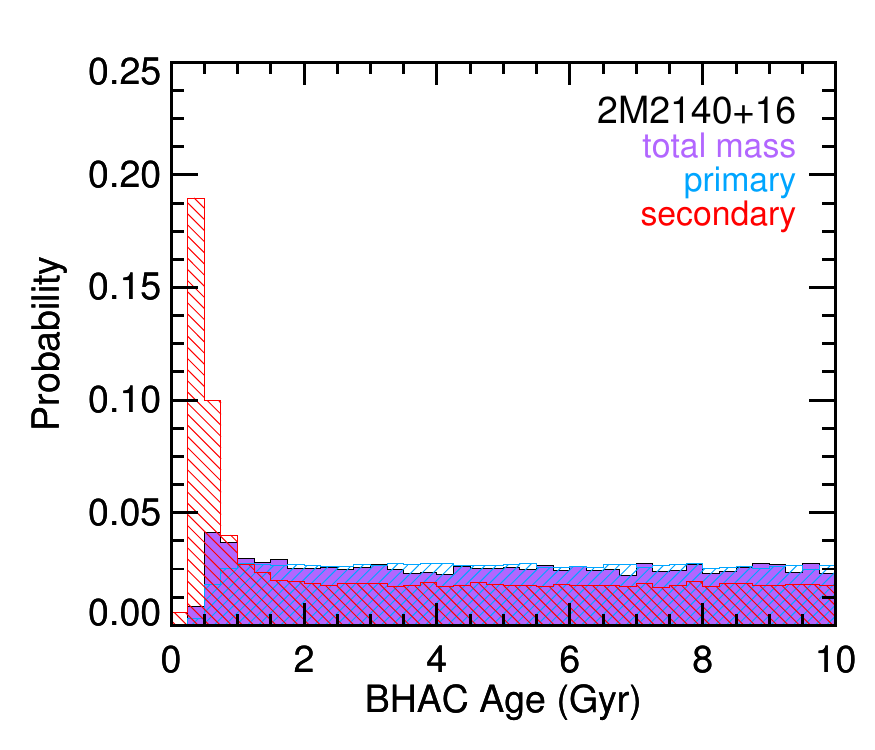}} 
\ContinuedFloat \caption{\normalsize (Continued)} \end{figure} 
\clearpage
\begin{figure} 
  \centerline{\includegraphics[width=3.0in,angle=0]{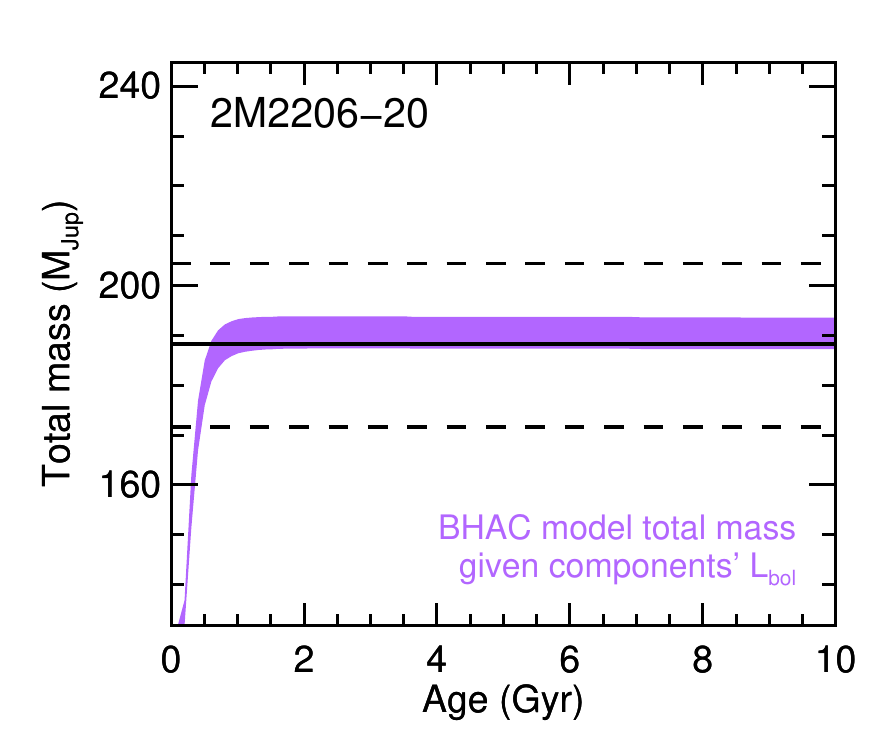} \includegraphics[width=3.0in,angle=0]{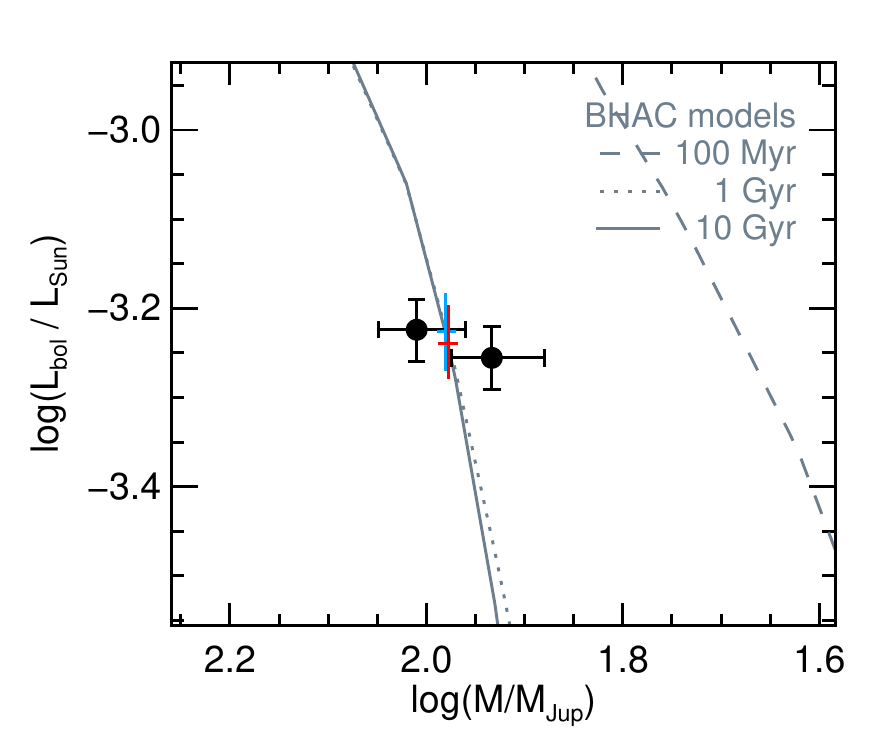} \includegraphics[width=3.0in,angle=0]{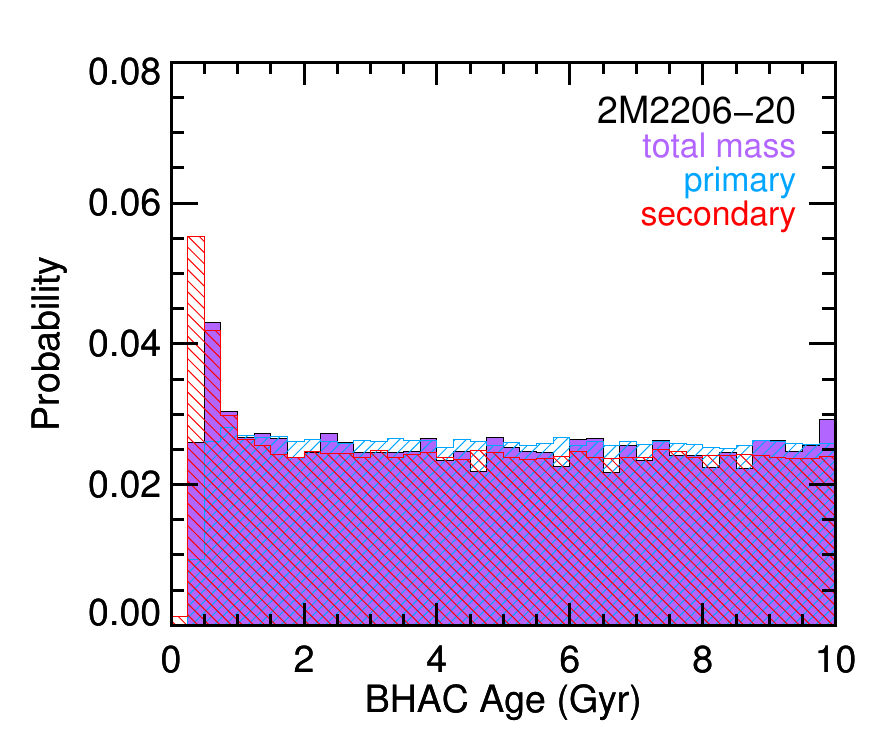}} 
\ContinuedFloat \caption{\normalsize (Continued)} \end{figure} 
\clearpage
\begin{figure} 
  \centerline{\includegraphics[width=3.0in,angle=0]{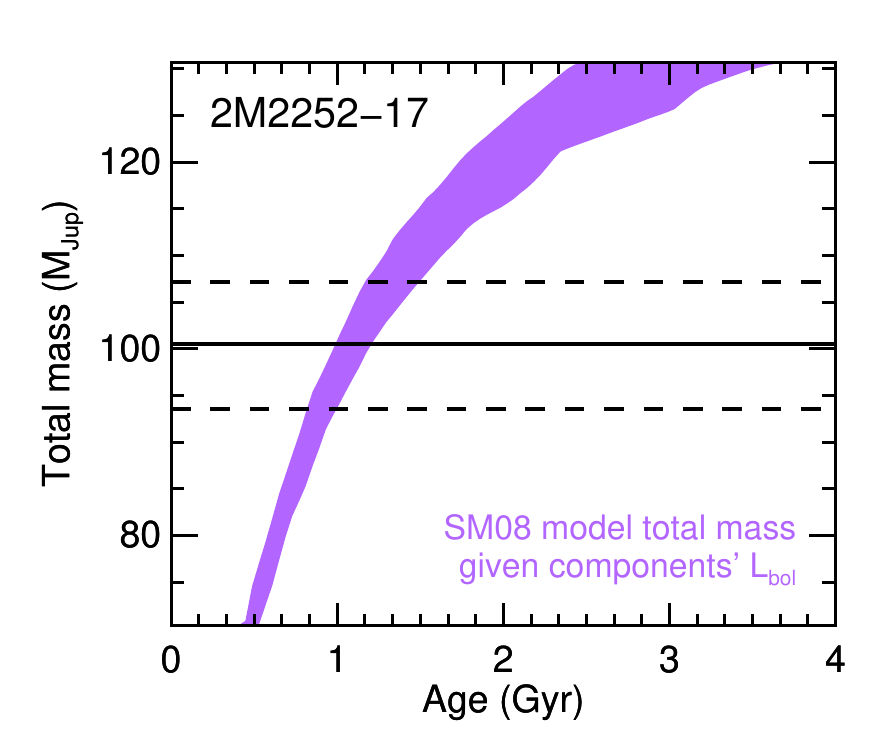} \includegraphics[width=3.0in,angle=0]{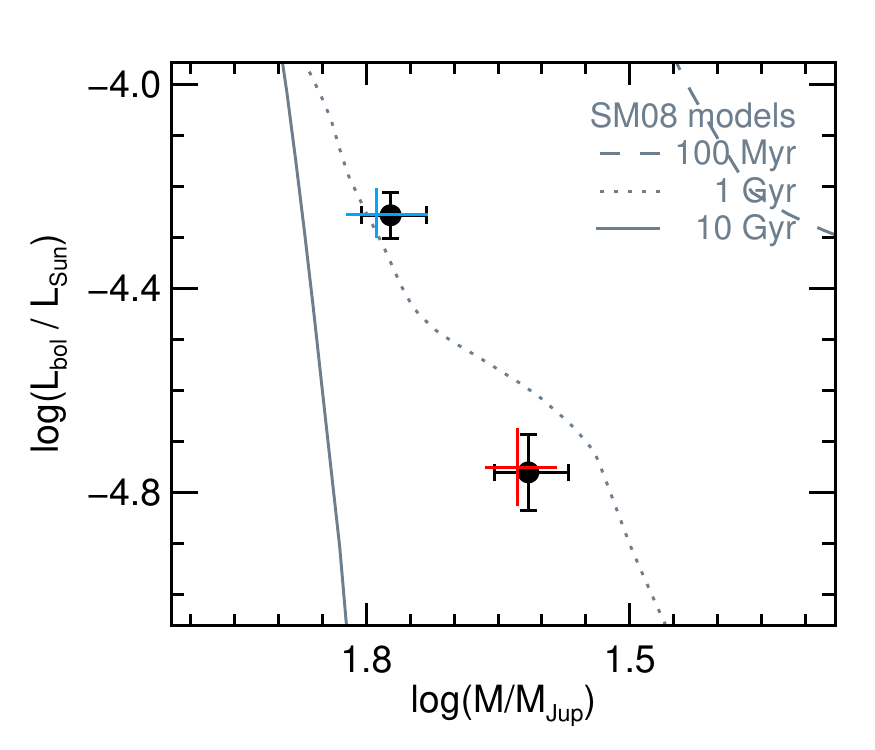} \includegraphics[width=3.0in,angle=0]{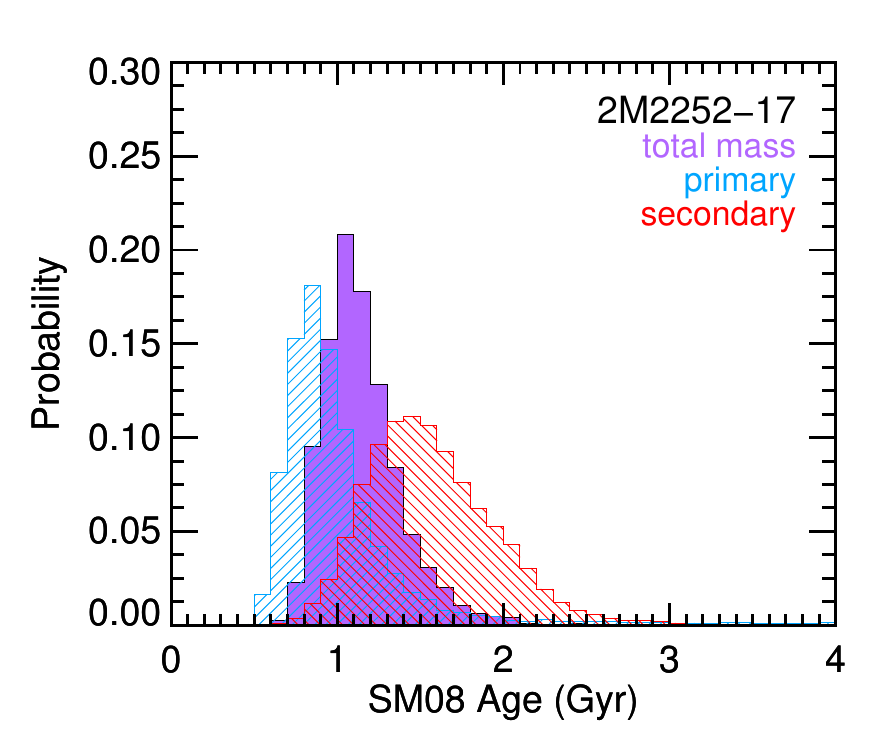}} 
  \centerline{\includegraphics[width=3.0in,angle=0]{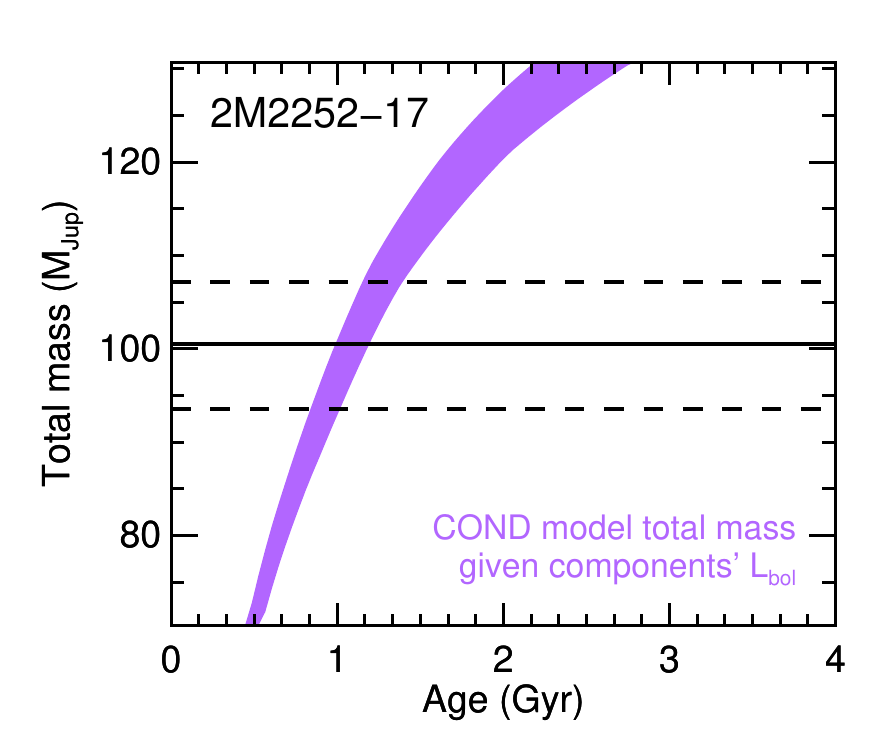} \includegraphics[width=3.0in,angle=0]{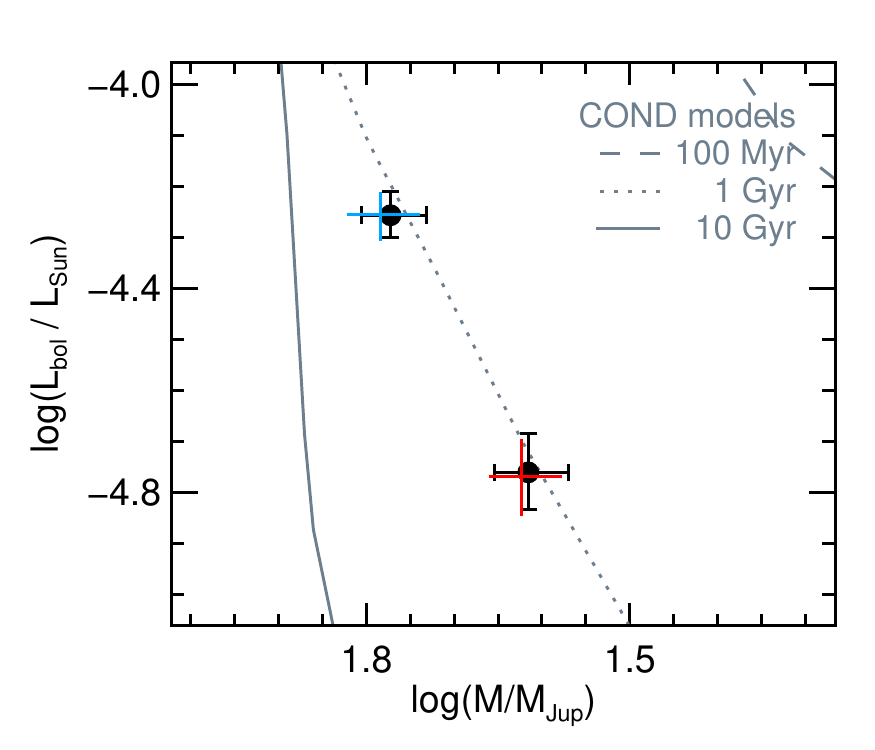} \includegraphics[width=3.0in,angle=0]{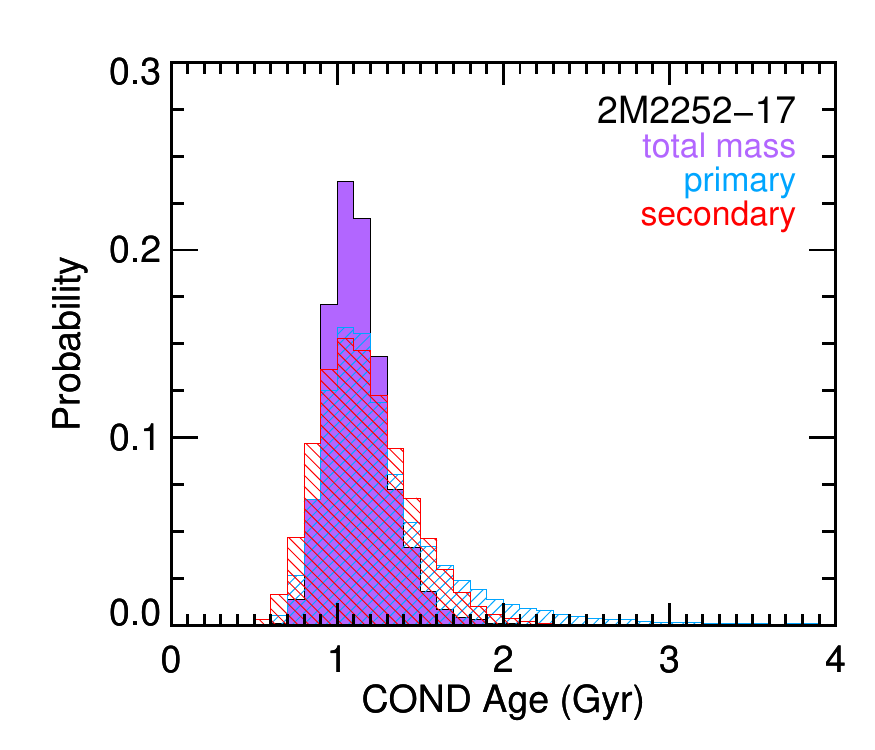}} 
\ContinuedFloat \caption{\normalsize (Continued)} \end{figure} 
\end{landscape}
\clearpage

\begin{landscape}
\begin{figure} 
  \vskip -0.4in
  \centerline{\includegraphics[width=4.2in,angle=0]{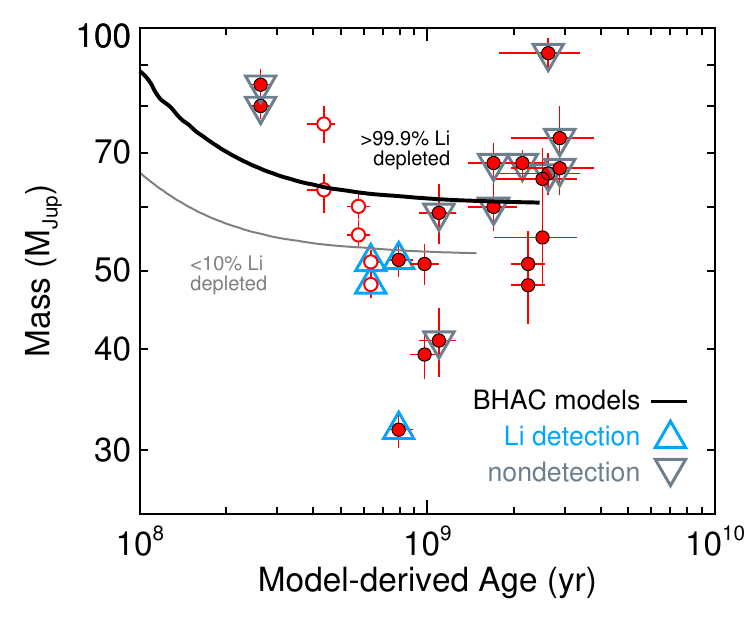} \includegraphics[width=4.2in,angle=0]{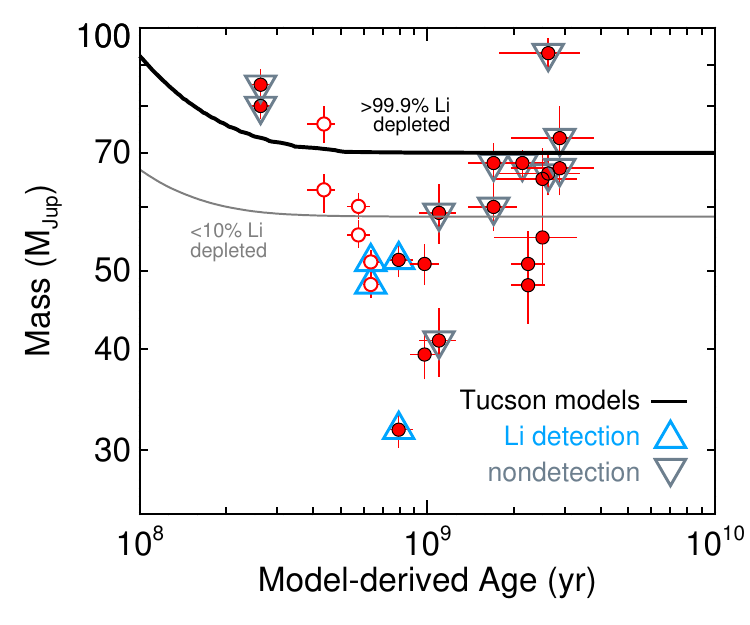}}
  \caption{\normalsize Individual masses plotted at their system ages
    for the 13 of our sample binaries that have well constrained ages
    according to evolutionary models.  Open symbols indicate systems
    where we use model-derived mass ratios to compute individual
    masses, including Gl~569Bab.  Up-pointing blue triangles indicate
    components of binaries where lithium is detected in integrated
    light, while down-pointing orange triangles indicate systems for
    which an optical spectrum is available and does not display
    lithium absorption.  Model predicted mass limits for strong
    ($>$99.9\%) and weak ($<$10\%) lithium depletion are shown as
    thick black and thin gray lines, respectively.  The models shown
    here are BHAC \citep{2015A&A...577A..42B} and Tucson
    \citep{1997ApJ...491..856B}. Lithium detections and nondetections
    seem to generally agree well with BHAC model-predicted lithium
    depletion, within the observational uncertainties on mass.  The
    mass limit for lithium depletion is higher in Tucson models
    (70\,\Mjup\ instead of 60\,\Mjup), making many nondetections of
    lithium seemingly discrepant with model predictions.  (Note that
    we use Cond model ages for plotting here because they are the only
    models that cover the full range of luminosities in our sample,
    and they tend to agree very well with the results from SM08 and
    BHAC models.  We use the ages from our total-mass analysis, except
    for LHS~2397aAB where we use the age from the individual-mass
    analysis of the secondary.  For clarity, the unresolved multiples
    2MASS~J0920+3517B and 2MASS~J0700+3157B are not shown.  The two
    young systems that lack lithium information are Gl~569Bab and
    HD~130948BC.)  \label{fig:m-age}}
\end{figure} 
\end{landscape}
\clearpage

\begin{figure} 
  \vskip -0.4in
  \centerline{\includegraphics[width=3.8in,angle=0]{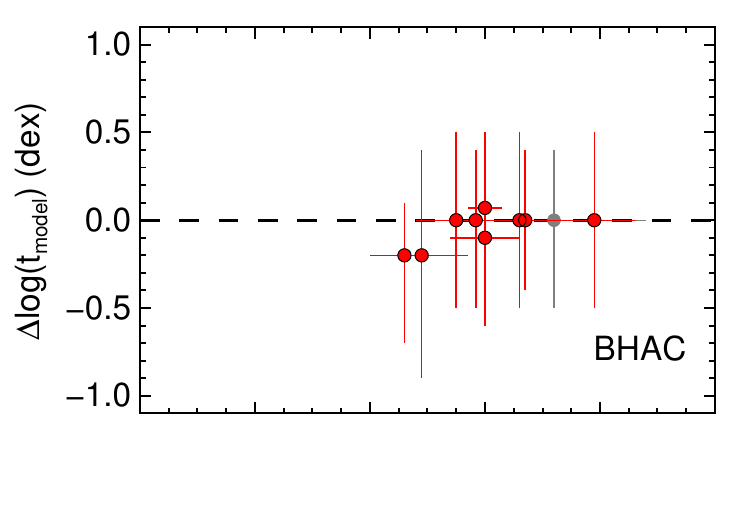}}
  \vskip -0.65in
  \centerline{\includegraphics[width=3.8in,angle=0]{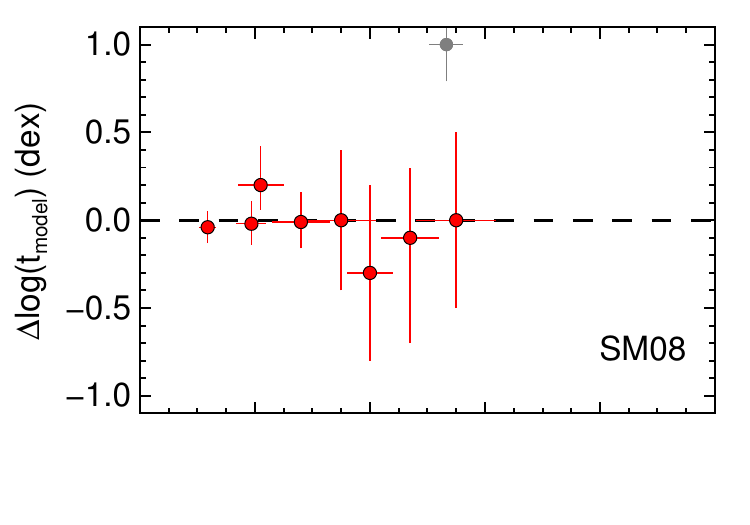}}
  \vskip -0.65in
  \centerline{\includegraphics[width=3.8in,angle=0]{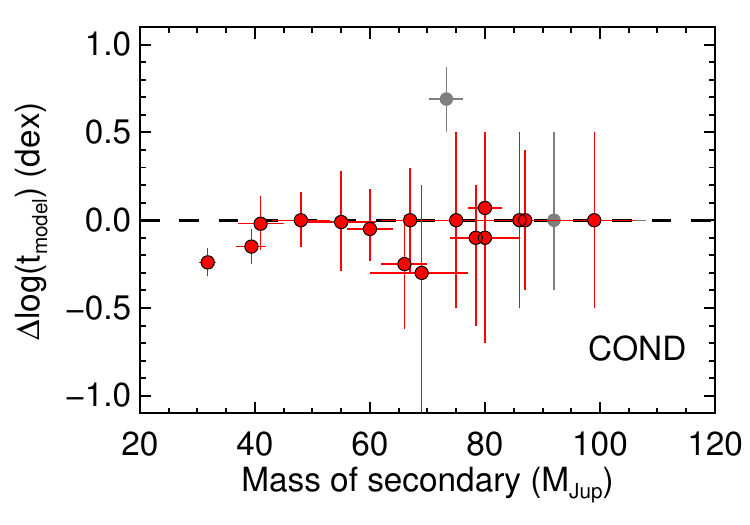}}
  \caption{\normalsize Coevality determinations plotted as a function
    of the mass of the secondary component.  For each system with
    directly measured individual masses and luminosities, we derive
    the age of each component from models ($t_1$ and $t_2$).  The data
    points here show the median and 1$\sigma$ intervals in the
    posterior distribution of the logarithm of the difference in age,
    computed as $\log{t_2}-\log{t_1}$.  Gray symbols indicate two of
    the likely unresolved multiples LP~415-20 and 2MASS~J0700+3157.
    Binaries composed of main-sequence stars (secondary masses
    $>$70\,\Mjup\ here) have essentially unconstrained ages and thus
    $\Delta\log{t} \approx 0.0\pm0.5$\,dex (i.e., both stars
    consistent with any main sequence age at 1$\sigma$).  Aside from
    the unresolved multiple 2MASS~J0700+3157, all but two systems give
    consistent ages for the primary and secondary at
    $\lesssim$1$\sigma$.  These are two binaries spanning the L/T
    transition that have the lowest mass secondaries.
    SDSS~J0423$-$0414AB and SDSS~J1052+4422AB are 3.0$\sigma$ and
    1.5$\sigma$, respectively, discrepant with coevality in the Cond
    models.  In contrast, the SM08 models that predict a shallower
    mass--luminosity relation in the L/T transition yield coeval ages
    for both systems.  \label{fig:mb-dt}}
\end{figure} 
\clearpage

\begin{landscape}
\begin{figure} 
  \centerline{\includegraphics[width=4.0in,angle=0]{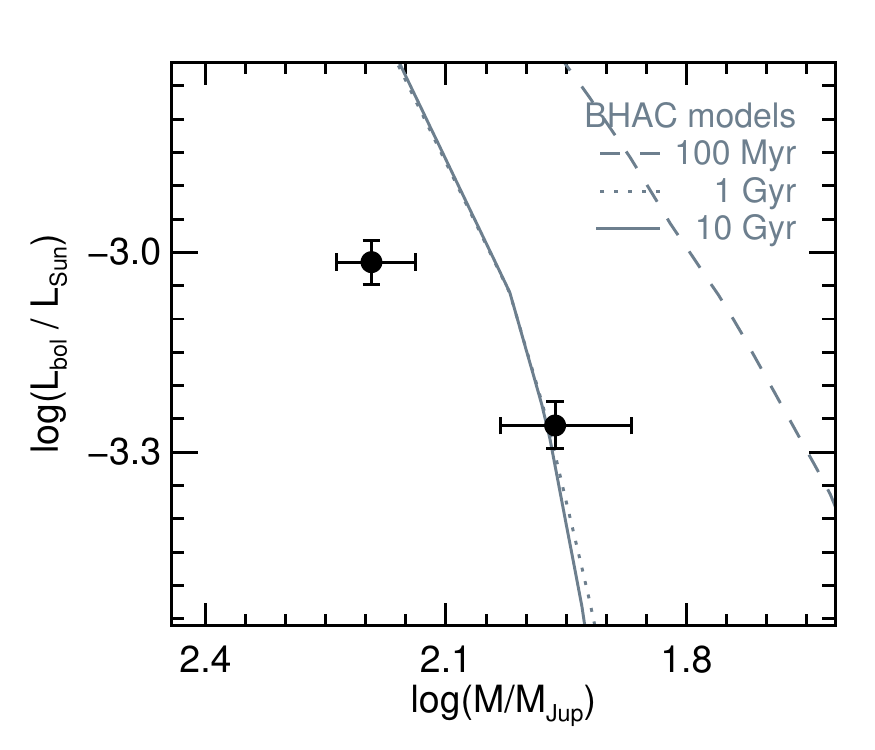}} 
  \caption{Individual mass and luminosity measurements shown alongside
    models.  These are all likely triples given that the data points
    significantly deviate from coevality (i.e., falling on a single
    isochrone).  In the cases of 2MASS~J0700+3157AB and
    2MASS~J0920+3517AB, the less luminous secondary is actually more
    massive than the more luminous primary, implying that the
    secondaries are in fact unresolved binaries.  In the case of
    LP~415-20AB, the primary is much less luminous than expected given
    its mass, implying that it is an unresolved pair of lower
    luminosity, lower mass objects. \label{fig:triple-ml}}
\end{figure}
\begin{figure} 
  \centerline{\includegraphics[width=4.0in,angle=0]{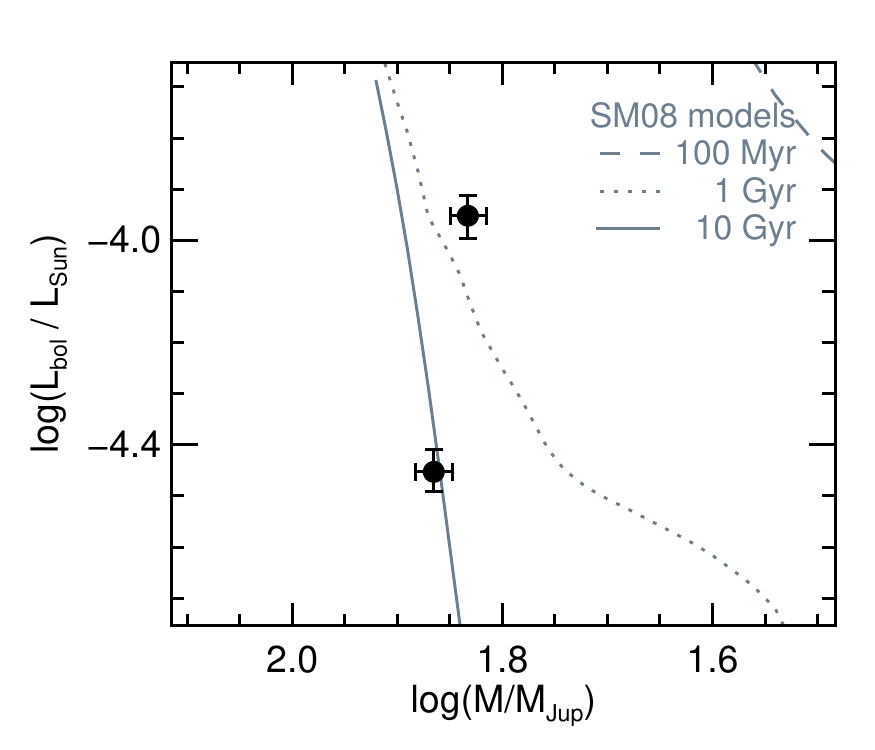} \includegraphics[width=4.0in,angle=0]{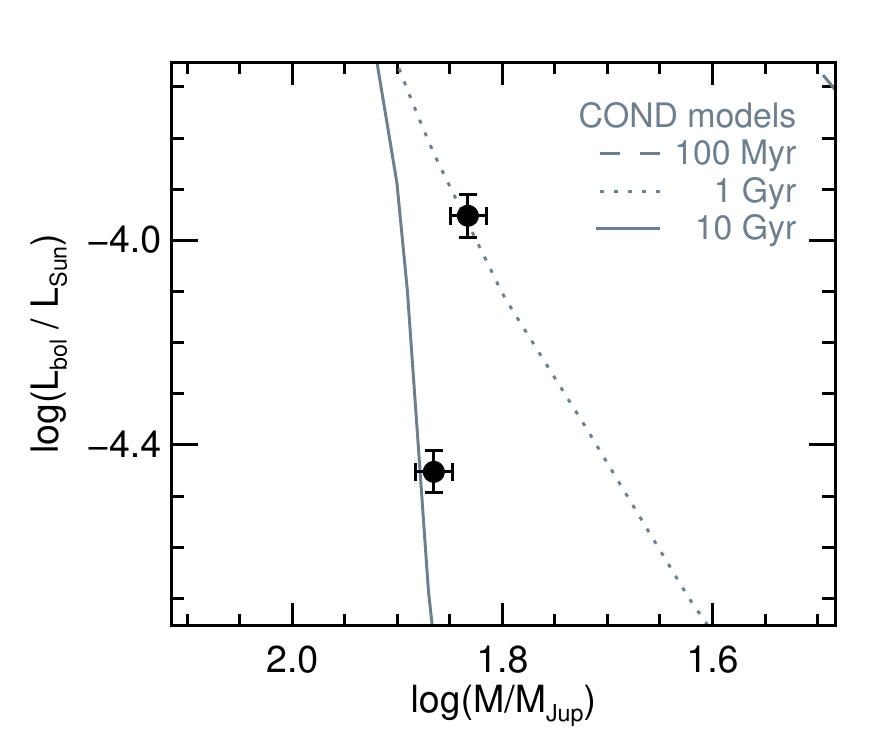}} 
\ContinuedFloat \caption{\normalsize (Continued)}
\end{figure} 
\begin{figure} 
  \centerline{\includegraphics[width=4.0in,angle=0]{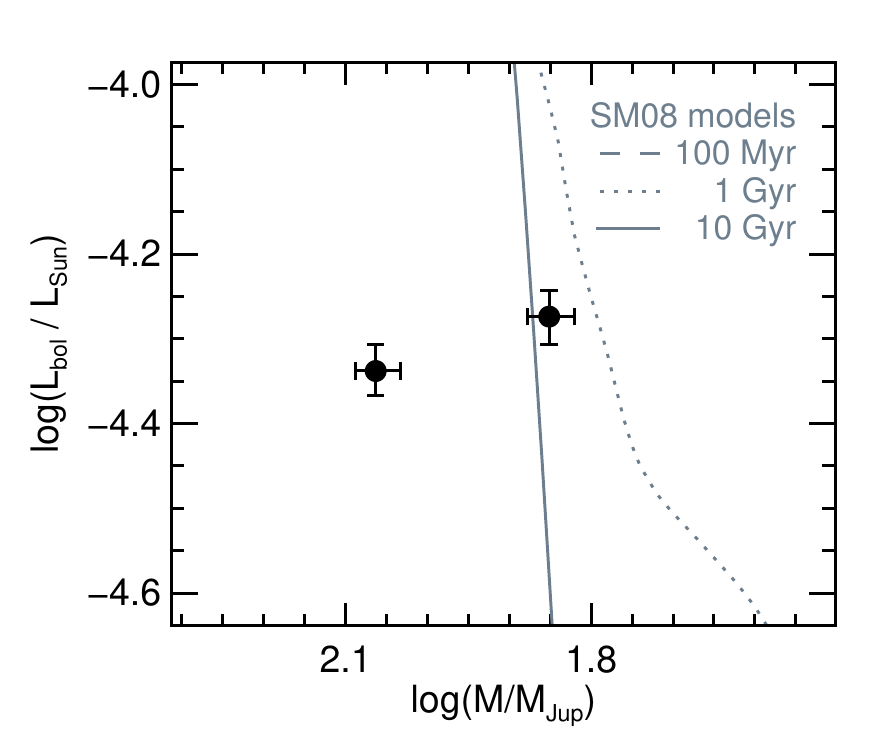} \includegraphics[width=4.0in,angle=0]{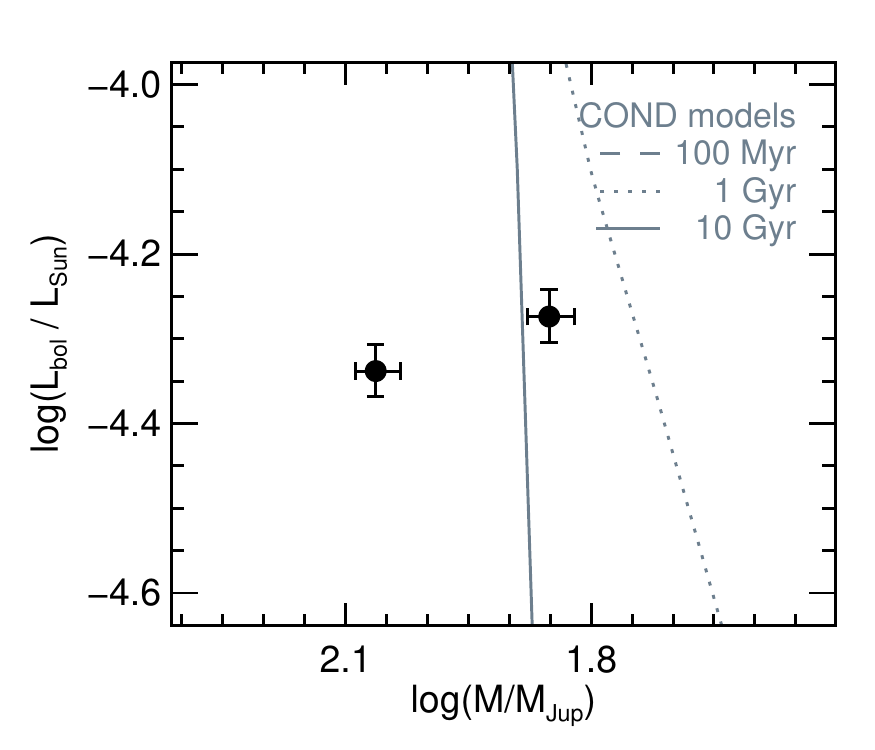}} 
\ContinuedFloat \caption{\normalsize (Continued)}
\end{figure} 
\end{landscape}
\clearpage

\begin{figure} 
  \centerline{\includegraphics[width=4.5in,angle=0]{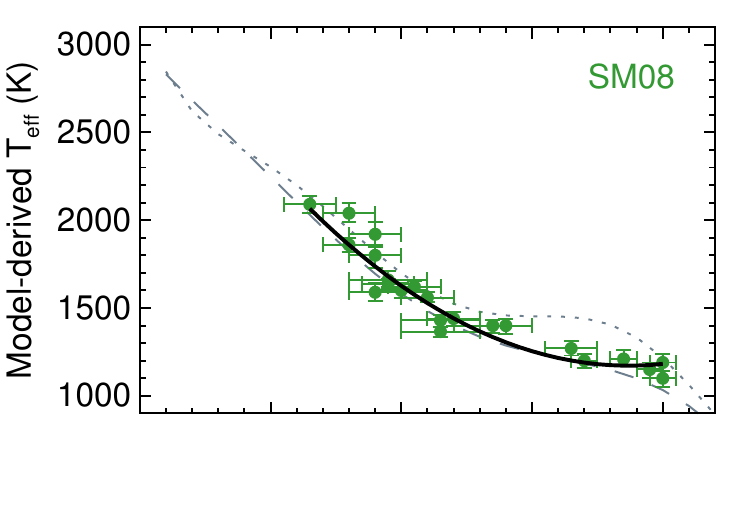}}
  \vskip -0.7in
  \centerline{\includegraphics[width=4.5in,angle=0]{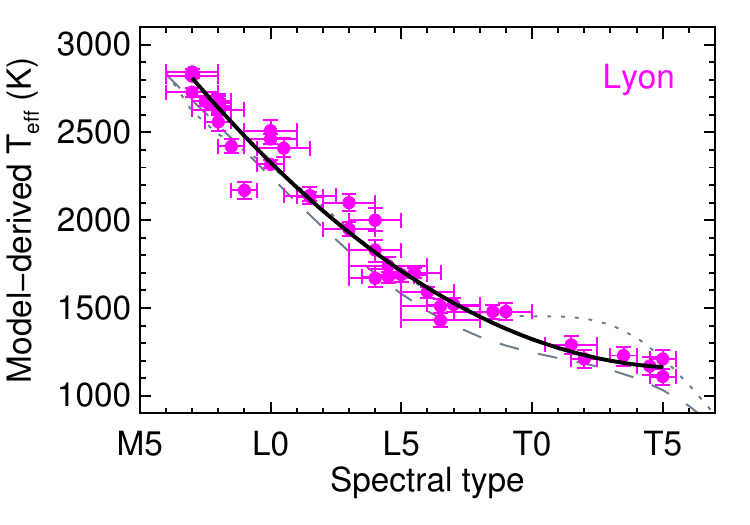}}
  \caption{\normalsize Effective temperatures derived from SM08 models
    (top) and from Lyon models (bottom; BHAC and Cond).  Solid black
    lines show our second-order polynomials to the relations, having
    rms scatter about the fit of 90\,K for the 40 objects with Lyon
    temperatures and 80\,K for the 22 objects with SM08 temperatures.
    Gray lines indicate previous literature polynomial relations from
    \citet[][dotted]{2004AJ....127.3516G} and
    \citet[][dashed]{2015ApJ...810..158F}.  \label{fig:spt-teff}}
\end{figure} 
\clearpage

\begin{figure} 
  \centerline{\includegraphics[width=6.5in,angle=0]{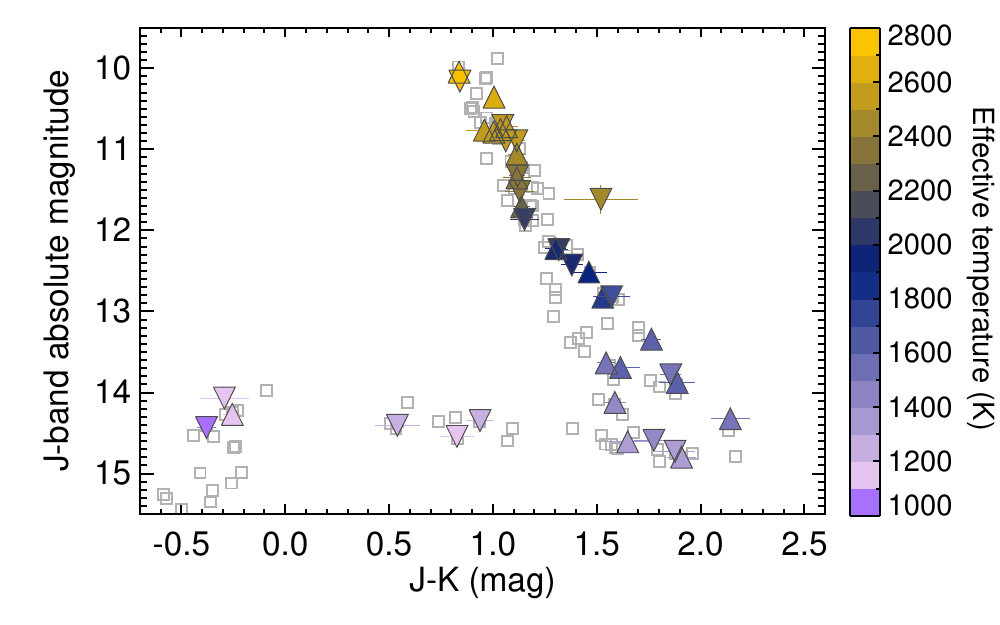}} 
  \caption{\normalsize Color--magnitude diagram (CMD) showing our
    dynamical mass sample with symbols colored according to their Lyon
    model-derived effective temperatures (BHAC when possible,
    otherwise Cond).  Up-pointing triangles indicate primary
    components, down-pointing triangles indicate secondary components,
    and the field sequence is shown for reference as gray squares
    using the latest compilation of ultracool dwarf parallaxes at
    \protect\url{http://www.as.utexas.edu/~tdupuy/plx}.  As expected,
    temperature correlates strongly with the location along the CMD
    sequence, consistent with temperature being a primary driver of
    the spectral energy distributions of ultracool dwarfs.  Objects at
    the faintest magnitudes ($M_J = 14$--15\,mag) span temperatures of
    $\approx$1500\,K to $\approx$1100\,K, where objects with bluer
    $J-K$ colors have cooler model-derived
    temperatures. \label{fig:cmd-teff}}
\end{figure} 
\clearpage

\begin{figure} 
  \centerline{\includegraphics[width=4.5in,angle=0]{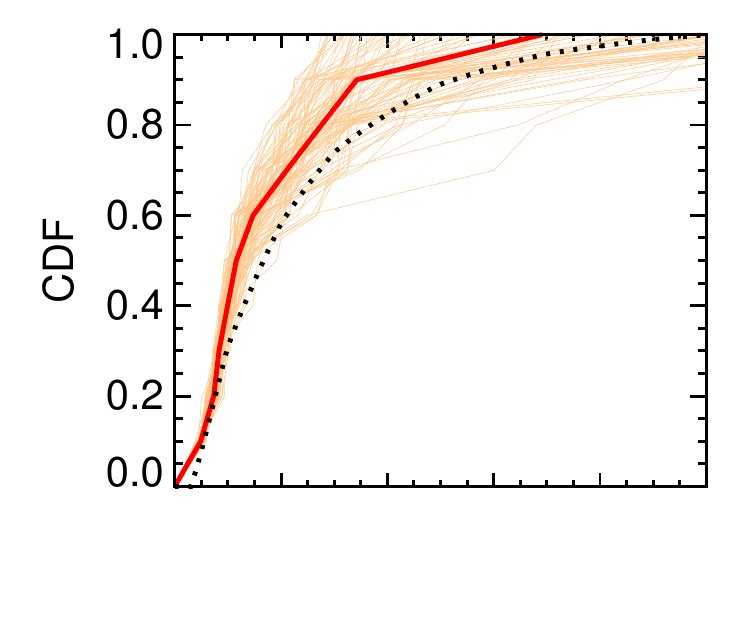}}
  \vskip -0.9in
  \centerline{\includegraphics[width=4.5in,angle=0]{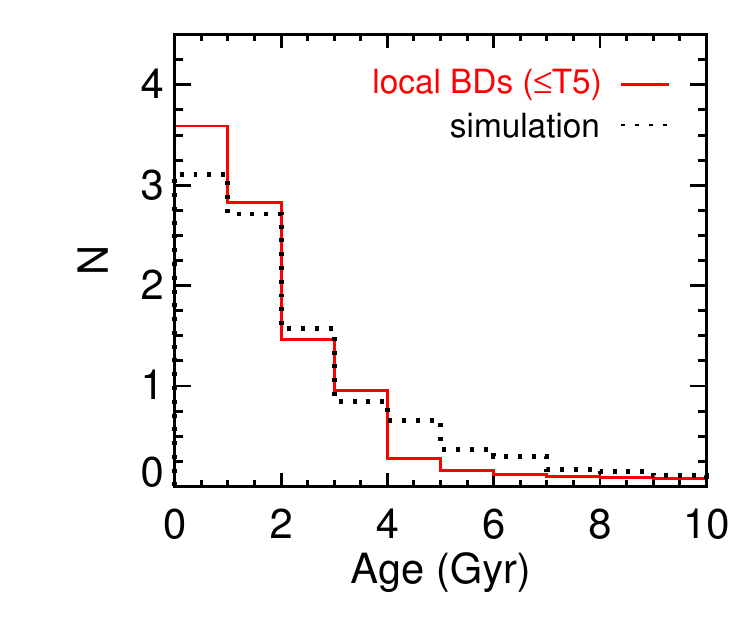}}
  \caption{\normalsize Bottom: age distribution for the systems with
    well determined ages (solid red line).  The histogram is a mean of
    $10^4$ posterior values for each system, resulting in noninteger
    values for each 1-Gyr-wide bin.  The dotted line shows the age
    distribution of a synthesized population of brown dwarfs with
    masses of 30--70\,\Mjup\ and assuming a constant star formation
    rate (but including the young-age skew due to dynamical heating
    that removes old objects from the Galactic midplane, according to
    the Besan\c{c}on model of the solar neighborhood).  Top:
    cumulative distribution functions computed for the median age of
    each system (thick red line) and for 100 randomly drawn posterior
    age values for each system (thin orange lines).  The observed
    distribution is remarkably consistent with the simplistic
    synthesized population that assumes a constant star formation rate
    for the thin disk and a spectral type cut of T5, as is the case
    for our sample of brown dwarfs here. \label{fig:age-hist}}
\end{figure} 
\clearpage

\begin{figure} 
  \centerline{\includegraphics[width=6.0in,angle=0]{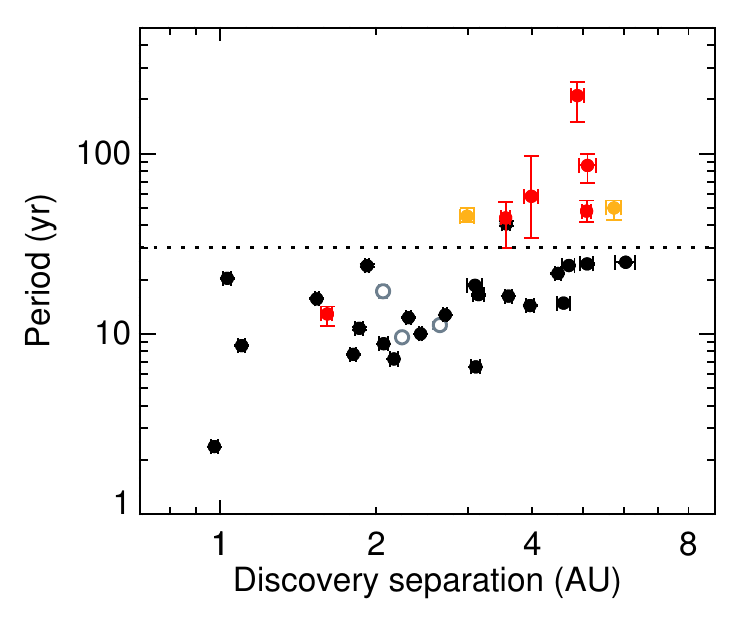}}
  \caption{\normalsize Orbital period of our sample binaries plotted
    as a function of projected separation at discovery.  Symbols are
    colored according to the quality of the orbit determination: good
    (black), marginal (orange), and poor (red).  Open gray symbols
    indicate results from the literature for LHS~1070BC,
    $\epsilon$~Ind~Bab, and LSPM~J1314+1320AB.  The quality of our
    orbit determinations are almost entirely determined by having
    sufficient observational time coverage, so the longest-period
    orbits tend to be poorly determined.  The only $P<30$\,yr binary
    with a poor orbit is 2MASS~J0926+5847AB, which is seen nearly edge
    on and lacks the phase coverage from our 3-year \HST\ program
    needed to robustly determine the orbit from such a challenging
    viewing angle.  Therefore, we conclude that a cut of $P<30$\,yr
    (dotted line) yields an effectively complete sample of orbit
    determinations to be used in a statistical study of orbital
    eccentricities.  Most orbits have error bars smaller than the
    plotting symbols. \label{fig:sep-p}}
\end{figure} 
\clearpage

\begin{figure} 
  \centerline{\includegraphics[width=6.0in,angle=0]{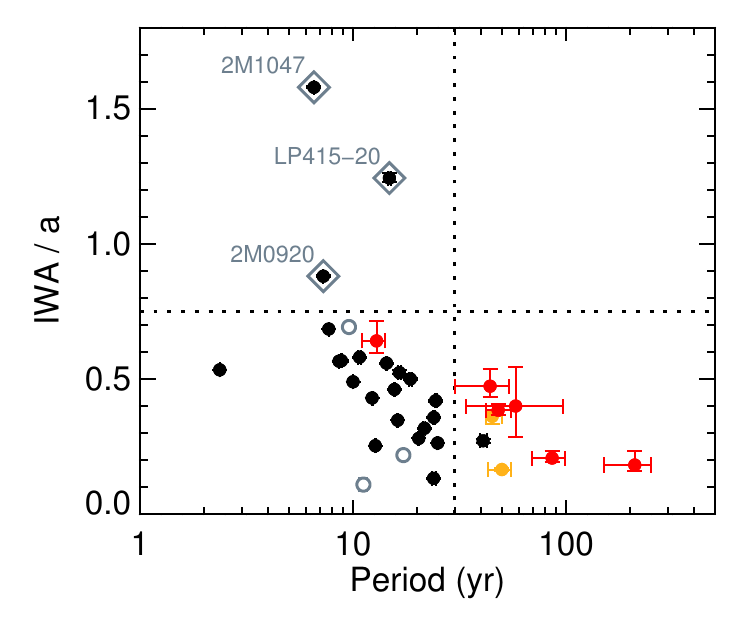}}
  \caption{\normalsize The ratio of the inner working angle (IWA) of
    the discovery data for each binary to its semimajor axis ($a$).
    For a given $a$, a survey with ${\rm IWA} \ll a$ is sensitive to
    binaries of all eccentricities, while a survey with ${\rm IWA} >
    a$ is strongly biased toward discovering eccentric binaries.
    According to simulations from \citet{2011ApJ...733..122D}, a
    cutoff of ${\rm IWA}/a < 0.75$ will result in minimal bias in the
    resulting eccentricity distribution.  Therefore, we exclude the
    binaries 2MASS~J1047+4026AB, LP~415-20AB, and 2MASS~J0920+3517AB
    (gray diamonds) from our statistical analysis.  The first two of
    these have IWA/$a > 1$ and indeed turn out to have quite eccentric
    ($\ge$0.7) orbits.  Most orbits have error bars smaller than the
    plotting symbols.  \label{fig:p-iwa}}
\end{figure} 
\clearpage

\begin{figure} 
  \centerline{\includegraphics[width=6.0in,angle=0]{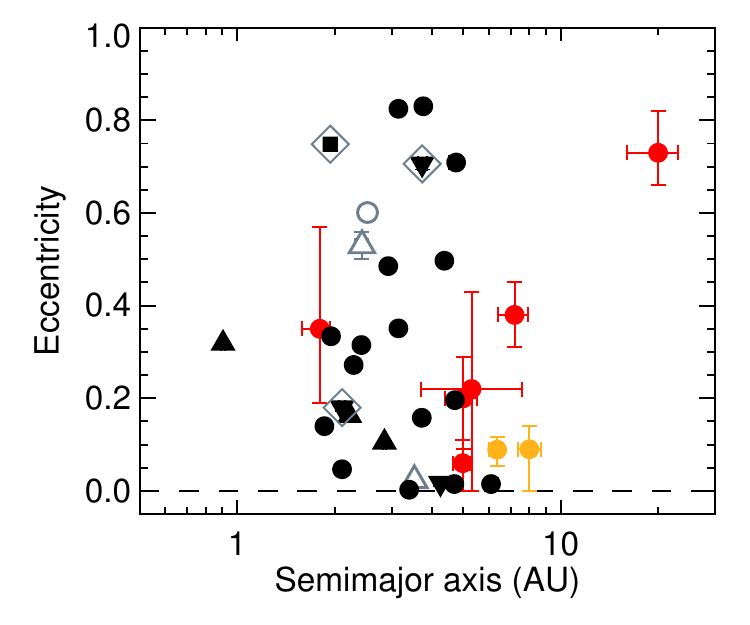}}
  \caption{\normalsize Eccentricity as a function of semimajor axis
    for all orbits in our sample.  Symbol shapes indicate orbits that
    are simple binaries (circles), inner pairs of hierarchical triples
    (up-pointing triangles), outer pairs of hierarchical triples
    (down-pointing triangles), or an inner pair in a quadruple
    (square).  Symbols are colored according to the quality of the
    orbit determination: good (black), marginal (orange), and poor
    (red).  Open gray symbols indicate results from the literature for
    LHS~1070BC, $\epsilon$~Ind~Bab, and LSPM~J1314+1320AB.  Objects
    expected to be impacted by discovery bias (IWA/$a > 0.75$) are
    enclosed by large gray diamonds.  No significant trends in
    eccentricity with semimajor axis are apparent.  Most orbits have
    error bars smaller than the plotting symbols. \label{fig:a-e}}
\end{figure} 
\clearpage

\begin{figure} 
  \centerline{\includegraphics[width=6.0in,angle=0]{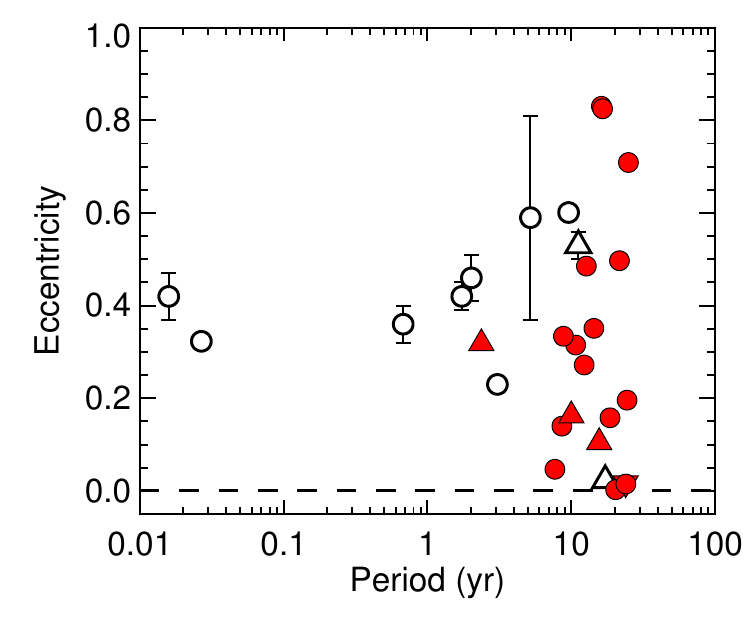}}
  \caption{\normalsize Eccentricity as a function of orbital period
    for binaries in our de-biased sample (filled red) and from the
    literature (open black).  Symbol shapes indicates orbits that are
    simple binaries (circles), inner pairs of a hierarchical triples
    (up-pointing triangles), or outer pairs of hierarchical triples
    (down-pointing triangles). Our sample and some published orbits
    have error bars smaller than the plotting symbols.  Literature
    results for spectroscopic, astrometric, and eclipsing binaries
    tend to have short periods and modest eccentricities (0.2--0.6).
    In contrast, visual binaries at longer periods have some very
    eccentric orbits ($\geq$0.7) and a substantial fraction of low-$e$
    orbits (10 of 22 have $e = 0.0$--0.2).  Whether this is indicative
    of an actual difference in orbital properties will require a
    larger sample of short-period binaries. \label{fig:p-e}}
\end{figure} 
\clearpage

\begin{figure} 
  \centerline{\includegraphics[width=4.5in,angle=0]{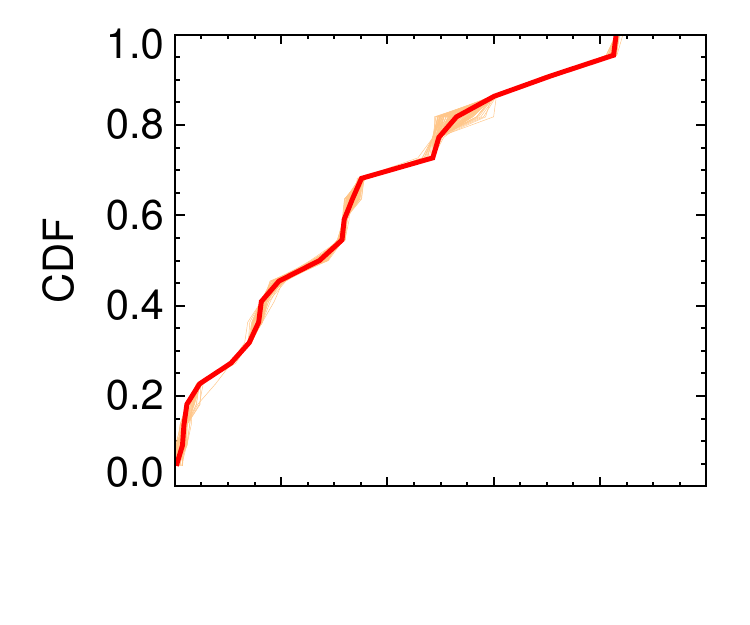}}
  \vskip -0.9in
  \centerline{\includegraphics[width=4.5in,angle=0]{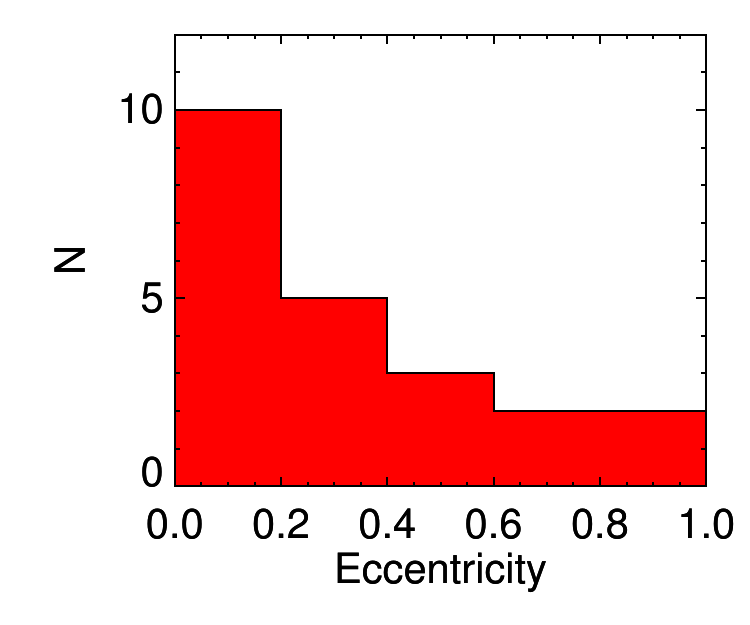}}
  \caption{\normalsize Bottom: eccentricity distribution of our
    de-biased visual binary sample ($P < 30$\,yr).  Top: cumulative
    distribution functions computed for the median eccentricity of
    each orbit (thick red line) and for 100 randomly drawn posterior
    values for each orbit (thin orange lines).  Almost half of the
    orbits (10 of 22) have low eccentricities
    (0.0--0.2). \label{fig:ecc-hist}}
\end{figure} 
\clearpage


\begin{figure} 
  \vskip -0.3in
  \centerline{\includegraphics[width=2.6in,angle=0]{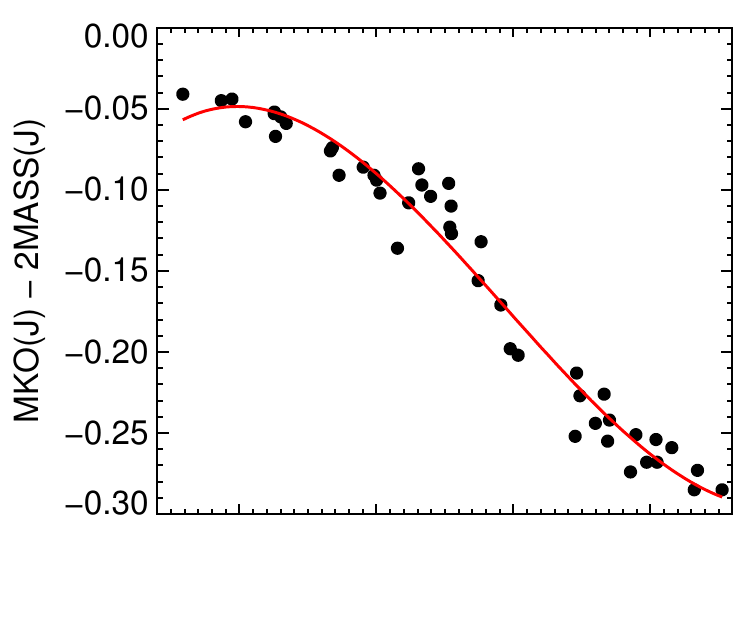} \includegraphics[width=2.6in,angle=0]{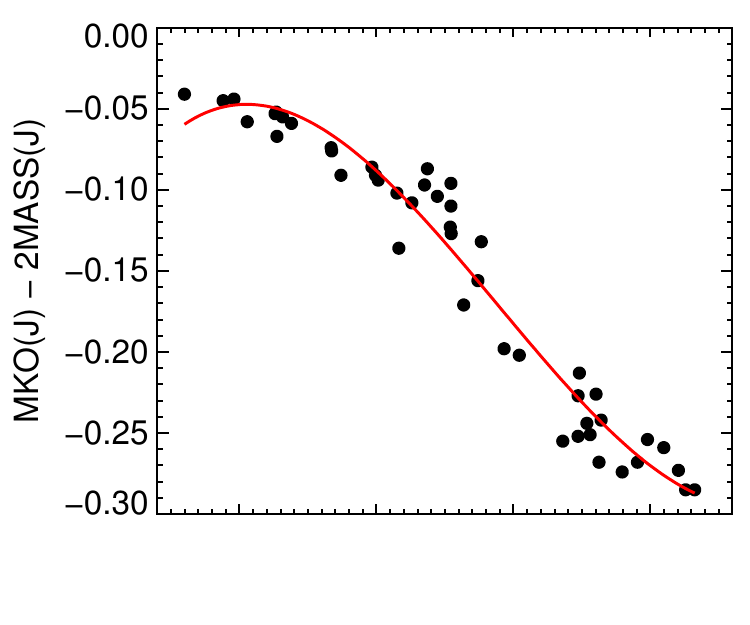}}  \vskip -0.4in
  \centerline{\includegraphics[width=2.6in,angle=0]{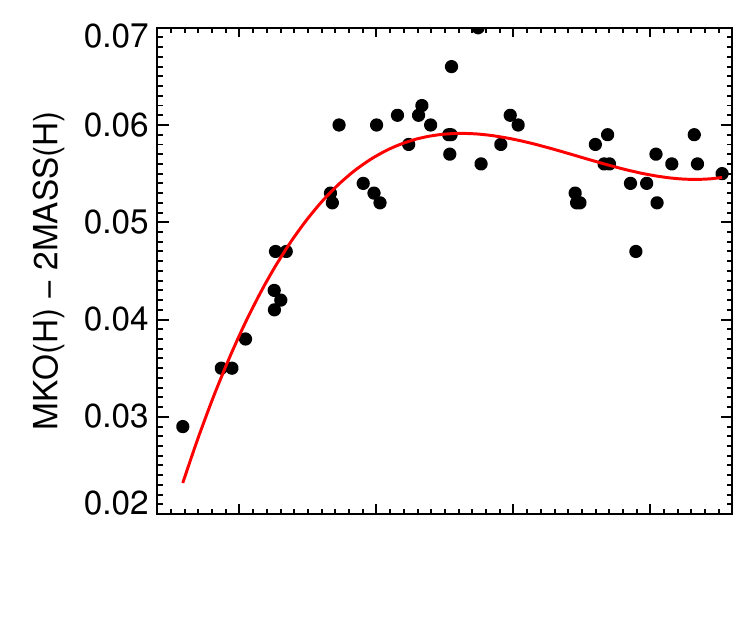} \includegraphics[width=2.6in,angle=0]{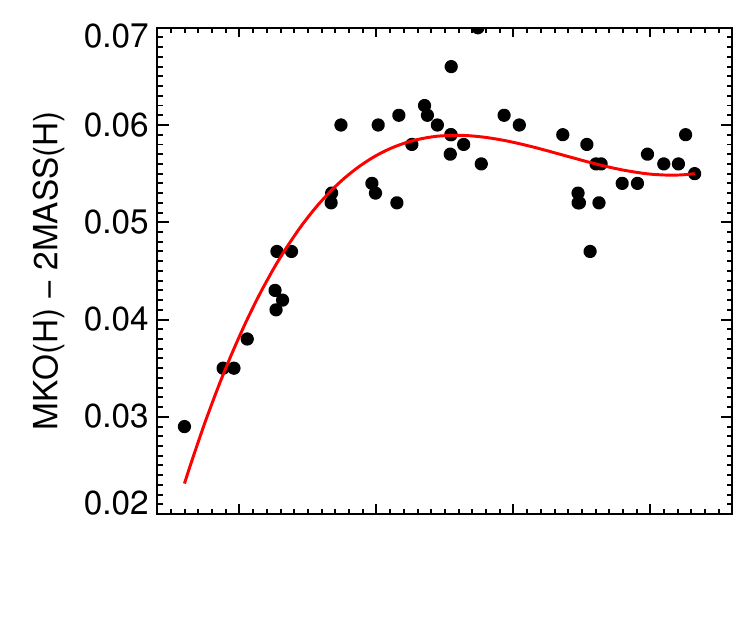}}  \vskip -0.4in
  \centerline{\includegraphics[width=2.6in,angle=0]{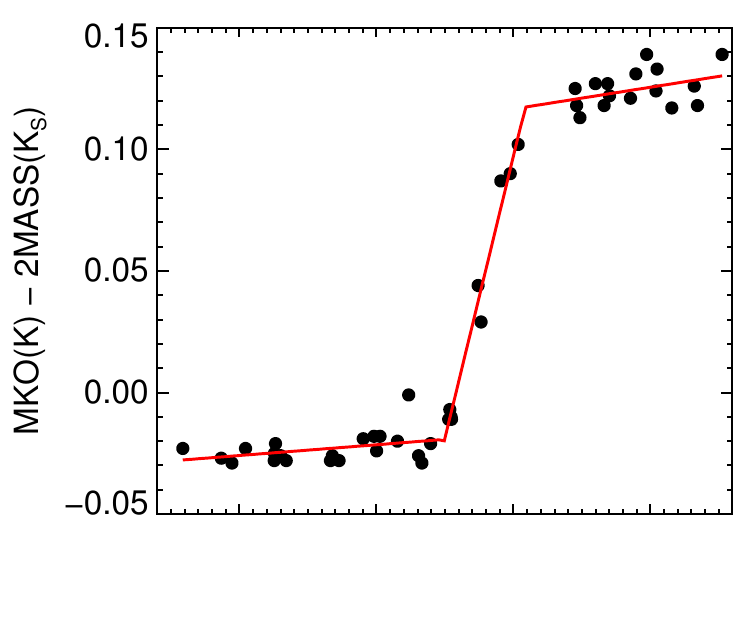} \includegraphics[width=2.6in,angle=0]{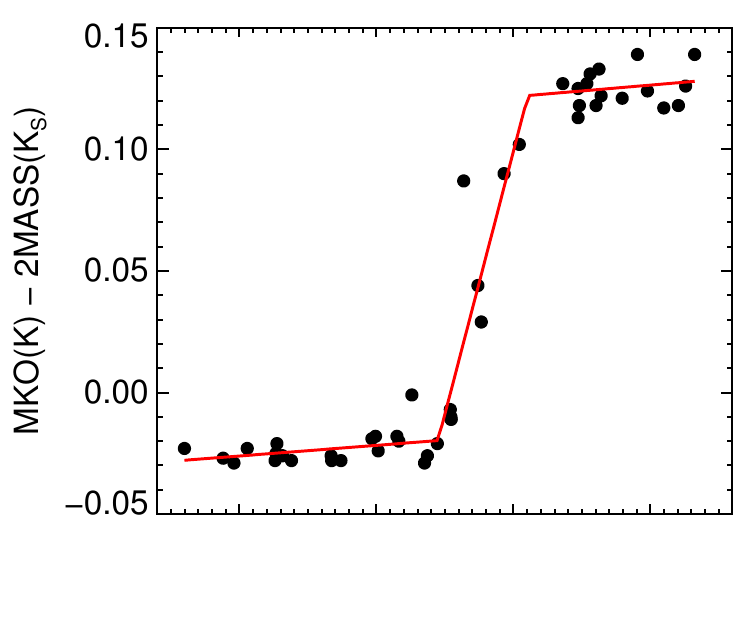}}  \vskip -0.4in
  \centerline{\includegraphics[width=2.6in,angle=0]{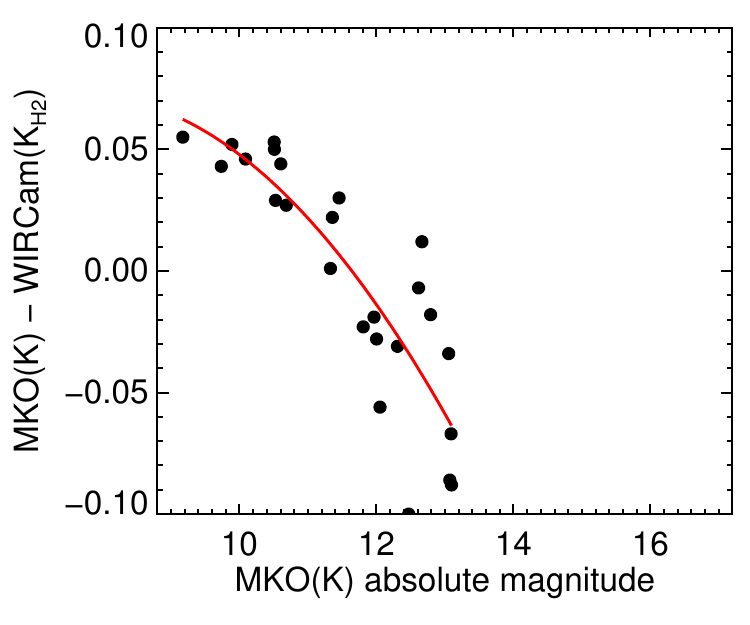} \includegraphics[width=2.6in,angle=0]{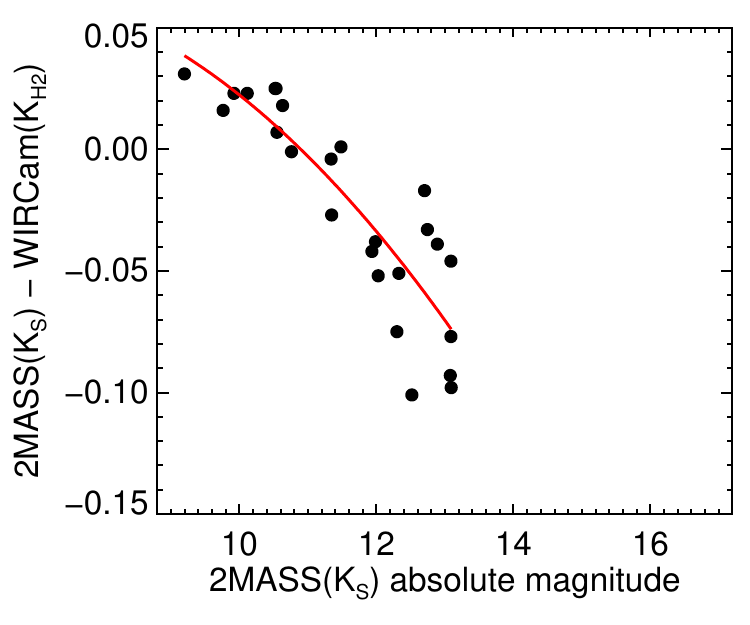}}
  \caption{\normalsize Polynomial relations between MKO-to-2MASS
    photometric conversions and absolute magnitude.  $K$-band is a
    three-part piecewise linear fit, while others are single
    polynomial fits.  The input data used to derive these fits (black
    points) are computed by synthetic photometry as described in
    Sections~\ref{app:phot-corr} and \ref{app:narr-corr}.  We use
    these relations to derive $JHK$ photometry on both 2MASS and MKO
    systems in cases where it is not measured
    directly. \label{fig:poly-corr}}
\end{figure}
\begin{figure} 
  \centerline{\includegraphics[width=2.6in,angle=0]{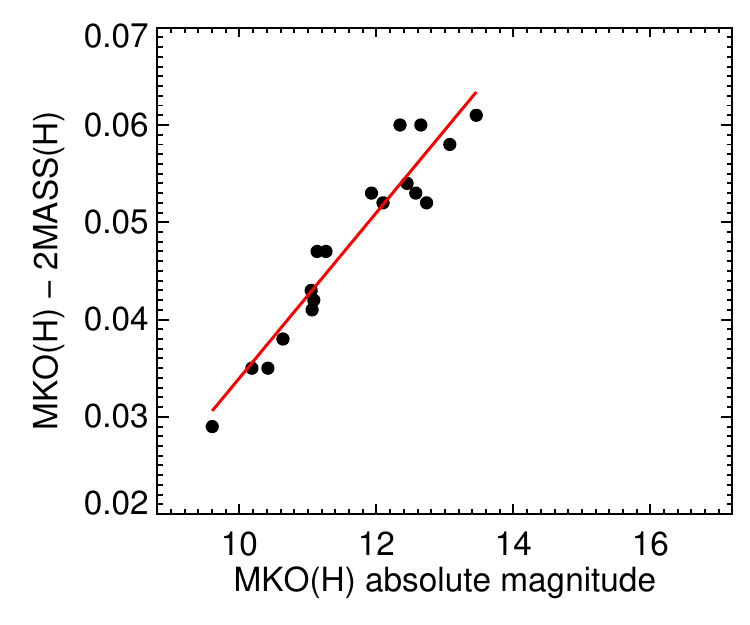} \includegraphics[width=2.6in,angle=0]{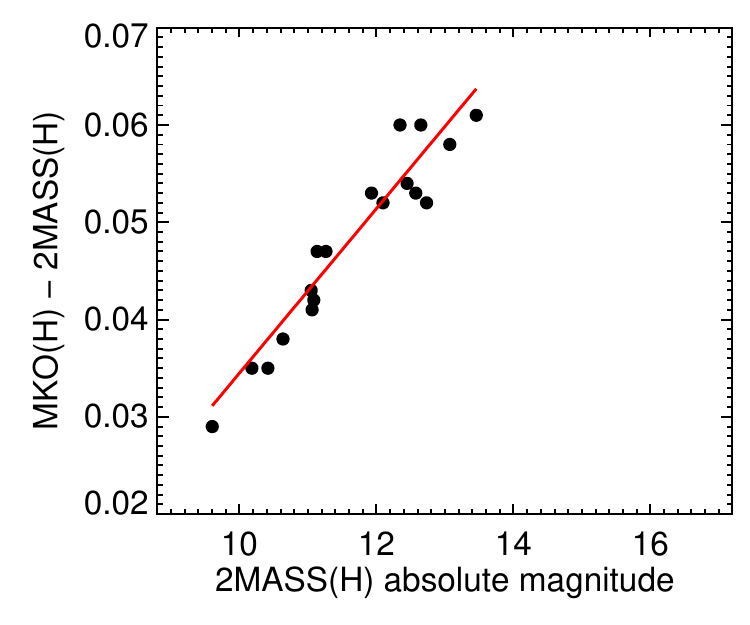}}
  \ContinuedFloat \caption{\normalsize (Continued)}
\end{figure}
\clearpage

\begin{figure} 
  \centerline{
    \includegraphics[width=3.25in,angle=0]{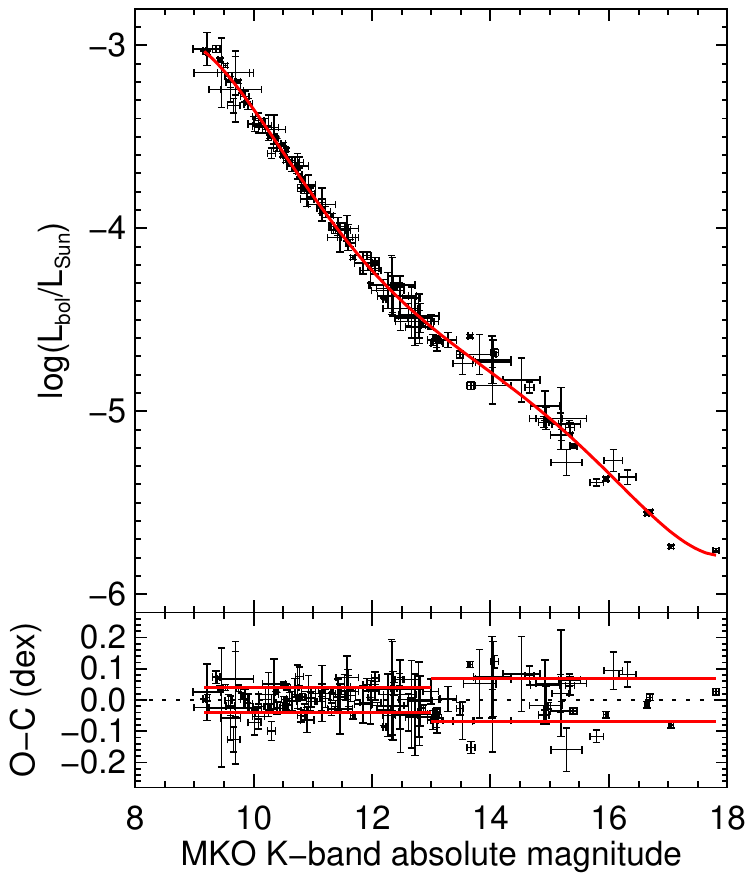}
    \includegraphics[width=3.25in,angle=0]{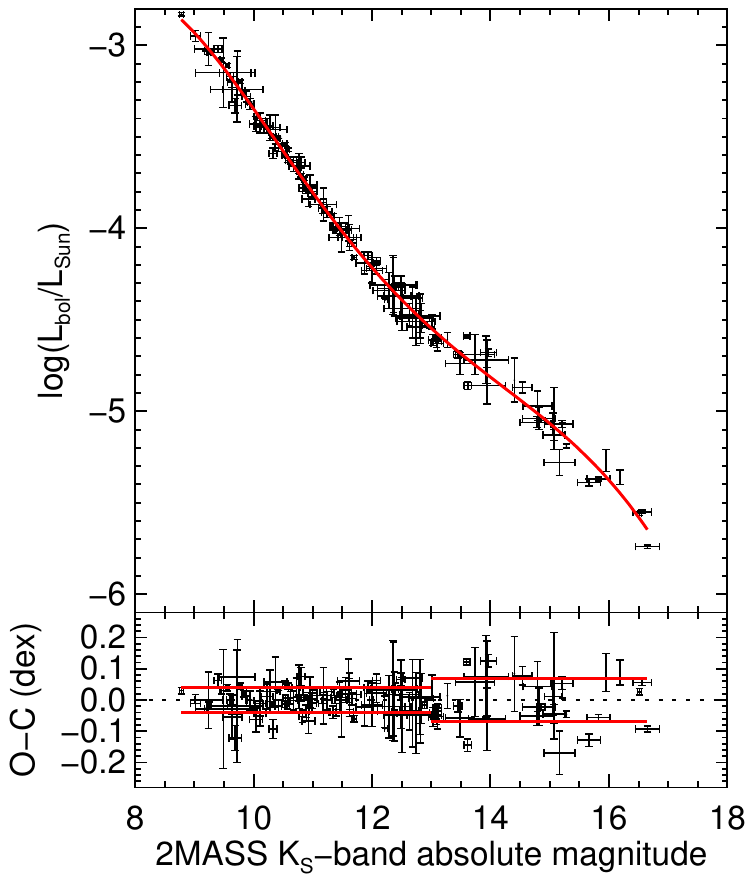}}  
  \caption{\normalsize Polynomial relations between bolometric
    luminosity and absolute magnitude (red) derived from input data
    from \citet{2015ApJ...810..158F} (black points) as described in
    Section~\ref{app:lbol-mag}.  Bottom panels show the residuals in
    the input data after subtracting the polynomials.  We use these
    relations to derive the component luminosities for our sample
    binaries.  When both components have $M_H < 13.3$\,mag, we use the
    $H$-band relations, otherwise we use $K$-band
    relations. \label{fig:poly-lbol}}
\end{figure}
\begin{figure} 
  \centerline{
    \includegraphics[width=3.25in,angle=0]{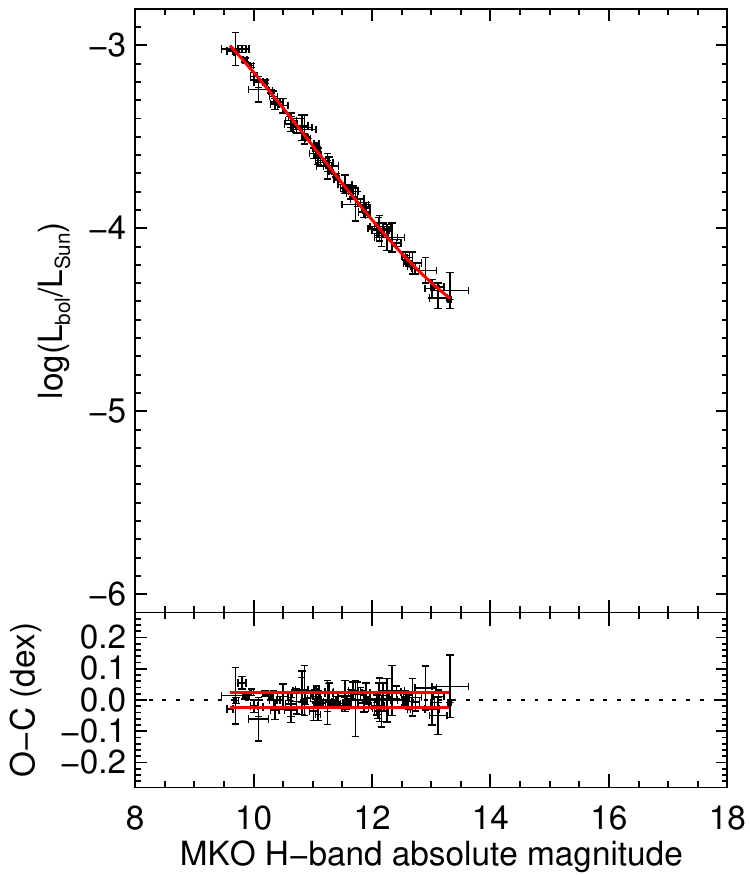}
    \includegraphics[width=3.25in,angle=0]{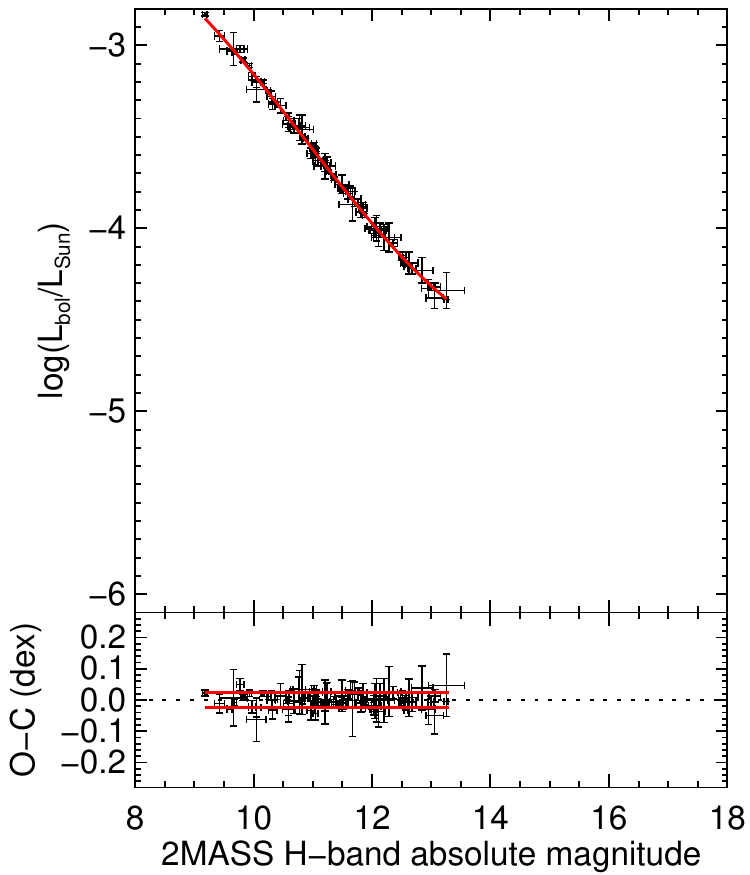}}
  \ContinuedFloat \caption{\normalsize (Continued)}
\end{figure}
\clearpage


\begin{deluxetable}{lcclccccccc}
\setlength{\tabcolsep}{0.065in}
\tabletypesize{\tiny}
\tablecaption{Orbit Monitoring Sample \label{tbl:sample}}
\tablewidth{0pt}
\tablehead{
\colhead{Name} &
\colhead{} &
\multicolumn{3}{c}{Binary Discovery} &
\colhead{} &
\multicolumn{2}{c}{Integrated-Light Spec.\ Type} &
\colhead{} &
\multicolumn{2}{c}{Component Spec.\ Types} \\

\cline{3-5}
\cline{7-8}
\cline{10-11}

\colhead{} &
\colhead{} &
\colhead{Sep.} &
\colhead{Epoch (UT)} &
\colhead{Ref.} &
\colhead{} &
\colhead{optical/IR} &
\colhead{Ref.} &
\colhead{} &
\colhead{A+B} &
\colhead{Ref.}
}
\startdata
LP~349-25AB                   & & 0$\farcs$125 & 2004-07-03 & Forv05  & & M8/M8~\intg\    & Gizi00, Dupu16b  & & M7:+M8:                     & Dupu12  \\
LP~415-20AB                   & & 0$\farcs$119 & 2002-02-07 & Sieg03  & & M7.5/\nodata    & Gizi00           & & M6:+M8                      & Dupu16b \\
SDSSp~J042348.57$-$041403.5AB & & 0$\farcs$164 & 2004-07-22 & Burg05  & & L7.5/T0         & Cruz03, Burg06a  & & L6.5:+T2                    & Dupu12  \\
2MASS~J05185995$-$2828372AB   & & 0$\farcs$051 & 2004-09-07 & Burg06b & & L7/T1p          & Kirk08, Burg06a  & & L6:+T4                      & Dupu12  \\
2MASS~J07003664+3157266AB     & & 0$\farcs$170 & 2004-12-29 & Reid06a & & L3.5/\nodata    & Thor03           & & L3:+L6.5:.                  & Dupu12  \\
LHS~1901AB                    & & 0$\farcs$275 & 2004-01-08 & Mont06  & & M7/M7           & Lepi09, Dupu10   & & M7:+M7:                     & Dupu12  \\
2MASSI~J0746425+200032AB      & & 0$\farcs$220 & 2000-04-25 & Reid01  & & L0.5/L1         & Kirk00, Knap04   & & L0+L1.5\tablenotemark{*}    & Bouy04  \\
2MASSs~J0850359+105716AB      & & 0$\farcs$160 & 2000-02-01 & Reid01  & & L6/\nodata      & Kirk99           & & L6.5:+L8.5:                 & Dupu12  \\
\enddata
\end{deluxetable}

\clearpage

\begin{deluxetable}{lccccccl}
\tabletypesize{\footnotesize}
\tablewidth{0pt}
\tablehead{
\multicolumn{2}{c}{Observation Date} &
\colhead{Separation}    &
\colhead{PA}      &
\colhead{$\Delta{m}$} &
\colhead{Bandpass}  &
\colhead{$N_{\rm frames}$}      &
\colhead{Notes}     \\
\colhead{(UT)}   &
\colhead{(MJD)}   &
\colhead{(mas)}    &
\colhead{(\degree)}      &
\colhead{(mag)} &
\colhead{} &
\colhead{} &
\colhead{} }
\tablecaption{Relative astrometry from Keck/NIRC2 Adaptive Optics Imaging and Masking \label{tbl:keck}}
\startdata
\multicolumn{8}{c}{LP~349-25AB ($N_{\rm ep} = 8, \Delta{t} =  3.99$\,yr)} \\ \cline{1-8}                                2008~Jan~16 & 54481.25 & $137.25\pm0.24  $ & $208.07\pm0.06  $ & $ 0.335\pm0.016$ & $\Ks       $ & 10 & I                    \\
2008~Jun~30 & 54647.55 & $114.88\pm0.25  $ & $194.94\pm0.12  $ & $ 0.285\pm0.011$ & $\Ks       $ &  9 & I                    \\
2008~Jun~30 & 54647.56 & $114.67\pm0.23  $ & $195.13\pm0.08  $ & $ 0.303\pm0.004$ & $H         $ &  7 & I                    \\
2008~Jun~30 & 54647.56 & $ 114.5\pm0.6   $ & $195.20\pm0.17  $ & $ 0.345\pm0.011$ & $J         $ & 11 & I                    \\
2008~Aug~20 & 54698.61 & $105.71\pm0.10  $ & $189.90\pm0.05  $ & $ 0.353\pm0.010$ & $\Ks       $ & 12 & I                    \\
2008~Sep~9  & 54718.58 & $102.17\pm0.14  $ & $187.62\pm0.09  $ & $ 0.243\pm0.005$ & $\Lp       $ & 12 & I                    \\
2009~Sep~28 & 55102.56 & $ 71.20\pm0.30  $ & $  98.4\pm0.6   $ & $ 0.230\pm0.030$ & $\Ks       $ & 14 & I                    \\
\enddata
\tablecomments{In the notes column, ``I'' indicates an observation done with direct imaging and ``M'' indicates non-redundant aperture masking.}
\tablenotetext{*}{This denotes a previously published data set that we obtained from the public NIRC2 archive.}
\tablenotetext{**}{This denotes a previously unpublished data set that we obtained from the public NIRC2 archive.}
\end{deluxetable}
\clearpage

\begin{deluxetable}{lccccl}
\tabletypesize{\scriptsize}
\tablewidth{0pt}
\tablehead{
\colhead{Date} &
\colhead{Separation}    &
\colhead{PA}      &
\colhead{$\Delta{m}$} &
\colhead{Bandpass}  &
\colhead{Notes}     \\
\colhead{(UT)}   &
\colhead{(mas)}    &
\colhead{(\degree)}      &
\colhead{(mag)} &
\colhead{} &
\colhead{} }
\tablecaption{Relative astrometry from \HST\ Imaging and the Literature \label{tbl:other}}
\startdata
\multicolumn{6}{c}{LP~349-25AB ($N_{\rm ep} = 2$)} \\ \cline{1-6}                                                       2004~Jul~3  & $   125\pm10    $ & $  12.7\pm2.0   $ &      $0.26\pm0.05$ & $\Kp       $ & CFHT/PUEO~(Forveille~et~al.~2005)                            \\
2004~Sep~26 & $   107\pm10    $ & $   7.1\pm0.5   $ &      $0.38\pm0.05$ & $H         $ & VLT/NACO~(Forveille~et~al.~2005)                             \\
 \cline{1-6} \multicolumn{6}{c}{} \\     
\multicolumn{6}{c}{LP~415-20AB ($N_{\rm ep} = 1$)} \\ \cline{1-6}                                                       2002~Feb~7  & $ 119.0\pm1.0   $ & $  91.2\pm0.7   $ &      $0.66\pm0.06$ & $\Kp       $ & Gemini/QUIRC~(Siegler~et~al.~2003)                           \\
 \cline{1-6} \multicolumn{6}{c}{} \\     
\multicolumn{6}{c}{SDSS~J0423$-$0414AB ($N_{\rm ep} = 8$)} \\
\enddata
\tablecomments{The measurements reported in this table are from our
  own analysis of archival data in this paper, unless another
  reference is given in the Notes column.}
\end{deluxetable}
\clearpage

\begin{deluxetable}{lccccccc}
\tabletypesize{\footnotesize}
\tablewidth{0pt}
\tablehead{
\multicolumn{2}{c}{Observation Date} &
\colhead{R.A.}    &
\colhead{Dec.}      &
\colhead{$\sigma_{\rm R.A.}\cos{\delta}$} &
\colhead{$\sigma_{\rm Dec.}$}  &
\colhead{Airmass}      &
\colhead{Seeing}     \\
\colhead{(UT)}   &
\colhead{(MJD)}   &
\colhead{(deg)}    &
\colhead{(deg)}      &
\colhead{(mas)} &
\colhead{(mas)} &
\colhead{} &
\colhead{(arcsec)} }
\tablecaption{Integrated-light astrometry from CFHT/WIRCam \label{tbl:cfht}}
\startdata
\multicolumn{8}{c}{LP~349-25AB ($N_{\rm ep} = 20, \Delta{t} =  8.26$\,yr)} \\ \cline{1-8}                               2008~Aug~9  & 54687.5662 & 006.98451510 & $+22.32547171$ &  2.4 &  2.6 & 1.001 & 0.61 \\
2008~Sep~6  & 54715.4857 & 006.98451751 & $+22.32546921$ &  3.0 &  8.6 & 1.001 & 0.48 \\
2008~Oct~8  & 54747.4233 & 006.98451811 & $+22.32545820$ &  2.8 &  3.3 & 1.008 & 0.53 \\
2008~Oct~9  & 54748.3920 & 006.98451801 & $+22.32545770$ &  2.7 &  3.5 & 1.002 & 0.75 \\
2008~Nov~16 & 54786.2900 & 006.98451972 & $+22.32544801$ &  1.9 &  3.2 & 1.002 & 0.90 \\
2009~Jul~29 & 55041.6143 & 006.98463346 & $+22.32542572$ &  2.6 &  4.0 & 1.006 & 0.84 \\
2009~Aug~4  & 55047.6397 & 006.98463611 & $+22.32542544$ &  3.1 &  3.6 & 1.063 & 0.56 \\
\enddata
\end{deluxetable}
\clearpage

\begin{landscape}

\begin{deluxetable}{lcccc}
\setlength{\tabcolsep}{0.10in}
\tabletypesize{\tiny}
\tablewidth{0pt}
\tablehead{
\colhead{Property}              &
\colhead{Median $\pm$1$\sigma$} &
\colhead{Best fit}              &
\colhead{95.4\% c.i.}           &
\colhead{Prior/Notes}           }
\tablecaption{MCMC Posteriors for the Orbit and Parallax of LP~349-25AB \label{tbl:mcmc-LP349-25}}
\startdata
\multicolumn{5}{c}{Fitted parameters} \\[1pt]
\cline{1-5}
\multicolumn{5}{c}{} \\[-5pt]
Orbital period $P$ (yr)                                                 & $7.698\pm0.014$                & 7.701         &        7.671, 7.727        & $1/P$ (log-flat)                           \\[3pt]
Semimajor axis $a$ (mas)                                                & $145.99_{-0.18}^{+0.17}$       & 145.98        &       145.65, 146.33       & $1/a$ (log-flat)                           \\[3pt]
Eccentricity $e$                                                        & $0.0468_{-0.0018}^{+0.0019}$   & 0.0470        &       0.0431, 0.0505       & uniform, $0 \leq e < 1$                    \\[3pt]
Inclination $i$ (\degree)                                               & $117.36_{-0.10}^{+0.11}$       & 117.36        &       117.16, 117.57       & $\sin(i)$, $0\degree < i < 180\degree$     \\[3pt]
PA of the ascending node $\Omega$ (\degree)                             & $36.64\pm0.10$                 & 36.62         &        36.43, 36.84        & uniform                                    \\[3pt]
Argument of periastron $\omega$ (\degree)                               & $262.2\pm1.8$                  & 262.2         &        258.6, 266.0        & uniform                                    \\
\enddata
\end{deluxetable}
   \clearpage
\begin{deluxetable}{lcccc}
\setlength{\tabcolsep}{0.10in}
\tabletypesize{\tiny}
\tablewidth{0pt}
\tablehead{
\colhead{Property}              &
\colhead{Median $\pm$1$\sigma$} &
\colhead{Best fit}              &
\colhead{95.4\% c.i.}           &
\colhead{Prior/Notes}           }
\tablecaption{MCMC Posteriors for the Orbit and Parallax of LP~415-20AB \label{tbl:mcmc-LP415-20}}
\startdata
\multicolumn{5}{c}{Fitted parameters} \\[1pt]
\cline{1-5}
\multicolumn{5}{c}{} \\[-5pt]
Orbital period $P$ (yr)                                                 & $14.82\pm0.24$                 & 14.81         &        14.34, 15.29        & $1/P$ (log-flat)                           \\[3pt]
Semimajor axis $a$ (mas)                                                & $96.5_{-1.4}^{+1.1}$           & 96.7          &         94.2, 99.4         & $1/a$ (log-flat)                           \\[3pt]
Eccentricity $e$                                                        & $0.706_{-0.012}^{+0.011}$      & 0.707         &        0.684, 0.731        & uniform, $0 \leq e < 1$                    \\[3pt]
Inclination $i$ (\degree)                                               & $62.4\pm1.6$                   & 62.4          &         59.2, 65.7         & $\sin(i)$, $0\degree < i < 180\degree$     \\[3pt]
PA of the ascending node $\Omega$ (\degree)                             & $82.1_{-0.9}^{+1.1}$           & 82.3          &         79.9, 84.0         & uniform                                    \\[3pt]
Argument of periastron $\omega$ (\degree)                               & $168\pm4$                      & 168           &          160, 176          & uniform                                    \\
\enddata
\end{deluxetable}
   \clearpage
\begin{deluxetable}{lcccc}
\setlength{\tabcolsep}{0.10in}
\tabletypesize{\tiny}
\tablewidth{0pt}
\tablehead{
\colhead{Property}              &
\colhead{Median $\pm$1$\sigma$} &
\colhead{Best fit}              &
\colhead{95.4\% c.i.}           &
\colhead{Prior/Notes}           }
\tablecaption{MCMC Posteriors for the Orbit and Parallax of SDSS~J0423$-$0414AB \label{tbl:mcmc-SD0423-04}}
\startdata
\multicolumn{5}{c}{Fitted parameters} \\[1pt]
\cline{1-5}
\multicolumn{5}{c}{} \\[-5pt]
Orbital period $P$ (yr)                                                 & $12.30\pm0.06$                 & 12.30         &        12.18, 12.41        & $1/P$ (log-flat)                           \\[3pt]
Semimajor axis $a$ (mas)                                                & $162.9\pm0.5$                  & 162.6         &        161.9, 163.9        & $1/a$ (log-flat)                           \\[3pt]
Eccentricity $e$                                                        & $0.272_{-0.007}^{+0.008}$      & 0.270         &        0.258, 0.287        & uniform, $0 \leq e < 1$                    \\[3pt]
Inclination $i$ (\degree)                                               & $65.4\pm0.3$                   & 65.2          &         64.7, 66.0         & $\sin(i)$, $0\degree < i < 180\degree$     \\[3pt]
PA of the ascending node $\Omega$ (\degree)                             & $34.35_{-0.23}^{+0.22}$        & 34.33         &        33.89, 34.78        & uniform                                    \\[3pt]
Argument of periastron $\omega$ (\degree)                               & $97.4\pm0.3$                   & 97.5          &         96.7, 98.1         & uniform                                    \\
\enddata
\end{deluxetable}
   \clearpage
\begin{deluxetable}{lcccc}
\setlength{\tabcolsep}{0.10in}
\tabletypesize{\tiny}
\tablewidth{0pt}
\tablehead{
\colhead{Property}              &
\colhead{Median $\pm$1$\sigma$} &
\colhead{Best fit}              &
\colhead{95.4\% c.i.}           &
\colhead{Prior/Notes}           }
\tablecaption{MCMC Posteriors for the Orbit and Parallax of 2MASS~J0700+3157AB \label{tbl:mcmc-2M0700+31}}
\startdata
\multicolumn{5}{c}{Fitted parameters} \\[1pt]
\cline{1-5}
\multicolumn{5}{c}{} \\[-5pt]
Orbital period $P$ (yr)                                                 & $23.9\pm0.5$                   & 23.9          &         22.7, 25.1         & $1/P$ (log-flat)                           \\[3pt]
Semimajor axis $a$ (mas)                                                & $377_{-6}^{+5}$                & 377           &          364, 389          & $1/a$ (log-flat)                           \\[3pt]
Eccentricity $e$                                                        & $0.017_{-0.007}^{+0.005}$      & 0.013         &        0.007, 0.040        & uniform, $0 \leq e < 1$                    \\[3pt]
Inclination $i$ (\degree)                                               & $88.143_{-0.024}^{+0.025}$     & 88.140        &       88.094, 88.193       & $\sin(i)$, $0\degree < i < 180\degree$     \\[3pt]
PA of the ascending node $\Omega$ (\degree)                             & $102.666_{-0.021}^{+0.022}$    & 102.666       &      102.623, 102.708      & uniform                                    \\[3pt]
Argument of periastron $\omega$ (\degree)                               & $70_{-40}^{+60}$               & 80            &            0, 140          & uniform                                    \\
\enddata
\end{deluxetable}
   \clearpage
\begin{deluxetable}{lcccc}
\setlength{\tabcolsep}{0.10in}
\tabletypesize{\tiny}
\tablewidth{0pt}
\tablehead{
\colhead{Property}              &
\colhead{Median $\pm$1$\sigma$} &
\colhead{Best fit}              &
\colhead{95.4\% c.i.}           &
\colhead{Prior/Notes}           }
\tablecaption{MCMC Posteriors for the Orbit and Parallax of LHS~1901AB \label{tbl:mcmc-LHS1901}}
\startdata
\multicolumn{5}{c}{Fitted parameters} \\[1pt]
\cline{1-5}
\multicolumn{5}{c}{} \\[-5pt]
Orbital period $P$ (yr)                                                 & $16.21\pm0.08$                 & 16.19         &        16.05, 16.37        & $1/P$ (log-flat)                           \\[3pt]
Semimajor axis $a$ (mas)                                                & $287.7_{-0.8}^{+0.7}$          & 287.4         &        286.2, 289.2        & $1/a$ (log-flat)                           \\[3pt]
Eccentricity $e$                                                        & $0.8304\pm0.0009$              & 0.8302        &       0.8287, 0.8322       & uniform, $0 \leq e < 1$                    \\[3pt]
Inclination $i$ (\degree)                                               & $71.97\pm0.05$                 & 71.98         &        71.86, 72.07        & $\sin(i)$, $0\degree < i < 180\degree$     \\[3pt]
PA of the ascending node $\Omega$ (\degree)                             & $2.27\pm0.05$                  & 2.28          &         2.17, 2.37         & uniform                                    \\[3pt]
Argument of periastron $\omega$ (\degree)                               & $44.49_{-0.15}^{+0.14}$        & 44.44         &        44.19, 44.78        & uniform                                    \\
\enddata
\end{deluxetable}
   \clearpage
\begin{deluxetable}{lcccc}
\setlength{\tabcolsep}{0.10in}
\tabletypesize{\tiny}
\tablewidth{0pt}
\tablehead{
\colhead{Property}              &
\colhead{Median $\pm$1$\sigma$} &
\colhead{Best fit}              &
\colhead{95.4\% c.i.}           &
\colhead{Prior/Notes}           }
\tablecaption{MCMC Posteriors for the Orbit and Parallax of 2MASS~J0746+2000AB \label{tbl:mcmc-2M0746+20}}
\startdata
\multicolumn{5}{c}{Fitted parameters} \\[1pt]
\cline{1-5}
\multicolumn{5}{c}{} \\[-5pt]
Orbital period $P$ (yr)                                                 & $12.736_{-0.029}^{+0.031}$     & 12.749        &       12.674, 12.796       & $1/P$ (log-flat)                           \\[3pt]
Semimajor axis $a$ (mas)                                                & $237.18\pm0.11$                & 237.16        &       236.96, 237.40       & $1/a$ (log-flat)                           \\[3pt]
Eccentricity $e$                                                        & $0.4854\pm0.0003$              & 0.4856        &       0.4848, 0.4861       & uniform, $0 \leq e < 1$                    \\[3pt]
Inclination $i$ (\degree)                                               & $138.56_{-0.21}^{+0.20}$       & 138.68        &       138.15, 138.96       & $\sin(i)$, $0\degree < i < 180\degree$     \\[3pt]
PA of the ascending node $\Omega$ (\degree)                             & $29.77_{-0.28}^{+0.30}$        & 29.79         &        29.19, 30.35        & uniform                                    \\[3pt]
Argument of periastron $\omega$ (\degree)                               & $355.6\pm0.6$                  & 355.8         &        354.5, 356.8        & uniform                                    \\
\enddata
\end{deluxetable}
   \clearpage
\begin{deluxetable}{lcccc}
\setlength{\tabcolsep}{0.10in}
\tabletypesize{\tiny}
\tablewidth{0pt}
\tablehead{
\colhead{Property}              &
\colhead{Median $\pm$1$\sigma$} &
\colhead{Best fit}              &
\colhead{95.4\% c.i.}           &
\colhead{Prior/Notes}           }
\tablecaption{MCMC Posteriors for the Orbit and Parallax of 2MASS~J0850+1057AB \label{tbl:mcmc-2M0850+10}}
\startdata
\multicolumn{5}{c}{Fitted parameters} \\[1pt]
\cline{1-5}
\multicolumn{5}{c}{} \\[-5pt]
Orbital period $P$ (yr)                                                 & $48_{-6}^{+7}$                 & 45            &           34, 94           & $1/P$ (log-flat)                           \\[3pt]
Semimajor axis $a$ (mas)                                                & $156_{-9}^{+8}$                & 152           &          140, 212          & $1/a$ (log-flat)                           \\[3pt]
Eccentricity $e$                                                        & $0.06_{-0.06}^{+0.05}$         & 0.02          &         0.00, 0.38         & uniform, $0 \leq e < 1$                    \\[3pt]
Inclination $i$ (\degree)                                               & $64.9_{-1.4}^{+1.5}$           & 65.7          &         60.7, 67.9         & $\sin(i)$, $0\degree < i < 180\degree$     \\[3pt]
PA of the ascending node $\Omega$ (\degree)                             & $113.3_{-1.7}^{+1.6}$          & 112.9         &        110.0, 119.0        & uniform                                    \\[3pt]
Argument of periastron $\omega$ (\degree)                               & $70_{-40}^{+150}$              & 270           &        $-$60, 270          & uniform                                    \\
\enddata
\end{deluxetable}
   \clearpage
\begin{deluxetable}{lcccc}
\setlength{\tabcolsep}{0.10in}
\tabletypesize{\tiny}
\tablewidth{0pt}
\tablehead{
\colhead{Property}              &
\colhead{Median $\pm$1$\sigma$} &
\colhead{Best fit}              &
\colhead{95.4\% c.i.}           &
\colhead{Prior/Notes}           }
\tablecaption{MCMC Posteriors for the Orbit and Parallax of 2MASS~J0920+3517AB \label{tbl:mcmc-2M0920+35}}
\startdata
\multicolumn{5}{c}{Fitted parameters} \\[1pt]
\cline{1-5}
\multicolumn{5}{c}{} \\[-5pt]
Orbital period $P$ (yr)                                                 & $7.258\pm0.009$                & 7.257         &        7.239, 7.277        & $1/P$ (log-flat)                           \\[3pt]
Semimajor axis $a$ (mas)                                                & $68.15\pm0.23$                 & 68.12         &        67.68, 68.62        & $1/a$ (log-flat)                           \\[3pt]
Eccentricity $e$                                                        & $0.180_{-0.007}^{+0.006}$      & 0.181         &        0.167, 0.193        & uniform, $0 \leq e < 1$                    \\[3pt]
Inclination $i$ (\degree)                                               & $87.15\pm0.17$                 & 87.18         &        86.81, 87.49        & $\sin(i)$, $0\degree < i < 180\degree$     \\[3pt]
PA of the ascending node $\Omega$ (\degree)                             & $66.73\pm0.12$                 & 66.76         &        66.49, 66.97        & uniform                                    \\[3pt]
Argument of periastron $\omega$ (\degree)                               & $317.5_{-1.5}^{+1.4}$          & 317.4         &        314.7, 320.6        & uniform                                    \\
\enddata
\end{deluxetable}
   \clearpage
\begin{deluxetable}{lcccc}
\setlength{\tabcolsep}{0.10in}
\tabletypesize{\tiny}
\tablewidth{0pt}
\tablehead{
\colhead{Property}              &
\colhead{Median $\pm$1$\sigma$} &
\colhead{Best fit}              &
\colhead{95.4\% c.i.}           &
\colhead{Prior/Notes}           }
\tablecaption{MCMC Posteriors for the Orbit and Parallax of SDSS~J0926+5847AB \label{tbl:mcmc-SD0926+58}}
\startdata
\multicolumn{5}{c}{Fitted parameters} \\[1pt]
\cline{1-5}
\multicolumn{5}{c}{} \\[-5pt]
Orbital period $P$ (yr)                                                 & $12.9_{-1.9}^{+1.3}$           & 11.1          &         10.4, 18.2         & $1/P$ (log-flat)                           \\[3pt]
Semimajor axis $a$ (mas)                                                & $78_{-8}^{+6}$                 & 80            &           68, 105          & $1/a$ (log-flat)                           \\[3pt]
Eccentricity $e$                                                        & $0.35_{-0.16}^{+0.22}$         & 0.48          &         0.00, 0.58         & uniform, $0 \leq e < 1$                    \\[3pt]
Inclination $i$ (\degree)                                               & $91.7_{-0.8}^{+0.6}$           & 91.3          &         90.5, 93.2         & $\sin(i)$, $0\degree < i < 180\degree$     \\[3pt]
PA of the ascending node $\Omega$ (\degree)                             & $133.6_{-0.5}^{+0.6}$          & 133.4         &        132.4, 134.6        & uniform                                    \\[3pt]
Argument of periastron $\omega$ (\degree)                               & $275_{-17}^{+11}$              & 277           &          210, 410          & uniform                                    \\
\enddata
\end{deluxetable}
   \clearpage
\begin{deluxetable}{lcccc}
\setlength{\tabcolsep}{0.10in}
\tabletypesize{\tiny}
\tablewidth{0pt}
\tablehead{
\colhead{Property}              &
\colhead{Median $\pm$1$\sigma$} &
\colhead{Best fit}              &
\colhead{95.4\% c.i.}           &
\colhead{Prior/Notes}           }
\tablecaption{MCMC Posteriors for the Orbit and Parallax of 2MASS~J1017+1308AB \label{tbl:mcmc-2M1017+13}}
\startdata
\multicolumn{5}{c}{Fitted parameters} \\[1pt]
\cline{1-5}
\multicolumn{5}{c}{} \\[-5pt]
Orbital period $P$ (yr)                                                 & $18.60_{-0.23}^{+0.22}$        & 18.62         &        18.16, 19.07        & $1/P$ (log-flat)                           \\[3pt]
Semimajor axis $a$ (mas)                                                & $120.0_{-1.2}^{+1.1}$          & 120.0         &        117.7, 122.3        & $1/a$ (log-flat)                           \\[3pt]
Eccentricity $e$                                                        & $0.158\pm0.010$                & 0.158         &        0.139, 0.178        & uniform, $0 \leq e < 1$                    \\[3pt]
Inclination $i$ (\degree)                                               & $35.4_{-0.9}^{+1.0}$           & 35.4          &         33.5, 37.2         & $\sin(i)$, $0\degree < i < 180\degree$     \\[3pt]
PA of the ascending node $\Omega$ (\degree)                             & $72.7_{-2.0}^{+1.8}$           & 72.3          &         69.0, 76.7         & uniform                                    \\[3pt]
Argument of periastron $\omega$ (\degree)                               & $5\pm5$                        & 7             &         $-$5, 16           & uniform                                    \\
\enddata
\end{deluxetable}
   \clearpage
\begin{deluxetable}{lcccc}
\setlength{\tabcolsep}{0.10in}
\tabletypesize{\tiny}
\tablewidth{0pt}
\tablehead{
\colhead{Property}              &
\colhead{Median $\pm$1$\sigma$} &
\colhead{Best fit}              &
\colhead{95.4\% c.i.}           &
\colhead{Prior/Notes}           }
\tablecaption{MCMC Posteriors for the Orbit and Parallax of SDSS~J1021$-$0304AB \label{tbl:mcmc-SD1021-03}}
\startdata
\multicolumn{5}{c}{Fitted parameters} \\[1pt]
\cline{1-5}
\multicolumn{5}{c}{} \\[-5pt]
Orbital period $P$ (yr)                                                 & $86_{-17}^{+13}$               & 75            &           60, 119          & $1/P$ (log-flat)                           \\[3pt]
Semimajor axis $a$ (mas)                                                & $241_{-28}^{+19}$              & 224           &          201, 294          & $1/a$ (log-flat)                           \\[3pt]
Eccentricity $e$                                                        & $0.38\pm0.07$                  & 0.31          &         0.24, 0.51         & uniform, $0 \leq e < 1$                    \\[3pt]
Inclination $i$ (\degree)                                               & $165_{-7}^{+6}$                & 161           &          154, 177          & $\sin(i)$, $0\degree < i < 180\degree$     \\[3pt]
PA of the ascending node $\Omega$ (\degree)                             & $260_{-90}^{+100}$             & 260           &           60, 400          & uniform                                    \\[3pt]
Argument of periastron $\omega$ (\degree)                               & $110_{-120}^{+70}$             & 70            &        $-$40, 300          & uniform                                    \\
\enddata
\end{deluxetable}
   \clearpage
\begin{deluxetable}{lcccc}
\setlength{\tabcolsep}{0.10in}
\tabletypesize{\tiny}
\tablewidth{0pt}
\tablehead{
\colhead{Property}              &
\colhead{Median $\pm$1$\sigma$} &
\colhead{Best fit}              &
\colhead{95.4\% c.i.}           &
\colhead{Prior/Notes}           }
\tablecaption{MCMC Posteriors for the Orbit and Parallax of 2MASS~J1047+4026AB~(a.k.a.~LP~213-68) \label{tbl:mcmc-2M1047+40}}
\startdata
\multicolumn{5}{c}{Fitted parameters} \\[1pt]
\cline{1-5}
\multicolumn{5}{c}{} \\[-5pt]
Orbital period $P$ (yr)                                                 & $6.562_{-0.026}^{+0.029}$      & 6.557         &        6.508, 6.619        & $1/P$ (log-flat)                           \\[3pt]
Semimajor axis $a$ (mas)                                                & $75.98_{-0.22}^{+0.21}$        & 75.94         &        75.55, 76.41        & $1/a$ (log-flat)                           \\[3pt]
Eccentricity $e$                                                        & $0.7485\pm0.0013$              & 0.7479        &       0.7458, 0.7511       & uniform, $0 \leq e < 1$                    \\[3pt]
Inclination $i$ (\degree)                                               & $30.0\pm0.5$                   & 29.9          &         29.0, 30.9         & $\sin(i)$, $0\degree < i < 180\degree$     \\[3pt]
PA of the ascending node $\Omega$ (\degree)                             & $85.5_{-1.0}^{+1.1}$           & 85.6          &         83.4, 87.5         & uniform                                    \\[3pt]
Argument of periastron $\omega$ (\degree)                               & $63.5\pm1.0$                   & 63.4          &         61.6, 65.6         & uniform                                    \\
\enddata
\end{deluxetable}
   \clearpage
\begin{deluxetable}{lcccc}
\setlength{\tabcolsep}{0.10in}
\tabletypesize{\tiny}
\tablewidth{0pt}
\tablehead{
\colhead{Property}              &
\colhead{Median $\pm$1$\sigma$} &
\colhead{Best fit}              &
\colhead{95.4\% c.i.}           &
\colhead{Prior/Notes}           }
\tablecaption{MCMC Posteriors for the Orbit and Parallax of SDSS~J1052+4422AB \label{tbl:mcmc-SD1052+44}}
\startdata
\multicolumn{5}{c}{Fitted parameters} \\[1pt]
\cline{1-5}
\multicolumn{5}{c}{} \\[-5pt]
Orbital period $P$ (yr)                                                 & $8.608_{-0.024}^{+0.025}$      & 8.603         &        8.559, 8.656        & $1/P$ (log-flat)                           \\[3pt]
Semimajor axis $a$ (mas)                                                & $70.67\pm0.24$                 & 70.61         &        70.19, 71.15        & $1/a$ (log-flat)                           \\[3pt]
Eccentricity $e$                                                        & $0.1399_{-0.0023}^{+0.0022}$   & 0.1410        &       0.1354, 0.1445       & uniform, $0 \leq e < 1$                    \\[3pt]
Inclination $i$ (\degree)                                               & $62.1\pm0.3$                   & 62.1          &         61.5, 62.7         & $\sin(i)$, $0\degree < i < 180\degree$     \\[3pt]
PA of the ascending node $\Omega$ (\degree)                             & $126.8\pm0.3$                  & 126.8         &        126.2, 127.5        & uniform                                    \\[3pt]
Argument of periastron $\omega$ (\degree)                               & $187.3\pm1.6$                  & 186.8         &        184.0, 190.6        & uniform                                    \\
\enddata
\end{deluxetable}
   \clearpage
\begin{deluxetable}{lcccc}
\setlength{\tabcolsep}{0.10in}
\tabletypesize{\tiny}
\tablewidth{0pt}
\tablehead{
\colhead{Property}              &
\colhead{Median $\pm$1$\sigma$} &
\colhead{Best fit}              &
\colhead{95.4\% c.i.}           &
\colhead{Prior/Notes}           }
\tablecaption{MCMC Posteriors for the Orbit and Parallax of Gl~417BC \label{tbl:mcmc-GL417B}}
\startdata
\multicolumn{5}{c}{Fitted parameters} \\[1pt]
\cline{1-5}
\multicolumn{5}{c}{} \\[-5pt]
Orbital period $P$ (yr)                                                 & $15.65\pm0.08$                 & 15.65         &        15.49, 15.81        & $1/P$ (log-flat)                           \\[3pt]
Semimajor axis $a$ (mas)                                                & $130.0\pm0.4$                  & 130.0         &        129.2, 130.9        & $1/a$ (log-flat)                           \\[3pt]
Eccentricity $e$                                                        & $0.105\pm0.003$                & 0.106         &        0.099, 0.112        & uniform, $0 \leq e < 1$                    \\[3pt]
Inclination $i$ (\degree)                                               & $102.9\pm0.5$                  & 102.9         &        101.9, 103.9        & $\sin(i)$, $0\degree < i < 180\degree$     \\[3pt]
PA of the ascending node $\Omega$ (\degree)                             & $101.01\pm0.22$                & 101.06        &       100.56, 101.45       & uniform                                    \\[3pt]
Argument of periastron $\omega$ (\degree)                               & $348_{-5}^{+4}$                & 348           &          339, 356          & uniform                                    \\
\enddata
\end{deluxetable}
   \clearpage
\begin{deluxetable}{lcccc}
\setlength{\tabcolsep}{0.10in}
\tabletypesize{\tiny}
\tablewidth{0pt}
\tablehead{
\colhead{Property}              &
\colhead{Median $\pm$1$\sigma$} &
\colhead{Best fit}              &
\colhead{95.4\% c.i.}           &
\colhead{Prior/Notes}           }
\tablecaption{MCMC Posteriors for the Orbit and Parallax of LHS~2397aAB \label{tbl:mcmc-LHS2397a}}
\startdata
\multicolumn{5}{c}{Fitted parameters} \\[1pt]
\cline{1-5}
\multicolumn{5}{c}{} \\[-5pt]
Orbital period $P$ (yr)                                                 & $14.37\pm0.05$                 & 14.36         &        14.27, 14.46        & $1/P$ (log-flat)                           \\[3pt]
Semimajor axis $a$ (mas)                                                & $214.8\pm0.8$                  & 214.6         &        213.2, 216.4        & $1/a$ (log-flat)                           \\[3pt]
Eccentricity $e$                                                        & $0.351\pm0.003$                & 0.353         &        0.345, 0.358        & uniform, $0 \leq e < 1$                    \\[3pt]
Inclination $i$ (\degree)                                               & $42.55_{-0.29}^{+0.28}$        & 42.41         &        41.98, 43.13        & $\sin(i)$, $0\degree < i < 180\degree$     \\[3pt]
PA of the ascending node $\Omega$ (\degree)                             & $74.7\pm0.8$                   & 74.4          &         73.2, 76.2         & uniform                                    \\[3pt]
Argument of periastron $\omega$ (\degree)                               & $221.2\pm1.3$                  & 221.5         &        218.5, 223.8        & uniform                                    \\
\enddata
\end{deluxetable}
   \clearpage
\begin{deluxetable}{lcccc}
\setlength{\tabcolsep}{0.10in}
\tabletypesize{\tiny}
\tablewidth{0pt}
\tablehead{
\colhead{Property}              &
\colhead{Median $\pm$1$\sigma$} &
\colhead{Best fit}              &
\colhead{95.4\% c.i.}           &
\colhead{Prior/Notes}           }
\tablecaption{MCMC Posteriors for the Orbit and Parallax of DENIS~J1228$-$1557AB \label{tbl:mcmc-DE1228-15}}
\startdata
\multicolumn{5}{c}{Fitted parameters} \\[1pt]
\cline{1-5}
\multicolumn{5}{c}{} \\[-5pt]
Orbital period $P$ (yr)                                                 & $50_{-7}^{+5}$                 & 47            &           40, 65           & $1/P$ (log-flat)                           \\[3pt]
Semimajor axis $a$ (mas)                                                & $303_{-11}^{+8}$               & 297           &          288, 331          & $1/a$ (log-flat)                           \\[3pt]
Eccentricity $e$                                                        & $0.089_{-0.035}^{+0.027}$      & 0.075         &        0.044, 0.197        & uniform, $0 \leq e < 1$                    \\[3pt]
Inclination $i$ (\degree)                                               & $142\pm4$                      & 140           &          135, 150          & $\sin(i)$, $0\degree < i < 180\degree$     \\[3pt]
PA of the ascending node $\Omega$ (\degree)                             & $60.0_{-2.0}^{+2.5}$           & 60.4          &         54.4, 64.3         & uniform                                    \\[3pt]
Argument of periastron $\omega$ (\degree)                               & $10_{-40}^{+50}$               & $-$30         &        $-$70, 60           & uniform                                    \\
\enddata
\end{deluxetable}
   \clearpage
\begin{deluxetable}{lcccc}
\setlength{\tabcolsep}{0.10in}
\tabletypesize{\tiny}
\tablewidth{0pt}
\tablehead{
\colhead{Property}              &
\colhead{Median $\pm$1$\sigma$} &
\colhead{Best fit}              &
\colhead{95.4\% c.i.}           &
\colhead{Prior/Notes}           }
\tablecaption{MCMC Posteriors for the Orbit and Parallax of Kelu-1AB \label{tbl:mcmc-KELU-1}}
\startdata
\multicolumn{5}{c}{Fitted parameters} \\[1pt]
\cline{1-5}
\multicolumn{5}{c}{} \\[-5pt]
Orbital period $P$ (yr)                                                 & $24.98\pm0.19$                 & 24.97         &        24.60, 25.36        & $1/P$ (log-flat)                           \\[3pt]
Semimajor axis $a$ (mas)                                                & $227.9_{-1.1}^{+0.9}$          & 227.7         &        226.0, 230.0        & $1/a$ (log-flat)                           \\[3pt]
Eccentricity $e$                                                        & $0.709\pm0.005$                & 0.709         &        0.699, 0.719        & uniform, $0 \leq e < 1$                    \\[3pt]
Inclination $i$ (\degree)                                               & $82.35\pm0.08$                 & 82.34         &        82.18, 82.51        & $\sin(i)$, $0\degree < i < 180\degree$     \\[3pt]
PA of the ascending node $\Omega$ (\degree)                             & $44.78\pm0.20$                 & 44.80         &        44.39, 45.20        & uniform                                    \\[3pt]
Argument of periastron $\omega$ (\degree)                               & $3.7\pm2.1$                    & 3.4           &       $-$0.5, 7.8          & uniform                                    \\
\enddata
\end{deluxetable}
   \clearpage
\begin{deluxetable}{lcccc}
\setlength{\tabcolsep}{0.10in}
\tabletypesize{\tiny}
\tablewidth{0pt}
\tablehead{
\colhead{Property}              &
\colhead{Median $\pm$1$\sigma$} &
\colhead{Best fit}              &
\colhead{95.4\% c.i.}           &
\colhead{Prior/Notes}           }
\tablecaption{MCMC Posteriors for the Orbit and Parallax of 2MASS~J1404$-$3159AB \label{tbl:mcmc-2M1404-31}}
\startdata
\multicolumn{5}{c}{Fitted parameters} \\[1pt]
\cline{1-5}
\multicolumn{5}{c}{} \\[-5pt]
Orbital period $P$ (yr)                                                 & $16.52_{-0.22}^{+0.21}$        & 16.51         &        16.10, 16.97        & $1/P$ (log-flat)                           \\[3pt]
Semimajor axis $a$ (mas)                                                & $133.7_{-2.3}^{+1.5}$          & 135.9         &        131.0, 138.8        & $1/a$ (log-flat)                           \\[3pt]
Eccentricity $e$                                                        & $0.825\pm0.005$                & 0.825         &        0.814, 0.836        & uniform, $0 \leq e < 1$                    \\[3pt]
Inclination $i$ (\degree)                                               & $164_{-7}^{+8}$                & 157           &          151, 177          & $\sin(i)$, $0\degree < i < 180\degree$     \\[3pt]
PA of the ascending node $\Omega$ (\degree)                             & $207_{-13}^{+176}$             & 208           &           23, 376          & uniform                                    \\[3pt]
Argument of periastron $\omega$ (\degree)                               & $52_{-17}^{+172}$              & 45            &        $-$60, 242          & uniform                                    \\
\enddata
\end{deluxetable}
   \clearpage
\begin{deluxetable}{lcccc}
\setlength{\tabcolsep}{0.10in}
\tabletypesize{\tiny}
\tablewidth{0pt}
\tablehead{
\colhead{Property}              &
\colhead{Median $\pm$1$\sigma$} &
\colhead{Best fit}              &
\colhead{95.4\% c.i.}           &
\colhead{Prior/Notes}           }
\tablecaption{MCMC Posteriors for the Orbit and Parallax of HD~130948BC \label{tbl:mcmc-HD130948B}}
\startdata
\multicolumn{5}{c}{Fitted parameters} \\[1pt]
\cline{1-5}
\multicolumn{5}{c}{} \\[-5pt]
Orbital period $P$ (yr)                                                 & $10.009\pm0.010$               & 10.008        &        9.990, 10.029       & $1/P$ (log-flat)                           \\[3pt]
Semimajor axis $a$ (mas)                                                & $122.52\pm0.08$                & 122.53        &       122.37, 122.68       & $1/a$ (log-flat)                           \\[3pt]
Eccentricity $e$                                                        & $0.1627\pm0.0017$              & 0.1622        &       0.1592, 0.1660       & uniform, $0 \leq e < 1$                    \\[3pt]
Inclination $i$ (\degree)                                               & $95.70\pm0.05$                 & 95.71         &        95.60, 95.81        & $\sin(i)$, $0\degree < i < 180\degree$     \\[3pt]
PA of the ascending node $\Omega$ (\degree)                             & $133.56\pm0.03$                & 133.56        &       133.50, 133.63       & uniform                                    \\[3pt]
Argument of periastron $\omega$ (\degree)                               & $68.39\pm0.24$                 & 68.35         &        67.93, 68.88        & uniform                                    \\
\enddata
\end{deluxetable}
   \clearpage
\begin{deluxetable}{lcccc}
\setlength{\tabcolsep}{0.10in}
\tabletypesize{\tiny}
\tablewidth{0pt}
\tablehead{
\colhead{Property}              &
\colhead{Median $\pm$1$\sigma$} &
\colhead{Best fit}              &
\colhead{95.4\% c.i.}           &
\colhead{Prior/Notes}           }
\tablecaption{MCMC Posteriors for the Orbit and Parallax of Gl~569Bab \label{tbl:mcmc-GL569B}}
\startdata
\multicolumn{5}{c}{Fitted parameters} \\[1pt]
\cline{1-5}
\multicolumn{5}{c}{} \\[-5pt]
Orbital period $P$ (yr)                                                 & $2.3707\pm0.0005$              & 2.3707        &       2.3696, 2.3718       & $1/P$ (log-flat)                           \\[3pt]
Semimajor axis $a$ (mas)                                                & $93.64\pm0.14$                 & 93.62         &        93.35, 93.93        & $1/a$ (log-flat)                           \\[3pt]
Eccentricity $e$                                                        & $0.3186\pm0.0010$              & 0.3184        &       0.3166, 0.3207       & uniform, $0 \leq e < 1$                    \\[3pt]
Inclination $i$ (\degree)                                               & $32.70\pm0.23$                 & 32.65         &        32.24, 33.15        & $\sin(i)$, $0\degree < i < 180\degree$     \\[3pt]
PA of the ascending node $\Omega$ (\degree)                             & $142.5\pm0.3$                  & 142.6         &        141.9, 143.2        & uniform                                    \\[3pt]
Argument of periastron $\omega$ (\degree)                               & $81.19_{-0.29}^{+0.28}$        & 81.18         &        80.63, 81.76        & uniform                                    \\
\enddata
\end{deluxetable}
   \clearpage
\begin{deluxetable}{lcccc}
\setlength{\tabcolsep}{0.10in}
\tabletypesize{\tiny}
\tablewidth{0pt}
\tablehead{
\colhead{Property}              &
\colhead{Median $\pm$1$\sigma$} &
\colhead{Best fit}              &
\colhead{95.4\% c.i.}           &
\colhead{Prior/Notes}           }
\tablecaption{MCMC Posteriors for the Orbit and Parallax of SDSS~J1534+1615AB \label{tbl:mcmc-SD1534+16}}
\startdata
\multicolumn{5}{c}{Fitted parameters} \\[1pt]
\cline{1-5}
\multicolumn{5}{c}{} \\[-5pt]
Orbital period $P$ (yr)                                                 & $58_{-24}^{+39}$               & 48            &           30, 454          & $1/P$ (log-flat)                           \\[3pt]
Semimajor axis $a$ (mas)                                                & $150_{-40}^{+60}$              & 130           &          100, 590          & $1/a$ (log-flat)                           \\[3pt]
Eccentricity $e$                                                        & $0.22_{-0.22}^{+0.21}$         & 0.11          &         0.00, 0.80         & uniform, $0 \leq e < 1$                    \\[3pt]
Inclination $i$ (\degree)                                               & $36_{-10}^{+15}$               & 28            &           13, 55           & $\sin(i)$, $0\degree < i < 180\degree$     \\[3pt]
PA of the ascending node $\Omega$ (\degree)                             & $146_{-18}^{+19}$              & 132           &          101, 189          & uniform                                    \\[3pt]
Argument of periastron $\omega$ (\degree)                               & $168_{-42}^{+13}$              & 128           &        $-$12, 234          & uniform                                    \\
\enddata
\end{deluxetable}
   \clearpage
\begin{deluxetable}{lcccc}
\setlength{\tabcolsep}{0.10in}
\tabletypesize{\tiny}
\tablewidth{0pt}
\tablehead{
\colhead{Property}              &
\colhead{Median $\pm$1$\sigma$} &
\colhead{Best fit}              &
\colhead{95.4\% c.i.}           &
\colhead{Prior/Notes}           }
\tablecaption{MCMC Posteriors for the Orbit and Parallax of 2MASS~J1534$-$2952AB \label{tbl:mcmc-2M1534-29}}
\startdata
\multicolumn{5}{c}{Fitted parameters} \\[1pt]
\cline{1-5}
\multicolumn{5}{c}{} \\[-5pt]
Orbital period $P$ (yr)                                                 & $20.35_{-0.06}^{+0.05}$        & 20.38         &        20.26, 20.48        & $1/P$ (log-flat)                           \\[3pt]
Semimajor axis $a$ (mas)                                                & $214.07_{-0.23}^{+0.27}$       & 213.92        &       213.54, 214.55       & $1/a$ (log-flat)                           \\[3pt]
Eccentricity $e$                                                        & $0.0027_{-0.0027}^{+0.0028}$   & 0.0058        &       0.0000, 0.0139       & uniform, $0 \leq e < 1$                    \\[3pt]
Inclination $i$ (\degree)                                               & $85.56_{-0.07}^{+0.08}$        & 85.55         &        85.40, 85.70        & $\sin(i)$, $0\degree < i < 180\degree$     \\[3pt]
PA of the ascending node $\Omega$ (\degree)                             & $13.61\pm0.05$                 & 13.61         &        13.51, 13.71        & uniform                                    \\[3pt]
Argument of periastron $\omega$ (\degree)                               & $80\pm40$                      & 80            &        $-$70, 260          & uniform                                    \\
\enddata
\end{deluxetable}
   \clearpage
\begin{deluxetable}{lcccc}
\setlength{\tabcolsep}{0.10in}
\tabletypesize{\tiny}
\tablewidth{0pt}
\tablehead{
\colhead{Property}              &
\colhead{Median $\pm$1$\sigma$} &
\colhead{Best fit}              &
\colhead{95.4\% c.i.}           &
\colhead{Prior/Notes}           }
\tablecaption{MCMC Posteriors for the Orbit and Parallax of 2MASS~J1728+3948AB \label{tbl:mcmc-2M1728+39}}
\startdata
\multicolumn{5}{c}{Fitted parameters} \\[1pt]
\cline{1-5}
\multicolumn{5}{c}{} \\[-5pt]
Orbital period $P$ (yr)                                                 & $40.8_{-1.2}^{+1.6}$           & 43.8          &         37.8, 47.9         & $1/P$ (log-flat)                           \\[3pt]
Semimajor axis $a$ (mas)                                                & $221_{-5}^{+6}$                & 232           &          210, 247          & $1/a$ (log-flat)                           \\[3pt]
Eccentricity $e$                                                        & $0.015_{-0.015}^{+0.013}$      & 0.055         &        0.000, 0.106        & uniform, $0 \leq e < 1$                    \\[3pt]
Inclination $i$ (\degree)                                               & $54.4\pm0.6$                   & 55.4          &         53.2, 56.8         & $\sin(i)$, $0\degree < i < 180\degree$     \\[3pt]
PA of the ascending node $\Omega$ (\degree)                             & $111.3\pm0.6$                  & 110.6         &        109.7, 112.3        & uniform                                    \\[3pt]
Argument of periastron $\omega$ (\degree)                               & $330_{-130}^{+50}$             & 340           &          100, 420          & uniform                                    \\
\enddata
\end{deluxetable}%
 \clearpage
\end{landscape}

\end{document}